\documentclass[preprint,trackchanges]{aastex63}

\usepackage{amsmath}
\usepackage{natbib}
\usepackage{multirow}  
\usepackage{booktabs}
\usepackage{verbatim}
\hypersetup{linkcolor=magenta, citecolor=cyan, filecolor=yellow, urlcolor=blue}

\received{2020 November 10}
\revised{2020 December 10}
\accepted{2021 January 11}
\published{2021 March 23}

\shorttitle{Testing High-latitude Curvature Effect of Gamma-Ray Bursts with {\it Fermi} Data}
\shortauthors{Li \& Zhang}

\begin{document}

\title{Testing High-latitude Curvature Effect of Gamma-Ray Bursts with {\it Fermi} Data: \\ Evidence of Bulk Acceleration in Prompt Emission}

\author[0000-0002-1343-3089]{Liang Li}
\affiliation{ICRANet, Piazza della Repubblica 10, 65122 Pescara, Italy; liang.li@icranet.org}
\affiliation{INAF -- Osservatorio Astronomico d'Abruzzo, Via M. Maggini snc, I-64100, Teramo, Italy}
\affiliation{ICRA, Dipartimento di Fisica, Sapienza Università di Roma, P.le Aldo Moro 5, I–00185 Rome, Italy}

\author{Bing Zhang}
\affiliation{Department of Physics and Astronomy, University of Nevada, Las Vegas, NV 89154, USA; zhang@physics.unlv.edu}

\correspondingauthor{Liang Li, Bing Zhang}
\email{liang.li@icranet.org, zhang@physics.unlv.edu}

\begin{abstract}
When a gamma-ray burst (GRB) emitter stops emission abruptly, the observer receives rapidly fading emission from high latitudes with respect to the line of sight, known as the ``curvature effect''. Identifying such emission from GRB prompt-emission lightcurves would constrain the radius of prompt emission from the central engine and the composition of GRB jets. We perform a dedicated search of high-latitude emission (HLE) through spectral and temporal analyses of a sample of single-pulse bursts detected by the Gamma-ray Burst Monitor on board the {\it Fermi} satellite. We identify HLE from a subsample of bursts and constrain the emission radius to be $R_{\rm GRB} \sim (10^{15}-10^{16})$ cm from the central engine. Some bursts have the HLE decay faster than predicted by a constant Lorentz factor jet, suggesting that the emission region is undergoing acceleration during prompt emission. This supports the Poynting-flux-dominated jet composition for these bursts. The conclusion is consistent with previous results drawn from spectral-lag modeling of prompt emission and HLE analysis of X-ray flares. 
\end{abstract}


\section{Introduction} \label{sec:intro}

Gamma-ray bursts (GRBs) are the most luminous explosions in the universe. While it is well established that the $\gamma$-ray emission originates from an internal site in a relativistic jet beaming toward Earth, the composition of the jet as well as the origin of $\gamma$-rays (energy-dissipation mechanism and radiation mechanism) are subject to intense debate \citep{Zhang2018}. The simplest model is the ``fireball'' model, which invokes a thermally accelerated, matter-dominated ejecta \citep{Goodman1986, Paczynski1986}. Within this framework, the outflow initially undergoes a rapid acceleration phase as the thermal energy of the fireball is quickly converted into the kinetic energy of the baryons at the coasting radius $\sim \Gamma (c t_{\rm pulse}) = 3\times 10^{12} \ {\rm cm} \Gamma_2 t_{\rm pulse}$ \citep{Shemi1990, meszaros93c, Piran1993, Kobayashi1999}, where $\Gamma$ is the Lorentz factor, and $t_{\rm pulse}$ is the duration of the GRB pulse in the source frame (the observed duration divided by the $(1+z)$ time dilation factor, where $z$ is the source redshift), and the convention $Q=10^n Q_n$ is adopted in cgs units throughout the text. Within this model, the $\gamma$-ray emission is released at the internal shock radius \citep{Rees1994} and the photospheric radius \citep{Meszaros2000}, both are typically smaller than $\sim 10^{14}$ cm from the central engine. The fireball is decelerated at $\sim 10^{17}$ cm by a pair of external shocks \citep{Rees1992, Meszaros1993b}.

An alternative scenario involves a Poynting-flux-dominated outflow to interpret GRBs. Within this model, the outflow initially has a magnetization parameter $\sigma_0 \gg 1$ (defined as the ratio between the Poynting flux and the plasma matter flux). The jet is accelerated gradually as the Poynting flux energy is converted to kinetic energy \citep[e.g.][]{Granot2011}. Since the majority of energy is not in the thermal form initially, the photosphere emission is suppressed \citep{Daigne2002,zhang2009}\footnote{If subphotosphere magnetic dissipation is significant such that $\sigma$ already drops to around unity at the photosphere, then the photosphere emission could be bright \citep[e.g.][]{Rees2005,Giannios2006,Peer2006a,Beloborodov2010,Levinson2012, Vurm2013,Begue2015}}. If the jet composition is still Poynting-flux dominated ($\sigma > 1$) at the traditional internal shock radius, the eventual energy-dissipation site would be at the location for internal collision-induced magnetic reconnection and turbulence (ICMART), which is typically beyond $10^{15}$ cm from the central engine \citep{Zhang2011b}. In reality, the jet composition may differ among different GRBs. Most likely the jet composition could be hybrid \citep{Gao2015,Li2020}, characterized by a relativistic outflow with a hot fireball component (defined by the dimensionless enthalpy $\eta$) and a cold Poynting-flux component (defined by magnetization $\sigma_{0}$ at the central engine). Indeed, observations show that GRB composition seems diverse. Whereas some GRBs indeed show the signature properties of a fireball with a dominant photospheric thermal spectral component \citep{Abdo2009,ryde2010,peer2012,Li2019a}, some others show evidence of a Poynting-flux-dominated flow \citep{Abdo2009a,zhang2009,Zhang2016,Zhang2018a}. The nondetection of high-energy neutrinos from GRBs disfavors the possibility that the majority of GRBs are matter dominated and is consistent with the hypothesis that most GRBs are Poynting-flux dominated \citep{Zhang2013b,Aartsen2017}.

For a relativistic jet, the observed emission does not stop immediately, even if the emission ceases abruptly. This is because the emission from higher latitudes with respect to the line of sight arrives at the observer later because of the extra path that photons travel. This high-latitude emission (HLE) ``curvature effect'' \citep[e.g.,][and references therein]{Fenimore1996, Ryde1999, Kumar2000, Zhang2006, Li2019} has some testable predictions. In particular, if the emitter Lorentz factor remains constant during the decaying wing of a pulse, the temporal index $\hat{\alpha}$ and the spectral index $\hat{\beta}$ should satisfy a simple closure relation \citep{Kumar2000}, i.e. 
\begin{equation}
\hat{\alpha} = 2+\hat{\beta},
\label{eq:ClosureRelation}
\end{equation}
where the convention $F_{\nu,t}\propto t^{-\hat{\alpha}} \nu^{-\hat{\beta}}$ is adopted, and the zero time to define the power-law temporal decay index is set to the beginning of the pulse \citep{Zhang2006}. If the emission region is accelerating or decelerating, the decay slope $\hat{\alpha}$ is steeper or shallower than this predicted relation \citep{Uhm2015}.

Testing the curvature effect using the data can bring clues to the unknown jet composition and GRB mechanism from two aspects. First, if a temporal segment during the decay phase of a GRB pulse is identified as HLE, one can immediately place a constraint on the GRB emission radius at
\begin{equation}
    R_{\rm GRB} \gtrsim \Gamma^2 c t_{\rm HLE} = (3\times 10^{14} \ {\rm cm}) \Gamma_2^2 \left(\frac{t_{\rm HLE}}{1 \ {\rm s}}\right),
    \label{eq:RGRB}
\end{equation}
where $t_{\rm HLE}$ is the duration of the HLE in the source frame (again the observed HLE duration divided by $(1+z)$). For seconds-duration pulses, a positive detection of HLE would immediately derive a GRB radius $R_{\rm GRB}$ much greater than the photosphere radius and the standard internal shock radius, lending support to Poynting-flux-dissipation models such as the ICMART model. Second, if GRB prompt emission is powered by dissipation of a Poynting flux, one would expect that about half of the dissipated magnetic energy goes to accelerate the ejecta while the other half powers the radiation. As a result, one would expect bulk acceleration in the emission region. An HLE curvature-effect test may help to find evidence of bulk acceleration and, hence, evidence of Poynting-flux dissipation in the GRB jet. 

Some attempts have been made to test the curvature effect using the GRB prompt-emission data \citep[e.g.,][]{Fenimore1996, Ryde1999}, but no firm conclusion has been drawn. This is because the prompt emission often has overlapping pulses that smear the curvature effect (if any). \cite{Uhm2016a} tested the HLE curvature effect in two X-ray flares with clean and extended decay tails and found convincing evidence of bulk acceleration in these two GRBs. \cite{Jia2016} extended the analysis to a large sample of GRB X-ray flares and found that bulk acceleration seems ubiquitous. Modeling of prompt-emission spectral lags by \cite{Uhm2016} also provided independent evidence of bulk acceleration in the GRB prompt-emission region. In all these analyses, the inferred GRB emission radius is $\sim (10^{15}-10^{16})$ cm from the central engine, again consistent with the physical picture of magnetic energy dissipation in a Poynting-flux-dominated flow.

Since its launch in 2008, {\it Fermi}-GBM has triggered more than 2000 GRBs and collected a large trove of prompt-emission data. Usually GRB prompt-emission lightcurves show a complicated and irregular temporal profile with overlapping pulses, suggesting an erratic central engine at work. Observationally, a small fraction of bursts only have one single pulse. Some other bursts may exhibit multiple pulses that are well separated. These bursts form a unique sample for testing the HLE curvature effect from the prompt-emission data. 

In this paper, we collect a sample of GRBs with single pulses and use the sample to test the curvature effect in the prompt-emission phase. The paper is organized as follows. In Section \ref{sec:sample}, we present our sample selection criteria and data reduction procedure. In Section \ref{sec:data}, we present the detailed data analysis methods. Our results are presented in Section \ref{sec:result}, and conclusions and discussions are summarized in Section \ref{sec:conclusion}.

\section{Sample Selection and Data Reduction}\label{sec:sample}

Since our primary interest concerns individual emission episodes, we pay special attention to single pulses. Our sample selection allows many smaller spikes on top of the main pulse structures. This is because for the specific large-radius magnetic-dissipation models (e.g. the ICMART) we are testing, rapid variability is expected to be superposed on the broad pulses, due to the existence of minijets from locally dissipated regions \citep{Zhang2011b,Zhang2014b}. We first visually inspected all of the time-tagged events (TTE) lightcurves to search for single-pulse bursts from the bursts detected by the Gamma-ray Burst Monitor (GBM; \citealt{Meegan2009}) on board the {\it Fermi} Gamma-ray Space Telescope during its first 10 years of mission. During this time period, GBM has triggered on at least 2000 bursts. After our initial checking, about 300 well-defined single-pulse bursts are selected as our initial  sample. 

Our next step is to use the Bayesian blocks (BBlocks; \citealt{Scargle2013}) method to rebin the TTE lightcurve of each individual burst from our initial sample. The significance ($S$; \citealt{Vianello2018a, Li1983}) for each individual time bin is calculated. In order to make the physical inferences trustworthy, high-quality data are required. In particular, the decay phase is our main interest. We therefore require at least five time bins with $S>15$ measured during the decay phase. Our final sample is reduced to 24 bursts that satisfy this criterion. The sample is listed in Table \ref{tab:property}, including 24 individual pulses from 23 long GRBs and one short GRB. Note that our sample selection is similar to that of \cite{Yu2019}. However, compared with the sample in \cite{Yu2019}, our sample is obtained with a higher selection criterion.

The prompt-emission properties of our sample are reported in Table \ref{tab:property}. We collect duration ($t_{90}$, Column 1) and 10-1000 keV fluence (Column 2) from the online {\it Fermi}-GBM GRB repository\footnote{\url{https://heasarc.gsfc.nasa.gov/W3Browse/fermi/fermigbrst.html}}. We also list the detectors used, the source and background intervals used in the analysis, the number of time bins using the BBlocks method across the source interval, and the number of time bins with statistical significance $S>15$ selected from the decay wing of the pulses. The detector in brackets is the brightest one, which is used for background and BBlock fits. 

\section{Methodology}\label{sec:data}

\subsection{Pulse Properties}\label{sec:FRED}

To delineate the characteristics of the pulses, several functional forms have been proposed \cite[e.g.,][]{Kocevski2003, Norris2005}. In order to adequately characterize a pulse shape, our next step is to employ an asymmetric fast-rising and exponential-decay function, the so-called the FRED model \citep{Kocevski2003}, to fit the entire lightcurve of that pulse (Figure \ref{fig:FRED}). The peak time of the pulse can be then determined. The function reads as
\begin{equation}
I(t)=I_{p} \left(\frac{t+t_{0}}{t_{p}+t_{0}}\right)^{r} \left[\frac{d}{r+d}+\frac{r}{r+d} \left(\frac{t+t_{0}}{t_{p}+t_{0}}\right)^{r+1} \right]^{-\frac{r+d}{r+1}}, 
\label{eq:FRED}
\end{equation}
where $I_{\rm p}$ is the amplitude, $t_{0}$ and $t_{\rm p}$ are the zero time and the peak time of the pulse, and $r$ and $d$ are the rise and decay time scale parameters, respectively. The model invokes a five parameters ($I_{\rm p}$, $t_{0}$, $t_{\rm p}$, $r$ and $d$). We also considered a broken power-law (BKPL) fit to the pulse (Appendix). In Figure \ref{fig:comparison} we present a comparison of the fitting results compared between the FRED model and the BKPL model.

In Table \ref{tab:FRED}, we list the best-fit parameters by adopting the FRED model for our sample. We list the used time resolution of the count rate (counts/sec) lightcurve used for each burst (Column 2), the start and stop times of the selected pulses (Column 3) and the corresponding significance $S$ (Column 4), as well as the best-fit parameters for the FRED model (Columns 5-9) including the normalization $I_{\rm p}$; the zero time $t_{\rm 0}$, which we fixed it to zero for each case; the peak time $t_{\rm p}$ of the pulses; and the rise $r$ and decay $d$ time scale parameters. The reduced chi-squared $\chi^{2}$/dof (Column 10), the Akaike Information Criterion (AIC) statistic (Column 11), and the Bayesian information criterion (BIC) statistic (Column 12) are also presented. Note that the goodness of fit (GOF) can be evaluated by calculating the reduced chi-squared statistic when the uncertainties in the data have been obtained. For a set of $N$ data points $ \lbrace x_{i},y_{i} \rbrace$ with the estimated uncertainties $\lbrace \sigma_{i} \rbrace$ in the $y_{i}$ values, one has chi-square $\chi^{2}=\Sigma^{N}_{i=1} \frac{(y_{i}-\hat{y_{i}})^{2}}{\sigma^{2}_{i}}$, and reduced chi-square $\chi^{2}_{\nu}=\chi^{2}/\rm dof$, where dof=($N-N_{\rm varys}$) is the degrees of freedom, $N$ is the number of data points, and $N_{\rm varys}$ is the number of variables in the fit. The bad fits (large $\chi^{2}_{\nu}$ values) indicate that these pulses cannot be well delineated by the FRED model. In Table \ref{tab:FRED}, AIC is calculated by $N \rm ln(\chi^{2}/N)+2N_{\rm varys}$, and BIC by $N \rm ln(\chi^{2}/N)+\rm ln(N) N_{\rm varys}$.

\subsection{Method to Measure Temporal Indices with a Simple Power-law Model}\label{sec:temporalfits}

We use the energy flux lightcurves to measure the temporal indices. This is because the indices thus defined can be better compared with model predictions. 

\begin{figure*}
\includegraphics[width=0.5\hsize,clip]{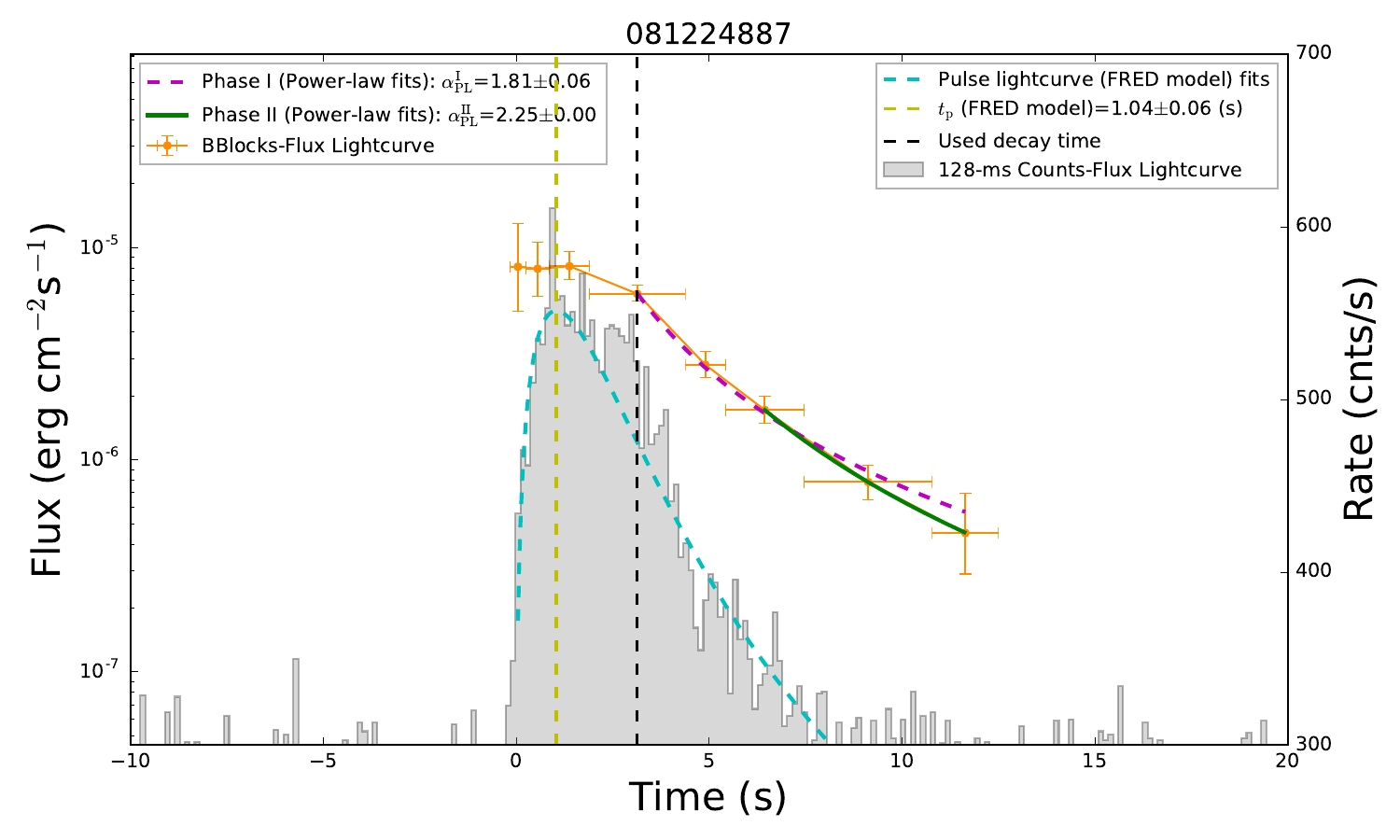}
\includegraphics[width=0.5\hsize,clip]{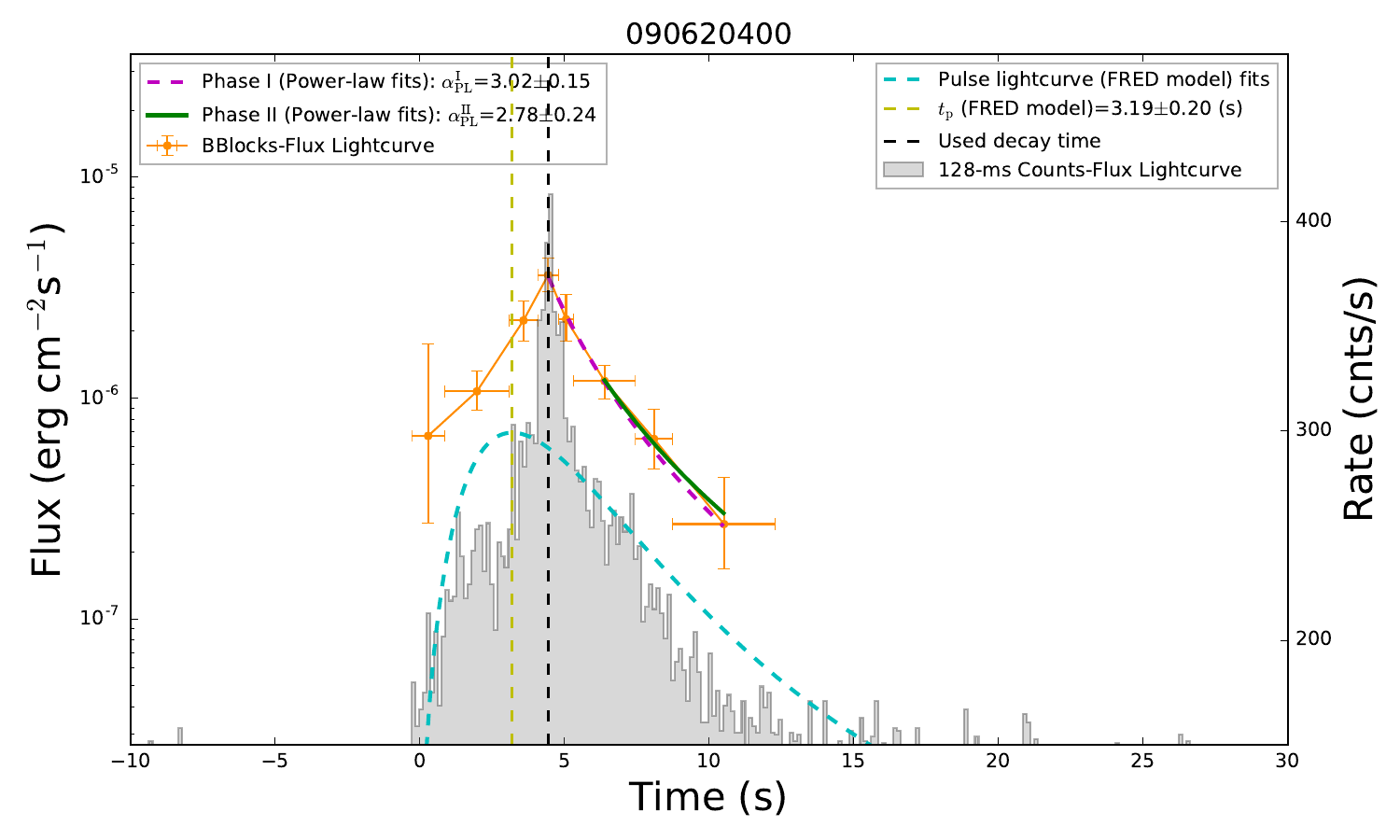}
\includegraphics[width=0.5\hsize,clip]{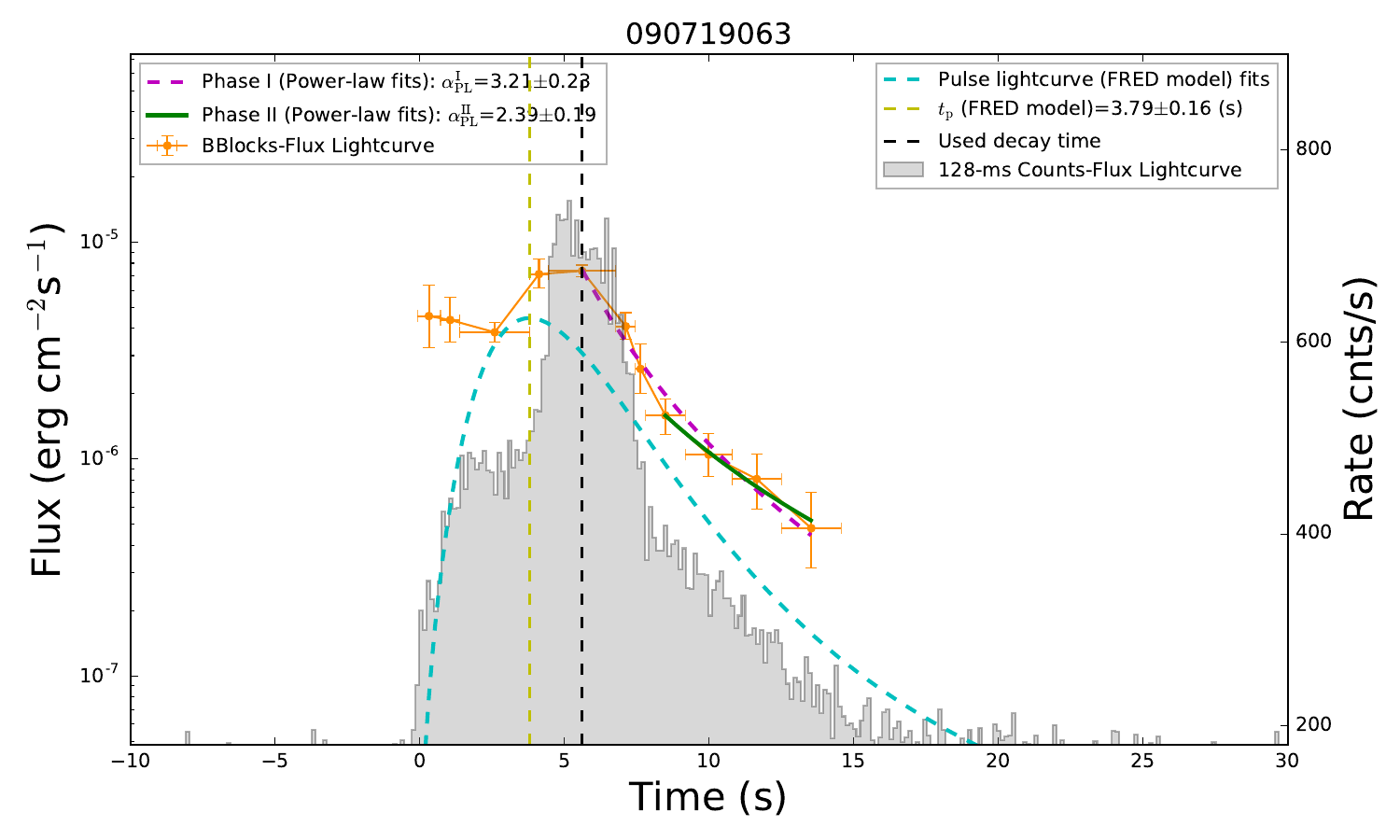}
\includegraphics[width=0.5\hsize,clip]{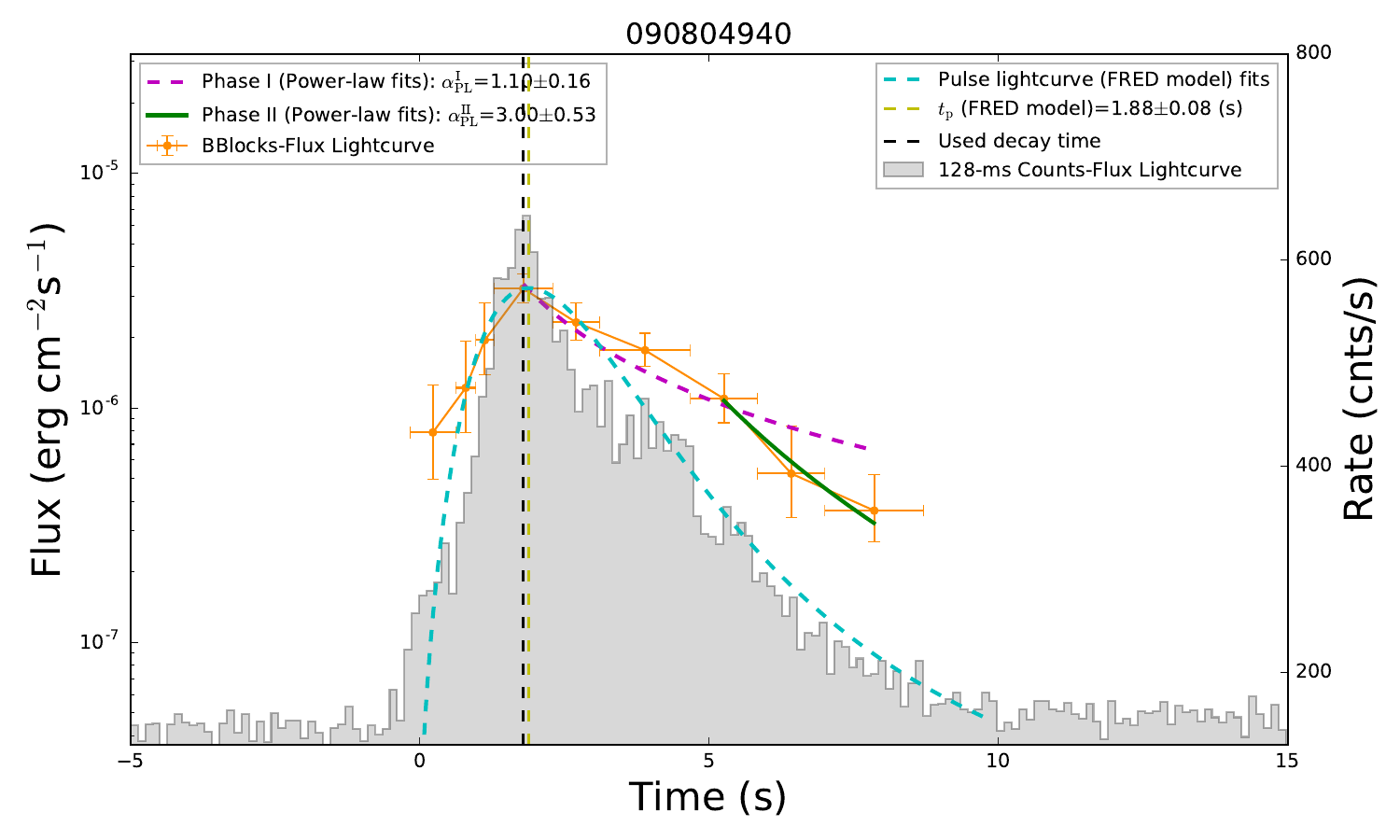}
\includegraphics[width=0.5\hsize,clip]{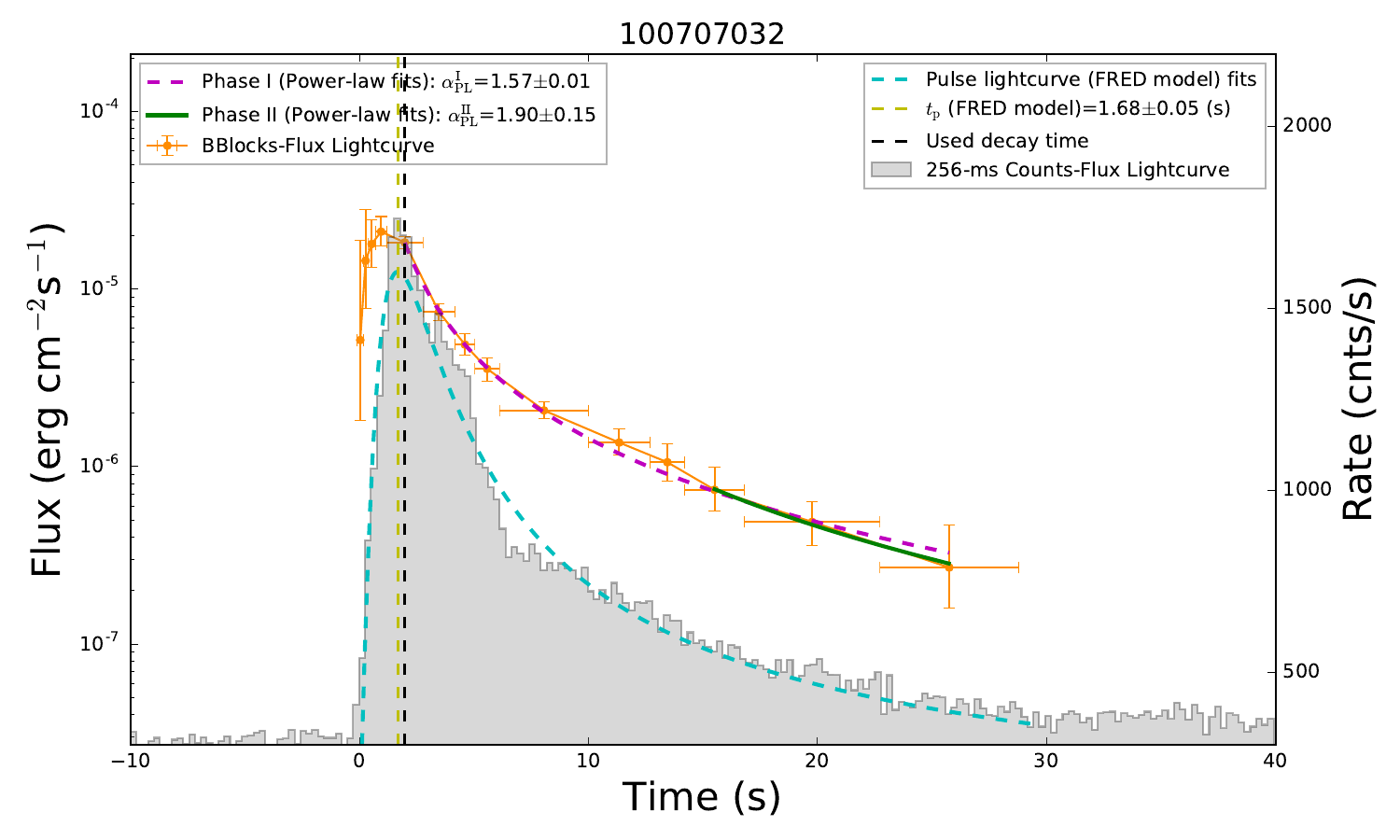}
\includegraphics[width=0.5\hsize,clip]{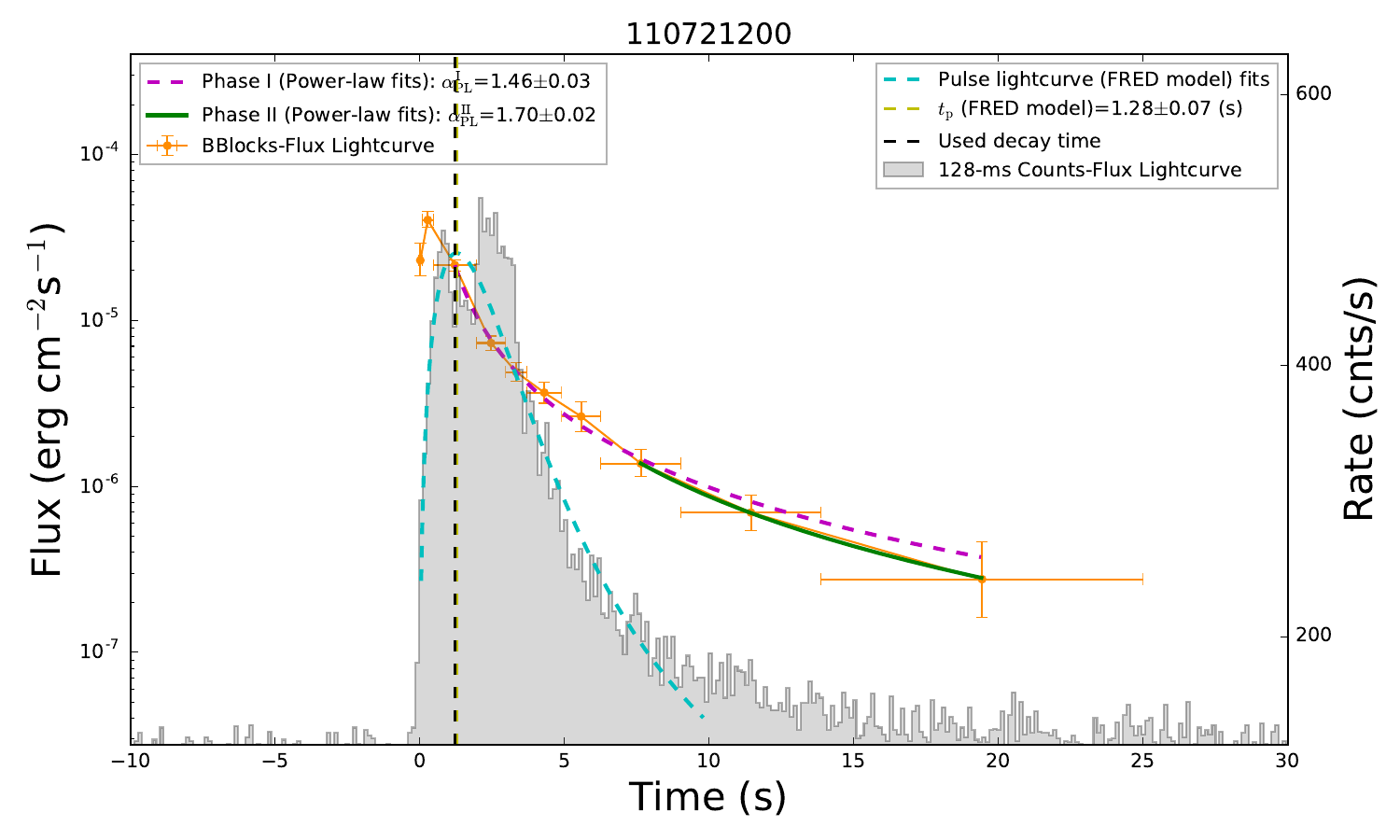}
\includegraphics[width=0.5\hsize,clip]{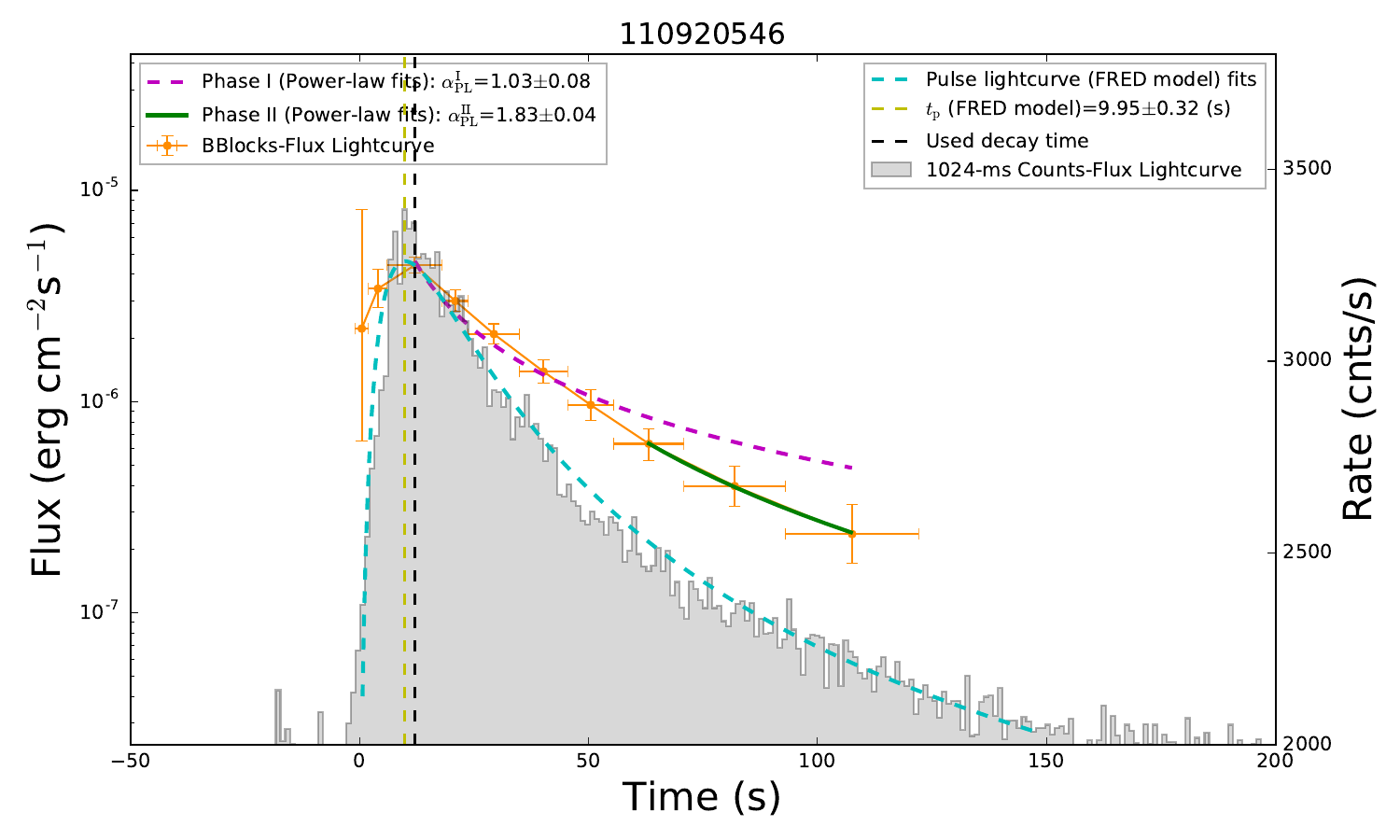}
\includegraphics[width=0.5\hsize,clip]{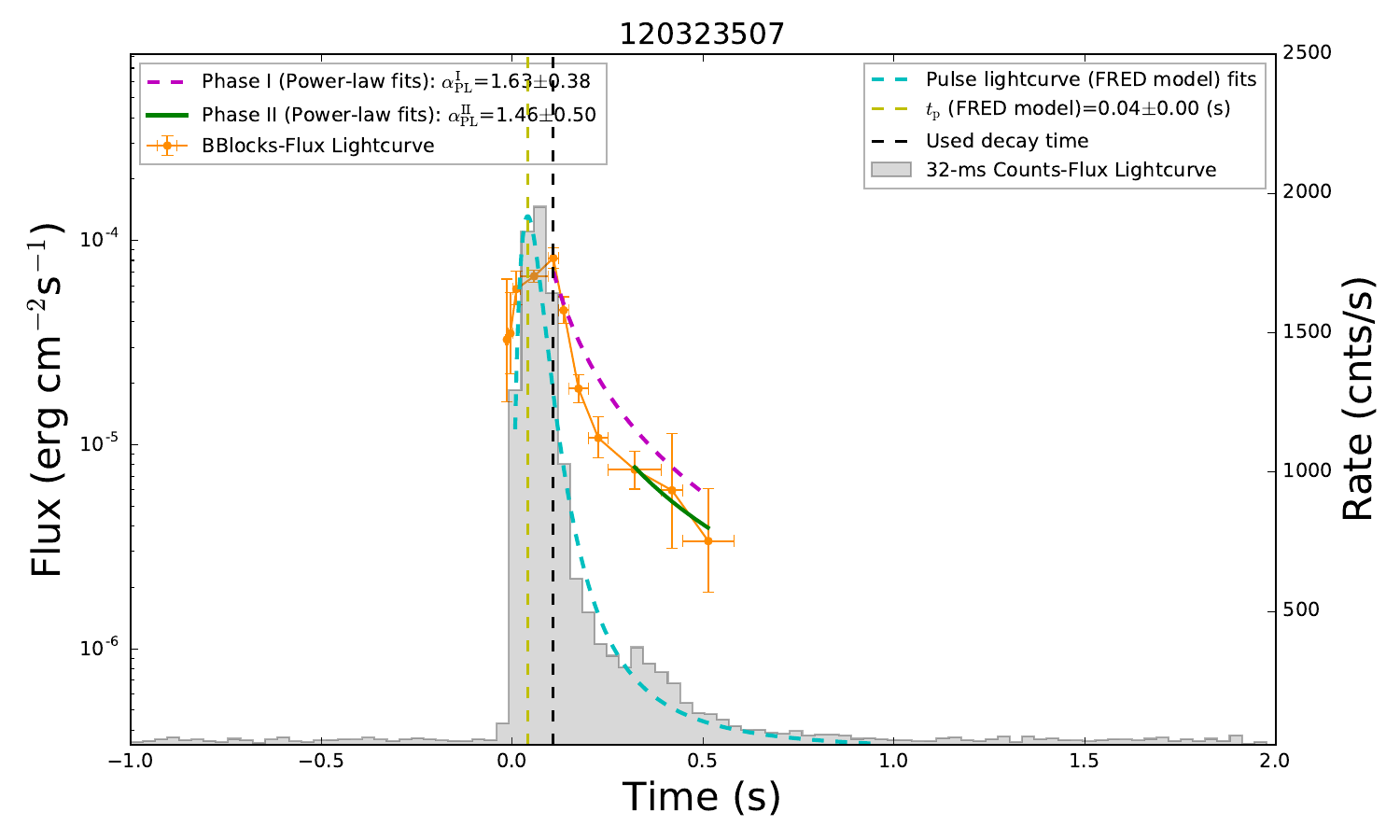}
\caption{Lightcurves of the pulses in our sample. For each panel, the left axis marks the energy flux. Its evolution is marked in orange. The best fits for Phase I are indicated with the dashed line with the purple dashed line, while those for Phase II are indicated with green solid lines. The right axis displays the count flux. The count lightcurves are in gray, overlaid with the best FRED model fits (cyan). The vertical yellow dashed line is the peak of the FRED fitting curve. The vertical black dashed line is the peak time identified by eye by inspecting the BBlock energy flux. }\label{fig:PL}
\end{figure*}

Our procedure to obtain the temporal indices includes the following steps:
\begin{enumerate}
\item Calculate the energy flux in each selected time bin. In order to obtain the energy flux, one needs to perform the spectral fits. For a given burst in our final sample, we therefore use the typical spectral model, called the Band function model \citep{Band1993} to fit the spectral data of each time bin ($S>15$) selected by the BBlocks method, and the best-fit parameters are evaluated by adopting the maximum-likelihood estimation (MLE) technique. The energy flux in such narrow time bins thus can be also calculated from the best fits, with a $k$-correction (1-10$^{4}$ keV) applied\footnote{Note that the energy flux obtained from different spectral models (Band and cutoff power law (CPL)) for the same time bin is very similar \citep{Li2019a,Li2020b}.}.
\item Determine the entire time interval of the decay wing of the pulses. In order to determine the entire time interval of the decay wing of the pulses, one needs to determine the peak times of the pulses. The peak times of the pulses can be roughly obtained by using the FRED model to fit their pulse lightcurves as we discussed in \S \ref{sec:FRED}. We find that the peak time determined by the FRED model for a good fraction of our sample can exactly match the true peaks of pulses (e.g., GRB 110920546). However, there are still some bursts whose peak times determined by the FRED model do not exactly describe the true peaks of the pulses\footnote{This is because some pulse lightcurves do not show an ``ideal'' asymmetric fast-rising and exponential-decay shape (e.g., GRB 090719063). In these cases, usually the true peak time of the pulse is apparently later than that derived from the FRED model.}. Therefore, we use two selection criteria. First, for the cases where the peak times determined by the FRED model can exactly match the true peaks of pulses, we use these values (see the vertical yellow color dashed lines in Fig. \ref{fig:PL}). That is, as long as the peak time ($t_{\rm p}$) of a certain pulse is obtained from the FRED model fits, the time window of the decaying wing of the pulse can be determined as $t_{\rm p}$-$t_{\rm stop}$, where $t_{\rm stop}$ is the end time of a pulse. The stop time of the decay wing of a certain pulse can be precisely determined by the stop time of the last time bin that satisfies $S>15$. Second, for the cases whose peak times determined by the FRED model do not exactly describe the true peaks of the pulses, we inspect the peak times from their lightcurves by eye (see the vertical black color dashed lines in Fig. \ref{fig:PL}). We define this phase as ``Phase I'' throughout the paper.
\item Determine the late-time interval of the decay wing of the pulses. Physically, the decay for prompt emission may not be fully controlled by the curvature effect. As shown in the theoretical modeling in \cite{Uhm2016} and \cite{Uhm2018}, the spectral lags are not caused by the curvature effect, and the temporal peaks of the pulses are often related to the time when the characteristic energy crosses the gamma-ray band as it decays with time. One possible test for this is to see whether the temporal peaks of the lightcurves for different GBM detectors that have different energy ranges occurr at different times. We therefore compare the NaI (8 keV-1 MeV) and BGO (200 keV-40 MeV) lightcurves for each individual burst, as shown in Figure \ref{fig:Bgo+NaI_LCs}. We find that in many cases in our sample the peak times are clearly shifted between two different detectors (GRB 081224887, GRB 110721200, GRB 120426090, GRB 160216801, GRB 170921168, and GRB 171210493), indicating that the peaks of the pulses are indeed related to crossing of a spectral break\footnote{Several other bursts, for example, GRBs 090620400, 090804940, 110920546, 130614997, 150510139, and 170114917, are consistent with having the same peak times in different bands. The HLE may come into play right after the peak time.}. For these bursts, the curvature effect does not kick in right after the peak. It may show up later in some bursts or would not show up at all in some others. When they show up, they may be related to the later part of the decay, usually not related to the decay right after the peak time. This brings an additional difficulty (other than the fact that the decay phase is usually short for prompt-emission pulses) in studying the curvature effect with the prompt-emission data. Besides testing the entire decay phase, we also adopt a more conservative approach by only testing the late-part time interval of the decay phase. Quantitatively, we only consider the last three time bins with $S>15$. In practice, when a certain model is used to fit the data, the number of data points $N$ should be greater than the number of variables $N_{\rm varys}$ of the model in order to get a good fitting result. The power-law model we use has two variables: amplitude and power-law index. This is why we include at least three data points in the fits. We define this phase as ``Phase II'' throughout the paper. 

\item After the time intervals are clearly defined in the aforementioned two cases, we then perform two fits\footnote{Note that we present the [log (Flux), time] plots in Figure \ref{fig:PL} since the count lightcurve before the GBM trigger relates to negative time. However, the power-law fits invoke the [log (Flux), log(time)] plots, so we give an example to show the [log (Flux), log(time)] plots (see Figure \ref{fig:loglog}).}. (see Figure \ref{fig:PL}): one uses a power-law model to fit the entire decay phase and the obtain a temporal decay index defined as $\hat{\alpha}_{\rm PL}^{\rm I}$; the other uses a power-law model to fit the later part of the decay to obtain a temporal decay index defined as $\hat{\alpha}_{\rm PL}^{\rm II}$. The power-law function we use to fit the lightcurves in order to obtain the $\hat{\alpha}$ indices is given by
\begin{equation}
 F_{t} = F_{t,0}\, (t+t_{0})^{-\hat{\alpha}},
 \label{eq:PL_Alpha}
\end{equation}
where $F_{t,0}$ is the amplitude and $\hat{\alpha}$ is the temporal slope. The $t_0$ parameter is fixed in the beginning of the pulse ($t_0=0$) for all cases in this task because this is physically more relevant \citep{Zhang2006,Uhm2015}. Note that the peak time $t_{\rm p}$ does not enter the problem of defining $\hat\alpha$, so that the inaccurate determination of $t_{\rm p}$ in the pulse lightcurve fitting does not noticeably affect our results. All these lightcurve fits are performed using a pure Python package called {\it lmfitt} \citep{Newville2016} by applying a nonlinear least-squares method using the Levenberg-Marquardt algorithm to fit a function to the data. Within {\it lmfitt} fits, we can set parameters with a varied or fixed value in the fit, or place an upper or lower bound on the value. The weight of parameter error is also easily taken into account in the fits. In Figure \ref{fig:comparison}, we also use GRB 131231198 as an example case to compare the fitting results obtained from different Python packages ($lmfit$ and $scipy.optimize.curve\_fit$).
\end{enumerate}

The start and stop times of each selected time interval (Column 2), the corresponding $S$ value (Column 3), the adopted zero time $t_{0}$ (Column 4), the best-fit parameters, include the normalization (Column 5), the power-law index (Column 6), and the AIC and BIC statistics (Column 7) are listed in Table \ref{tab:relation}. For each burst, the entire decay phase is marked with the symbol (1) and the late-part decay phase is marked with the symbol (2). 

\subsection{Method to Measure Spectral Indices with a Simple Power-law Model}\label{sec:spectralfits}
 
The GRB prompt-emission spectra are likely curved. However, since the simplest curvature-effect model (Eq. \ref{eq:ClosureRelation}) applies to single power-law spectral models, we first apply a simple power-law fit to the time bins where the curvature effect is tested:
\begin{equation}
 F_{\nu} = F_{\nu,0}\, \nu^{-\hat{\beta}},
  \label{eq:PL_Beta}
\end{equation}
where $F_{\nu,0}$ is the amplitude and $\hat{\beta}$ is the spectral index. The spectral analysis is performed using a pure Python package called the Multi-Mission Maximum Likelihood Framework (3ML; \citealt{Vianello2015}). The best model parameters can be evaluated using a given model to fit the data by applying either the MLE technique or the full Bayesian approach. Usually the best-fit results obtained from both methods are the same\footnote{There are some unexpected cases. For example, the prior range for the Bayesian inference is not included in the real solution; namely, the prior settings are not very informative, or the analyzed time bin has a low significance (e.g., $S<15$) or low peak energy (e.g., $E_{\rm p}<$20 keV). We refer to \cite{Li2019a, Li2019b, Li2020} and \cite{Li2020b} for the details of the data reduction procedure.}. 

We attempt two fits using the simple power-law model. One is to select the entire decay phase as the time interval to perform the spectral fit. The spectral index obtained this way is defined as $\hat{\beta}_{\rm PL}^{\rm I}$. The other is to select the later part of the decay as the time interval. The spectral index thus obtained is defined as $\hat{\beta}_{\rm PL}^{\rm II}$. 

For each spectral fit, we employ a fully Bayesian approach to explore the best parameter space and to obtain the best-fit parameters. The best-fit parameters, including the normalization (Column 8) and the power-law index (Column 9), as well as the deviance information criterion (DIC; \citealt{Moreno2013}; Column 10) and $p_{\rm DIC}$ (\citealt{Gelman2014}; Column 10), are tabulated in Table \ref{tab:relation}. 

\subsection{Method to Measure Spectral Indices with a General Non–power-law Spectral Model}

The aforementioned discussion invokes the simplest curvature-effect model, which assumes that the instantaneous spectrum of the prompt-emission tail is a simple power law. In this case, the predicted temporal decay and the spectral indices satisfy with the simplest closure relation (Eq.\ref{eq:ClosureRelation}).
However, the instantaneous spectrum upon the cessation of prompt emission is likely not to be a simple power law, but it may follow a non-power-law model such as the Band function \citep[e.g.][]{Band1993}. The characteristic frequency $\nu_{\rm c}$ may not be far outside the GBM spectral window. In this case, testing the curvature effect would become more complicated. 

We also test the curvature effect using the more complicated model as described in \cite{Zhang2009b}. We consider that for each time bin the photon flux can be described by a power-law spectrum with an exponential cutoff. This spectrum has one parameter less than the Band function and is found to be  adequate to describe the GRB spectra during the decay phase\footnote{Previous studies show that the CPL model is a sufficient model for the majority of GRB spectra \citep[e.g.][]{Yu2019, Li2020b}. On the other hand, GRBs usually exhibit strong spectral evolution. In order to best characterize the spectral shape, one needs to introduce an evolving spectral model within a burst or even within a pulse \citep{Li2020b}. For simplicity, we perform the HLE test only considering the CPL model. We also notice that there are clear predictions for $\alpha$ evolution for HLE if the emergent spectrum is indeed described by the Band function, which has been studied by some authors \cite[e.g.,][]{Genet2009}.}, i.e.
\begin{equation}
N(E,t)=N_{0}(t) \left(\frac{E}{E_{\rm piv}}\right)^{-\hat{\Gamma}} {\rm exp}\left\lbrace-\left[\frac{E}{E_{c}(t)}\right]\right\rbrace,
\label{eq:CPL}
\end{equation}
where $\hat{\Gamma}=\hat{\beta}+1$ is the power-law photon index, and $E_{\rm piv}$ is the pivot energy fixed at 100 keV, and $N_{0}(t)=N_{0,\rm p}[(t-t_{0})/(t_{\rm p}-t_{0})]^{-(1+\hat{\Gamma})}$ is the time-dependent photon flux (in units of photons keV$^{-1}$ cm$^{-2}$ s$^{-1}$) at 100 keV (see also Eq.7 in \citealt{Zhang2009b}). For such a spectrum, the standard curvature effect predicts 
\begin{equation}
E_{\rm c}(t)=E_{\rm c,p} \left(\frac{t-t_{0}}{t_{\rm p}-t_{0}}\right)^{-1}
\label{eq:Ec}
\end{equation}
where $E_{\rm c,p}=E_{c}(t_{\rm p})$, $t_{0}$ is fixed to zero, and $t_{\rm p}$ is the beginning of the decay of the pulses; and
\begin{equation}
F_{\nu,c}(t)=F_{\nu,c,p} \left(\frac{t-t_{0}}{t_{\rm p}-t_{0}}\right)^{-2}
\label{eq:Fnu}
\end{equation}
where $F_{\nu,c}(t) = E_{\rm c}(t)N_{\rm c}(t)$, and $F_{\nu,c,p} = E_{\rm c,p} N_{\rm c,p}$, where $N_{\rm c}(t)=N(E_{\rm c},t)=N_{0}(t) (E_{\rm c}/E_{\rm piv})^{-\hat{\Gamma}} {\rm exp}(-1)$, which is calculated using Eq.(\ref{eq:CPL}) when $E$ is at cutoff energy $E_{\rm c}$, and $N_{\rm c,p}=N(E_{\rm c},t_{\rm p})=N_{0}(t_{\rm p}) (E_{\rm c}/E_{\rm piv})^{-\hat{\Gamma}} {\rm exp}(-1)$, which is calculated at time $t_{\rm p}$ and cutoff energy $E_{\rm c}$.

With Eq.(\ref{eq:Ec}) and Eq.(\ref{eq:Fnu}), one can also get a direct relation between $F_{\nu,c}(t)$ and $E_{c}(t)$:
\begin{equation}
F_{\nu,c}(t)= \frac{N_{\rm c, p}} {E_{\rm c,p}} E^{2}_{\rm c}(t)
\label{eq:FnuEc}
\end{equation}

From the data, the time-dependent parameters $E_c(t)$ and $F_{\nu,c}(t)$ can be directly measured. One can then directly compare the data against the model predictions in Equations (\ref{eq:Ec})-(\ref{eq:FnuEc}).

\section{Results}\label{sec:result} 

\subsection{The Case of Power-law Spectra}

\begin{figure*}
\centering
\includegraphics[width=0.9\hsize,clip]{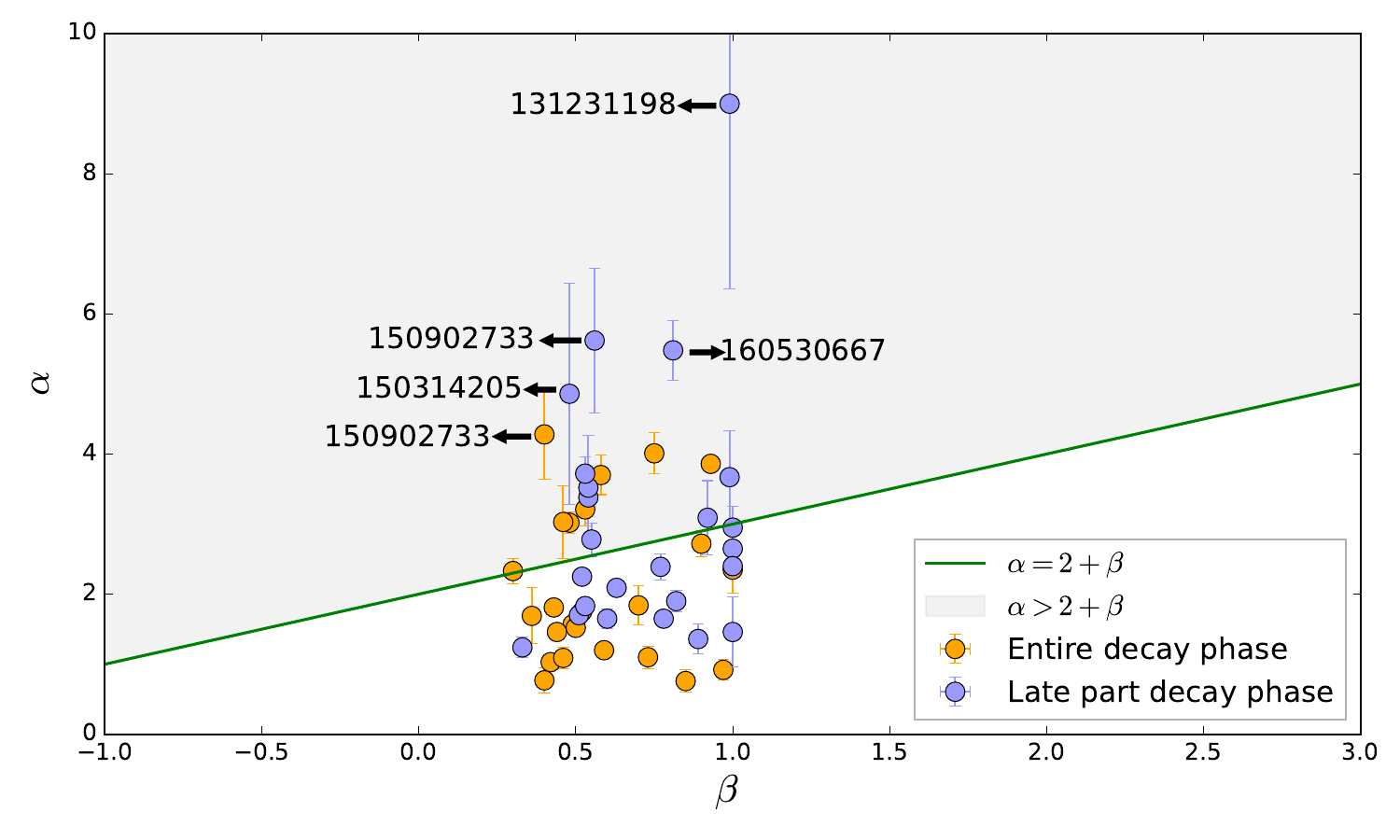}
\caption{Testing the closure relation of the curvature effect in the decaying wing using prompt-emission data. The closure relation between the temporal index $\hat\alpha$ and the spectral index $\hat\beta$ \citep{Kumar2000}, that is, $\hat\alpha \geq$ 2+$\hat\beta$, is marked as the solid green line, with the convention $F^{\rm obs}_{\nu_{\rm obs}} \propto t^{-\hat\alpha}_{\rm obs} \nu^{-\hat\beta}_{\rm obs}$. The orange and blue colors indicate different decay phases, Phase I and Phase II, respectively, as defined in the text. The shaded area stands for $\hat\alpha > 2+\hat\beta$, which requires bulk acceleration in the emission region. }\label{fig:relation}
\end{figure*}

For the case of power-law spectra, as discussed above, we measure the temporal indices for two phases (Phase I and II) and their corresponding spectral indices (using a time-integrated spectrum throughout the decay phase). The results are as follows:
\begin{itemize}
\item Entire decay phase (Phase I): The parameter set ($\hat{\alpha}_{\rm PL}^{\rm I}$-$\hat{\beta}_{\rm PL}^{\rm I}$) is presented as orange dots in Figure \ref{fig:relation}. Eight out of 24 cases satisfy satisfy the inequality $\hat{\alpha} \geq 2+\hat{\beta}$. These bursts are GRB 090620400, GRB 090719063, GRB 130305486, GRB 131231198, GRB 141028455, GRB 150213001, GRB 150902733, and GRB 160530667. Other bursts are below the line, suggesting that not the entire decay segment can be attributed to the curvature effect for these bursts, which is quite reasonable in view of the modeling presented in \cite{Uhm2016} and \cite{Uhm2018}.
\item Late-part decay phase (Phase II): The parameter set ($\hat{\alpha}_{\rm PL}^{\rm II}$-$\hat{\beta}_{\rm PL}^{\rm II}$) is presented as blue dots in Figure \ref{fig:relation}. Upward 11 out of 24 cases now satisfy  the inequality $\hat{\alpha} \geq 2+\hat{\beta}$. These bursts include GRB 090620400, GRB 090804940, GRB 120426090, GRB 131231198, GRB 141028455, GRB 150314205, GRB 150510139, GRB 150902733, GRB  160530667, GRB 170921168, and GRB 180305393. This suggests that three additional bursts have the curvature effect showing up during the last three data points, while the remaining 13 bursts still do not have the HLE turned on by the end of the observed pulse. 
\end{itemize}

One immediate observation is that a good fraction of our sample has entered the $\hat{\alpha} > 2+\hat{\beta}$ regime. Since the HLE curvature effect defines the steepest decay index allowed in a GRB pulse, the results strongly suggest that the emission region is undergoing bulk acceleration in the region where prompt emission is released. We calculated the distance of this region from the central engine, $R_{\rm GRB}$, using Eq.\ref{eq:RGRB}, and found that they are typically $\sim 10^{15}-10^{16}$ cm for a typical Lorentz factor $\Gamma \sim 100$ (Table \ref{tab:RGRB}). In this region, it is impossible to have thermally driven bulk acceleration. The only possibility is that the jet is Poynting-flux dominated in the region, and the GRB emission is powered by the dissipation of a Poynting flux \citep{Zhang2011b}. About one-half of the dissipated energy is released as GRB emission while the other half is used to accelerate the ejecta. This conclusion is consistent with previous results from prompt-emission spectral-lag analysis \citep{Uhm2016} and the curvature-effect test of X-ray flares \citep{Jia2016,Uhm2016a}. 

A few bursts (GRB 081224887, GRB 090719063, GRB 100707032, GRB 110721200, and GRB 110920546) have been reported in some previous studies \citep{Iyyani2013, Iyyani2015, Iyyani2016, Li2019b} to require an additional thermal component in order to produce acceptable spectral fits. The thermal component is also included in our analysis for these bursts. For a self-consistency test, it is worth noting that these GRBs do not qualify for our Phase II sample and only one burst (GRB 090719063) is included in our Phase I sample. The results imply that the emission in these bursts may be dominated by other mechanisms (e.g., photosphere emission). The existence of a thermal component is consistent with a lower magnetization in the jet \citep{Gao2015}.

We notice that six cases (GRB 090804940, GRB 120426090, GRB 150314205, GRB 150510139, GRB 170921168, and GRB 180305393) are not included in the Phase I sample but are included in the Phase II sample, indicating that the curvature effect may only dominate the later part of emission for these bursts. It is also interesting to note that three cases (GRB 090719063, GRB 130305486, and GRB 150213001) are included in the Phase I sample but not in the Phase II sample. These may be spurious cases, which may have contamination from another emission episode. Our analysis below confirms this speculation.

\subsection{The Case of Cutoff Power-law Spectra}

In total, 14 bursts (including eight cases in the Phase I sample and 11 cases in the Phase II sample, noticing that some cases appear in both samples) meet the HLE-dominated criterion based on the power-law spectral analysis. These bursts are our primary interest. Our next step is to study these bursts in detail by investigating their compliance with the curvature-effect predictions in the more complicated cutoff power-law model using a time-dependent analysis. 

\begin{figure*}
\includegraphics[width=0.5\hsize,clip]{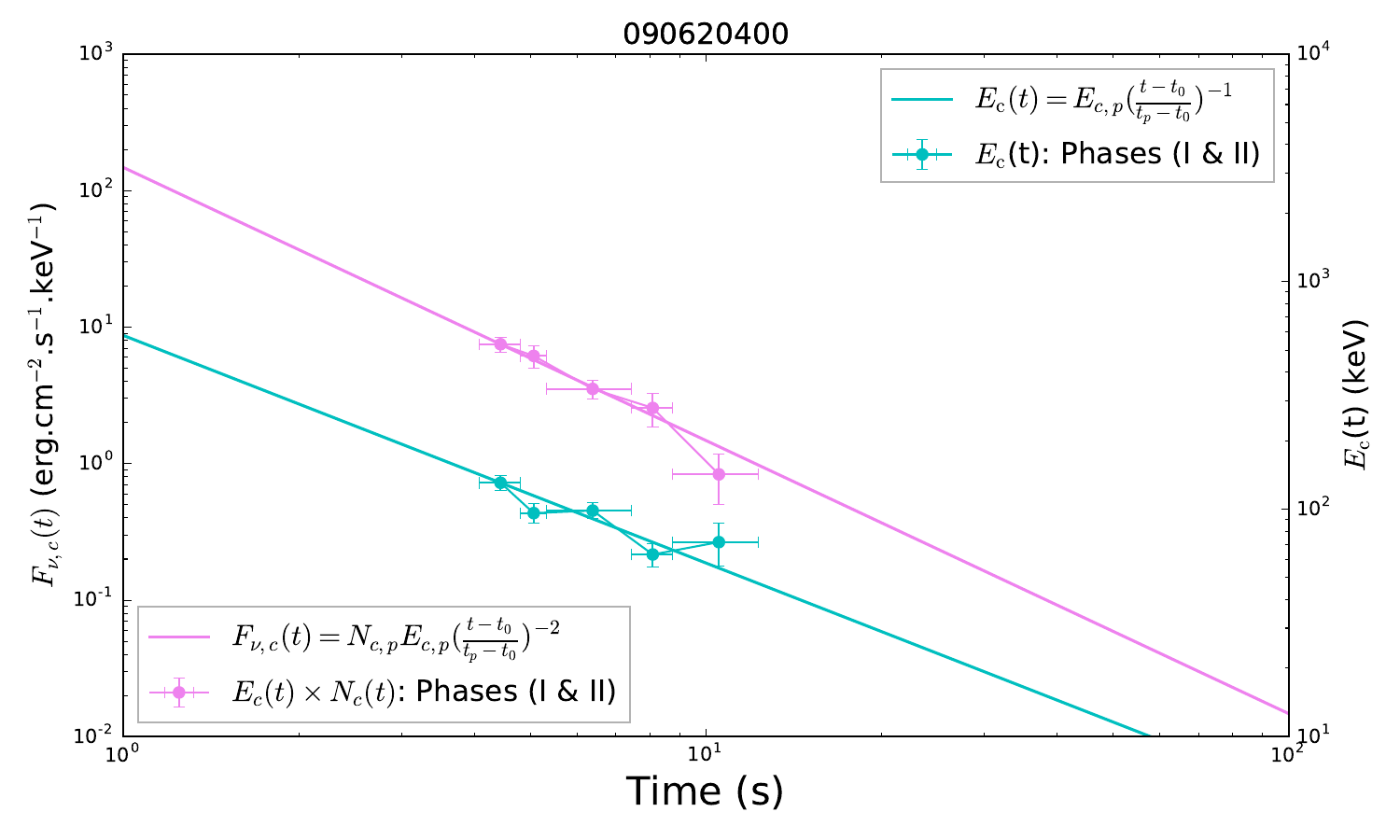}
\includegraphics[width=0.5\hsize,clip]{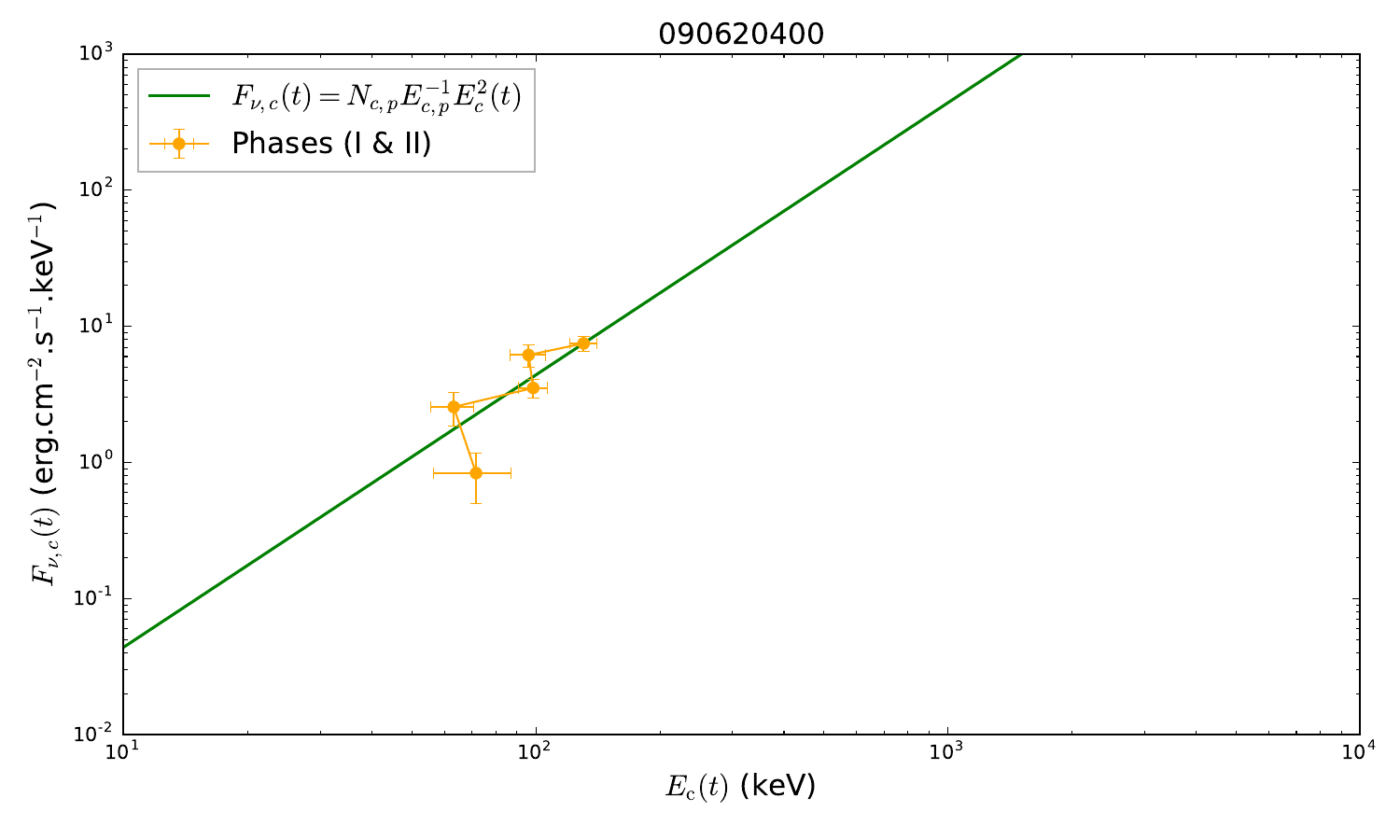}
\includegraphics[width=0.5\hsize,clip]{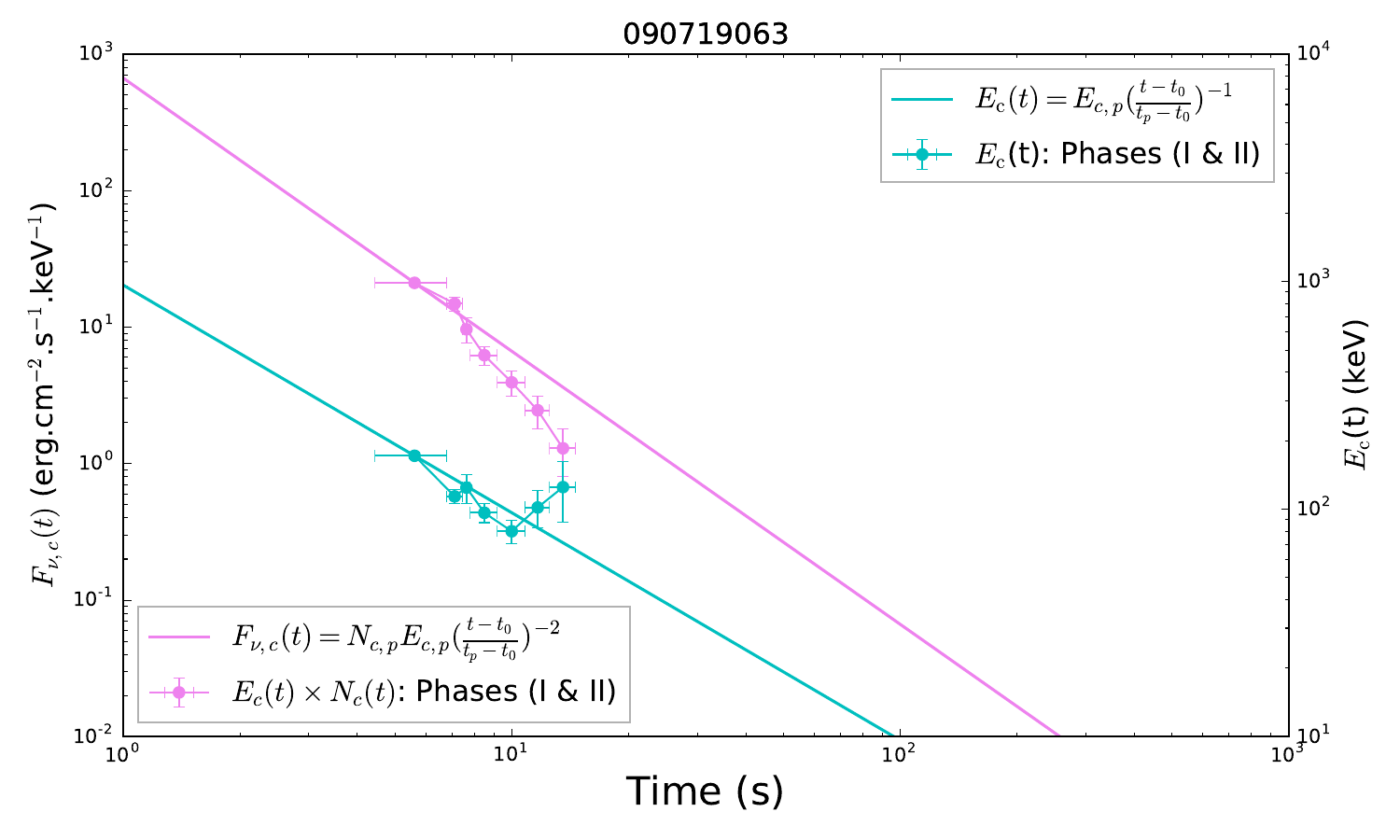}
\includegraphics[width=0.5\hsize,clip]{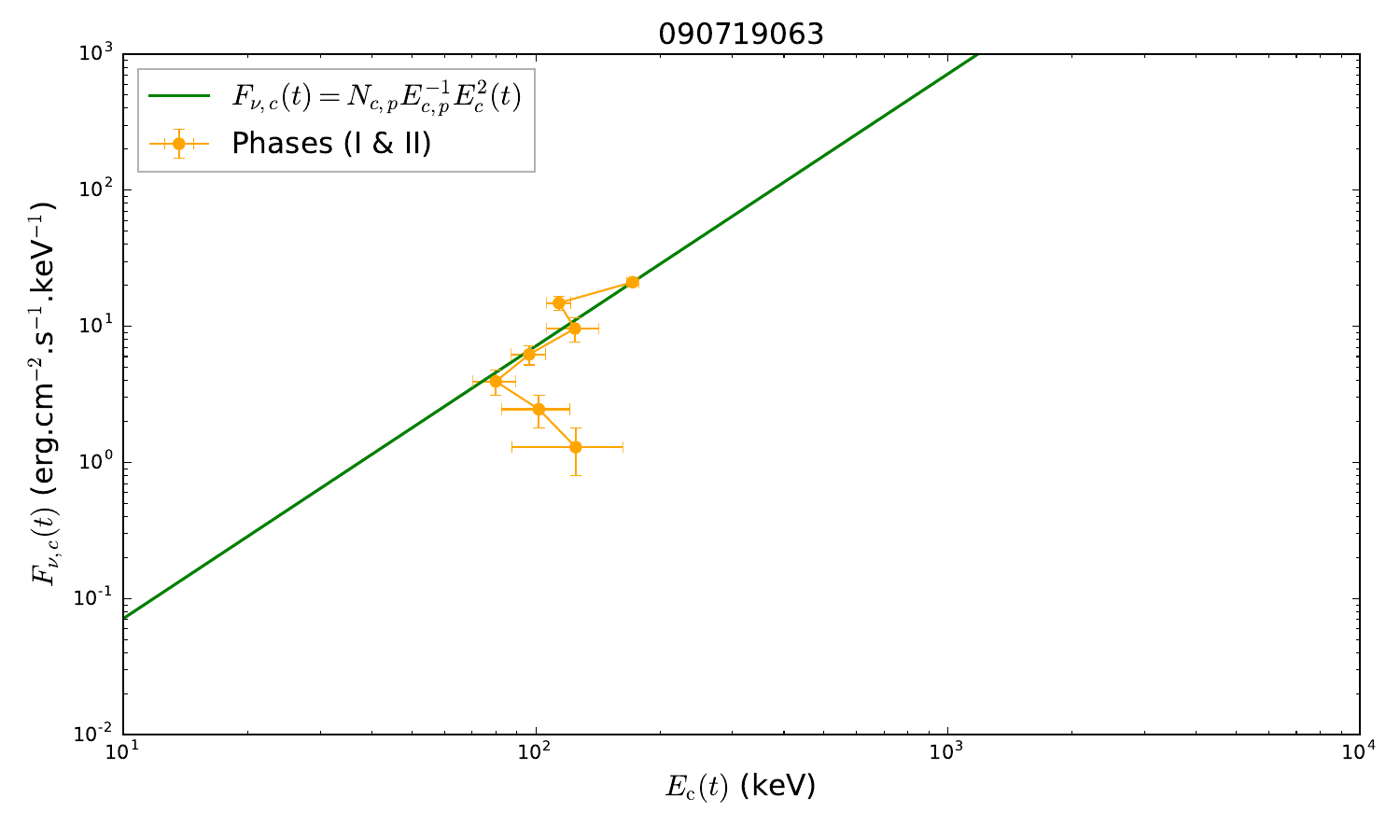}
\includegraphics[width=0.5\hsize,clip]{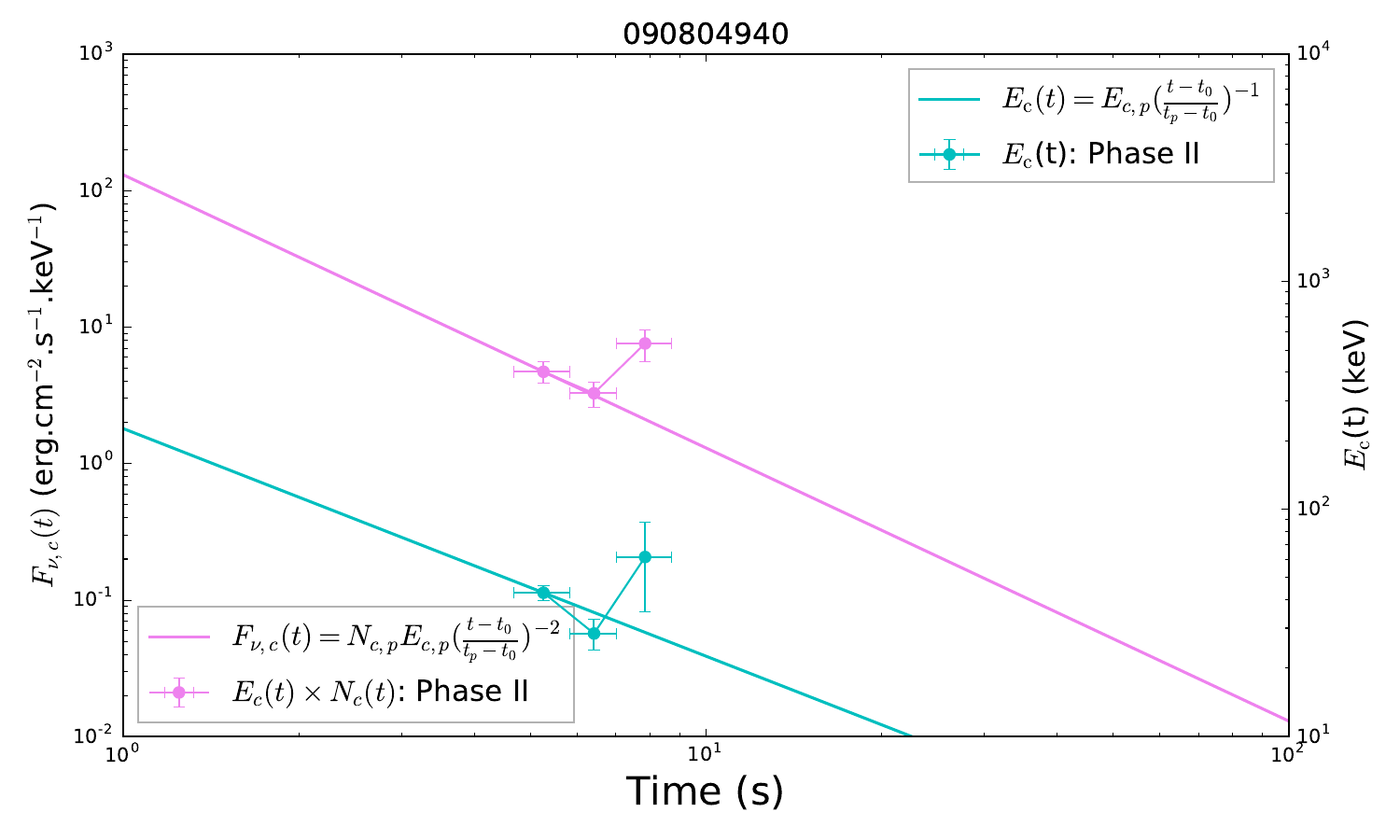}
\includegraphics[width=0.5\hsize,clip]{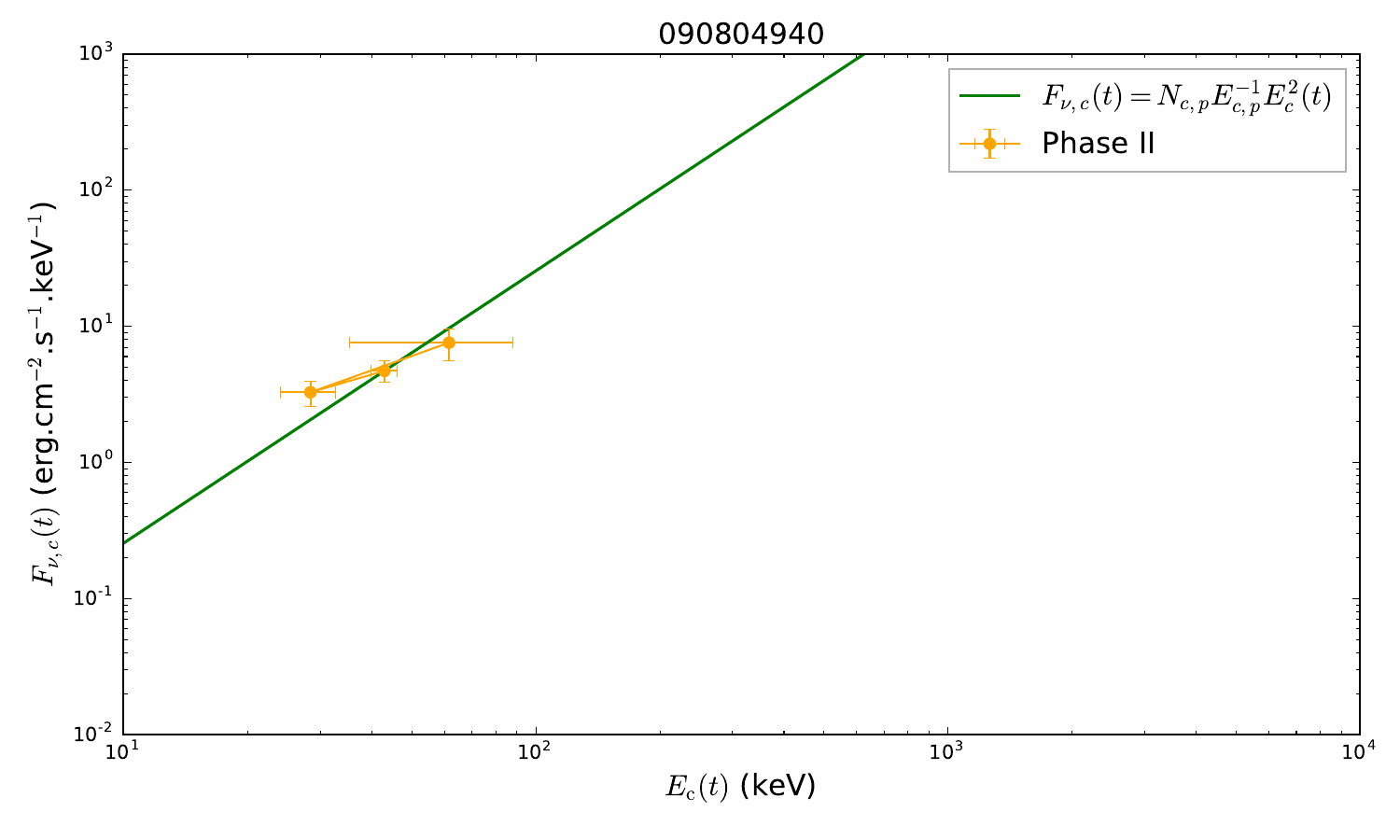}
\includegraphics[width=0.5\hsize,clip]{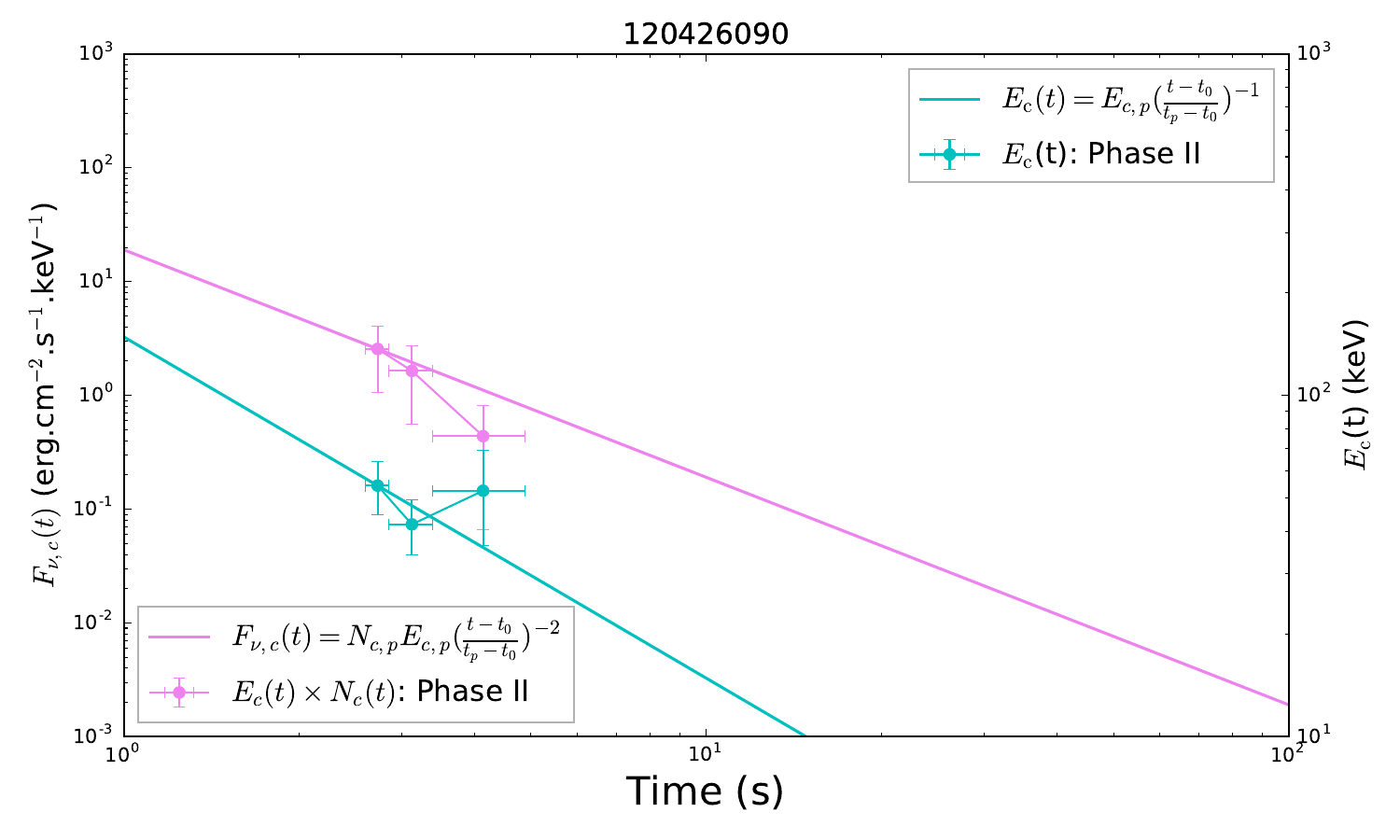}
\includegraphics[width=0.5\hsize,clip]{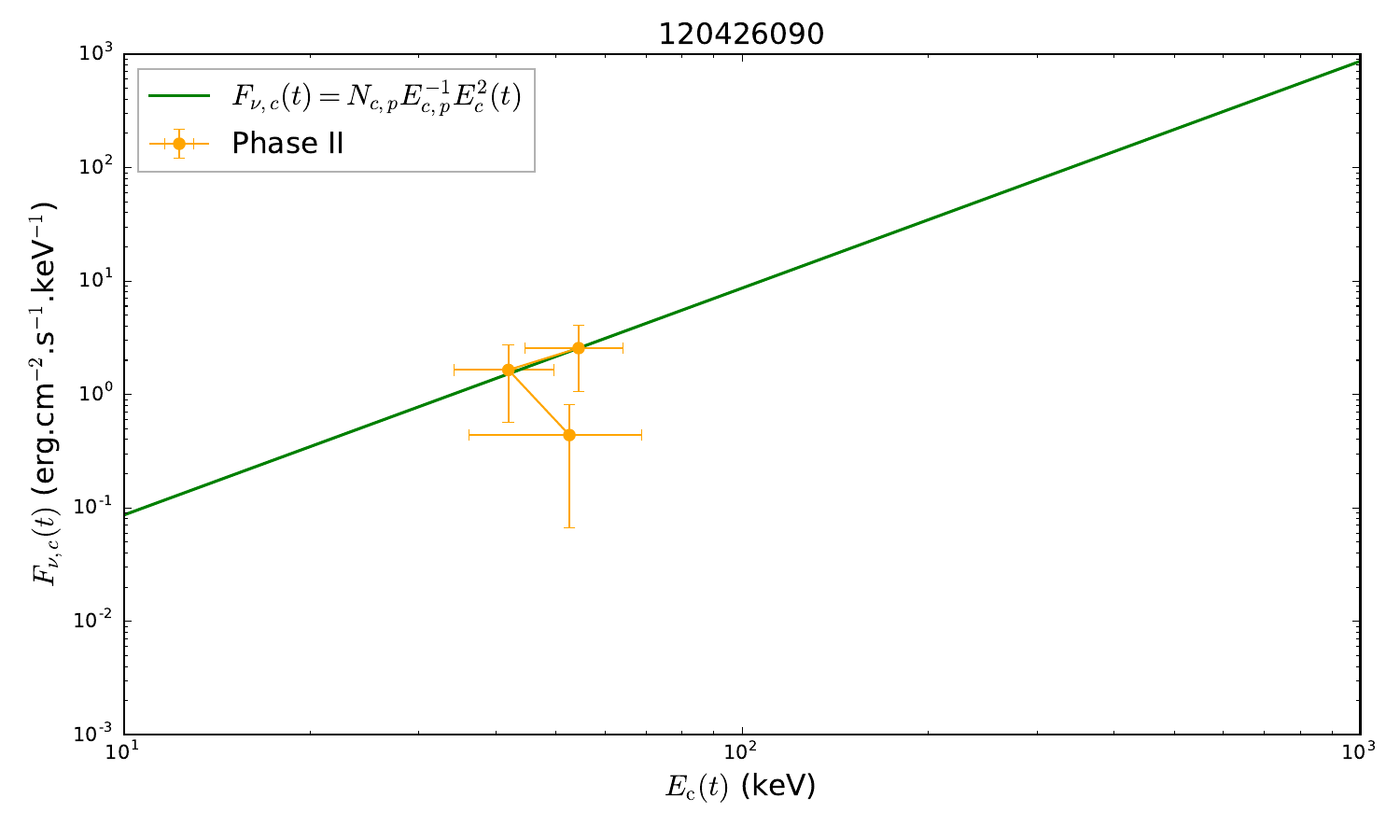}
\caption{Testing the non-power-law curvature-effect model developed in \cite{Zhang2009b} with observed data. The two panels in each row represent one individual pulse. Left panels: the cyan data points indicate the temporal evolution of the flux density $F_{\nu,c}(t)$ at the the characteristic energy $E_{\rm c}(t)$, while the pink data points indicate the evolution of the characteristic energy $E_{\rm c}(t)$. The cyan and pink solid lines represent the relevant theoretical predictions. Right panels: the orange data points indicate the data observed in the [$F_{\nu,c}(t)$, $E_{\rm c}(t)$] plane, while the green line represents the theoretical prediction between the two parameters.}\label{fig:CPL}
\end{figure*}

To test whether the CPL can account for the observed data as well, we adopt the following procedures:
\begin{enumerate}
\item We first apply the CPL model to fit the spectral data for these cases using the same episodes as the PL model to check whether the CPL model can improve the spectral fit results compared with the PL model. We find that the CPL fits are much better than the PL fits for all these cases by comparing the DIC statistic. We report our results in Table \ref{tab:relation}. For each individual fit, we fix $t_{0}$ to zero and $t_{\rm p}$ to the starting time of the Phase I or the Phase II. The best-fit parameters, including $t_{0}$ (fixed, Column 4), $N_{0,p}$ (Column 11), $t_{\rm p}$ (fixed, Column 12), $\Gamma$ index (Column 13), and  cutoff energy $E_{\rm c}$ (Column 14), as well as the DIC (Column 15) and $p_{\rm DIC}$ statistics (Column 15), are listed in Table \ref{tab:relation}.
\item Theoretically, we consider the evolution of $E_c$ and $F_{\rm \nu,c}$ according to Equations \ref{eq:Ec} and Eq. \ref{eq:Fnu} as predicted by the HLE curvature-effect theory (for a constant $\Gamma$). The predicted parameter evolution curves for both $F_{\nu,c}(t)$ and $E_{\rm c}(t)$ are plotted in the left panel of Figure \ref{fig:CPL} for each case to be directly compared with the data. In the right panel of Figure \ref{fig:CPL}, we plot the theoretically predicted $E_{\rm c}-F_{\nu,c}$ relation for each case to be directly compared with the observations.
\item The observed parameters for each time slice, including $N_{0}(t)$, $\Gamma$, and $E_{\rm c}(t)$, have been obtained by applying Step (1) in Section \ref{sec:temporalfits}. Since we consider the case at the characteristic energy $E_{\rm c}$, one needs to obtain $F_{\nu,c}(t)$ and $E_{\rm c}(t)$. The characteristic energy $E_{\rm c}$ is straightforwardly obtained, and $F_{\nu,c}(t)$ is derived using Eq. \ref{eq:Fnu}. For this step, $N_{\rm c,p}$ is calculated at peak time $t_{\rm p}$ with characteristic energy $E_{\rm c}$ using Eq. \ref{eq:CPL}.
\item Test the model with observed data. Through Step (3), the observed data points are available in the forms of [$F_{\nu,c}(t)$, t], [$E_{\rm c}(t)$, t], and [$F_{\nu,c}(t)$, $E_{\rm c}(t)$]. The [$F_{\nu,c}(t)$, t], [$E_{\rm c}(t)$, t] data points are plotted in the left panel of Figure \ref{fig:CPL}, and the [$F_{\nu,c}(t)$, $E_{\rm c}(t)$] data points are plotted in the right panel of Figure \ref{fig:CPL} for each burst. They are directly compared with the model predictions.
\end{enumerate}

From the left panel in Figure \ref{fig:CPL}, we can see that, except for several apparent cases that violate the predictions (090719063, 090804940, 130305486, 150213001, 150902733, 170921168), all the other data points are generally consistent with the model predictions. The data of some bursts (090620400, 120426090, 150510139) match the constant $\Gamma$ predictions well, suggesting that they are consistent with HLE emission with no significant acceleration. Some other cases (131231198, 141028455, 150314205, 160530667, 180305393) have either $E_{\rm c}(t)$ or $F_{\nu,c}(t)$ below the model prediction lines, consistent with the bulk acceleration in the emission region. For both cases, the [$F_{\nu,c}(t)$, $E_{\rm c}(t)$] test generally satisfies the model prediction (Eq.\ref{eq:FnuEc}) within error. This is consistent with Uhm et al. (2018, preprint) and Tak et al. (2020, in preparation) who first performed such a test and showed that Eq. \ref{eq:FnuEc} is generally valid regardless of bulk Lorentz factor evolution in the emission region. 

It is interesting to note that the three cases (GRB 090719063, GRB 130305486, and GRB 150213001) that are in the Phase I sample but not in the Phase II sample indeed do not satisfy the simple model predictions in the [$F_{\nu,c}(t)$, $E_{\rm c}(t)$] test, supporting that the cases are spurious.

\section{Conclusions and Discussions}\label{sec:conclusion} 

In this paper, we have tested the HLE curvature effect using the prompt-emission data. We selected 24 single-pulse GRBs detected by {\it Fermi} that are ideal for performing such a test. In order to avoid the $t_{0}$ effect and the overlapping effect, we focused on the single-pulse cases. In order to make the physical inferences trustworthy, we only selected the bursts with high statistical significance. In order to determine the temporal peaks ($t_{\rm p}$) of the pulses so that the starting time of the decay phase can be estimated, we employed the FRED model to fit the count-rate lightcurves for our sample. The time window of the entire decay phase is thus determined. Since the curvature effect is more likely to dominate the late-part emission of the decay phase, we are also concerned with such late-time segments. For the most conservative approach, we only selected the time intervals of the last three time bins with $S>15$ to conduct the HLE test. 

We then used two methods to measure the temporal indices and corresponding spectral indices: $\hat{\alpha}_{\rm PL}^{\rm I}$ and $\hat{\beta}_{\rm PL}^{\rm I}$ as derived from the entire decay phase, and $\hat{\alpha}_{\rm PL}^{\rm II}$ and $\hat{\beta}_{\rm PL}^{\rm II}$ as derived from the late-time decay phase. We perform the HLE curvature effect during these two phases. Using the simple power-law spectral analysis, we tested the $\hat{\alpha}_{\rm PL}$-$\hat{\beta}_{\rm PL}$ relation. We found that five out of 24 pulses for Phase I (except for three spurious cases as we discussed in Sec.\ref{sec:result}) and 11 out of 24 pulses for Phase II are consistent with the curvature effect. Some fall into the regime that requires bulk acceleration in the emission region.

We further test these candidate HLE-dominated pulses using a more complicated HLE model \citep{Zhang2009b} invoking a cutoff power-law fits to the time-dependent spectra. We confirm that the HLE effect is still valid for most of the cases, and that some of them indeed showed the evidence of bulk acceleration in the emission region.

Based on the duration of the HLE-dominated emission, we estimated the radius of the emission region from the central engine. For a typical bulk Lorentz factor, the radius $R_{\rm GRB}$ is typically of the order of $10^{15}-10^{16}$ cm, which is much greater than the photosphere radius and the standard internal shock radius. 

The evidence of bulk acceleration and a large emission radius in these bursts is fully consistent with the GRB prompt-emission models invoking direct dissipation of a Poynting flux to power $\gamma$-ray emission \citep[e.g.][]{Zhang2011b}. This suggests that at least for some GRBs, the jet composition is Poynting-flux dominated at the central engine and even in the emission region. This conclusion is consistent with previous independent modeling of GRB spectral lags \citep{Uhm2016} and $E_{\rm p}$ evolution patterns \citep{Uhm2018}, the HLE test for a sample of X-ray flares \citep{Jia2016,Uhm2016a}, and the nondetection of high-energy neutrinos from GRBs \citep{Zhang2013b,Aartsen2017}. Our analysis is also consistent with the recent investigations of Uhm et al. (2018, preprint) and Tak et al. (2020, in preparation).

\acknowledgments

We appreciate the valuable comments from the anonymous referee, and we thank Dr. Yu Wang for useful discussions on $lmfit$. This research made use of the High Energy Astrophysics Science Archive Research Center (HEASARC) Online Service at the NASA/Goddard Space Flight Center (GSFC). 

\facilities{{\it Fermi}/GBM}
\software{3ML \citep{Vianello2015}, matplotlib \citep{Hunter2007}}, $lmfit$ \citep{Newville2016}
\bibliography{Myreferences.bib}

\clearpage
\begin{deluxetable*}{cccccccc}
\tablewidth{0pt}
\tabletypesize{\scriptsize}
\tablecaption{Properties of Prompt Emission of Our Sample}
\tablehead{
\colhead{GRB}
&\colhead{$t_{90}$}
&\colhead{Fluence}
&\colhead{Detectors}
&\colhead{$\Delta T_{\rm src}$}
&\colhead{$[\Delta T_{\rm (bkg,1)},\Delta T_{\rm (bkg,2)}]$}
&\colhead{$N_{\rm tot}$}
&\colhead{$N_{(\rm S\geq 15)}$} 
\\
&(s)&(erg cm$^{-2}$)&&(s)&(s)&(Number)&(Number)
}
\colnumbers
\startdata
081224887&16.448$\pm$1.159&(3.76$\pm$0.02)$\times$10$^{-5}$&(n6)n7n9b1&-1$\sim$20&[-20$\sim$-10,40$\sim$60]&9&5\\
090620400&13.568$\pm$0.724&(1.33$\pm$0.01)$\times$10$^{-5}$&n6(n7)nab1&-1$\sim$30&[-20$\sim$-10,40$\sim$60]&11&5\\
090719063&11.392$\pm$0.896&(4.68$\pm$0.02)$\times$10$^{-5}$&n7(n8)b1&-1$\sim$20&[-20$\sim$-10,40$\sim$60]&13&7\\
090804940&5.568$\pm$0.362&(1.42$\pm$0.02)$\times$10$^{-5}$&n3n4(n5)b0&-1$\sim$15&[-25$\sim$-10,40$\sim$60]&11&6\\
100707032&81.793$\pm$1.218&(8.77$\pm$0.02)$\times$10$^{-5}$&n7(n8)b1&-1$\sim$20&[-50$\sim$-10,80$\sim$100]&16&10\\
110721200&21.822$\pm$0.572&(3.70$\pm$0.01)$\times$10$^{-5}$&(n6)n7n9b1&-1$\sim$25&[-20$\sim$10,40$\sim$60]&10&8\\
110920546&160.771$\pm$5.221&(1.72$\pm$0.01)$\times$10$^{-4}$&(n0)n1n3b0&-1$\sim$160&[-20$\sim$-10,180$\sim$190]&11&8\\
120323507&0.384$\pm$0.036&(1.04$\pm$0.01)$\times$10$^{-5}$&n0(n3)b0&-1$\sim$5&[-20$\sim$-10,10$\sim$20]&12&7\\
120426090&2.688$\pm$0.091&(2.10$\pm$0.01)$\times$10$^{-5}$&(n2)nab1&-1$\sim$10&[-20$\sim$-10,40$\sim$60]&15&7\\
130305486&25.600$\pm$1.557&(4.65$\pm$0.01)$\times$10$^{-5}$&n6(n9)nab1&-1$\sim$35&[50-70]&11&6\\
130614997&9.280$\pm$1.972&(6.72$\pm$0.10)$\times$10$^{-6}$&(n0)n1n3b0&-1$\sim$10&[-25$\sim$-10,20$\sim$45]&8&5\\
131231198&31.232$\pm$0.572&(1.52$\pm$0.01)$\times$10$^{-4}$&n0(n3)n4b0&0.064$\sim$60&[-50$\sim$-10,80$\sim$100]&31&17\\
141028455&31.489$\pm$2.429&(3.48$\pm$0.01)$\times$10$^{-5}$&(n6)n7n9b1&-1$\sim$40&[-30$\sim$-10,50$\sim$100]&15&8\\
150213001&4.096$\pm$0.091&(2.88$\pm$0.01)$\times$10$^{-5}$&n6n7(n8)b1&-1$\sim$10&[-25$\sim$-10,20-40]&23&11\\
150314205&10.688$\pm$0.143&(8.16$\pm$0.01)$\times$10$^{-5}$&n1(n9)b1&-1$\sim$15&[-25$\sim$-10,30$\sim$50]&16&11\\
150510139&51.904$\pm$0.384&(9.86$\pm$0.01)$\times$10$^{-5}$&n0(n1)n5b0&-1$\sim$50&[-25$\sim$-10,100$\sim$130]&22&16\\
150902733&13.568$\pm$0.362&(8.32$\pm$0.01)$\times$10$^{-5}$&(n0)n1n3b0&-1$\sim$25&[-25$\sim$-10,30$\sim$60]&17&9\\
151021791&7.229$\pm$0.602&(1.23$\pm$0.01)$\times$10$^{-5}$&n9(na)b1&-1$\sim$10&[-25$\sim$-10,30$\sim$50]&9&5\\
160216801&7.677$\pm$0.571&(9.90$\pm$0.02)$\times$10$^{-6}$&(n9)nanbb1&-1$\sim$15&[-20$\sim$-10,40$\sim$60]&13&6\\
160530667&9.024$\pm$3.584&(9.19$\pm$0.01)$\times$10$^{-5}$&n1(n2)n5b0&-1$\sim$25&[-40$\sim$-10,40$\sim$100]&21&12\\
170114917&12.032$\pm$1.305&(1.82$\pm$0.01)$\times$10$^{-5}$&n1(n2)nab0&-1$\sim$15&[-20$\sim$10,80$\sim$100]&11&7\\
170921168&39.361$\pm$4.481&(6.56$\pm$0.03)$\times$10$^{-5}$&(n1)n2n5b0&-1$\sim$40&[-20$\sim$-10,40$\sim$60]&8&6\\
171210493&143.107$\pm$2.573&(8.08$\pm$0.01)$\times$10$^{-5}$&n0(n1)n2b0&-1$\sim$100&[-30$\sim$-10,210$\sim$240]&13&9\\
180305393&13.056$\pm$0.810&(5.80$\pm$0.01)$\times$10$^{-5}$&n1(n2)nab0&-1$\sim$20&[-20$\sim$-10,40$\sim$60]&12&5\\
\enddata
\tablecomments{A sample of 23 long GRBs and one short GRB including 24 individual pulses used in this study. Column (1) lists GRB name, Column (2) lists the corresponding duration, Column (3) lists the fluence at 10-1000 keV, Column (4) lists the detectors used, and Columns (5) and (6) list the source and background intervals used in the analysis. Columns (7) and (8) list the number of time bins using the BBlocks method across the source interval, and the number of time bins with statistical significance $S>15$ selected from the decay wing of the pulses. The detector in brackets is the brightest one, used for background and BBlock fits.}
\label{tab:property}
\end{deluxetable*}

\clearpage
\begin{deluxetable*}{cccccccccccc}
\tablewidth{0pt}
\tabletypesize{\scriptsize}
\tablecaption{Results of Lightcurve (Pulses) Fitting of Our Sample with FRED model}\label{tab:FRED}
\tablehead{
\colhead{GRB}
&\colhead{Time Res} 
&\colhead{$t_{\rm start}$$\sim$$t_{\rm stop}$}
&\colhead{$S$} 
&\colhead{$I_{\rm p}$}
&\colhead{$t_{\rm 0}$}
&\colhead{$t_{\rm p}$}
&\colhead{$r$}
&\colhead{$d$}
&\colhead{$\chi^{2}$/dof}
&\colhead{AIC}
&\colhead{BIC}
}
\colnumbers
\startdata
081224887&0.128-s&0$\sim$10&100.96&4413$\pm$59&0&1.04$\pm$0.06&0.18$\pm$0.03&1.10$\pm$0.24&33/73&-1241&-1232\\
090620400&0.128-s&0$\sim$20&46.40&2216$\pm$45&0&3.19$\pm$0.20&0.38$\pm$0.05&1.45$\pm$0.38&324/151&-2144&-2132\\
090719063&0.128-s&0$\sim$25&117.04&4629$\pm$99&0&3.79$\pm$0.16&0.56$\pm$0.06&2.25$\pm$0.43&774/190&-3137&-3124\\
090804940&0.128-s&0$\sim$10&97.93&4245$\pm$84&0&1.88$\pm$0.08&0.56$\pm$0.06&2.27$\pm$0.51&117/73&-1270&-1260\\
100707032&0.256-s&0$\sim$30&138.83&6407$\pm$83&0&1.68$\pm$0.05&0.86$\pm$0.06&0.70$\pm$0.02&66/112&-2118&-2107\\
110721200&0.128-s&0$\sim$10&112.92&3865$\pm$68&0&1.28$\pm$0.07&0.28$\pm$0.03&2.62$\pm$0.86&77/73&-1269&-1260\\
110920546&1.024-s&0$\sim$150&54.53&3172$\pm$16&0&9.95$\pm$0.32&0.28$\pm$0.02&0.28$\pm$0.01&80/141&-2242&-2230\\
120323507&0.032-s&0$\sim$1&177.24&63949$\pm$2469&0&0.04$\pm$0.002&0.52$\pm$0.07&2.40$\pm$0.42&191/26&-710&-704\\
120426090&0.064-s&0$\sim$6&145.48&8927$\pm$182&0&1.04$\pm$0.03&0.87$\pm$0.07&3.65$\pm$0.61&726/89&-1759&-1749\\
130305486&0.128-s&0$\sim$20&54.24&2901$\pm$72&0&4.63$\pm$0.23&0.81$\pm$0.10&1.78$\pm$0.41&684/151&-2233&-2221\\
130614997&0.128-s&0$\sim$10&59.80&3158$\pm$57&0&0.22$\pm$0.09&0.04$\pm$0.02&1.89$\pm$0.73&49/73&-1260&-1251\\
131231198&0.512-s&0$\sim$60&324.86&5324$\pm$169&0&24.76$\pm$0.57&3.34$\pm$0.37&3.17$\pm$0.50&1875/112&-1878&-1867\\
141028455&0.256-s&0$\sim$50&68.31&2085$\pm$45&0&11.57$\pm$0.57&0.77$\pm$0.09&1.46$\pm$0.30&784/190&-2613&-2600\\
150213001&0.064-s&0$\sim$6&295.19&17545$\pm$570&0&2.08$\pm$0.05&1.93$\pm$0.19&10.00$\pm$3.76&1692/89&-1805&-1795\\
150314205&0.128-s&0$\sim$20&177.73&7426$\pm$133&0&1.85$\pm$0.06&0.72$\pm$0.06&1.41$\pm$0.10&386/151&-2813&-2801\\
150510139&0.256-s&0$\sim$50&96.98&5796$\pm$242&0&0.08$\pm$0.01&0.57$\pm$0.15&0.26$\pm$0.01&296/190&-2904&-2891\\
150902733&0.128-s&0$\sim$25&137.63&4538$\pm$121&0&8.44$\pm$0.23&1.67$\pm$0.16&3.72$\pm$0.80&1794/190&-3069&-3056\\
151021791&0.128-s&0$\sim$10&63.15&3672$\pm$83&0&0.80$\pm$0.05&0.51$\pm$0.07&0.82$\pm$0.07&96/73&-1242&-1233\\
160216801&0.128-s&0$\sim$15&98.56&4676$\pm$139&0&3.97$\pm$0.14&1.37$\pm$0.15&3.05$\pm$0.63&1064/112&-1865&-1854\\
160530667&0.128-s&0$\sim$20&228.04&12390$\pm$148&0&5.93$\pm$0.04&3.83$\pm$0.15&3.01$\pm$0.12&1671/151&-3119&-3107\\
170114917&0.128-s&0$\sim$10&76.96&3269$\pm$100&0&2.05$\pm$0.14&0.75$\pm$0.13&1.33$\pm$0.33&261/73&-1131&-1122\\
170921168&0.256-s&0$\sim$50&68.47&2975$\pm$41&0&4.35$\pm$0.25&0.21$\pm$0.03&1.11$\pm$0.17&241/190&-2929&-2916\\
171210493&0.512-s&0$\sim$100&93.34&2798$\pm$24&0&5.24$\pm$0.17&0.61$\pm$0.04&0.36$\pm$0.01&58/190&-2973&-2960\\
180305393&0.128-s&0$\sim$20&95.60&3941$\pm$82&0&4.65$\pm$0.18&0.84$\pm$0.09&2.04$\pm$0.39&647/151&-2395&-2383\\
\enddata
\tablecomments{Column (1) lists GRB name; Column (2) lists the time resolution used (Time Res) of the count-rate lightcurve of each burst; Column (3) lists the start and stop times of the pulses, in units of s; Column (4) lists the significance $S$ of the entire pulse; Columns (5)-(9) list the best-fit parameters for the FRED model: normalization $I_{\rm p}$, the zero time $t_{\rm 0}$, and the peak time $t_{\rm p}$ of pulses, and the rise $r$ and decay $d$ time scale parameters; Column (10) lists the reduced chi-squared $\chi^{2}$/dof; Column (11) lists the AIC statistic; Column (12) lists the BIC statistic.} 
\end{deluxetable*}

\clearpage
\begin{longrotatetable}
\begin{deluxetable*}{cccc|ccc|ccc|ccccc}
\centering
\rotate
\setlength{\tabcolsep}{0.15em}
\tabletypesize{\scriptsize}
\tablecaption{Results of Lightcurve and Spectral Fitting of the Decaying Wing of the Pulses}\label{tab:relation}
\tablehead{
\specialrule{0em}{3pt}{3pt}
\multirow{3}{*}{GRB}
&\multirow{3}{*}{$t_{\rm start}$$\sim$$t_{\rm stop}$} 
&\multirow{3}{*}{$S$} 
&\multirow{3}{*}{$t_{0}$} 
&\multicolumn{3}{c}{Lightcurve Power-law Fitting}
&\multicolumn{3}{c}{Spectral Power-law Fitting}
&\multicolumn{5}{c}{Spectral Cutoff Power-law Fitting}\\
\cmidrule(lr){5-7} \cmidrule(lr){8-10}  \cmidrule(lr){11-15}   
&&&
&\colhead{$F_{\rm t,0}$}
&\colhead{$\hat{\alpha}$}
&\colhead{AIC/BIC}
&\colhead{$F_{\rm \nu,0}$}
&\colhead{$\hat{\beta}$}
&\colhead{DIC/$p_{\rm DIC}$}
&\colhead{$N_{\rm 0,p}$}
&\colhead{$t_{\rm p}$}
&\colhead{$\hat{\Gamma}$}
&\colhead{$E_{\rm c}$}
&\colhead{DIC/$p_{\rm DIC}$}\\
\hline
}
\colnumbers
\startdata
081224887(1)&1.896$\sim$12.502&88.10&0&(2.62$\pm$0.06)$\times$10$^{-6}$&1.81$\pm$0.06&-159/-160&(1.62$^{+0.04}_{-0.04}$)$\times$10$^{1}$&0.43$^{+0.00}_{-0.00}$&8255/1.98&\nodata&\nodata&\nodata&\nodata&\nodata\\
081224887(2)&5.424$\sim$12.502&47.56&0&(8.16$\pm$0.01)$\times$10$^{-7}$&2.25$\pm$0.00&-121/-123&(1.43$^{+0.06}_{-0.06}$)$\times$10$^{1}$&0.52$^{+0.01}_{-0.01}$&5533/2.00&\nodata&\nodata&\nodata&\nodata&\nodata\\
090620400(1)&4.076$\sim$12.289&47.80&0&(2.46$\pm$0.05)$\times$10$^{-6}$&3.02$\pm$0.15&-161/-162&(1.27$^{+0.05}_{-0.05}$)$\times$10$^{1}$&0.48$^{+0.01}_{-0.01}$&6135/2.02&(5.43$\pm$0.34)$\times$10$^{-2}$&4.076&0.51$\pm$0.04&117.3$\pm$6.8&5116/2.83\\
090620400(2)&5.319$\sim$12.289&35.56&0&(4.91$\pm$0.34)$\times$10$^{-7}$&2.78$\pm$0.24&-101/-103&(1.34$^{+0.07}_{-0.07}$)$\times$10$^{1}$&0.55$^{+0.01}_{-0.01}$&5301/1.99&(5.25$\pm$0.53)$\times$10$^{-2}$&5.319&0.55$\pm$0.06&87.9$\pm$6.8&4827/2.57\\
090719063(1)&4.443$\sim$14.562&128.19&0&(3.68$\pm$0.17)$\times$10$^{-6}$&3.21$\pm$0.23&-207/-207&(3.73$^{+0.08}_{-0.08}$)$\times$10$^{1}$&0.53$^{+0.00}_{-0.00}$&6860/1.98&(8.73$\pm$0.25)$\times$10$^{-2}$&4.443&0.79$\pm$0.02&182.8$\pm$6.9&4292/2.95\\
090719063(2)&7.810$\sim$14.562&66.34&0&(8.21$\pm$0.34)$\times$10$^{-7}$&2.39$\pm$0.19&-132/-133&(5.05$^{+0.24}_{-0.24}$)$\times$10$^{1}$&0.77$^{+0.01}_{-0.01}$&4085/1.98&\nodata&\nodata&\nodata&\nodata&\nodata\\
090804940(1)&1.279$\sim$8.705&98.14&0&(1.09$\pm$0.14)$\times$10$^{-6}$&1.10$\pm$0.16&-179/-179&(7.29$^{+0.19}_{-0.20}$)$\times$10$^{1}$&0.73$^{+0.01}_{-0.01}$&7458/1.99&\nodata&\nodata&\nodata&\nodata&\nodata\\
090804940(2)&4.678$\sim$8.705&40.50&0&(5.12$\pm$0.58)$\times$10$^{-7}$&3.09$\pm$0.53&-97/-99&(7.21$^{+0.48}_{-0.48}$)$\times$10$^{1}$&0.92$^{+0.02}_{-0.02}$&4486/1.98&(54.40$\pm$2.66)$\times$10$^{-2}$&4.678&0.72$\pm$0.09&49.4$\pm$4.0&4167/2.73\\
100707032(1)&1.631$\sim$28.780&131.02&0&(4.27$\pm$0.05)$\times$10$^{-6}$&1.57$\pm$0.01&-320/-319&(2.42$^{+0.04}_{-0.04}$)$\times$10$^{1}$&0.49$^{+0.00}_{-0.00}$&9384/1.98&\nodata&\nodata&\nodata&\nodata&\nodata\\
100707032(2)&14.210$\sim$28.78&47.30&0&(4.01$\pm$0.17)$\times$10$^{-7}$&1.90$\pm$0.15&-105/-107&(3.23$^{+0.23}_{-0.23}$)$\times$10$^{1}$&0.82$^{+0.02}_{-0.02}$&4646/1.97&\nodata&\nodata&\nodata&\nodata&\nodata\\
110721200(1)&0.470$\sim$25.000&76.49&0&(2.72$\pm$0.09)$\times$10$^{-6}$&1.46$\pm$0.03&-243/-242&(1.06$^{+0.02}_{-0.02}$)$\times$10$^{1}$&0.44$^{+0.00}_{-0.00}$&7613/1.99&\nodata&\nodata&\nodata&\nodata&\nodata\\
110721200(2)&6.252$\sim$25.000&28.610&0&(4.07$\pm$0.05)$\times$10$^{-7}$&1.70$\pm$0.02&-112/-114&5.51$^{+0.34}_{-0.34}$&0.51$^{+0.01}_{-0.01}$&6119/1.99&\nodata&\nodata&\nodata&\nodata&\nodata\\
110920546(1)&9.966$\sim$122.091&59.68&0&(2.75$\pm$0.12)$\times$10$^{-6}$&1.03$\pm$0.08&-241/-241&8.71$^{+0.20}_{-0.20}$&0.42$^{+0.00}_{-0.00}$&12386/2.00&\nodata&\nodata&\nodata&\nodata&\nodata\\
110920546(2)&55.534$\sim$122.091&34.09&0&(3.39$\pm$0.04)$\times$10$^{-7}$&1.83$\pm$0.04&-113/-115&8.00$^{+0.37}_{-0.37}$&0.53$^{+0.01}_{-0.01}$&9532/1.99&\nodata&\nodata&\nodata&\nodata&\nodata\\
120323507(1)&0.094$\sim$0.581&130.70&0&(15.32$\pm$1.63)$\times$10$^{-6}$&2.72$\pm$0.19&-176/-177&(8.57$^{+0.33}_{-0.33}$)$\times$10$^{2}$&0.90$^{+0.01}_{-0.01}$&1643/2.00&\nodata&\nodata&\nodata&\nodata&\nodata\\
120323507(2)&0.252$\sim$0.581&66.33&0&(5.31$\pm$0.57)$\times$10$^{-6}$&1.46$\pm$0.50&-83/-85&(6.13$^{+0.13}_{-0.13}$)$\times$10$^{2}$&1.00$^{+0.00}_{-0.00}$&1161/1.00&\nodata&\nodata&\nodata&\nodata&\nodata\\
120426090(1)&1.044$\sim$4.882&125.87&0&(4.94$\pm$0.53)$\times$10$^{-6}$&1.84$\pm$0.28&-190/-190&(1.41$^{+0.04}_{-0.04}$)$\times$10$^{2}$&0.70$^{+0.01}_{-0.01}$&5338/2.04&\nodata&\nodata&\nodata&\nodata&\nodata\\
120426090(2)&2.600$\sim$4.882&35.11&0&(6.14$\pm$1.14)$\times$10$^{-7}$&3.67$\pm$0.66&-94/-96&(1.25$^{+0.05}_{-0.05}$)$\times$10$^{2}$&0.99$^{+0.01}_{-0.01}$&2529/1.14&(9.19$\pm$2.13)$\times$10$^{-2}$&2.600&1.06$\pm$0.12&52.9$\pm$7.4&2408/-0.66\\
130305486(1)&4.632$\sim$32.212&36.88&0&(3.12$\pm$0.26)$\times$10$^{-6}$&2.33$\pm$0.18&-173/-173&4.02$^{+0.12}_{-0.12}$&0.30$^{+0.01}_{-0.01}$&8462/1.99&(1.53$\pm$0.04)$\times$10$^{-2}$&4.632&0.67$\pm$0.03&545.7$\pm$44.0&7110/2.92\\
130305486(2)&8.849$\sim$32.212&16.99&0&(9.97$\pm$0.87)$\times$10$^{-7}$&1.24$\pm$0.15&-96/-98&2.20$^{+0.15}_{-0.15}$&0.33$^{+0.01}_{-0.01}$&6953/2.00&\nodata&\nodata&\nodata&\nodata&\nodata\\
130614997(1)&0.457$\sim$6.210&64.91&0&(0.90$\pm$0.10)$\times$10$^{-6}$&0.76$\pm$0.16&-123/-124&(7.23$^{+0.32}_{-0.32}$)$\times$10$^{1}$&0.85$^{+0.01}_{-0.01}$&4828/2.00&\nodata&\nodata&\nodata&\nodata&\nodata\\
130614997(2)&2.030$\sim$6.210&44.82&0&(6.37$\pm$0.51)$\times$10$^{-7}$&1.36$\pm$0.21&-98/-100&(6.41$^{+0.40}_{-0.40}$)$\times$10$^{1}$&0.89$^{+0.02}_{-0.02}$&4217/1.97&\nodata&\nodata&\nodata&\nodata&\nodata\\
131231198(1)&22.406$\sim$59.114&298.10&0&(1.43$\pm$0.20)$\times$10$^{-6}$&4.01$\pm$0.29&-555/-553&(1.22$^{+0.01}_{-0.01}$)$\times$10$^{2}$&0.75$^{+0.00}_{-0.00}$&12654/2.01&(7.56$\pm$0.11)$\times$10$^{-2}$&22.406&1.34$\pm$0.01&239.8$\pm$6.1&8716/2.97\\
131231198(2)&47.97$\sim$59.114&31.39&0&(2.87$\pm$0.60)$\times$10$^{-7}$&9.00$\pm$2.64&-98/-100&(4.10$^{+0.13}_{-0.13}$)$\times$10$^{1}$&0.99$^{+0.01}_{-0.01}$&5371/1.08&(0.98$\pm$0.18)$\times$10$^{-2}$&47.970&1.62$\pm$0.09&132.7$\pm$30.7&5314/0.75\\
141028455(1)&11.565$\sim$40.000&69.79&0&(2.96$\pm$0.24)$\times$10$^{-6}$&3.03$\pm$0.52&-258/-257&8.35$^{+0.23}_{-0.23}$&0.46$^{+0.01}_{-0.01}$&7260/2.02&(1.73$\pm$0.05)$\times$10$^{-2}$&11.565&1.00$\pm$0.02&364.8$\pm$25.8&6401/2.92\\
141028455(2)&22.335$\sim$40.000&21.28&0&(2.27$\pm$0.04)$\times$10$^{-7}$&3.38$\pm$0.07&-114/-116&4.17$^{+0.37}_{-0.38}$&0.54$^{+0.02}_{-0.02}$&5685/1.96&(0.68$\pm$0.08)$\times$10$^{-2}$&22.335&1.08$\pm$0.08&269.6$\pm$60.7&5620/1.80\\
150213001(1)&2.227$\sim$6.661&198.47&0&(2.56$\pm$0.05)$\times$10$^{-6}$&3.86$\pm$0.05&-342/-341&(4.17$^{+0.08}_{-0.08}$)$\times$10$^{2}$&0.93$^{+0.01}_{-0.01}$&6143/2.04&(17.50$\pm$0.59)$\times$10$^{-2}$&2.227&1.33$\pm$0.02&114.0$\pm$4.0&4426/2.94\\
150213001(2)&4.085$\sim$6.661&49.81&0&(8.86$\pm$0.50)$\times$10$^{-7}$&2.95$\pm$0.30&-130/-131&(1.29$^{+0.03}_{-0.03}$)$\times$10$^{2}$&1.00$^{+0.00}_{-0.00}$&3350/1.03&\nodata&\nodata&\nodata&\nodata&\nodata\\
150314205(1)&1.846$\sim$14.999&176.97&0&(8.70$\pm$0.73)$\times$10$^{-6}$&1.09$\pm$0.15&-287/-287&(3.93$^{+0.06}_{-0.06}$)$\times$10$^{1}$&0.46$^{+0.00}_{-0.00}$&11811/2.01&\nodata&\nodata&\nodata&\nodata&\nodata\\
150314205(2)&7.847$\sim$14.999&71.42&0&(1.86$\pm$0.51)$\times$10$^{-6}$&4.86$\pm$1.58&-111/-112&(1.95$^{+0.07}_{-0.07}$)$\times$10$^{1}$&0.48$^{+0.01}_{-0.01}$&4482/1.98&(3.44$\pm$0.11)$\times$10$^{-2}$&7.847&1.03$\pm$0.02&499.7$\pm$44.1&3734/2.94\\
150510139(1)&0.889$\sim$49.997&90.65&0&(8.14$\pm$1.96)$\times$10$^{-6}$&0.77$\pm$0.18&-390/-389&9.19$^{+0.17}_{-0.17}$&0.40$^{+0.00}_{-0.00}$&10870/1.99&\nodata&\nodata&\nodata&\nodata&\nodata\\
150510139(2)&28.736$\sim$49.997&34.51&0&(5.90$\pm$0.94)$\times$10$^{-7}$&3.52$\pm$0.75&-95/-97&7.31$^{+0.43}_{-0.43}$&0.54$^{+0.01}_{-0.01}$&6605/1.99&(0.90$\pm$0.06)$\times$10$^{-2}$&28.736&1.21$\pm$0.04&575.0$\pm$122.1&6447/2.29\\
150902733(1)&8.934$\sim$25.000&112.07&0&(2.51$\pm$0.71)$\times$10$^{-6}$&4.28$\pm$0.64&-260/-260&(1.65$^{+0.03}_{-0.03}$)$\times$10$^{1}$&0.40$^{+0.00}_{-0.00}$&11345/2.01&(4.68$\pm$0.07)$\times$10$^{-2}$&8.934&0.77$\pm$0.01&375.8$\pm$13.5&6656/2.97\\
150902733(2)&14.609$\sim$25.000&32.82&0&(3.49$\pm$0.64)$\times$10$^{-7}$&5.62$\pm$1.03&-97/-99&9.79$^{+0.61}_{-0.61}$&0.56$^{+0.01}_{-0.01}$&5601/2.01&(2.05$\pm$0.20)$\times$10$^{-2}$&14.609&0.87$\pm$0.07&161.8$\pm$19.8&5366/2.51\\
151021791(1)&0.797$\sim$7.923&62.48&0&(8.56$\pm$1.15)$\times$10$^{-7}$&1.52$\pm$0.12&-151/-152&(1.62$^{+0.06}_{-0.06}$)$\times$10$^{1}$&0.50$^{+0.01}_{-0.01}$&4361/1.99&\nodata&\nodata&\nodata&\nodata&\nodata\\
151021791(2)&2.286$\sim$7.923&36.55&0&(4.81$\pm$0.39)$\times$10$^{-7}$&1.65$\pm$0.14&-100/-102&(1.47$^{+0.10}_{-0.10}$)$\times$10$^{1}$&0.60$^{+0.02}_{-0.02}$&3488/2.00&\nodata&\nodata&\nodata&\nodata&\nodata\\
160216801(1)&5.031$\sim$14.999&53.76&0&(1.18$\pm$0.21)$\times$10$^{-6}$&2.35$\pm$0.33&-175/-175&(8.11$^{+0.12}_{-0.12}$)$\times$10$^{1}$&1.00$^{+0.00}_{-0.00}$&6954/1.00&\nodata&\nodata&\nodata&\nodata&\nodata\\
160216801(2)&6.876$\sim$14.999&21.46&0&(7.69$\pm$0.27)$\times$10$^{-7}$&2.65$\pm$0.10&-103/-104&(3.40$^{+0.12}_{-0.12}$)$\times$10$^{1}$&1.00$^{+0.00}_{-0.00}$&5288/1.02&\nodata&\nodata&\nodata&\nodata&\nodata\\
160530667(1)&6.661$\sim$20.442&168.89&0&(4.36$\pm$0.31)$\times$10$^{-6}$&3.70$\pm$0.28&-336/-335&(6.19$^{+0.09}_{-0.09}$)$\times$10$^{1}$&0.58$^{+0.00}_{-0.00}$&13223/2.00&(14.30$\pm$0.33)$\times$10$^{-2}$&6.661&0.77$\pm$0.01&130.3$\pm$3.1&6883/2.99\\
160530667(2)&12.961$\sim$20.442&37.55&0&(3.15$\pm$0.24)$\times$10$^{-7}$&5.48$\pm$0.43&-136/-137&(3.03$^{+0.22}_{-0.21}$)$\times$10$^{1}$&0.81$^{+0.02}_{-0.02}$&5215/2.00&(4.44$\pm$0.68)$\times$10$^{-2}$&12.961&0.86$\pm$0.08&66.6$\pm$6.9&5001/1.85\\
170114917(1)&2.047$\sim$14.999&57.52&0&(1.01$\pm$0.08)$\times$10$^{-6}$&1.75$\pm$0.09&-217/-217&(1.28$^{+0.05}_{-0.05}$)$\times$10$^{1}$&0.52$^{+0.01}_{-0.01}$&6211/2.02&\nodata&\nodata&\nodata&\nodata&\nodata\\
170114917(2)&4.702$\sim$14.999&30.73&0&(2.94$\pm$0.12)$\times$10$^{-7}$&2.09$\pm$0.09&-107/-109&(1.01$^{+0.09}_{-0.09}$)$\times$10$^{1}$&0.63$^{+0.02}_{-0.02}$&5404/2.01&\nodata&\nodata&\nodata&\nodata&\nodata\\
170921168(1)&4.353$\sim$25.654&92.75&0&(2.03$\pm$0.13)$\times$10$^{-6}$&0.92$\pm$0.15&-150/-150&(2.57$^{+0.06}_{-0.06}$)$\times$10$^{-2}$&0.97$^{+0.01}_{-0.01}$&8470/1.97&\nodata&\nodata&\nodata&\nodata&\nodata\\
170921168(2)&15.707$\sim$25.654&43.76&0&(1.38$\pm$0.03)$\times$10$^{-6}$&2.40$\pm$0.14&-101/-103&(1.75$^{+0.02}_{-0.02}$)$\times$10$^{-2}$&1.00$^{+0.00}_{-0.00}$&6527/0.99&(3.81$\pm$0.43)$\times$10$^{-2}$&15.707&1.73$\pm$0.05&105.0$\pm$13.1&6147/2.05\\
171210493(1)&5.237$\sim$137.109&74.10&0&(2.28$\pm$0.02)$\times$10$^{-6}$&1.20$\pm$0.02&-310/-309&(1.11$^{+0.03}_{-0.03}$)$\times$10$^{1}$&0.59$^{+0.01}_{-0.01}$&10380/2.03&\nodata&\nodata&\nodata&\nodata&\nodata\\
171210493(2)&64.334$\sim$137.109&27.51&0&(1.32$\pm$0.04)$\times$10$^{-7}$&1.65$\pm$0.11&-113/-114&9.59$^{+0.73}_{-0.72}$&0.78$^{+0.02}_{-0.02}$&8145/2.02&\nodata&\nodata&\nodata&\nodata&\nodata\\
180305393(1)&3.449$\sim$16.537&101.51&0&(2.72$\pm$0.86)$\times$10$^{-6}$&1.69$\pm$0.40&-181/-181&(1.60$^{+0.03}_{-0.03}$)$\times$10$^{1}$&0.36$^{+0.00}_{-0.00}$&11378/1.98&\nodata&\nodata&\nodata&\nodata&\nodata\\
180305393(2)&8.933$\sim$16.537&35.63&0&(4.78$\pm$0.33)$\times$10$^{-7}$&3.72$\pm$0.24&-101/-103&(1.27$^{+0.07}_{-0.07}$)$\times$10$^{1}$&0.53$^{+0.01}_{-0.01}$&5710/2.00&(6.05$\pm$0.73)$\times$10$^{-2}$&8.933&0.47$\pm$0.07&80.5$\pm$7.0&5312/2.31\\
\hline
\enddata
\tablecomments{Column (1) lists the GRB name; Column (2) lists the start and stop times of the decay phases (in units of s); Column (3) lists the statistical significance $S$; Column (4) lists the model parameter $t_{0}$ as described in Equations.(\ref{eq:PL_Alpha}), Eq.(\ref{eq:Ec}), and Eq.(\ref{eq:Fnu}), which we fixed it to zero; Columns (5)-(7) list the best-fit parameters for the power-law model in Eq.(\ref{eq:PL_Alpha}): the normalization $F_{\rm t,0}$ (in units of erg cm$^{-2}$ s$^{-1}$), the power-law index $\hat{\alpha}$, and the AIC and BIC statistics; Columns (8)-(10) list the best-fit parameters for the power-law model as shown in Eq.(\ref{eq:PL_Beta}): normalization $F_{\rm \nu,0}$ (in units of phs cm$^{-2}$ s$^{-1}$ keV$^{-1}$), the power-law index $\hat{\beta}$, and the DIC and $p_{\rm DIC}$ statistics; Columns (11)-(15) list the best-fit parameters for the cutoff power-law model as presented in Eqs.(\ref{eq:CPL})-(\ref{eq:Fnu}): normalization $N_{\rm 0,p}$ (in units of phs cm$^{-2}$ s$^{-1}$ keV$^{-1}$), parameter $t_{\rm p}$, which we fixed it at the beginning of the decay phases, the cutoff power-law index $\Gamma$, and the $\hat{\beta}$ index derived from $\Gamma$, and the AIC and BIC statistics. Note that (1) marks the entire decay phase of the pulses and (2) marks the later-part decay phase of the pulses. Note that we did not apply the chi-squared test for our sample, because the sample size in our selected bursts is not large enough; the chi-squared test usually requires a relatively large sample size.}
\end{deluxetable*}
\end{longrotatetable}

\clearpage
\begin{deluxetable*}{ccc|cc|cc}
\tablewidth{0pt}
\tabletypesize{\scriptsize}
\tablecaption{Estimation of GRB Emission Radius Using High-latitude Emission}
\tablehead{
\colhead{GRB}
&\colhead{$\Gamma_{2}$}
&\colhead{$z$}
&\colhead{$t^{\rm I}_{\rm HLE}$}
&\colhead{$R^{\rm I}_{\rm GRB}$}
&\colhead{$t^{\rm II}_{\rm HLE}$}
&\colhead{$R^{\rm II}_{\rm GRB}$}
\\
&(used)&(used)&(s)&(cm)&(s)&(cm)
}
\colnumbers
\startdata
090620400&1.0&1.0&4.11&1.2$\times$10$^{15}$&3.48&1.0$\times$10$^{15}$\\
090719063&1.0&1.0&5.06&1.5$\times$10$^{15}$&\nodata&\nodata\\
090804940&1.0&1.0&\nodata&\nodata&2.01&0.6$\times$10$^{15}$\\
120426090&1.0&1.0&\nodata&\nodata&1.14&0.3$\times$10$^{15}$\\
130305486&1.0&1.0&13.79&4.1$\times$10$^{15}$&\nodata&\nodata\\
131231198&1.0&0.642&22.36&6.7$\times$10$^{15}$&6.79&2.0$\times$10$^{15}$\\
141028455&1.0&2.33&8.54&2.6$\times$10$^{15}$&5.30&1.6$\times$10$^{15}$\\
150213001&1.0&1.0&2.22&0.7$\times$10$^{15}$&\nodata&\nodata\\
150314205&1.0&1.758&\nodata&\nodata&2.59&0.8$\times$10$^{15}$\\
150510139&1.0&1.0&\nodata&\nodata&10.63&3.2$\times$10$^{15}$\\
150902733&1.0&1.0&8.03&2.4$\times$10$^{15}$&5.20&1.6$\times$10$^{15}$\\
160530667&1.0&1.0&6.89&2.1$\times$10$^{15}$&3.74&1.1$\times$10$^{15}$\\
170921168&1.0&1.0&\nodata&\nodata&4.97&1.5$\times$10$^{15}$\\
180305393&1.0&1.0&\nodata&\nodata&3.80&1.1$\times$10$^{15}$\\
\enddata
\tablecomments{Column (1) lists the GRB name; Column (2) lists the $\Gamma$ values used, where we adopted a typical value ($\Gamma_{2}$=1) for all the cases. Column (3) lists the redshift used, a majority of bursts in our sample have no redshift observations, so we adopt a typical value ($z$=1) instead. Column (4) lists the duration of the HLE in the source frame for ``Phase I'', which is calculated using the observed HLE duration divided by (1+z). Column (5) lists the GRB emission radius $R_{\rm GRB}$ for ``Phase I'', derived using Eq. \ref{eq:RGRB}. Again, Column (6) lists the duration of the HLE in the source frame for ``Phase II'', and Column (5) lists the GRB emission radius $R_{\rm GRB}$ for ``Phase II''.}
\label{tab:RGRB}
\end{deluxetable*}

\clearpage
\begin{figure*}
\includegraphics[width=0.5\hsize,clip]{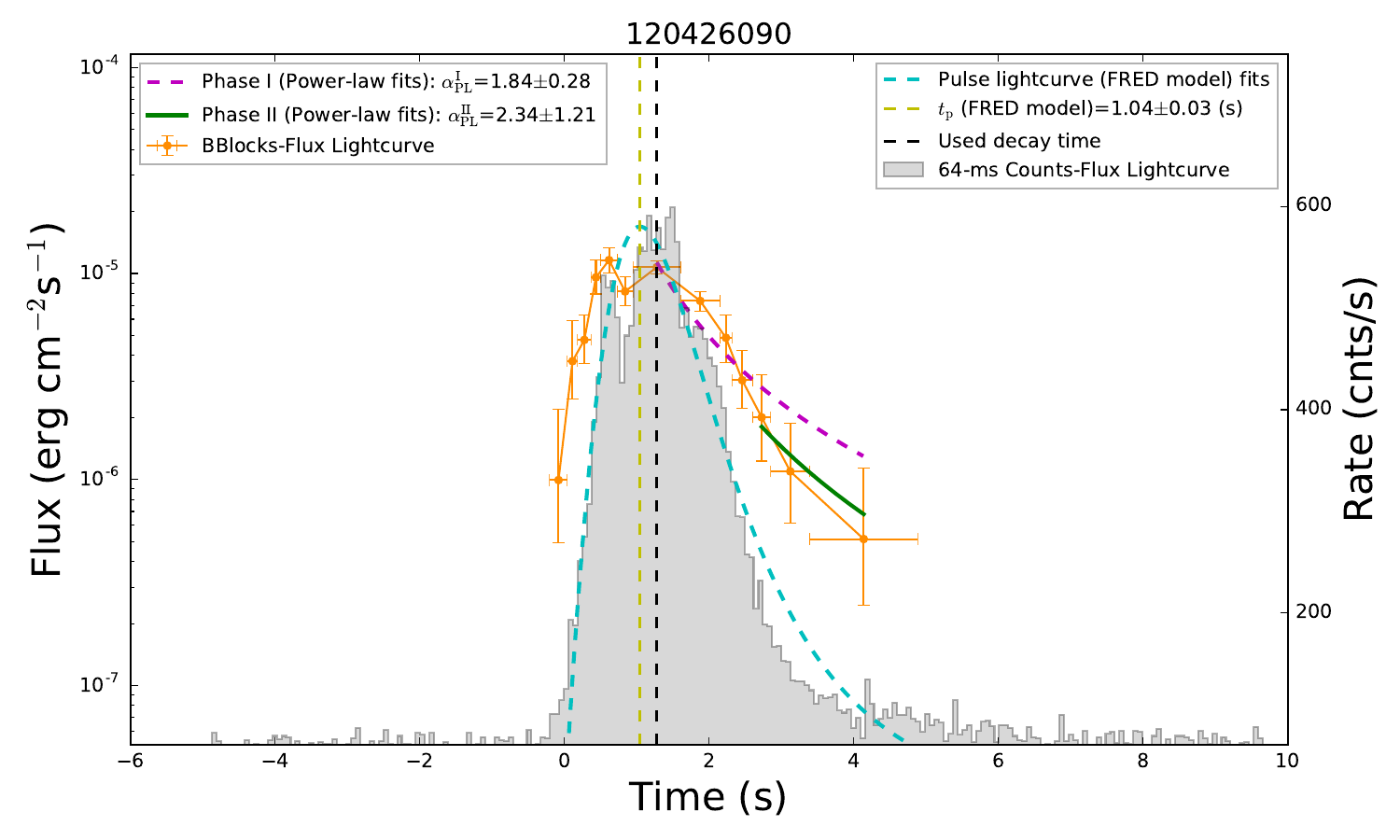}
\includegraphics[width=0.5\hsize,clip]{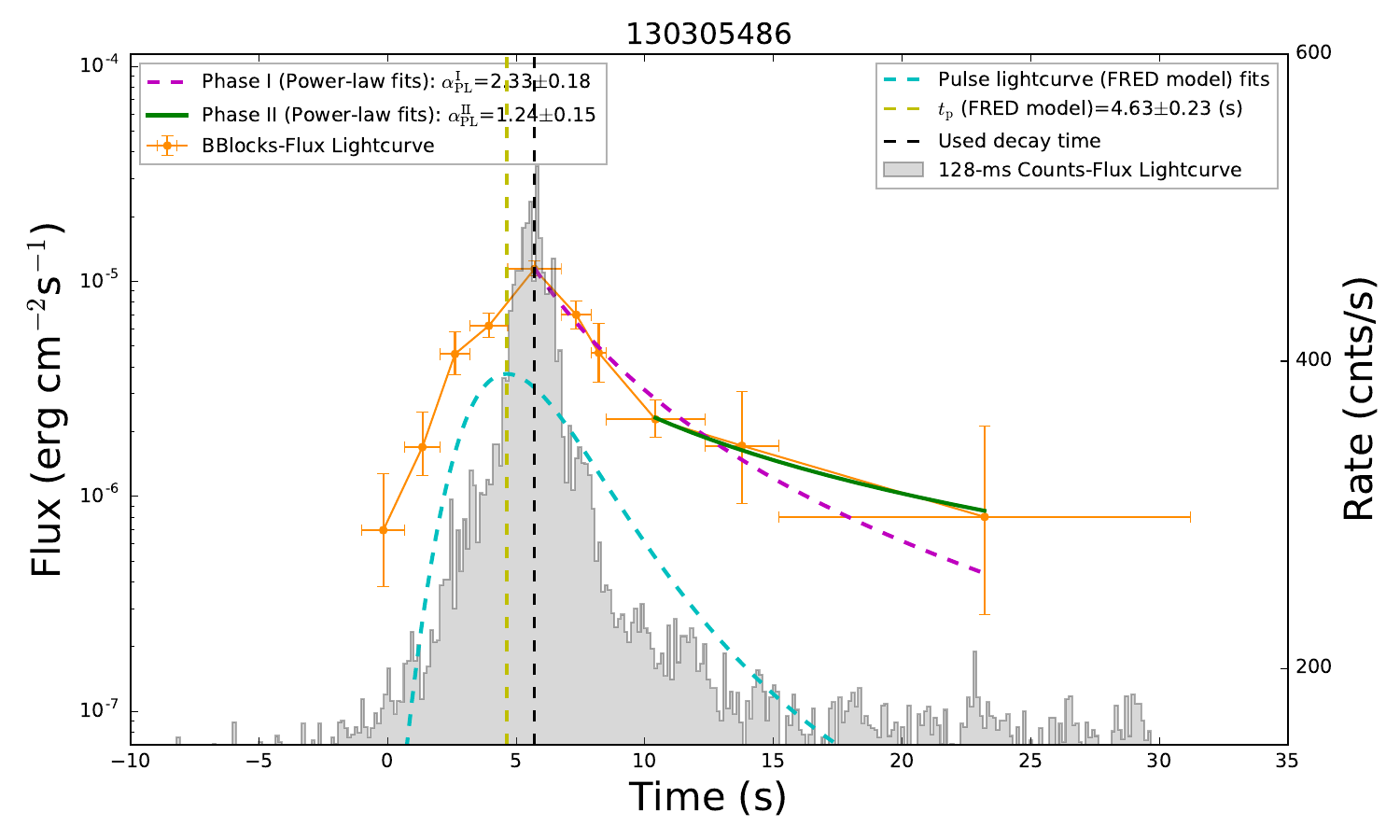}
\includegraphics[width=0.5\hsize,clip]{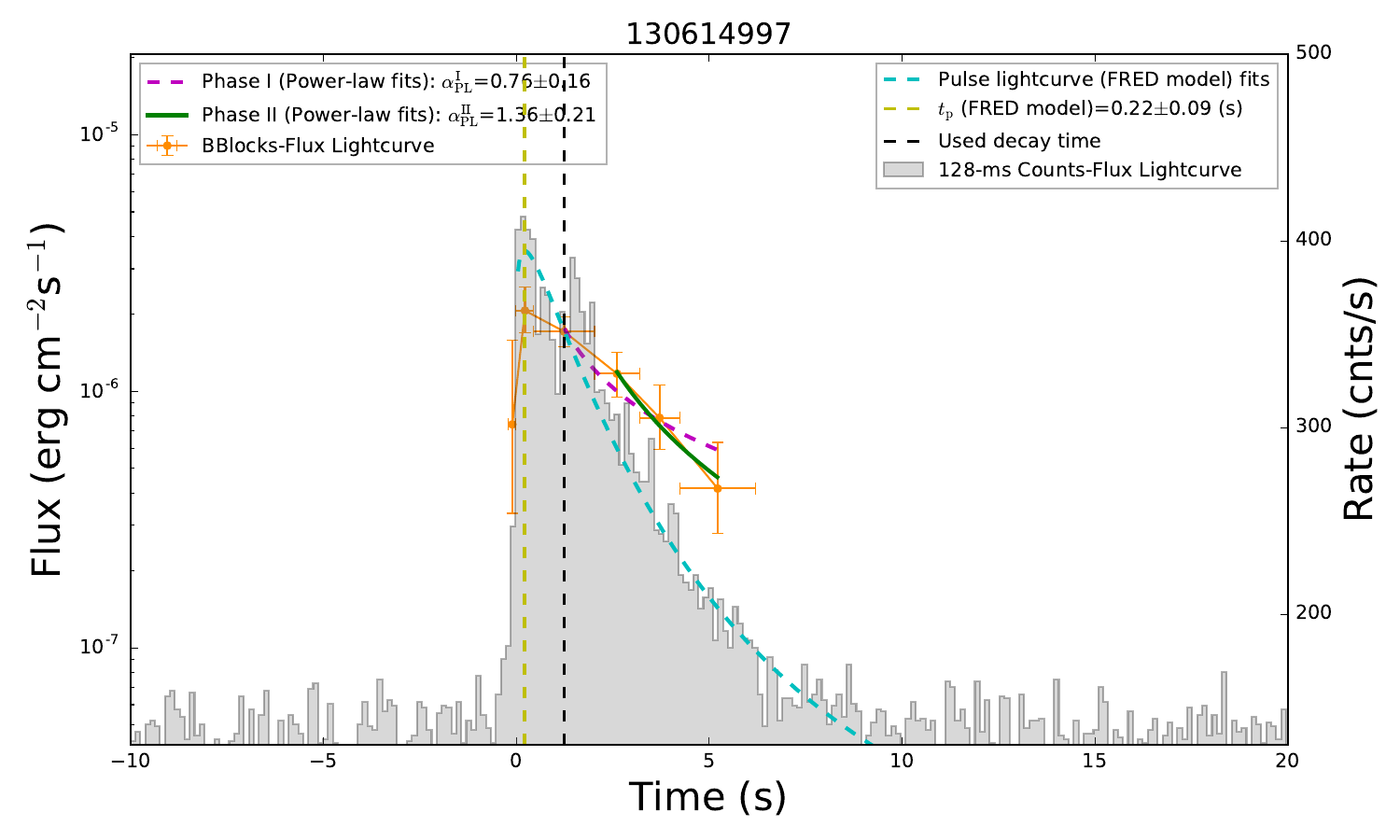}
\includegraphics[width=0.5\hsize,clip]{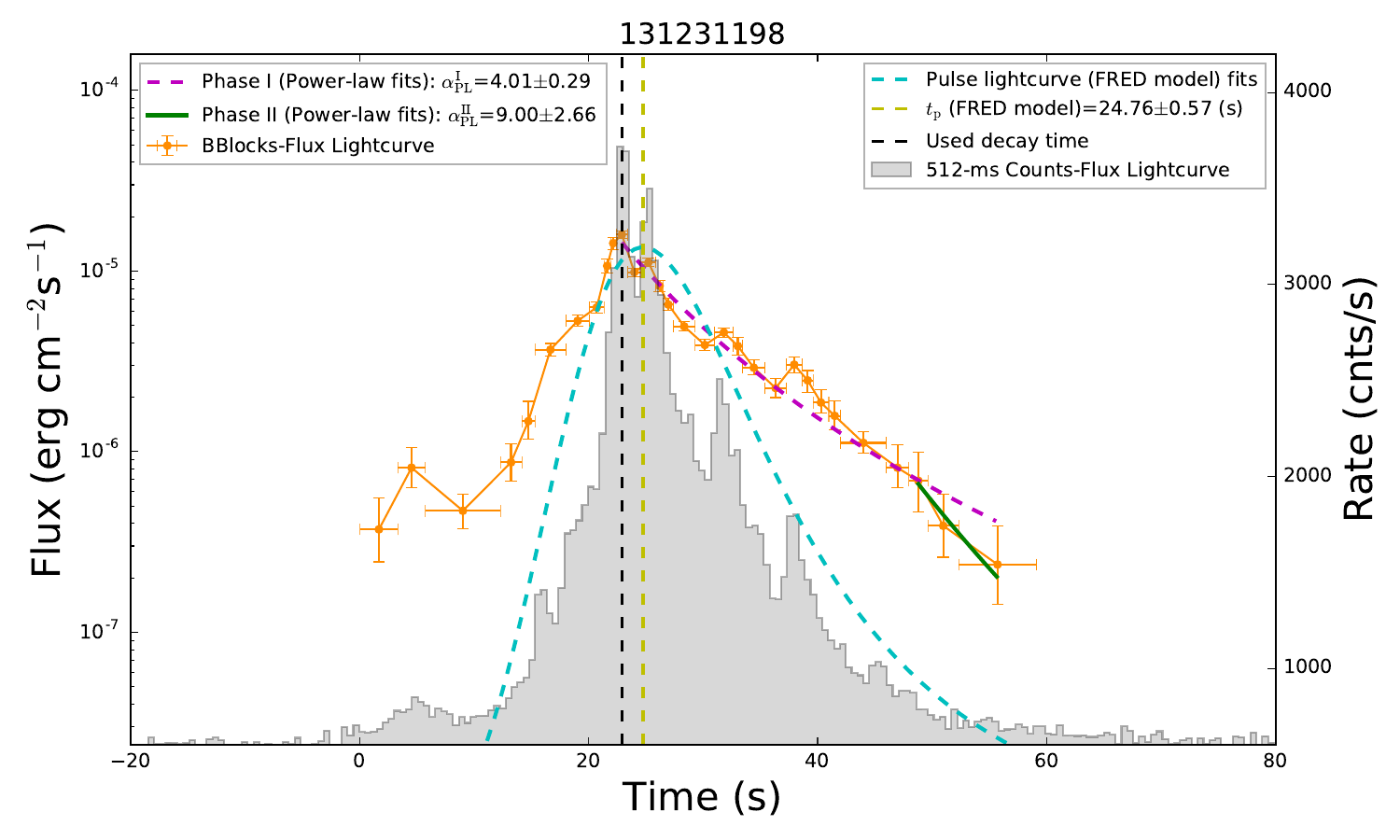}
\includegraphics[width=0.5\hsize,clip]{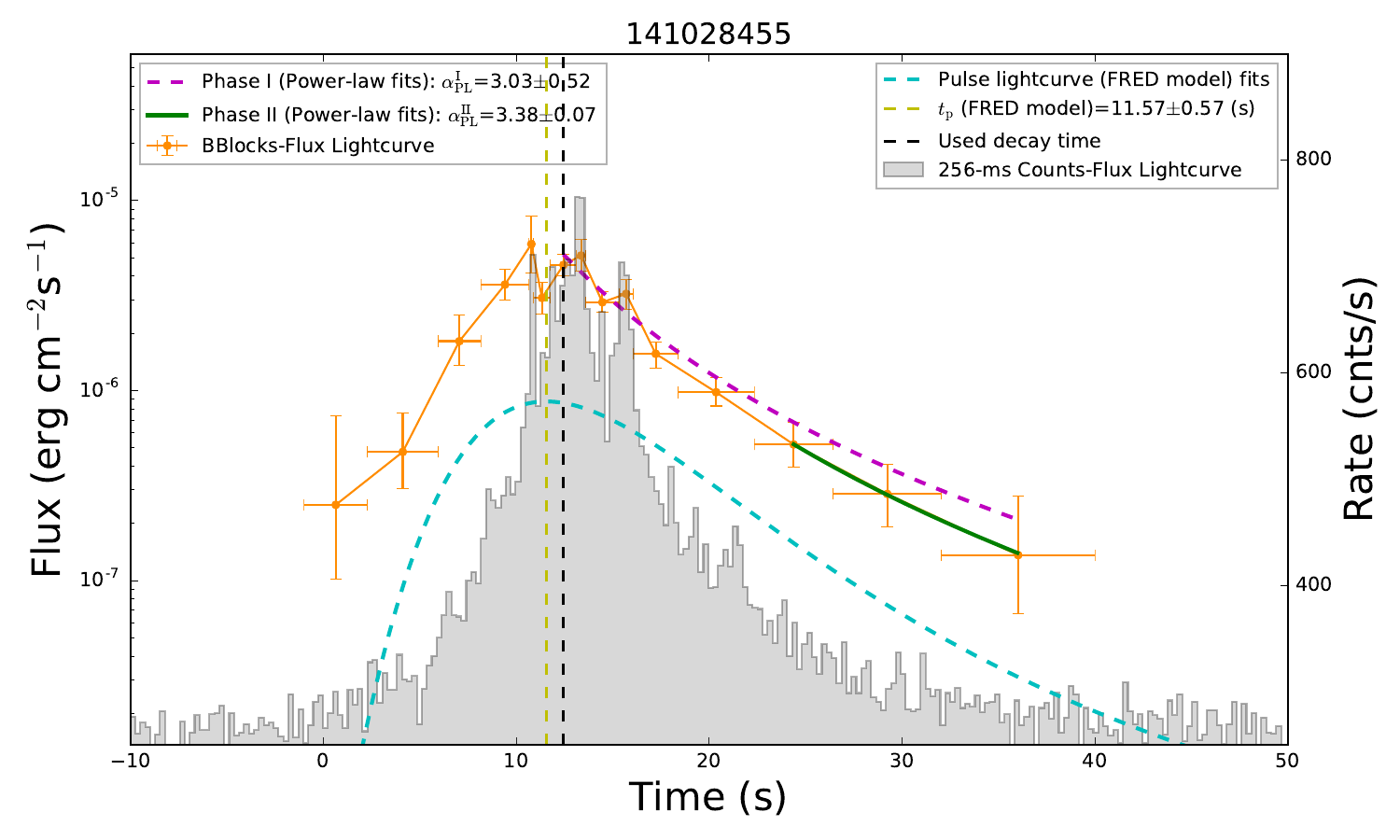}
\includegraphics[width=0.5\hsize,clip]{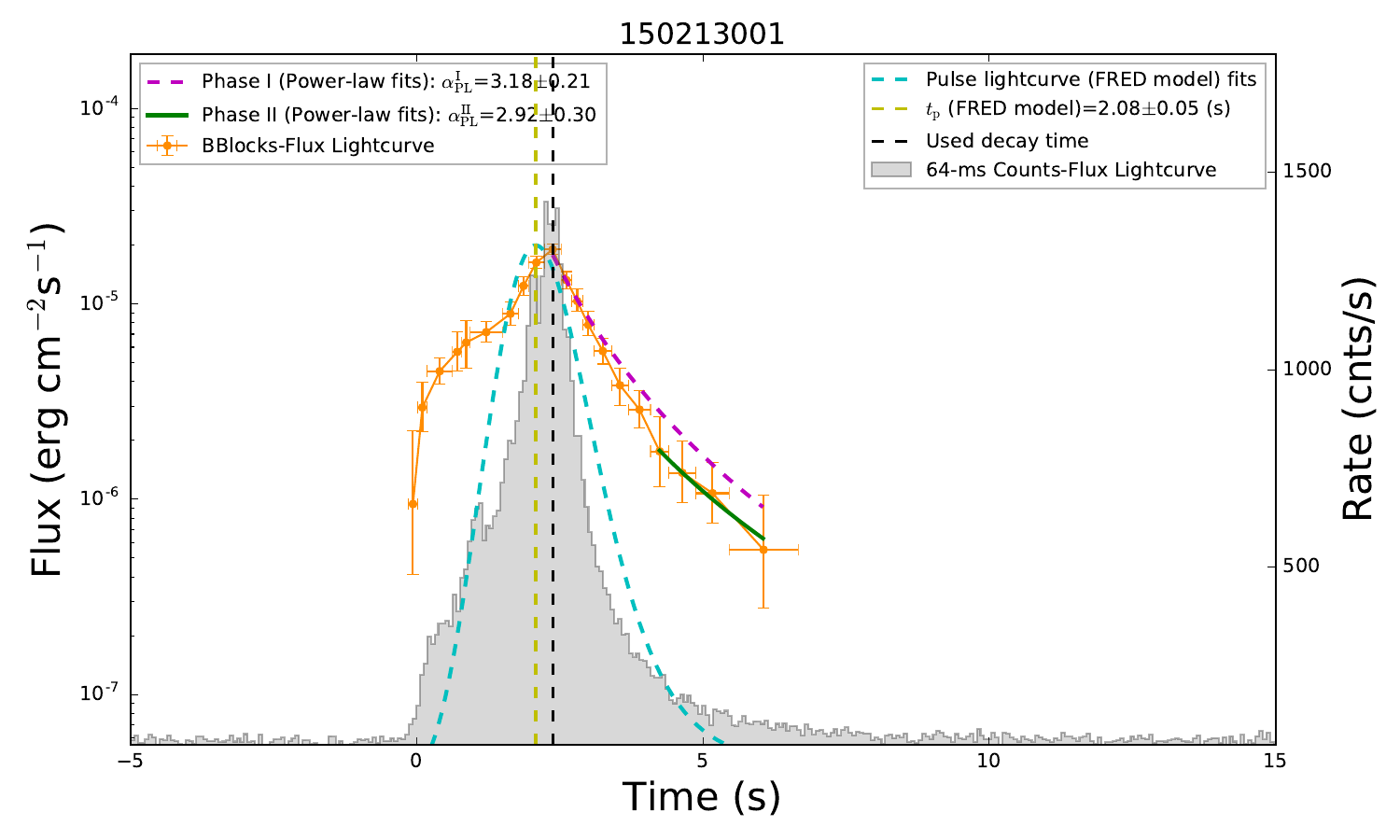}
\includegraphics[width=0.5\hsize,clip]{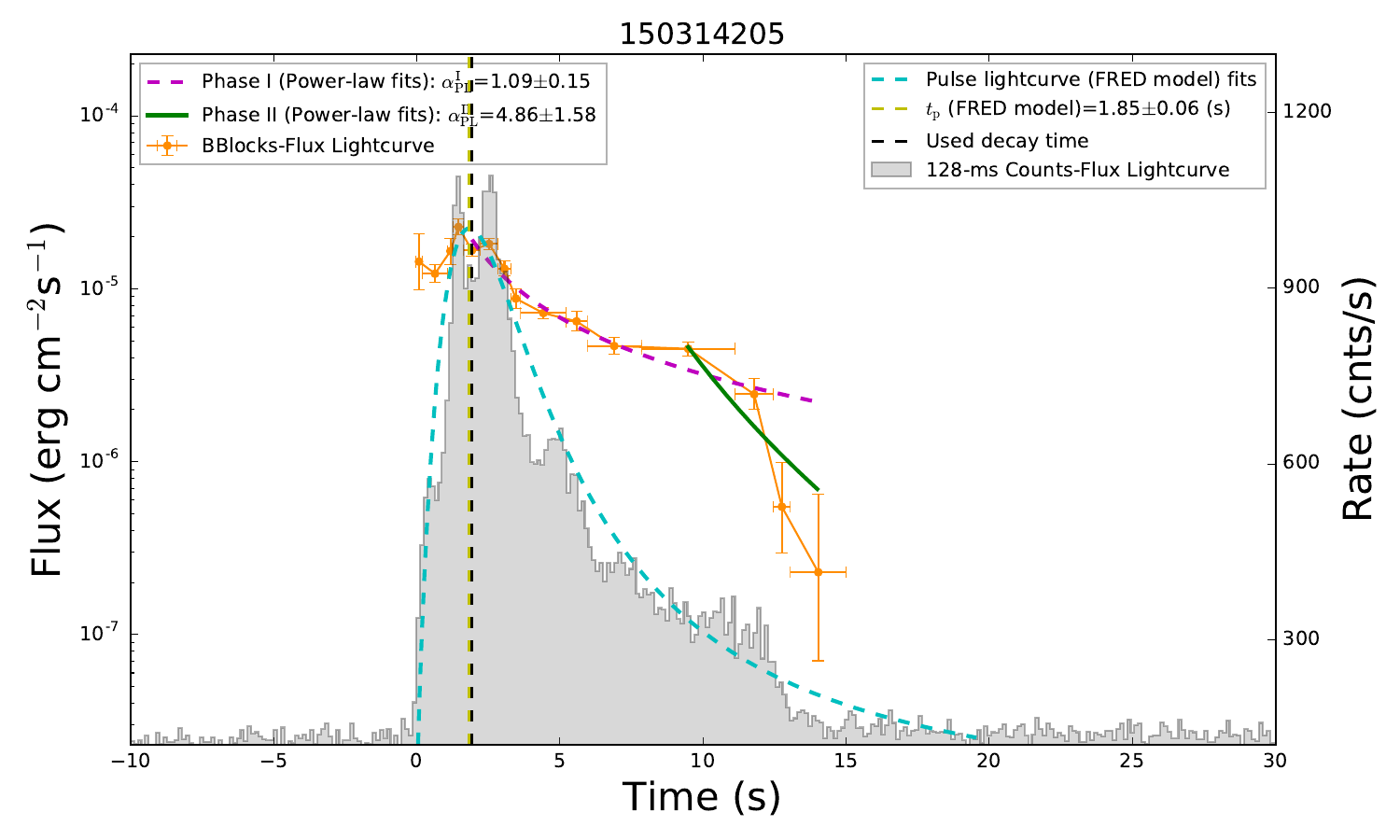}
\includegraphics[width=0.5\hsize,clip]{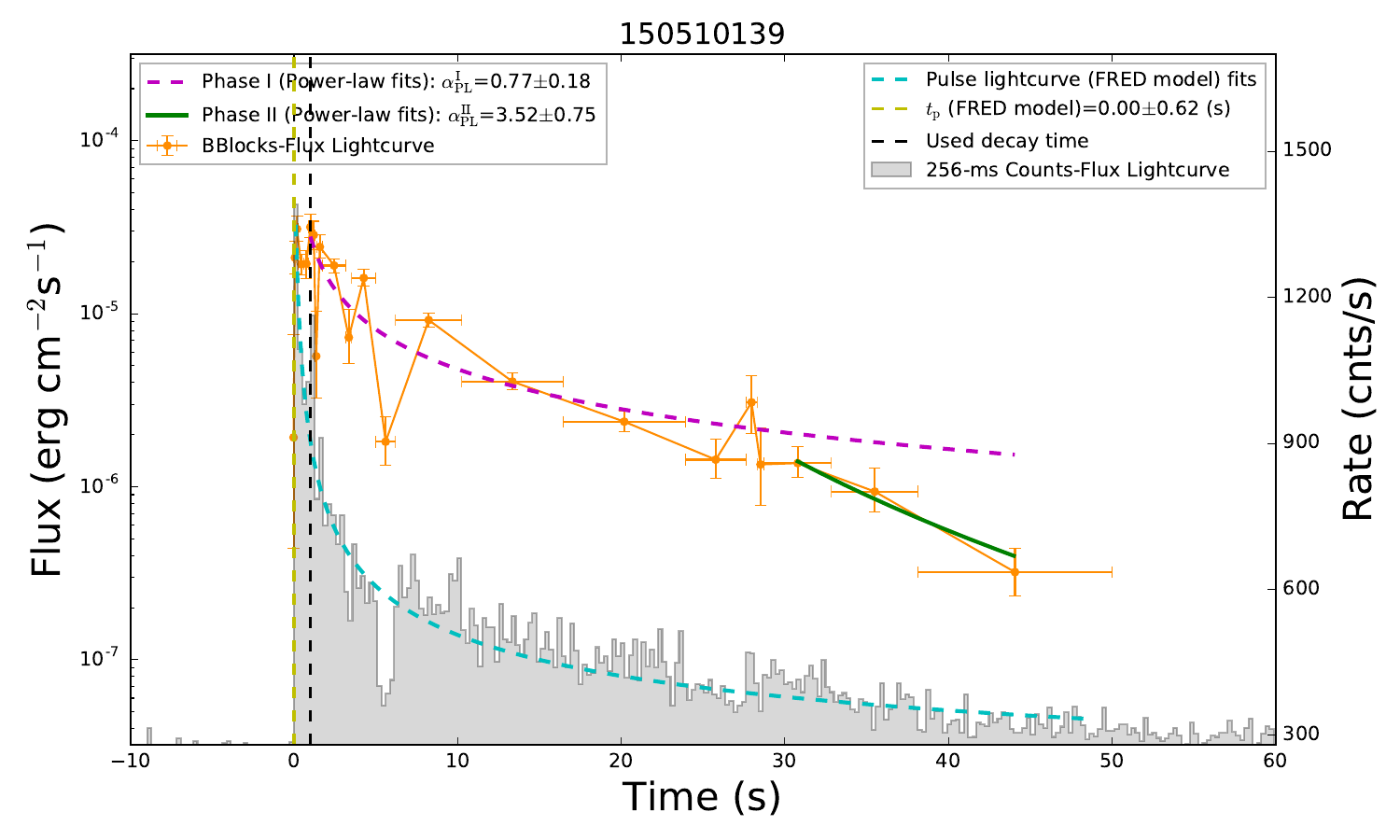}
\center{Fig. \ref{fig:PL}--- Continued}
\end{figure*}
\begin{figure*}
\includegraphics[width=0.5\hsize,clip]{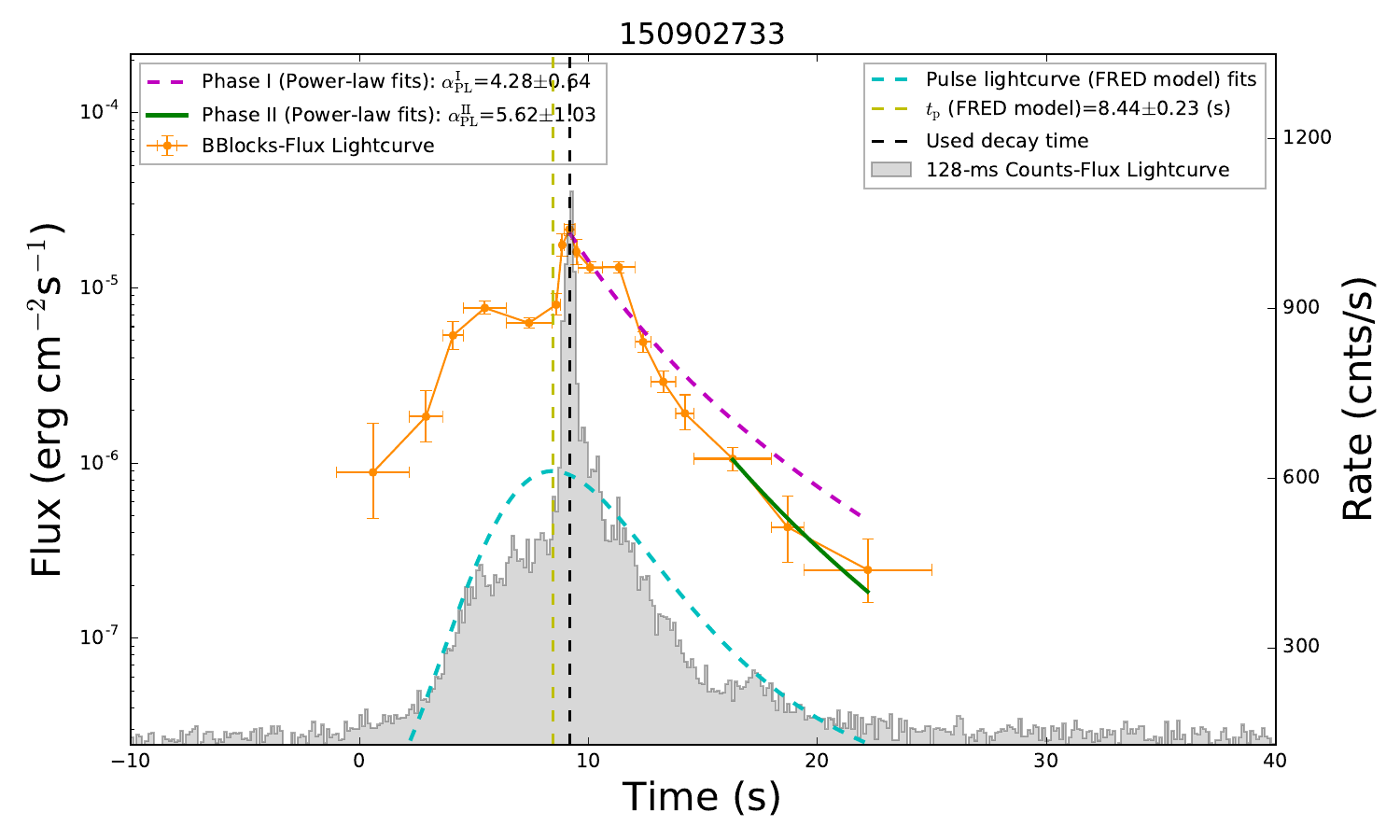}
\includegraphics[width=0.5\hsize,clip]{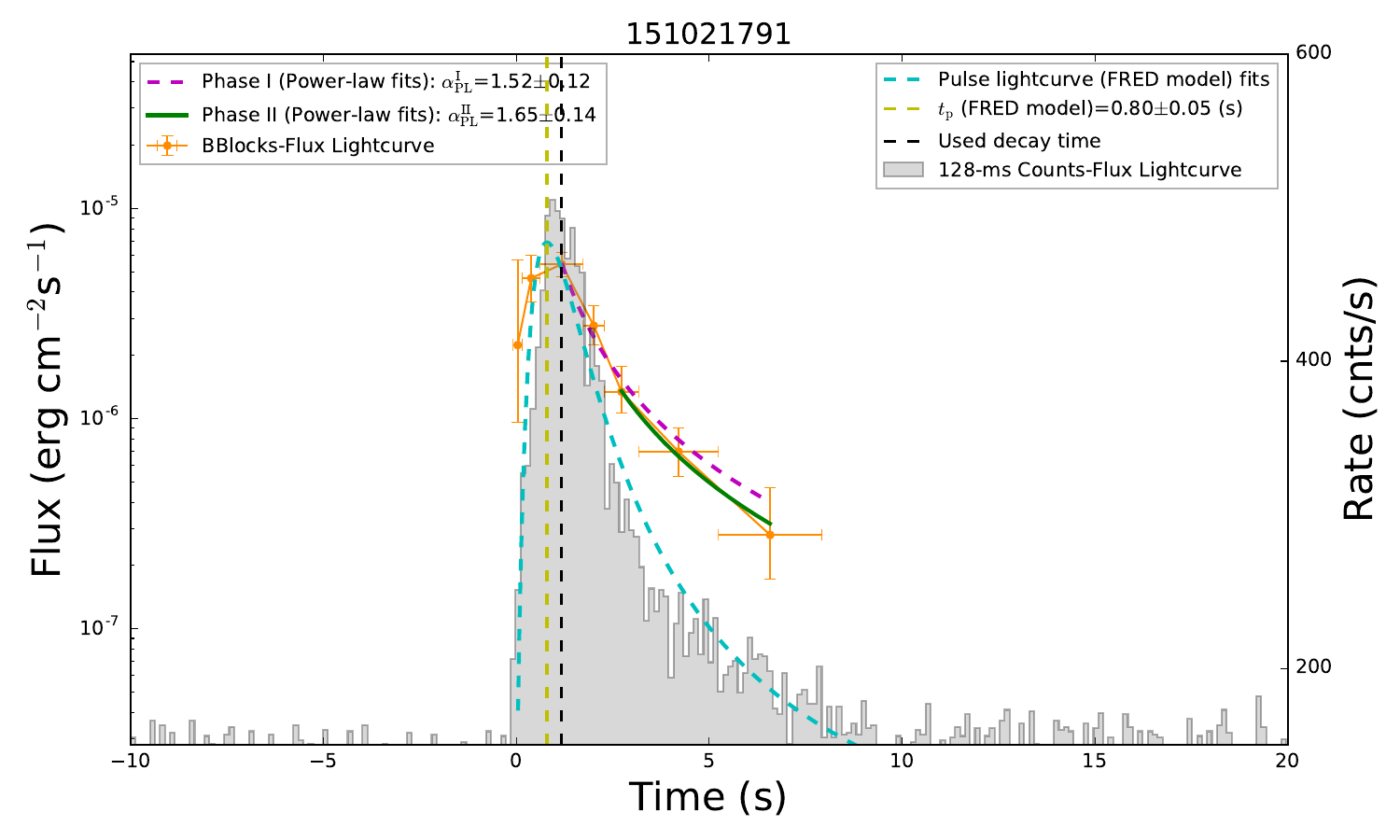}
\includegraphics[width=0.5\hsize,clip]{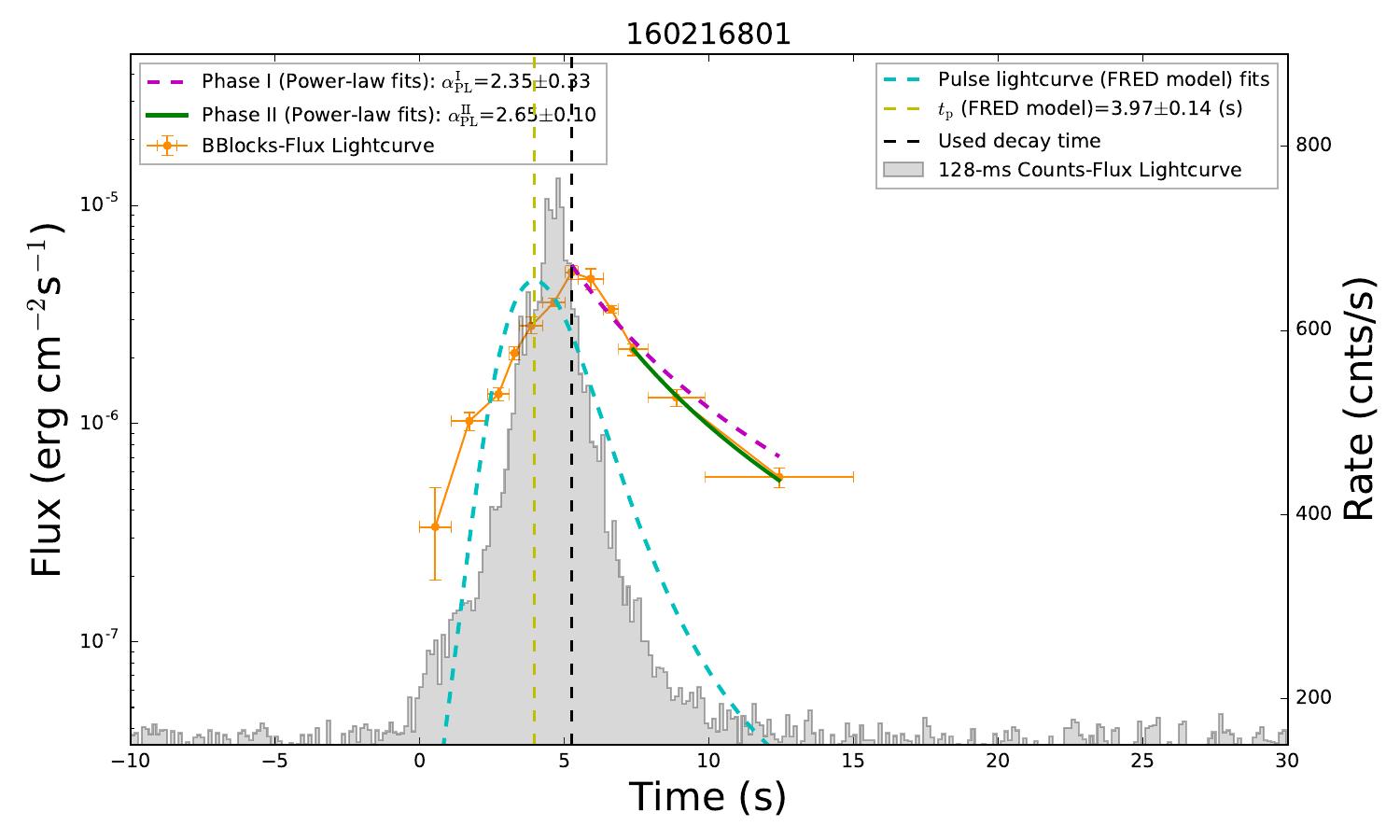}
\includegraphics[width=0.5\hsize,clip]{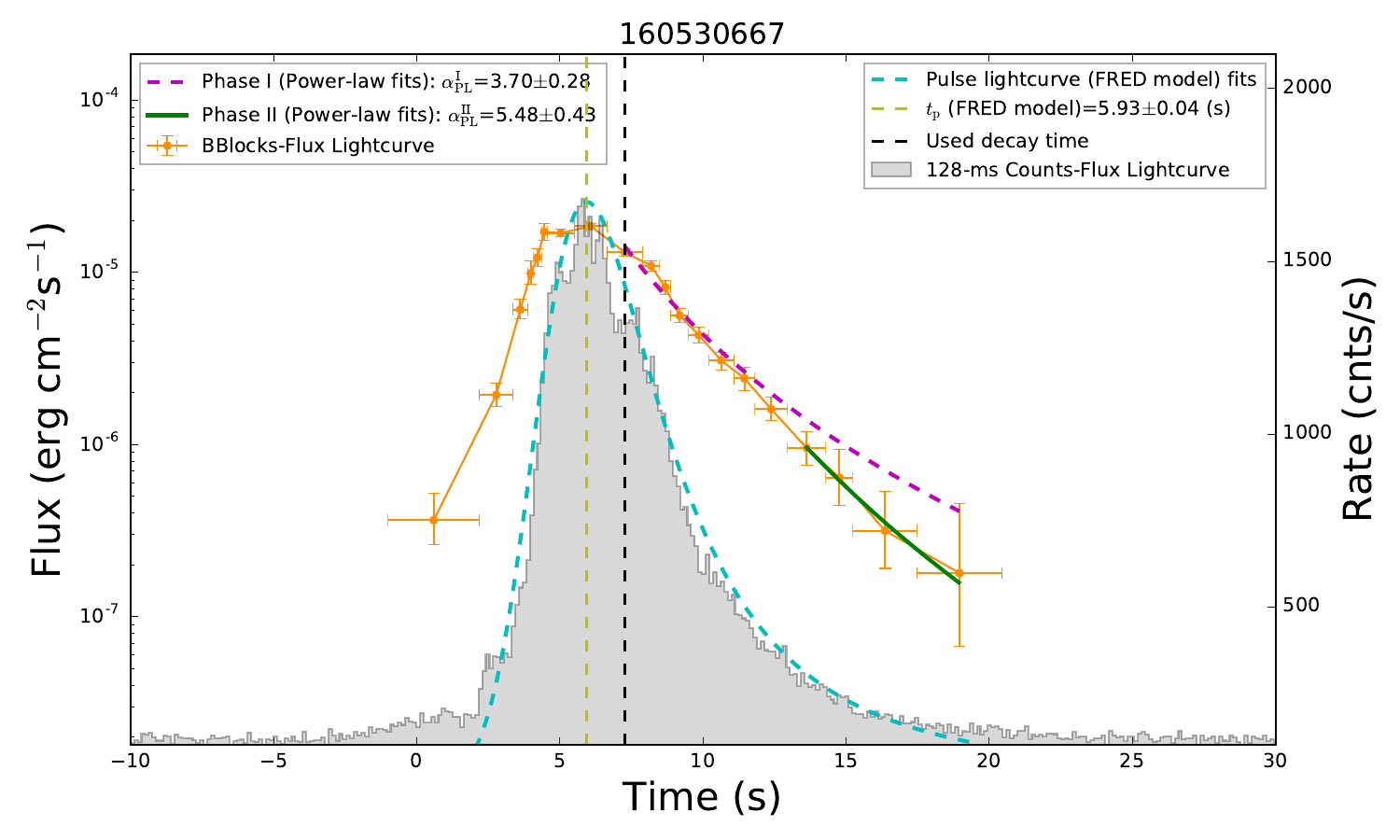}
\includegraphics[width=0.5\hsize,clip]{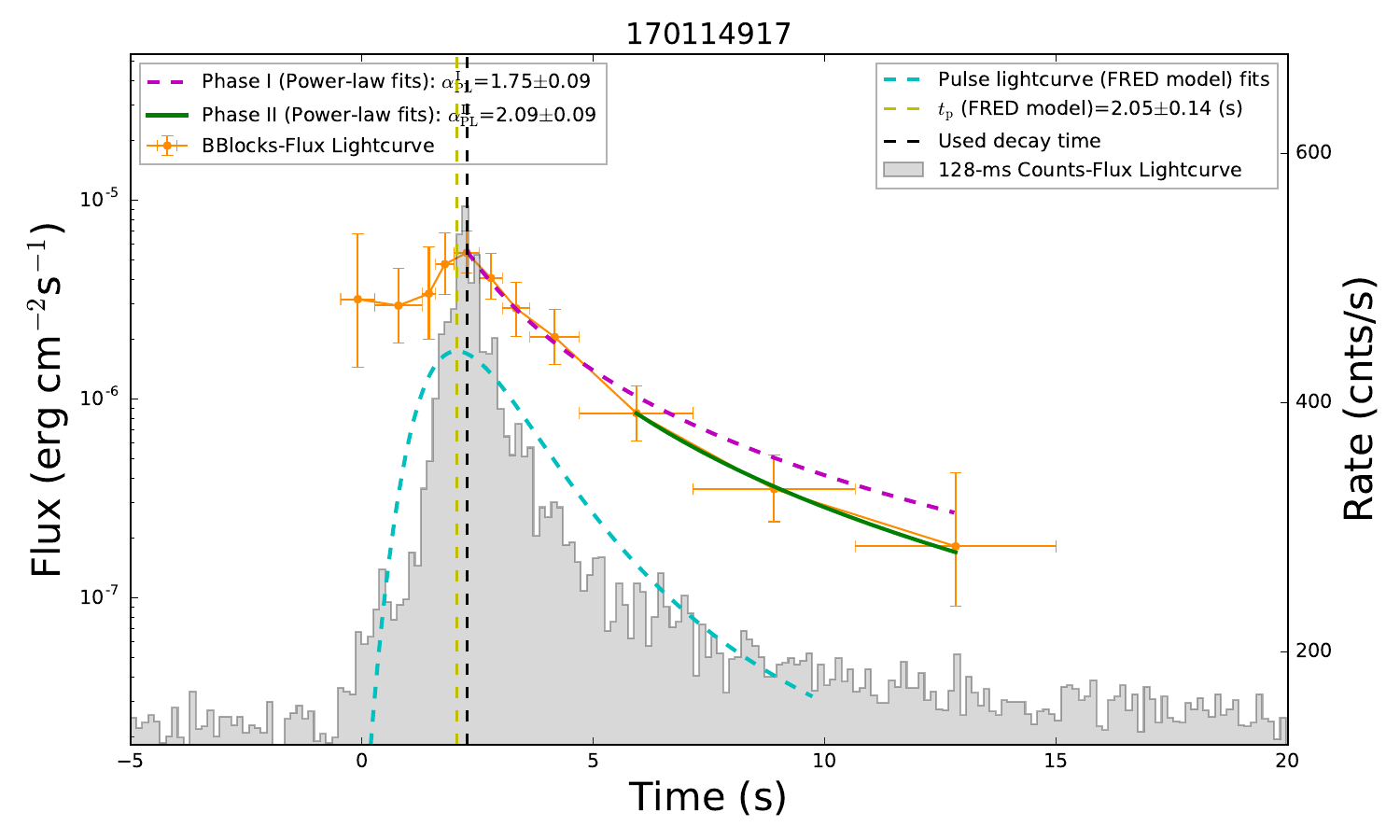}
\includegraphics[width=0.5\hsize,clip]{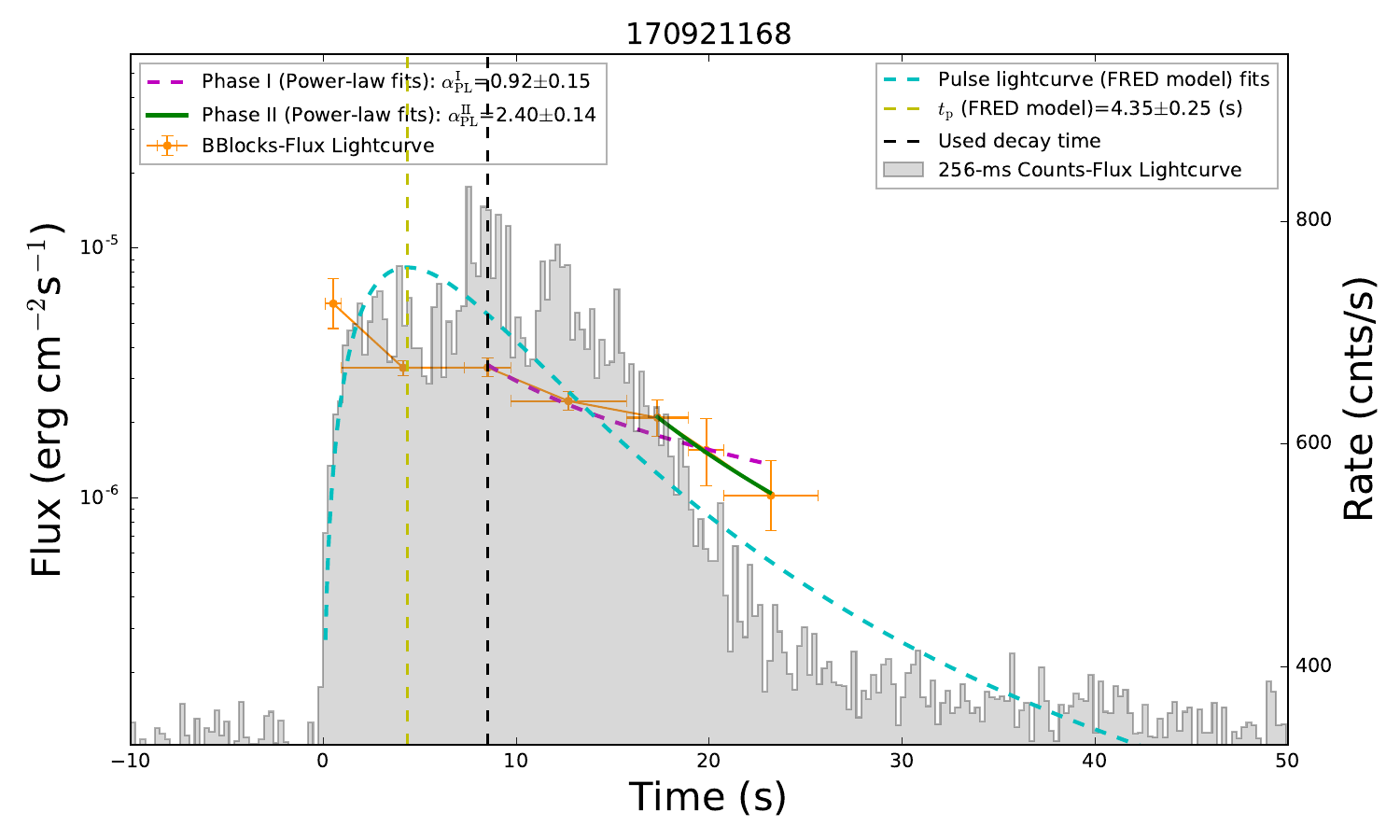}
\includegraphics[width=0.5\hsize,clip]{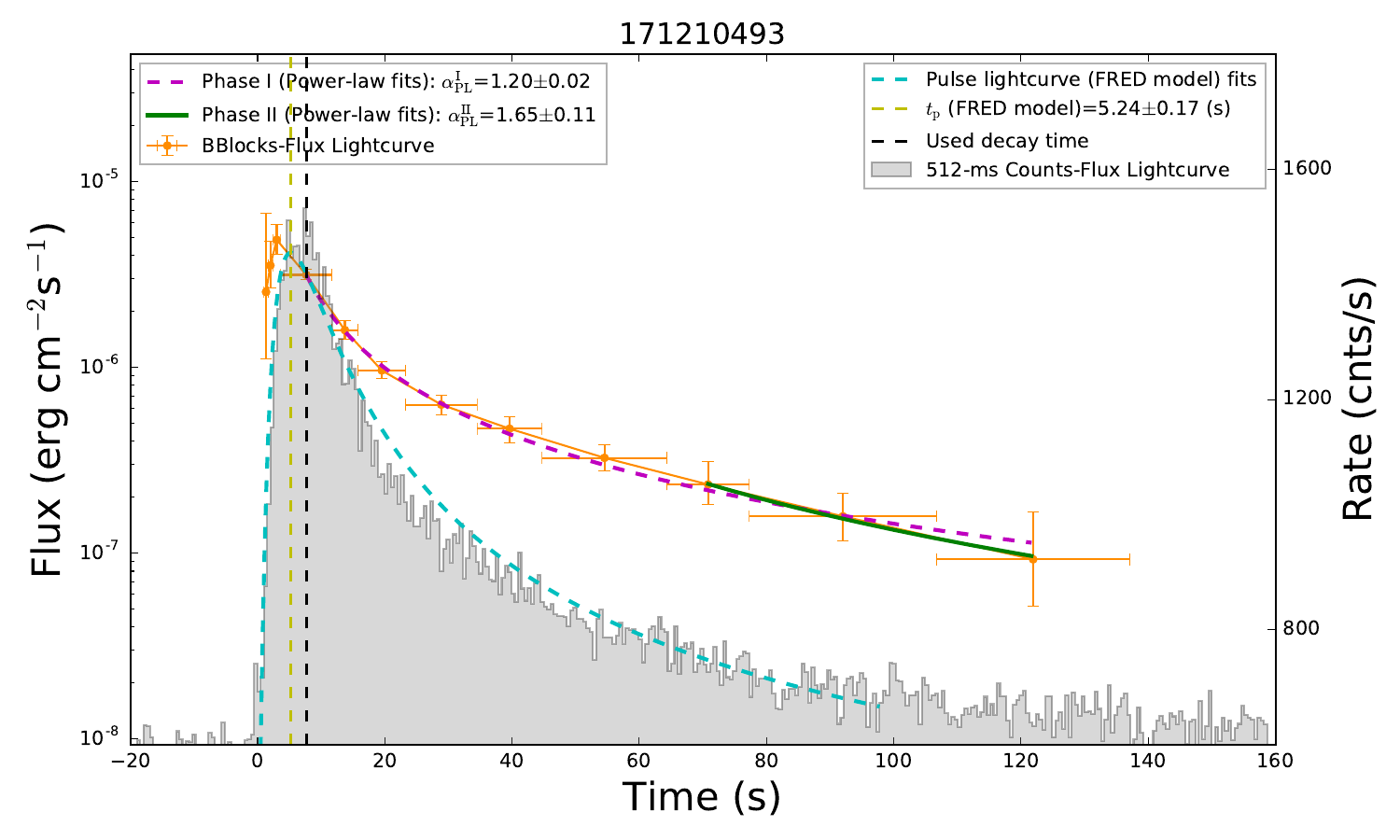}
\includegraphics[width=0.5\hsize,clip]{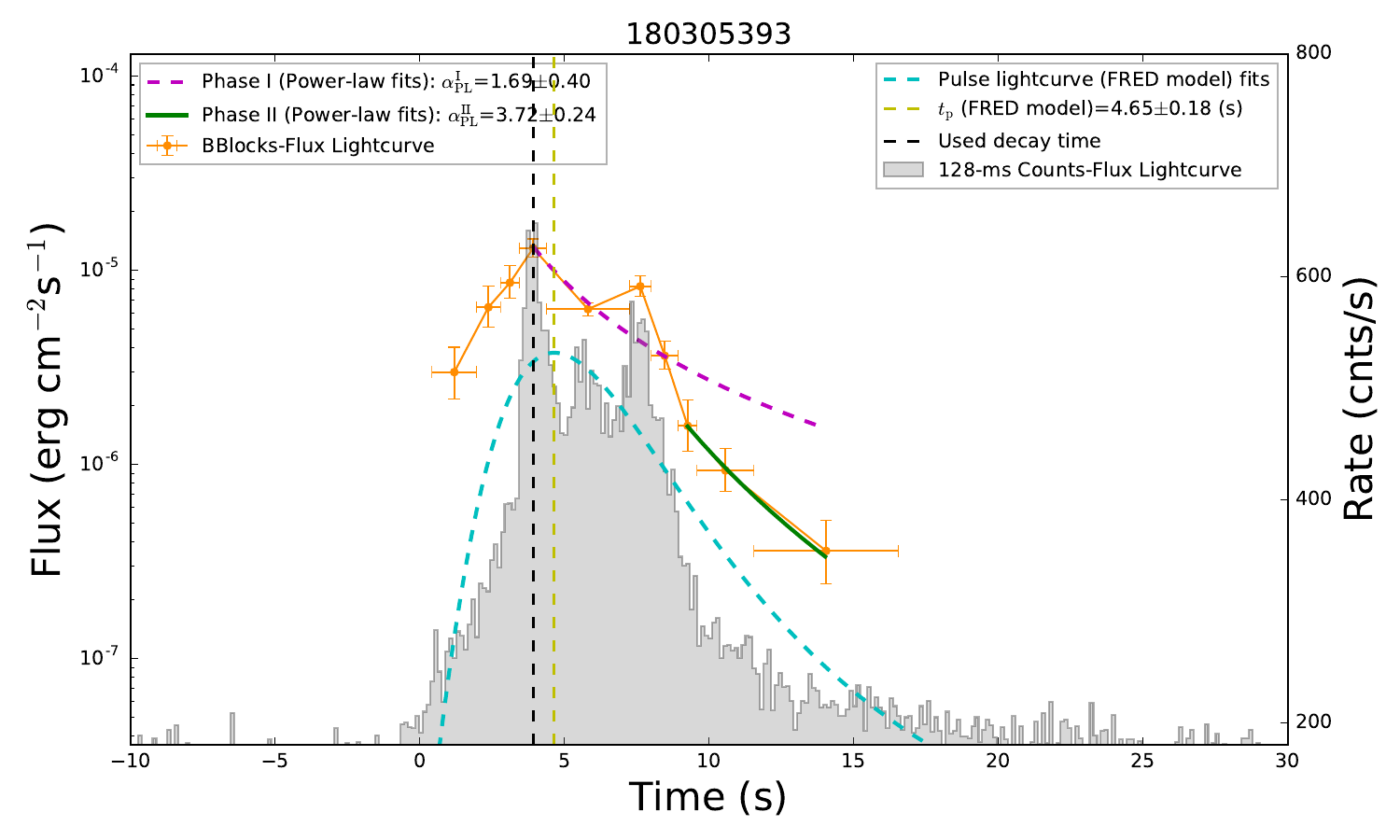}
\center{Fig. \ref{fig:PL}--- Continued}
\end{figure*}

\clearpage

\clearpage
\begin{figure*}
\includegraphics[width=0.5\hsize,clip]{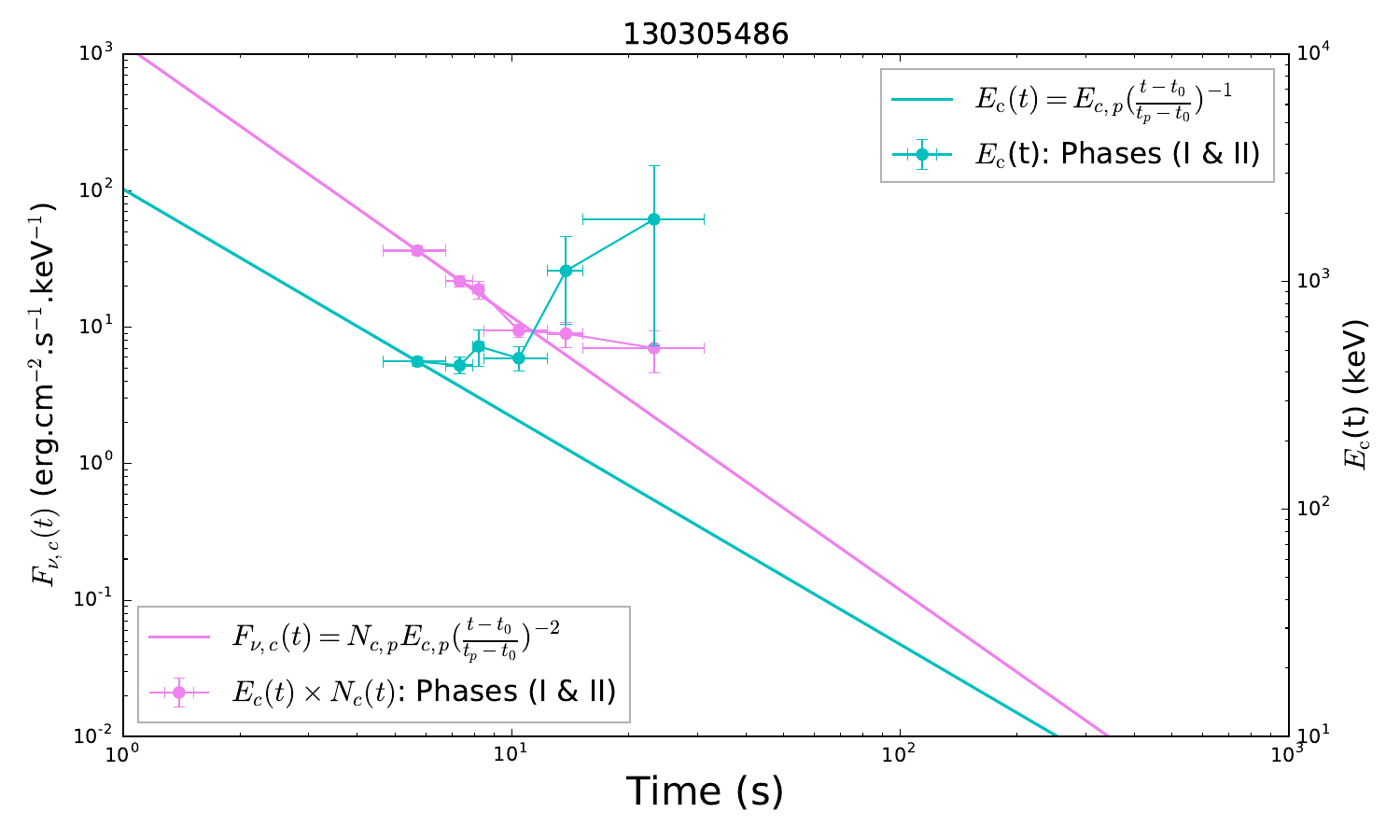}
\includegraphics[width=0.5\hsize,clip]{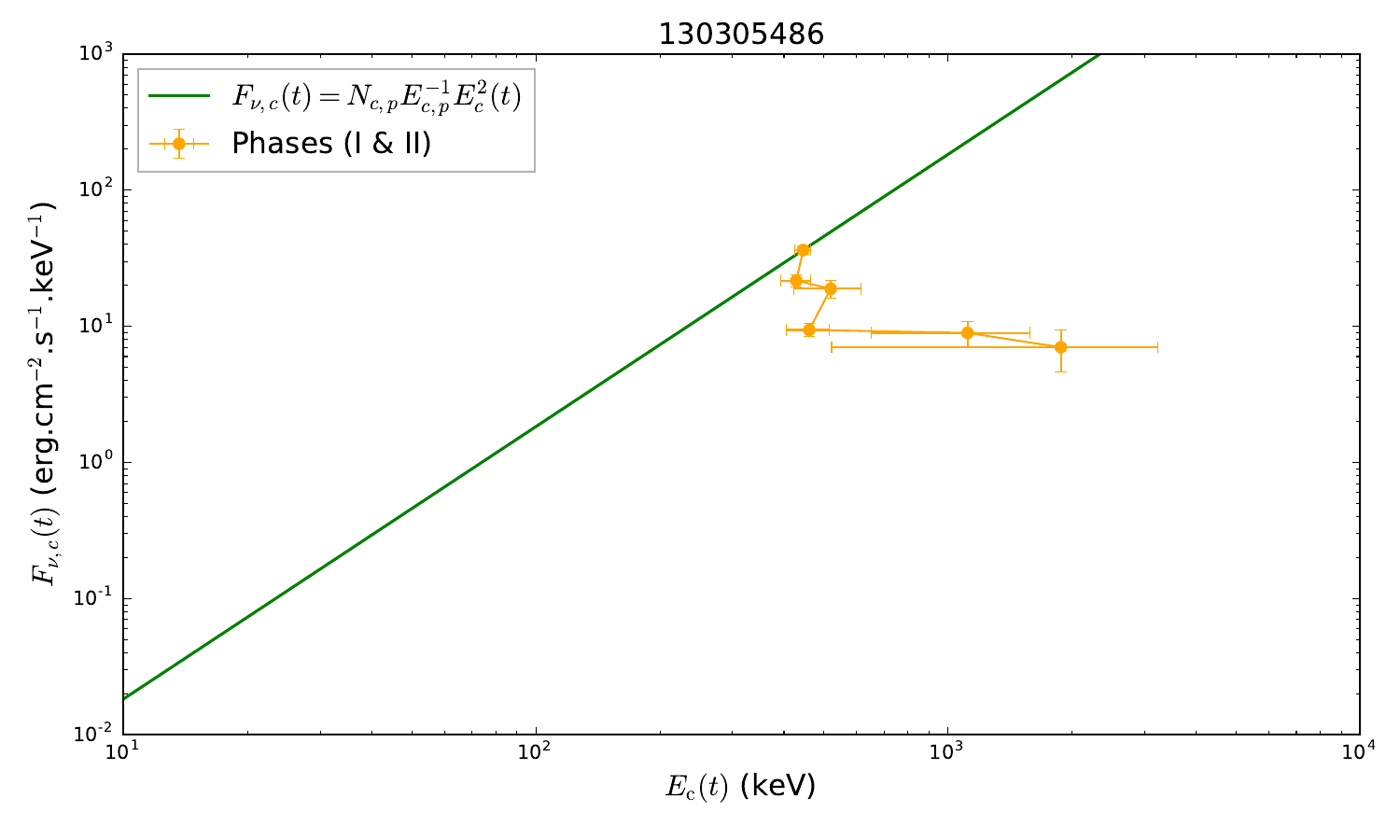}
\includegraphics[width=0.5\hsize,clip]{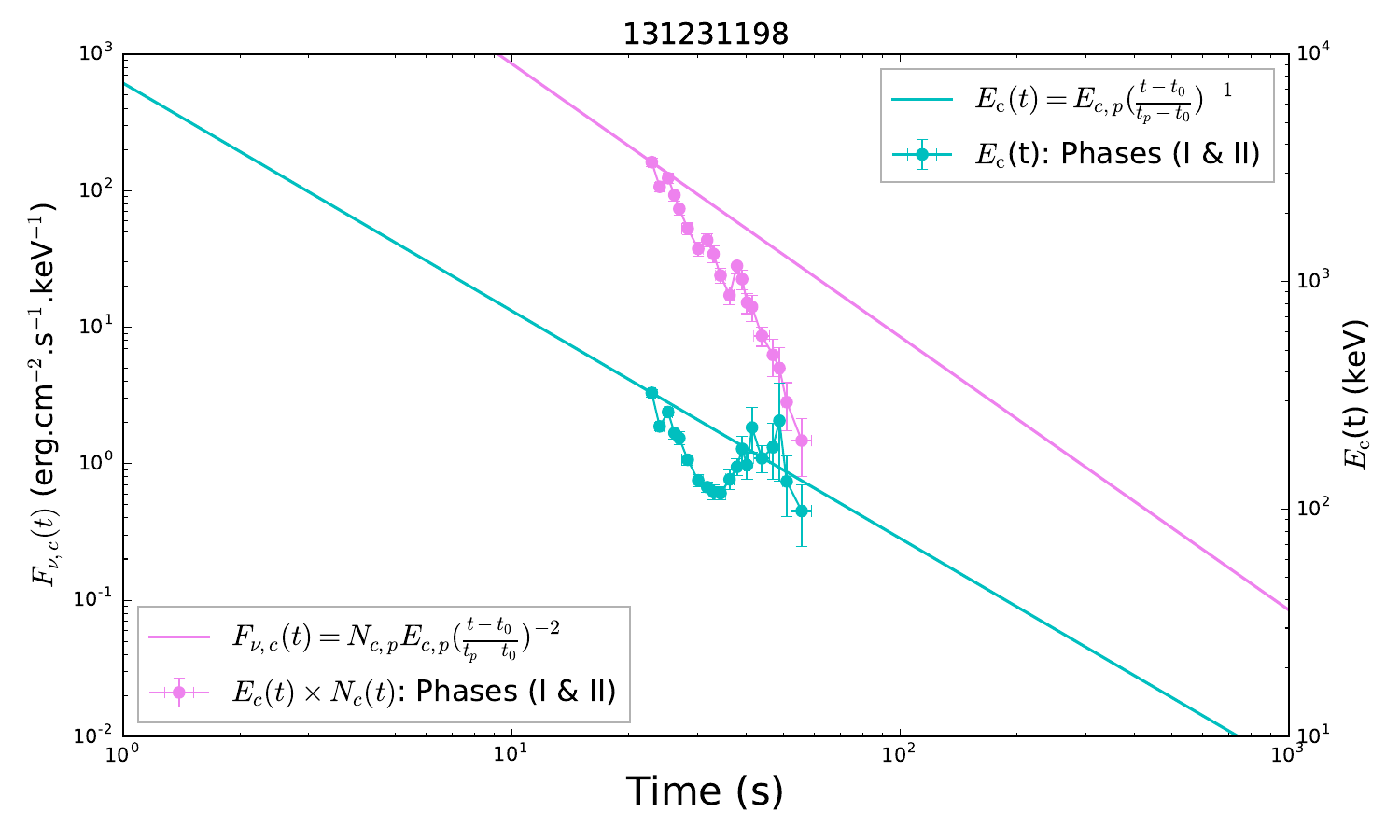}
\includegraphics[width=0.5\hsize,clip]{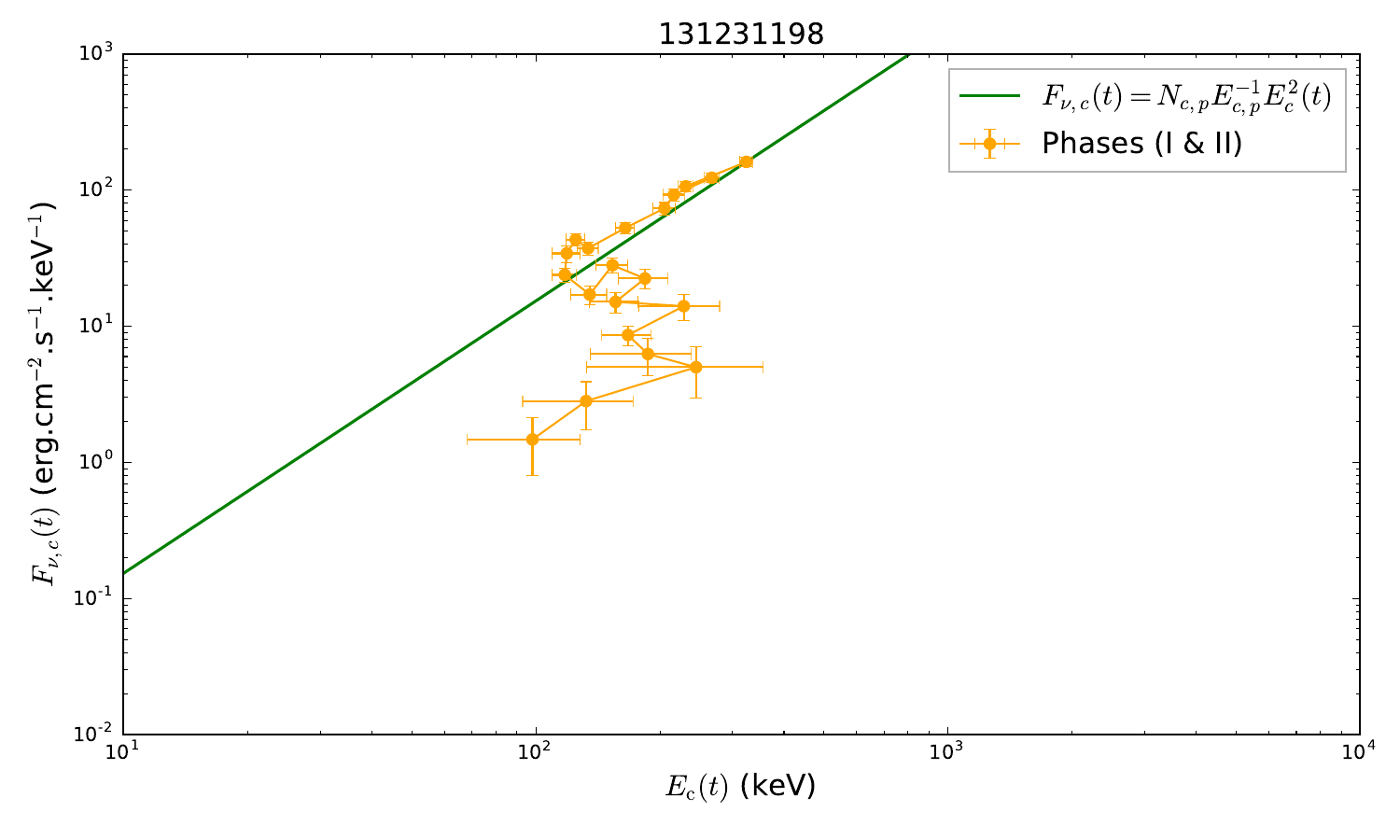}
\includegraphics[width=0.5\hsize,clip]{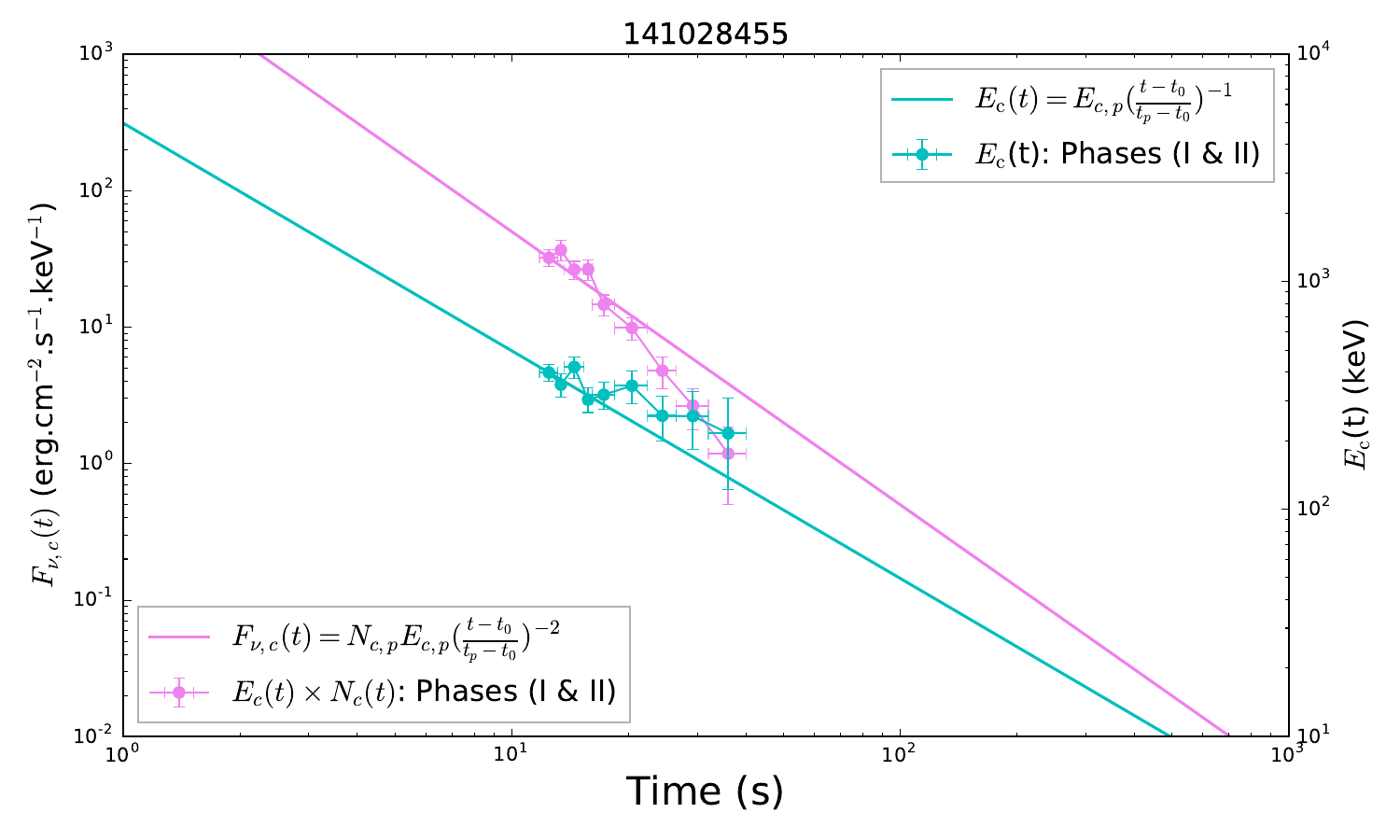}
\includegraphics[width=0.5\hsize,clip]{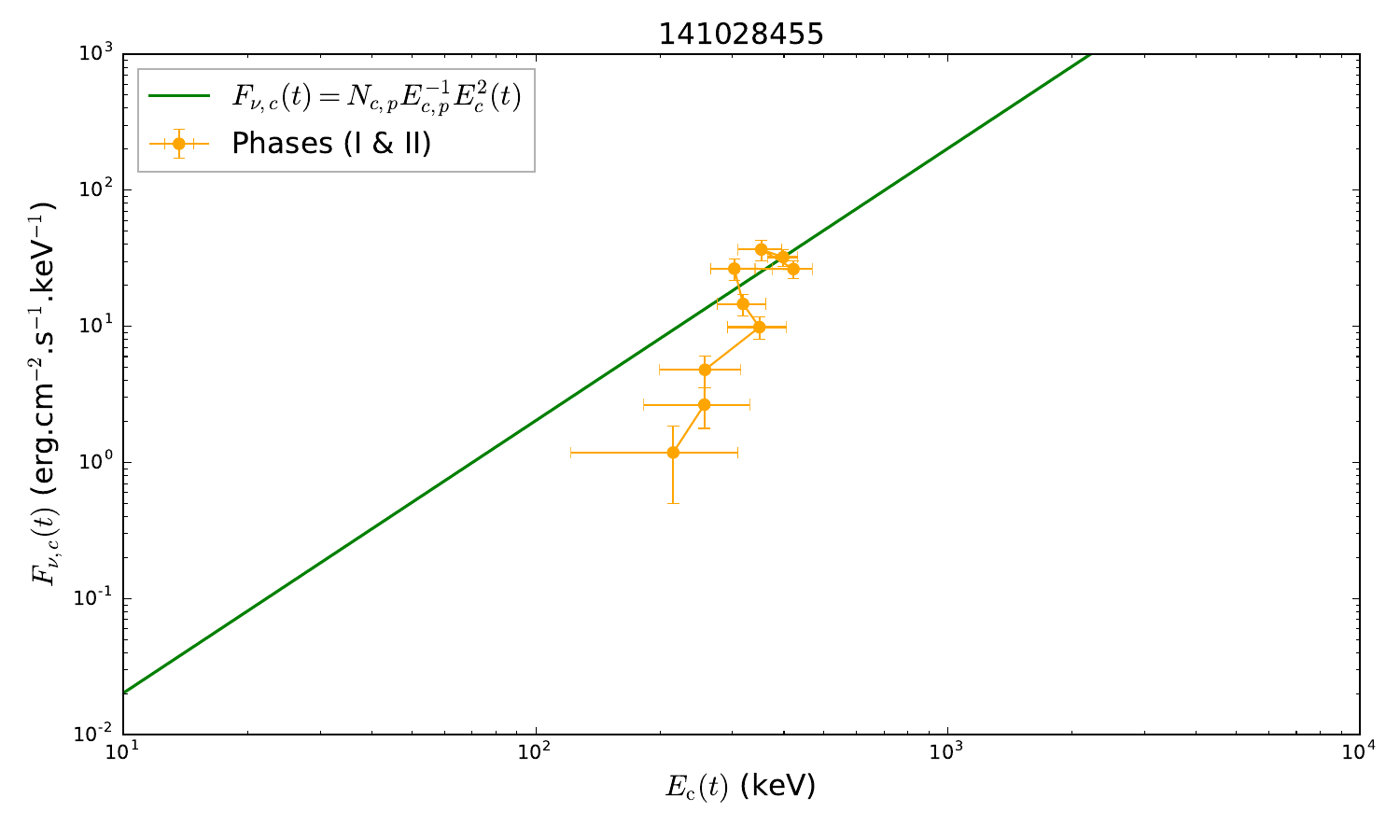}
\includegraphics[width=0.5\hsize,clip]{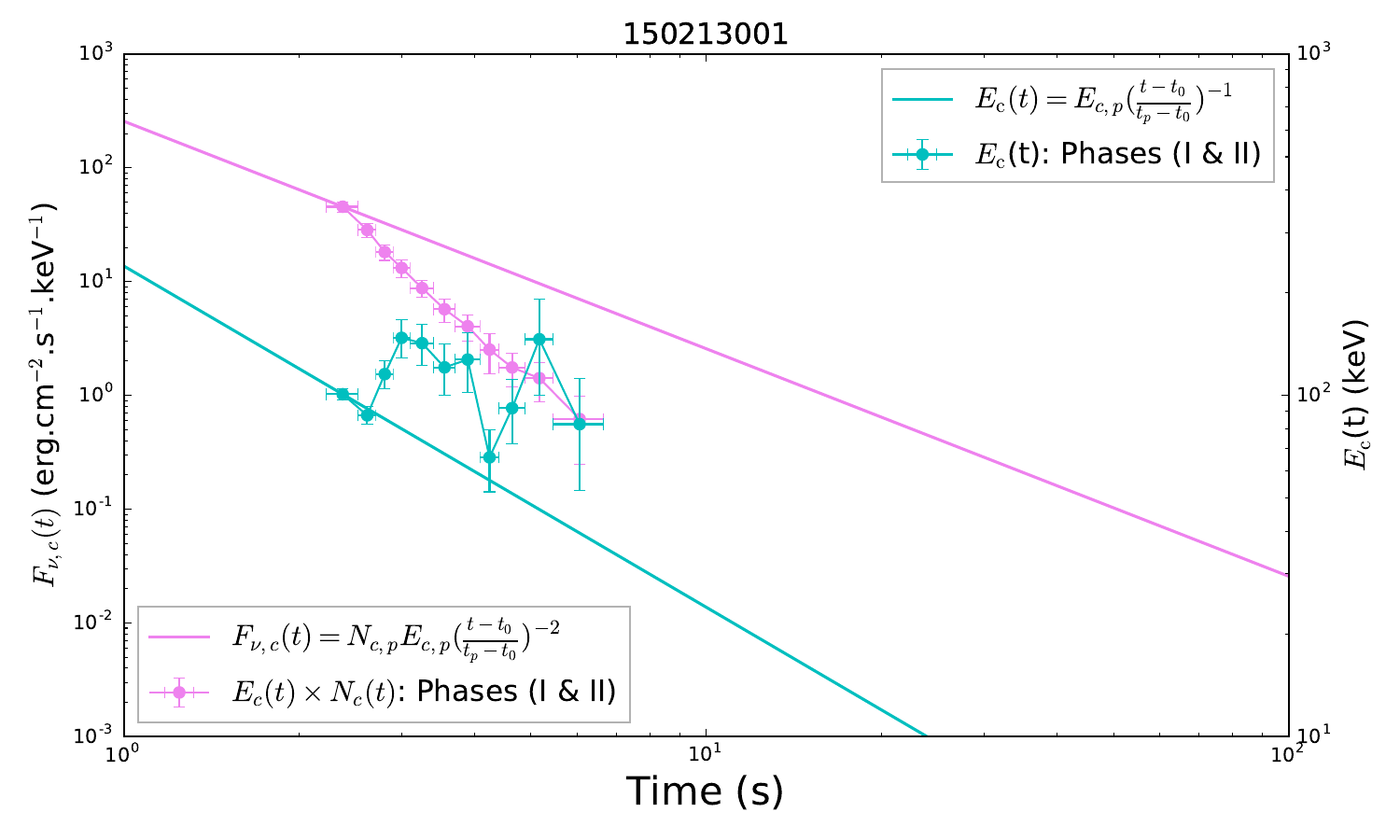}
\includegraphics[width=0.5\hsize,clip]{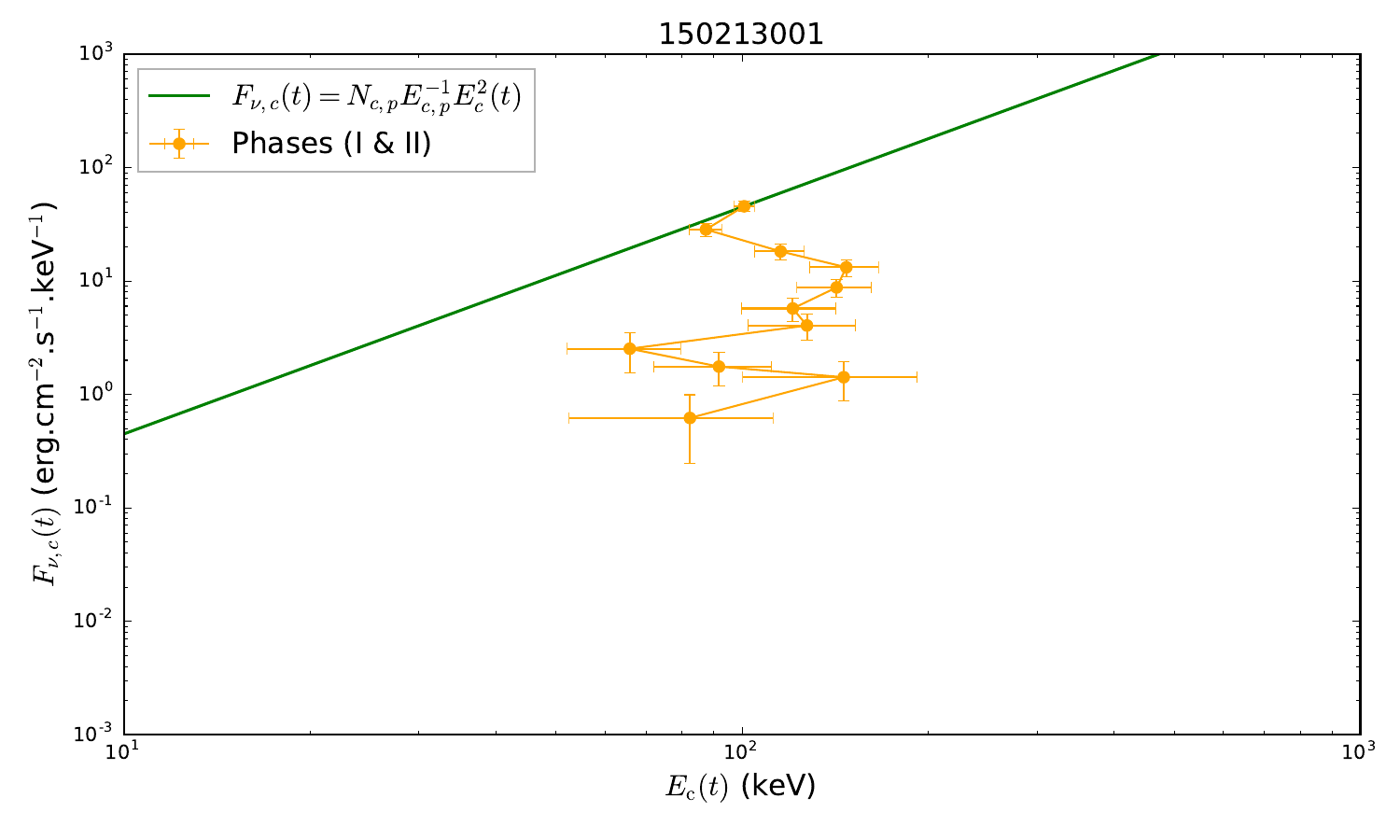}
\center{Fig. \ref{fig:CPL}--- Continued}
\end{figure*}
\begin{figure*}
\includegraphics[width=0.5\hsize,clip]{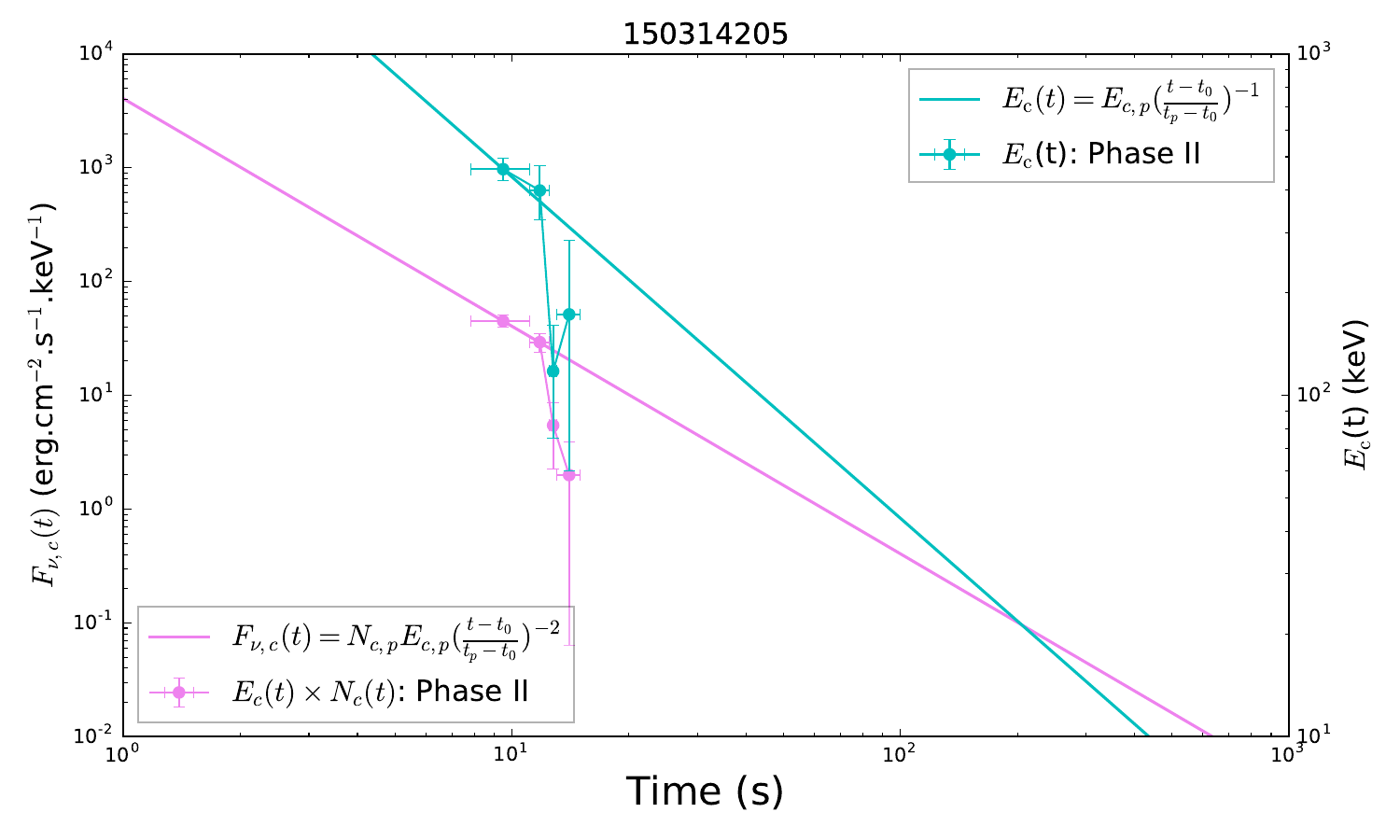}
\includegraphics[width=0.5\hsize,clip]{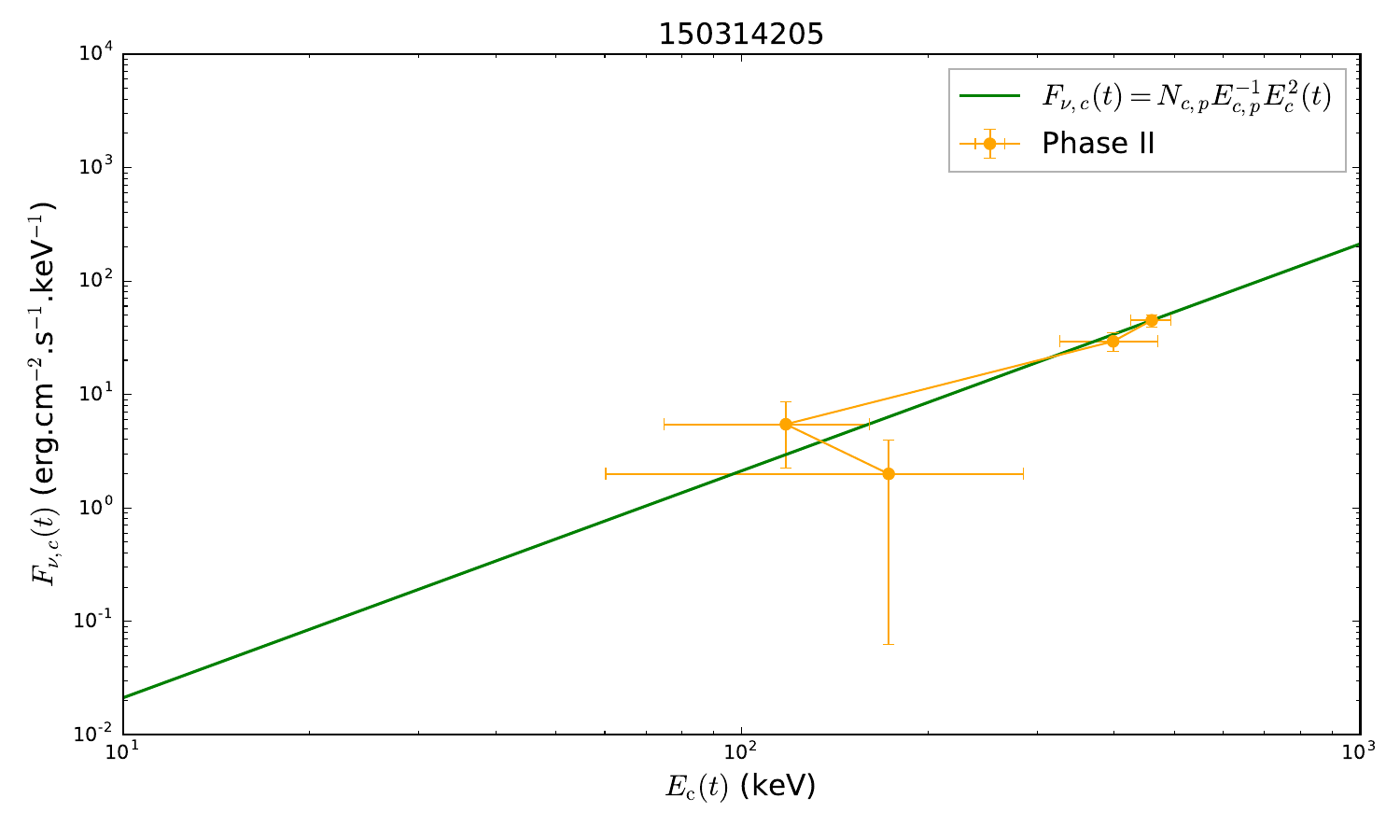}
\includegraphics[width=0.5\hsize,clip]{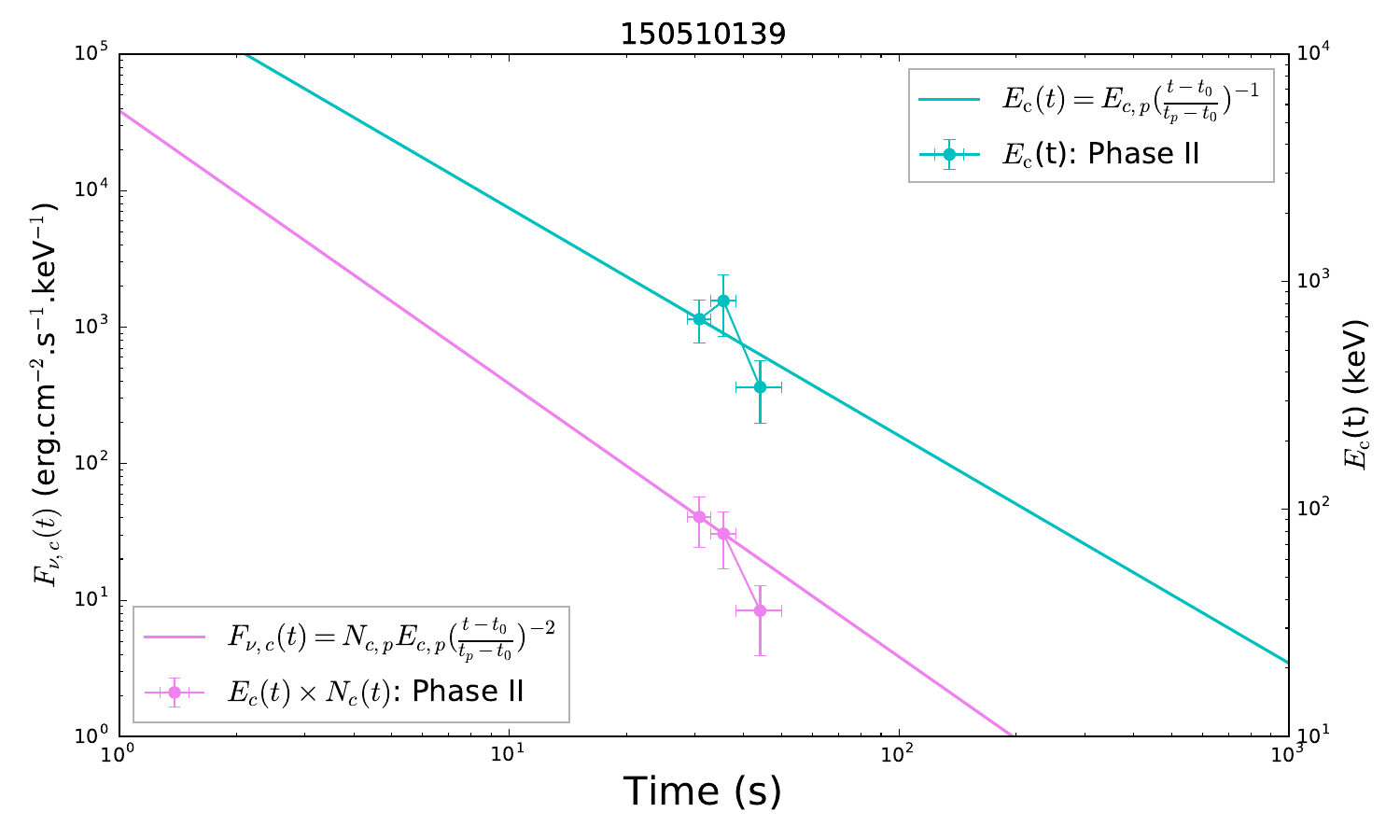}
\includegraphics[width=0.5\hsize,clip]{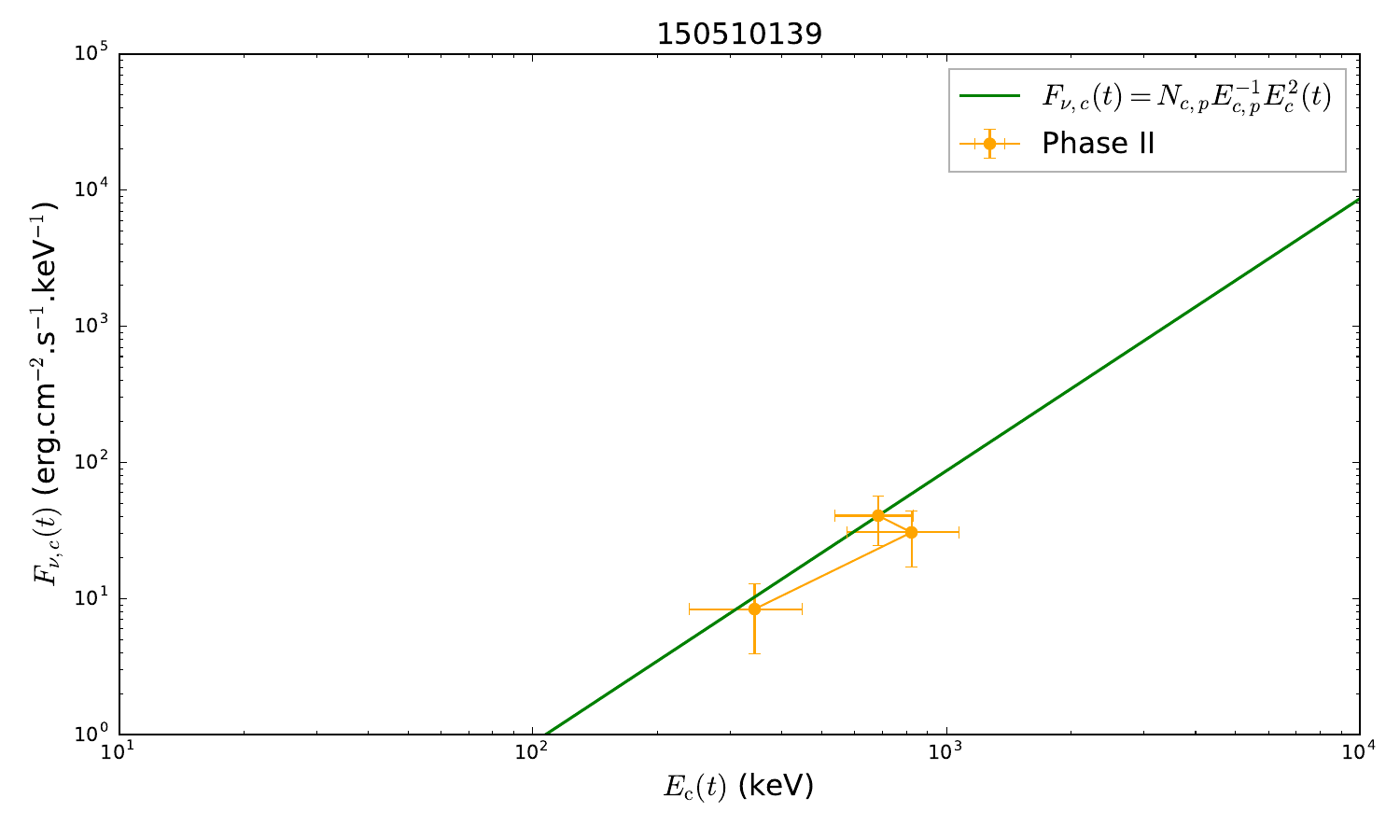}
\includegraphics[width=0.5\hsize,clip]{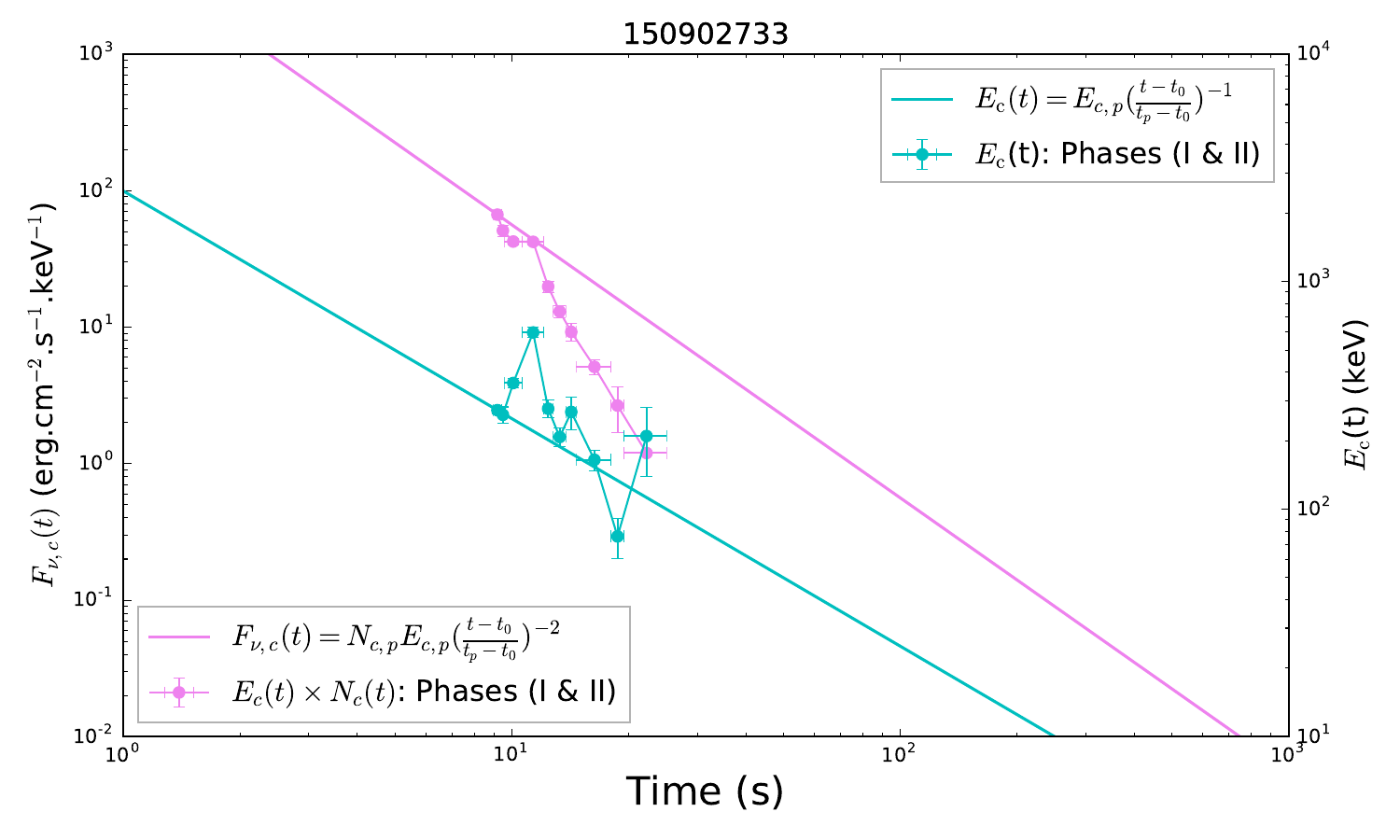}
\includegraphics[width=0.5\hsize,clip]{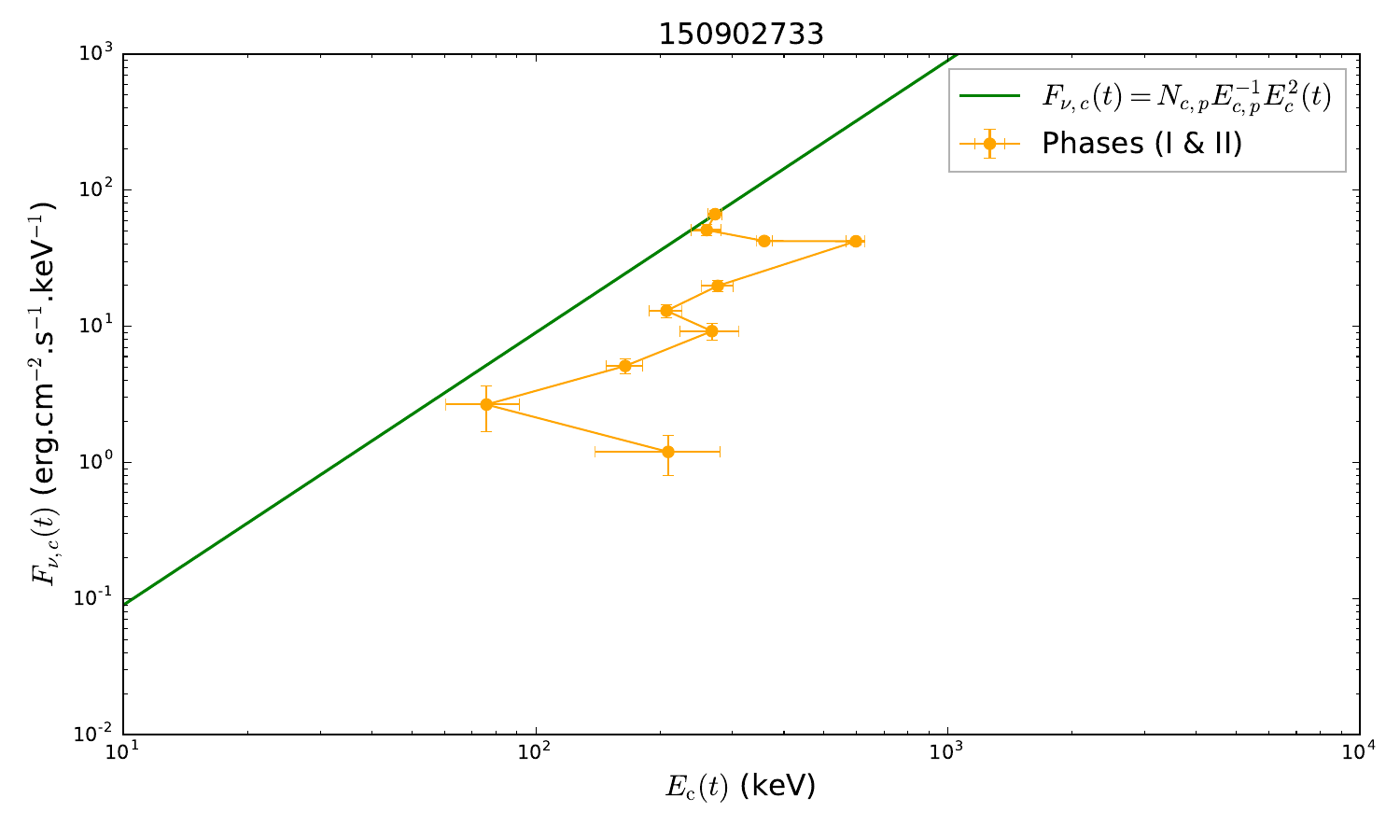}
\includegraphics[width=0.5\hsize,clip]{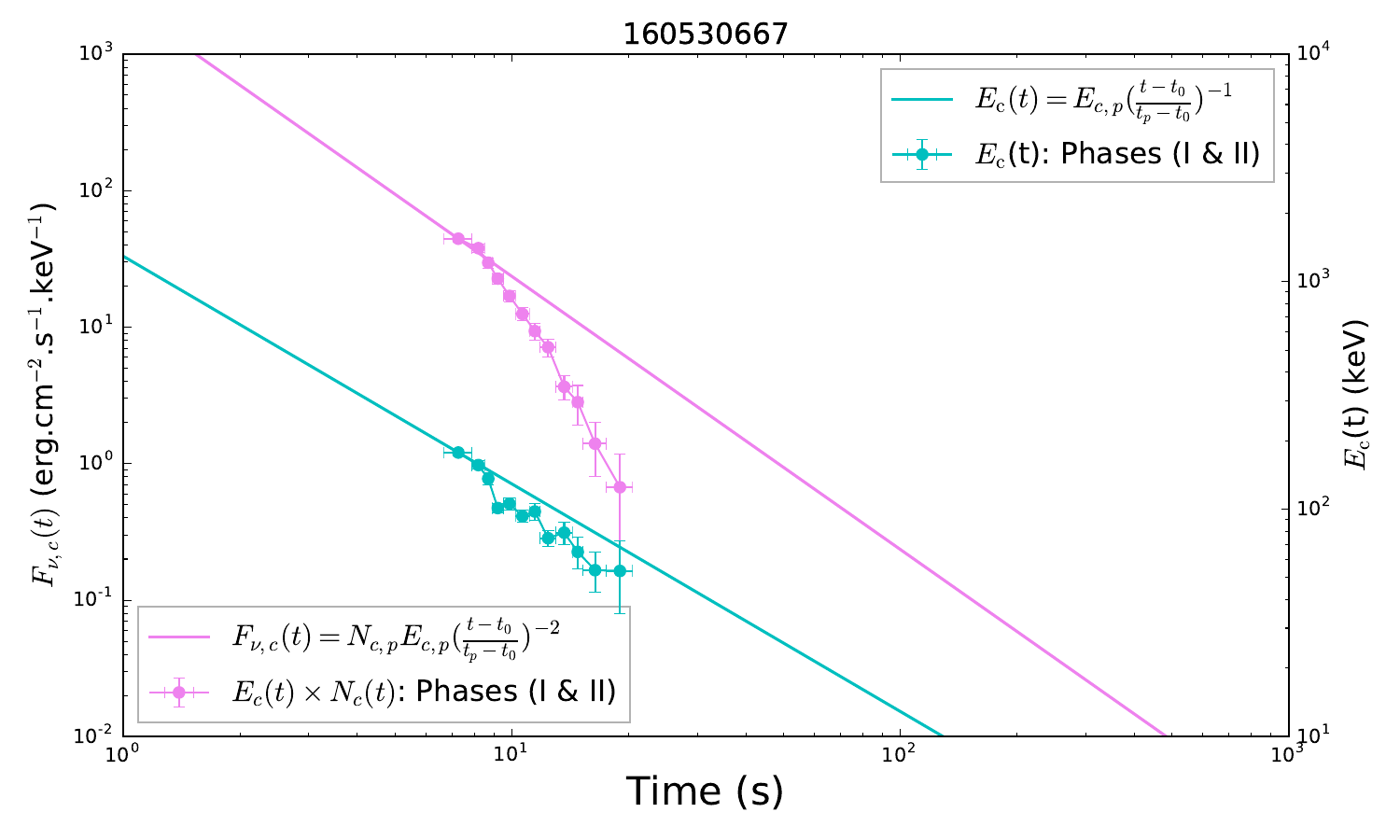}
\includegraphics[width=0.5\hsize,clip]{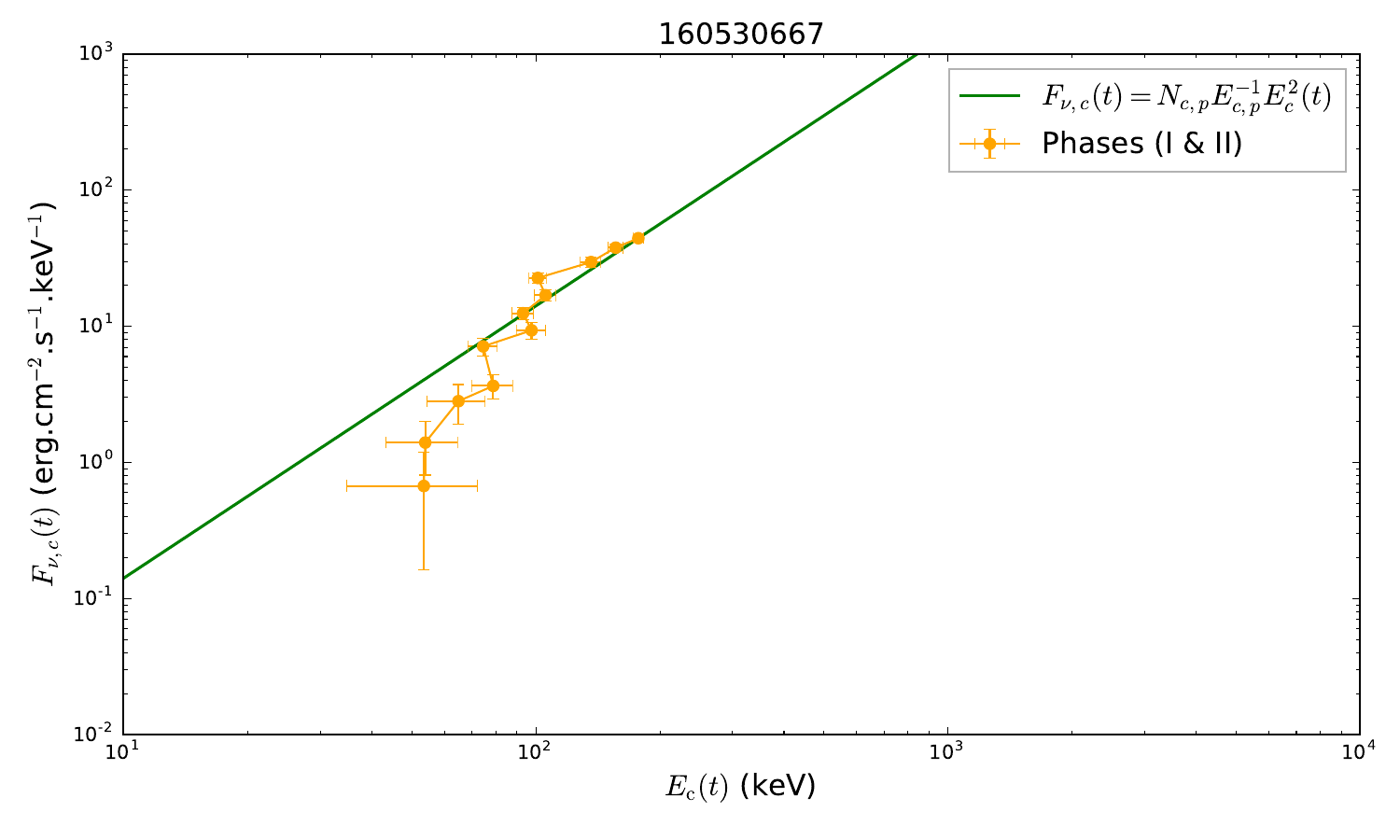}
\center{Fig. \ref{fig:CPL}--- Continued}
\end{figure*}
\begin{figure*}
\includegraphics[width=0.5\hsize,clip]{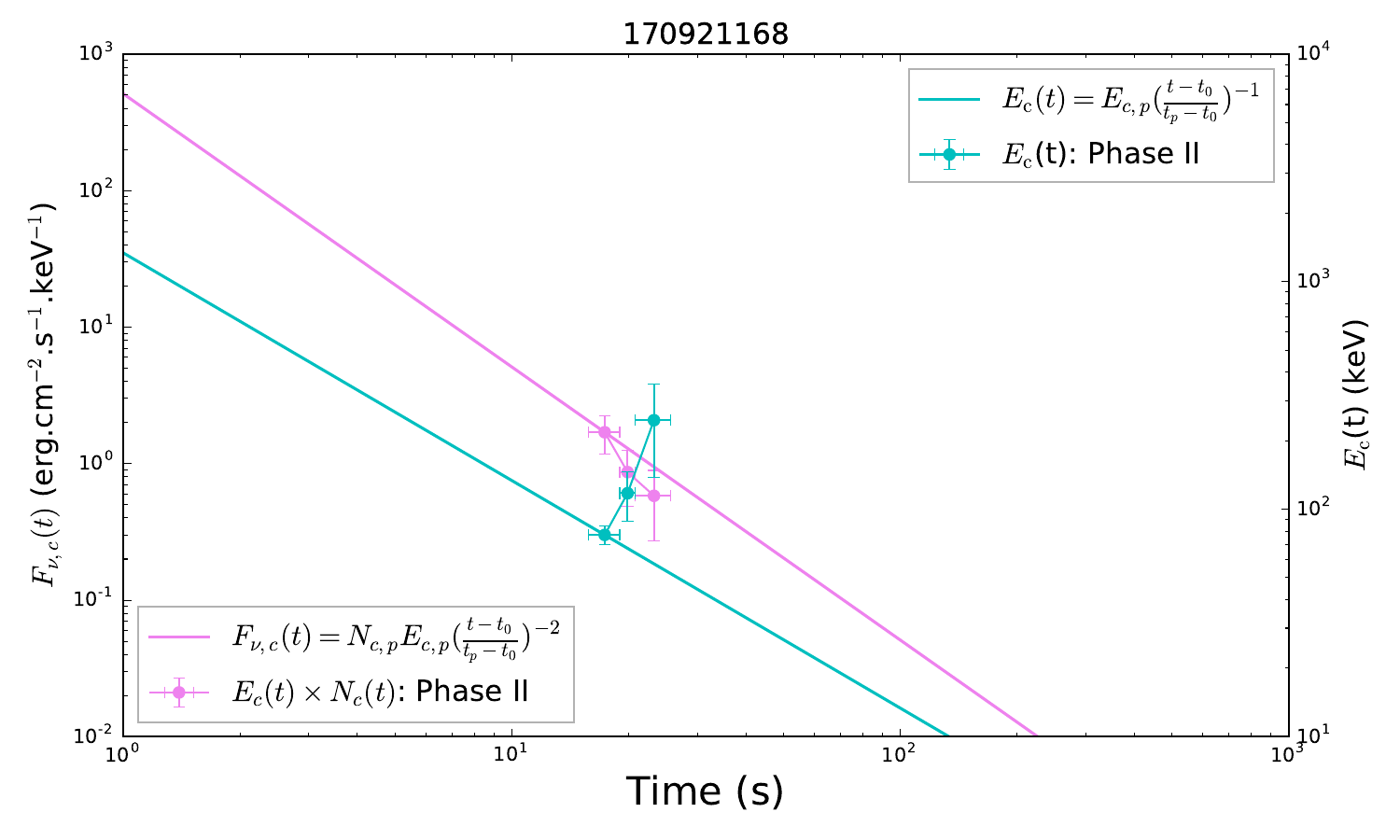}
\includegraphics[width=0.5\hsize,clip]{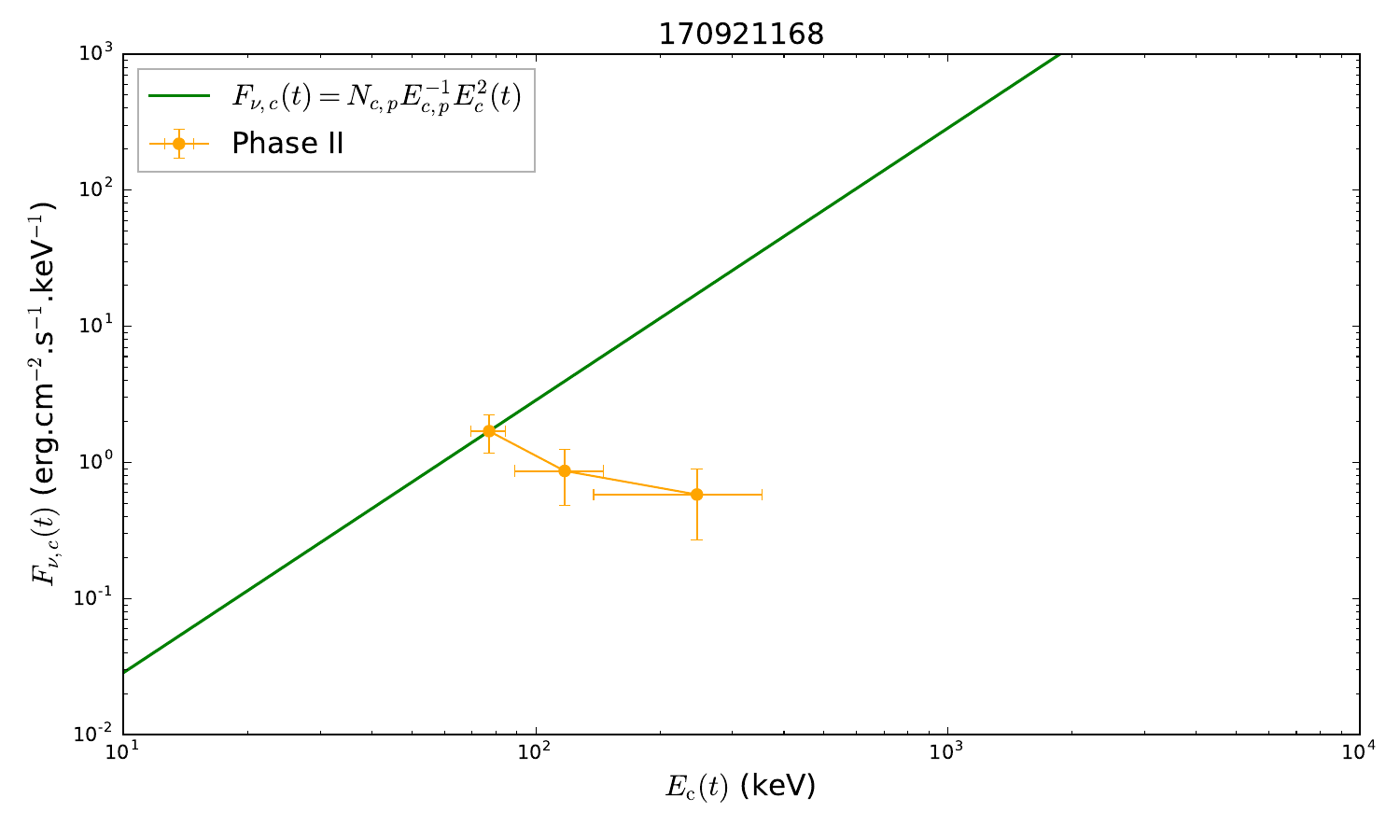}
\includegraphics[width=0.5\hsize,clip]{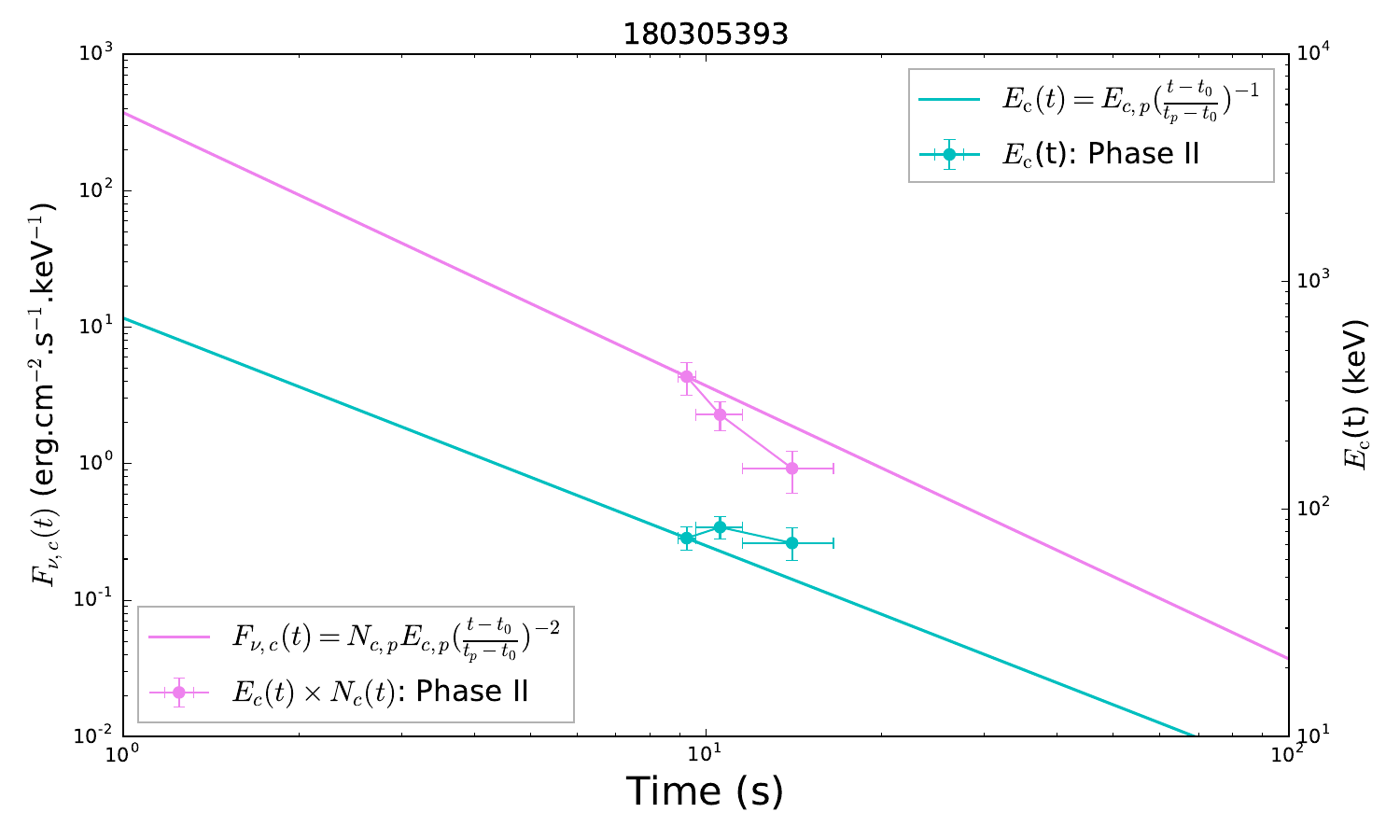}
\includegraphics[width=0.5\hsize,clip]{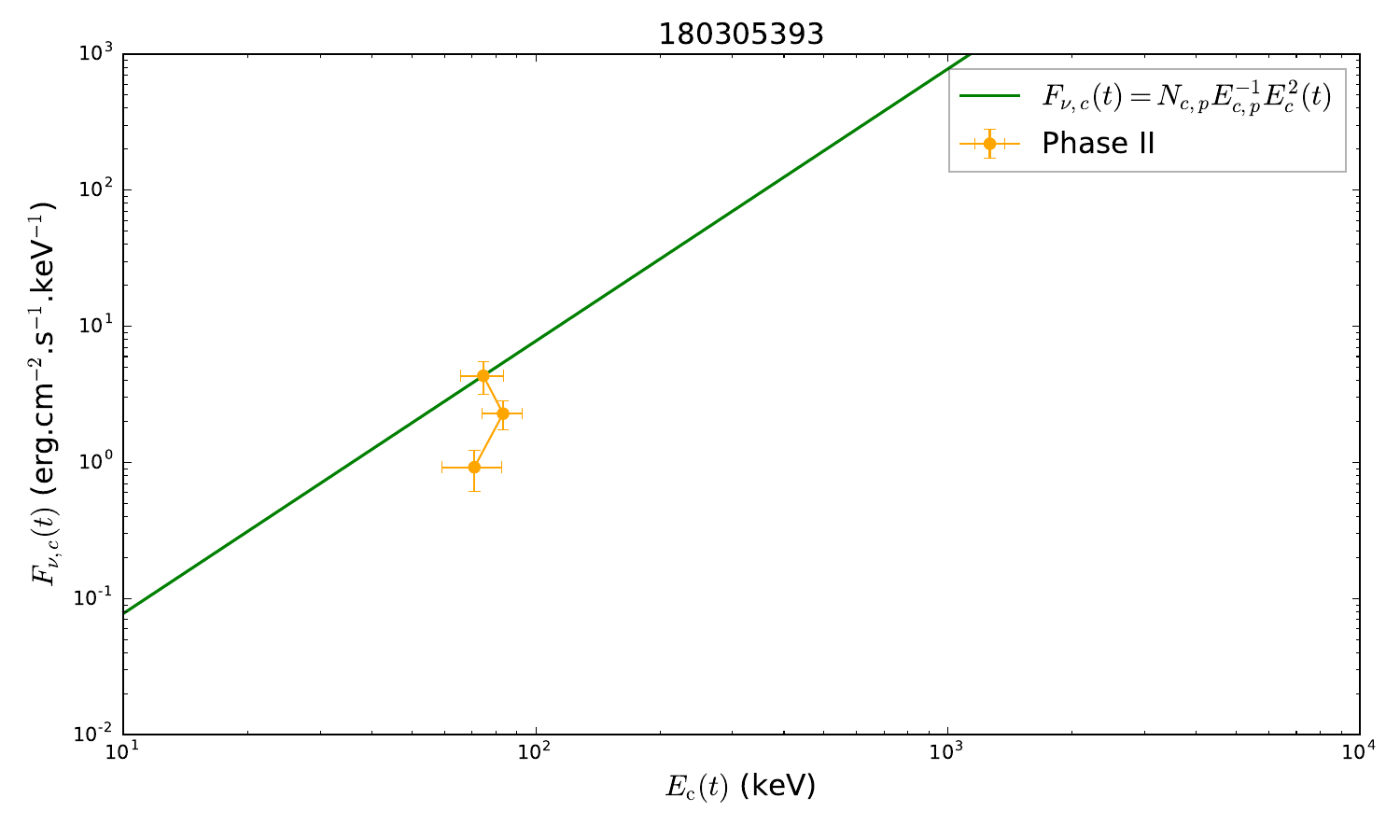}
\center{Fig. \ref{fig:CPL}--- Continued}
\end{figure*}

\clearpage
\appendix
\setcounter{figure}{0}    
\setcounter{section}{0}
\setcounter{table}{0}
\renewcommand{\thesection}{A\arabic{section}}
\renewcommand{\thefigure}{A\arabic{figure}}
\renewcommand{\thetable}{A\arabic{table}}
\renewcommand{\theequation}{A\arabic{equation}}

In this appendix, we provide additional figures.
\begin{figure*}
\includegraphics[width=0.5\hsize,clip]{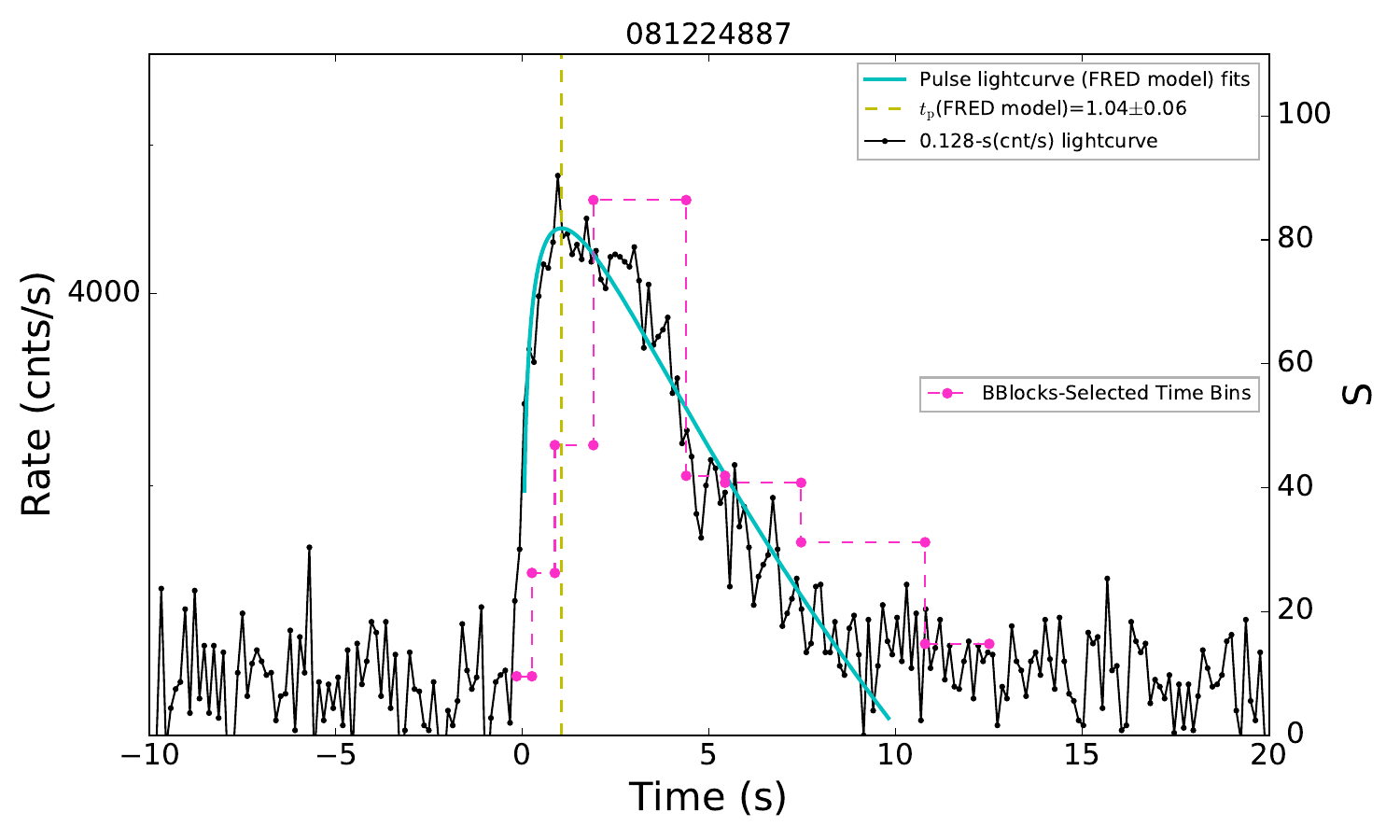}
\includegraphics[width=0.5\hsize,clip]{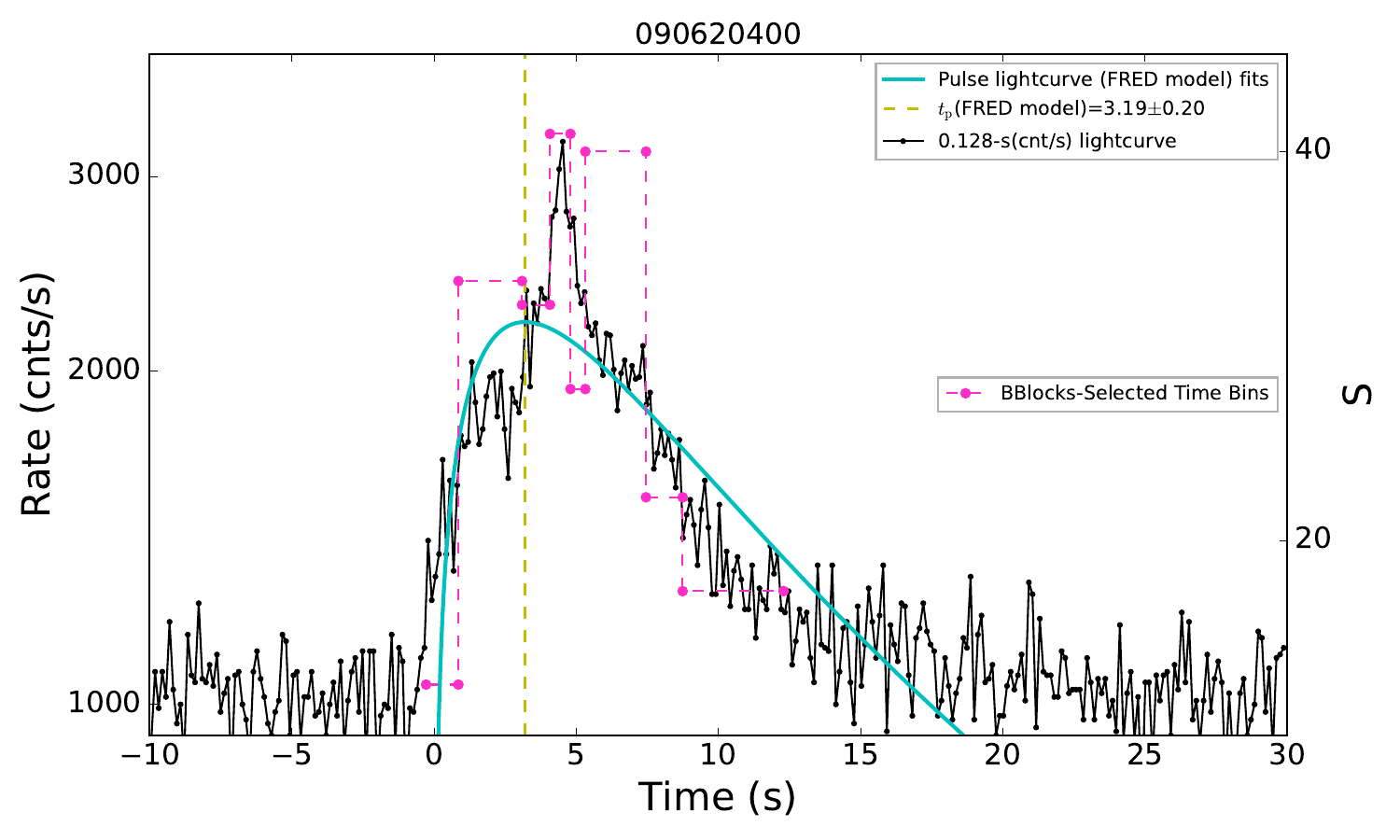}
\includegraphics[width=0.5\hsize,clip]{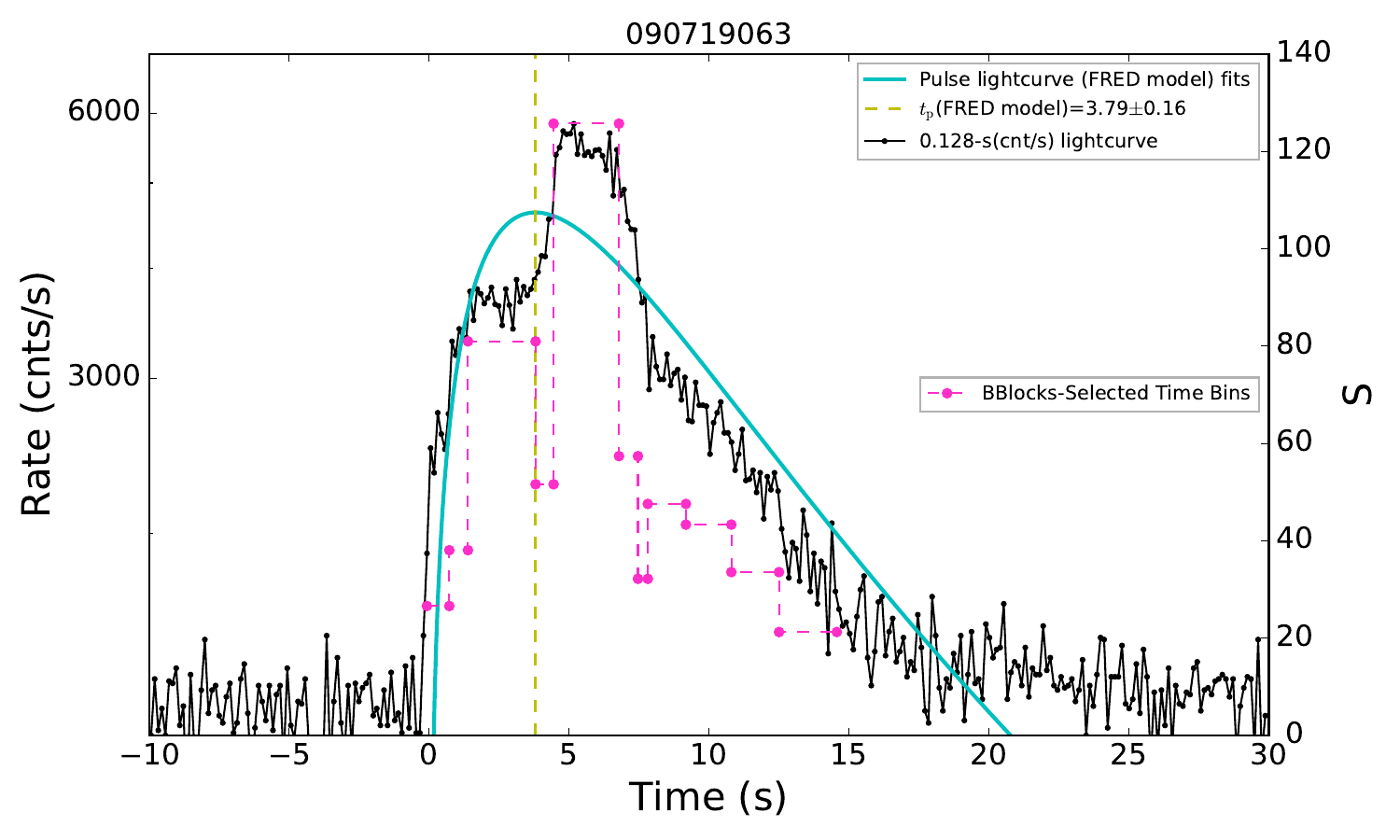}
\includegraphics[width=0.5\hsize,clip]{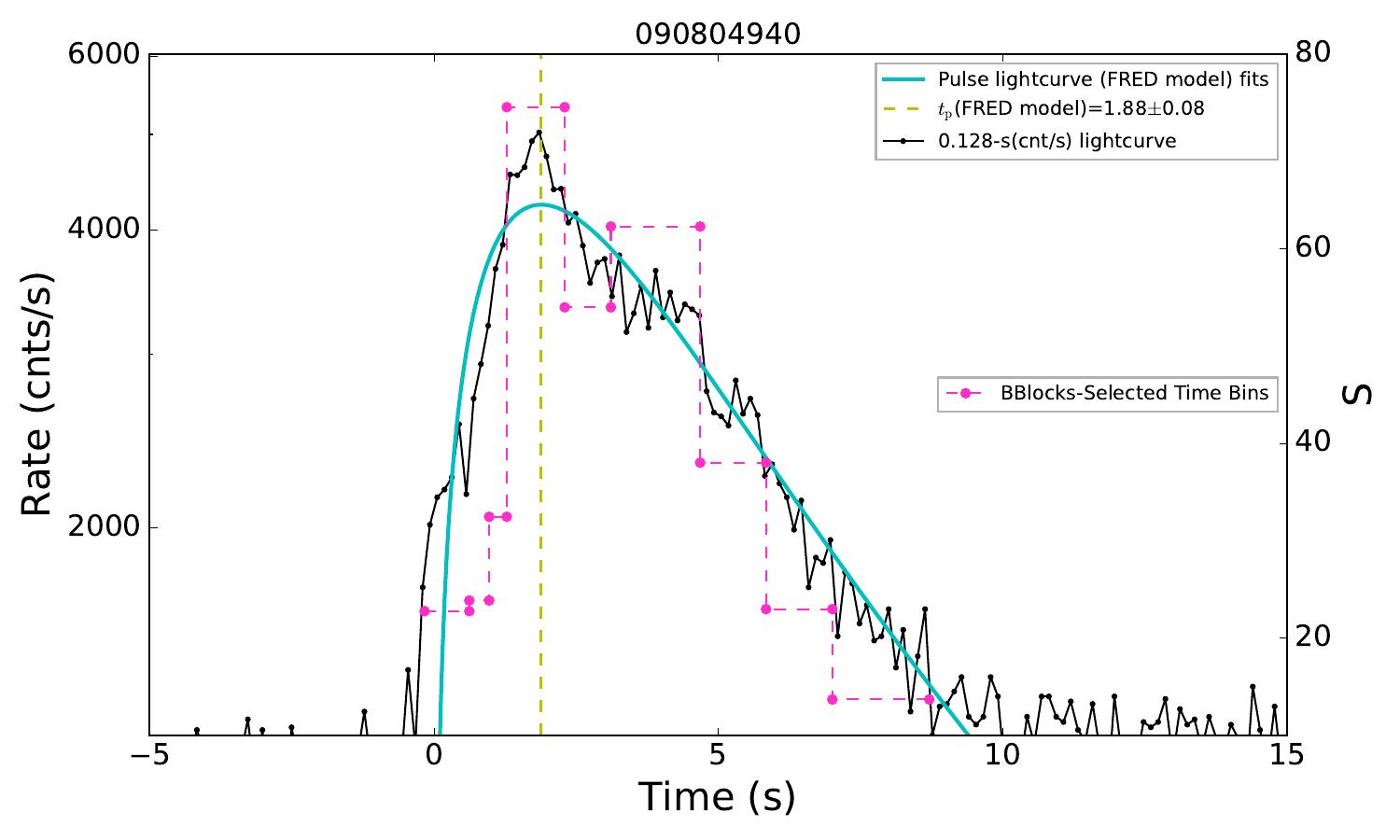}
\includegraphics[width=0.5\hsize,clip]{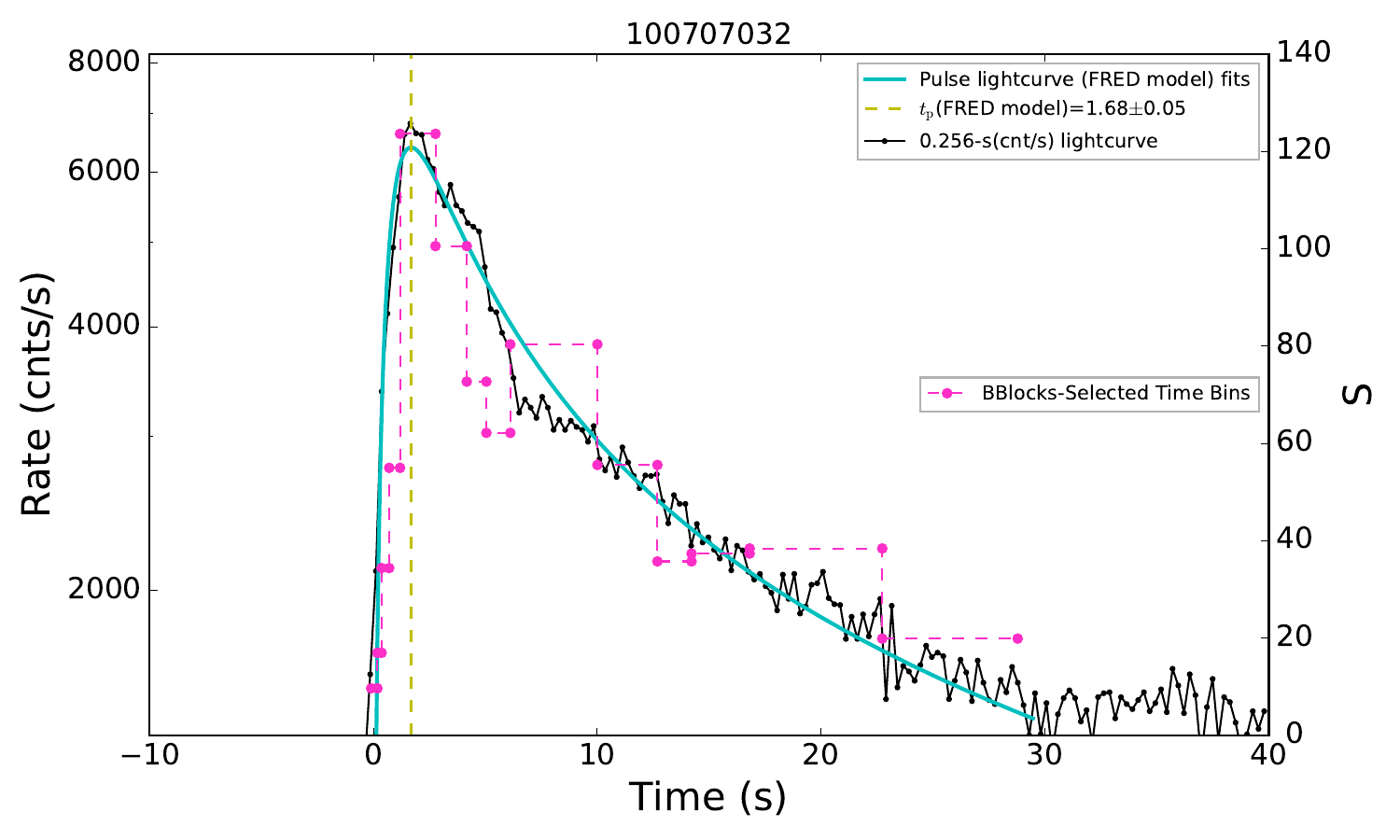}
\includegraphics[width=0.5\hsize,clip]{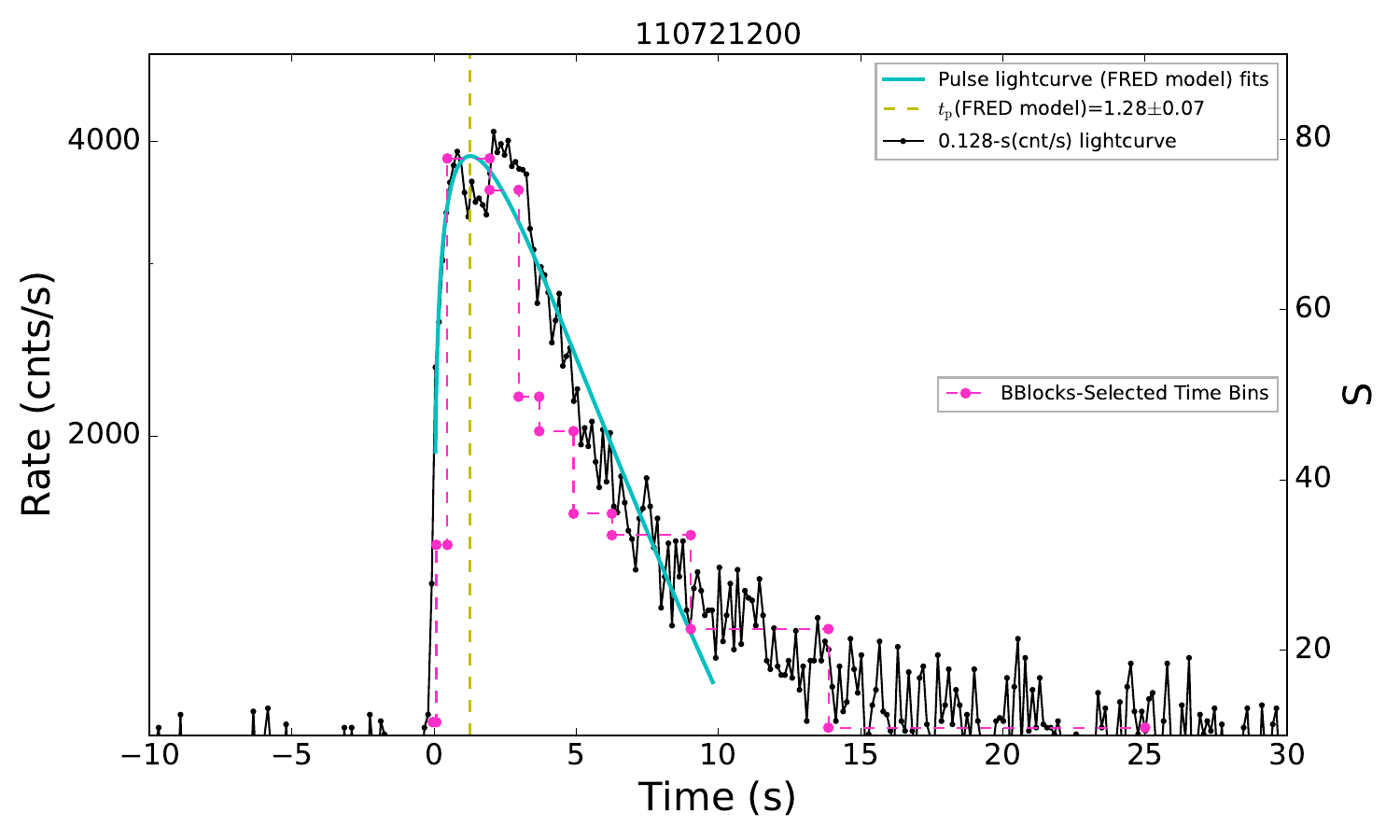}
\includegraphics[width=0.5\hsize,clip]{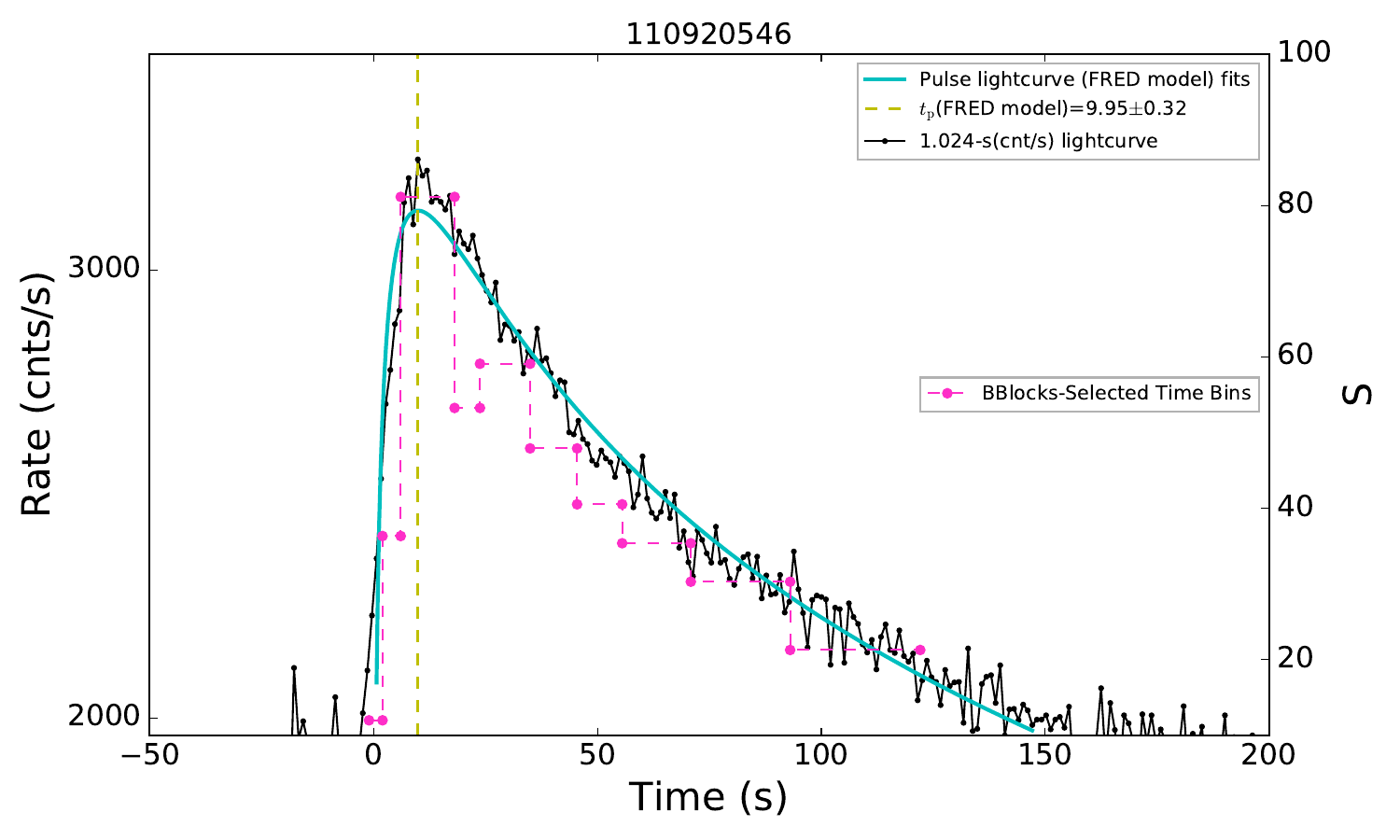}
\includegraphics[width=0.5\hsize,clip]{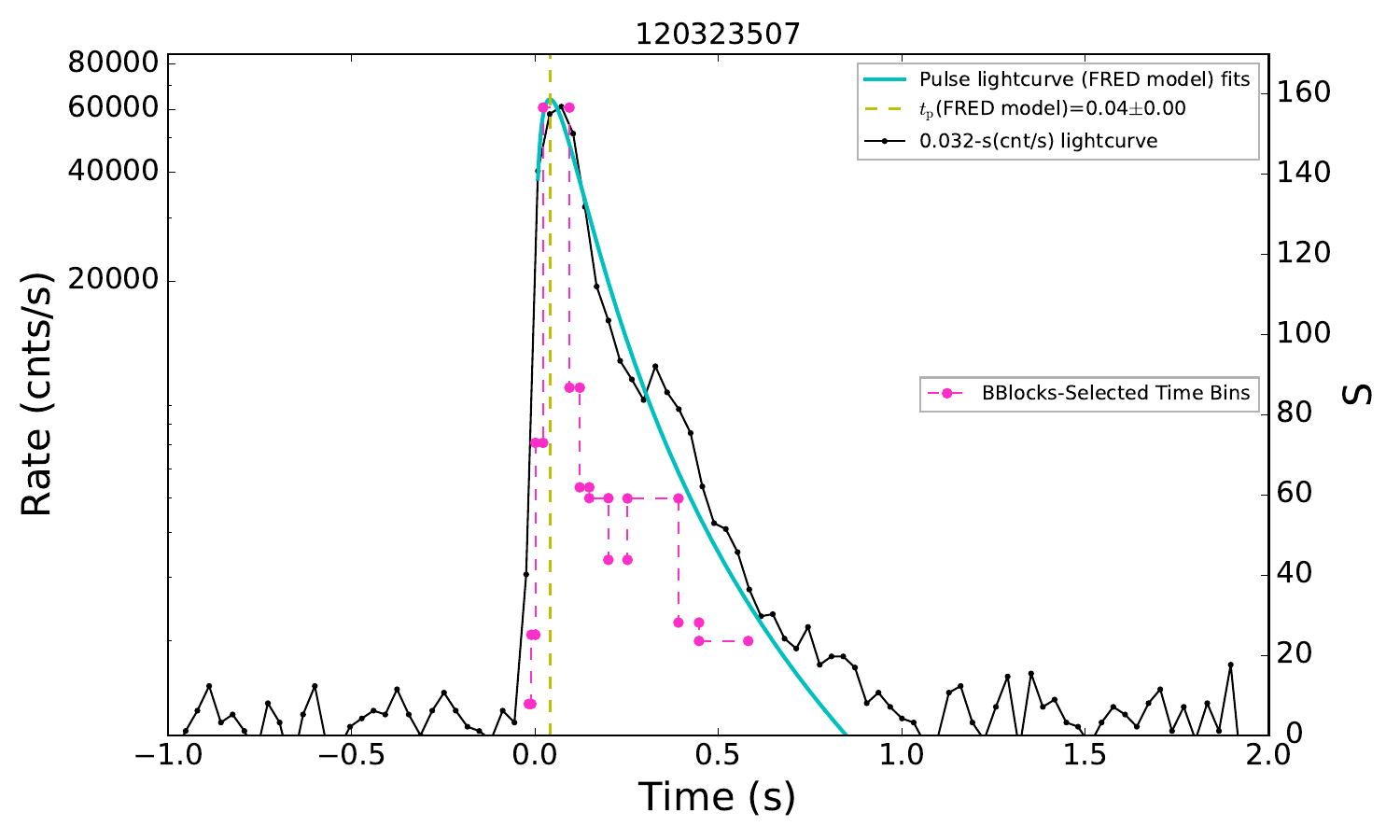}
\caption{Count rate lightcurves, as well as their best-fit results using the FRED model. Solid points connected by the solid line in  black color represent the lightcurve, while the solid lines marked with cyan color are the best FRED model fits. The peak times obtained from the best-fit FRED model are indicated by the yellow vertical dashed line. Solid points connected by the dashed line in pink color represent the time bins selected using the BBlocks method.}\label{fig:FRED}
\end{figure*}
\begin{figure*}
\includegraphics[width=0.5\hsize,clip]{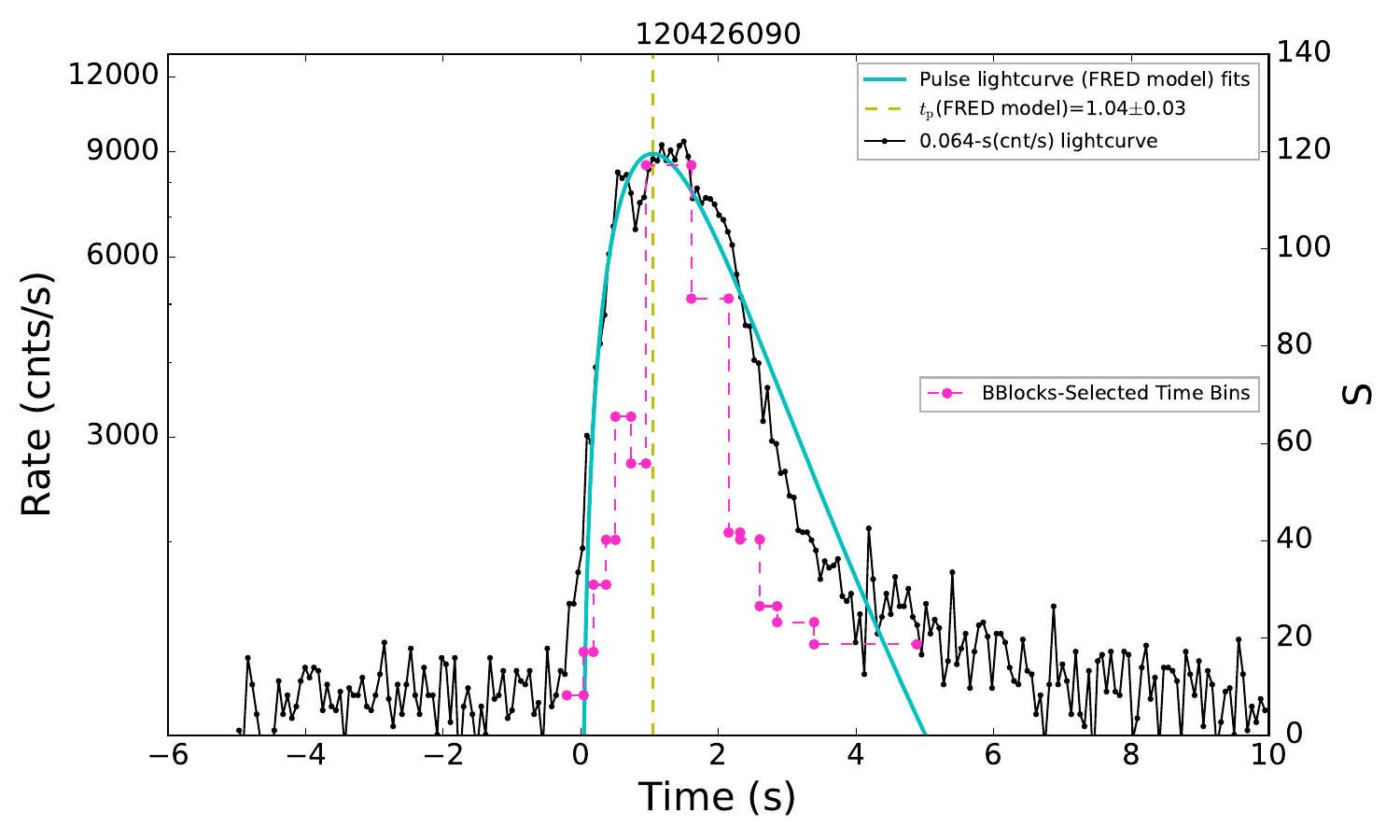}
\includegraphics[width=0.5\hsize,clip]{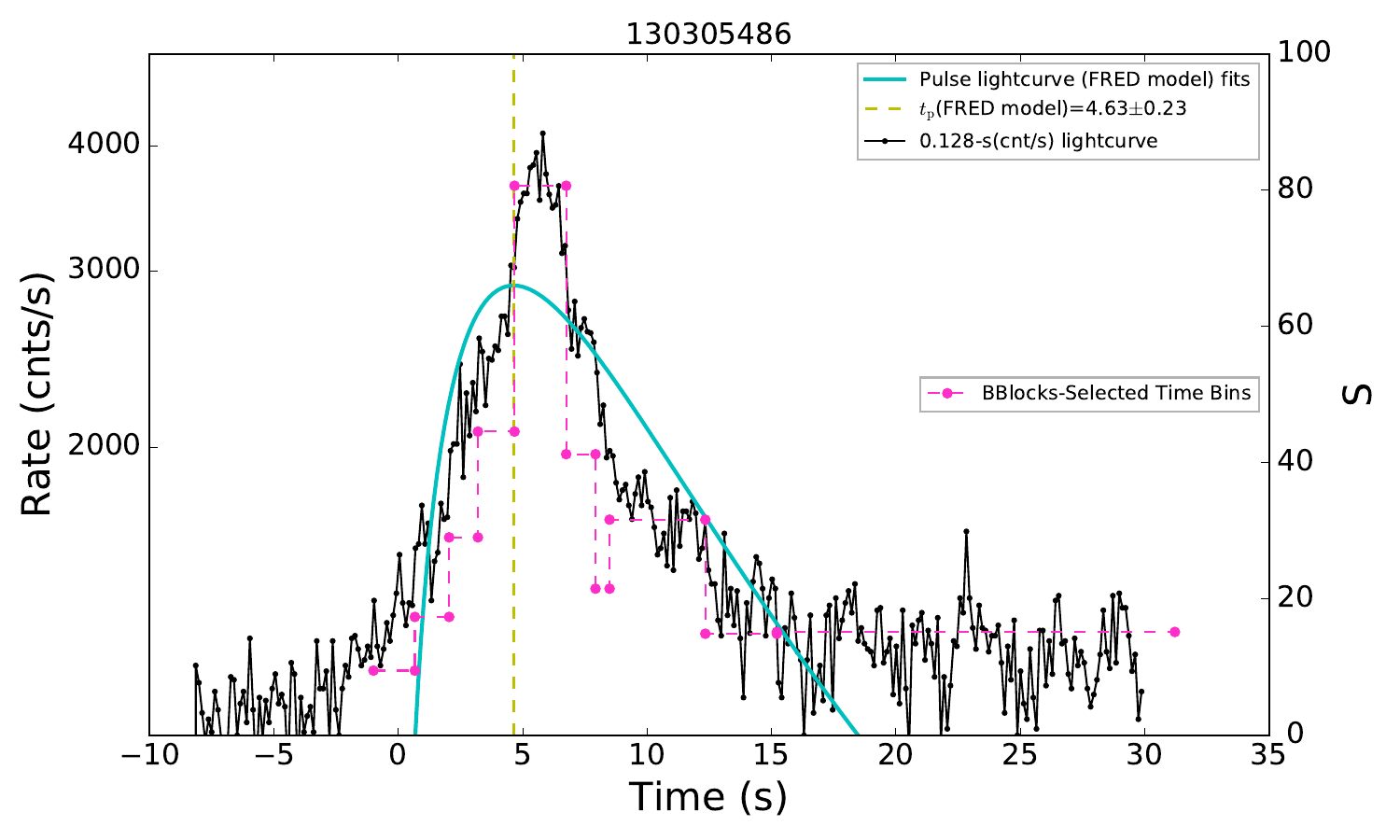}
\includegraphics[width=0.5\hsize,clip]{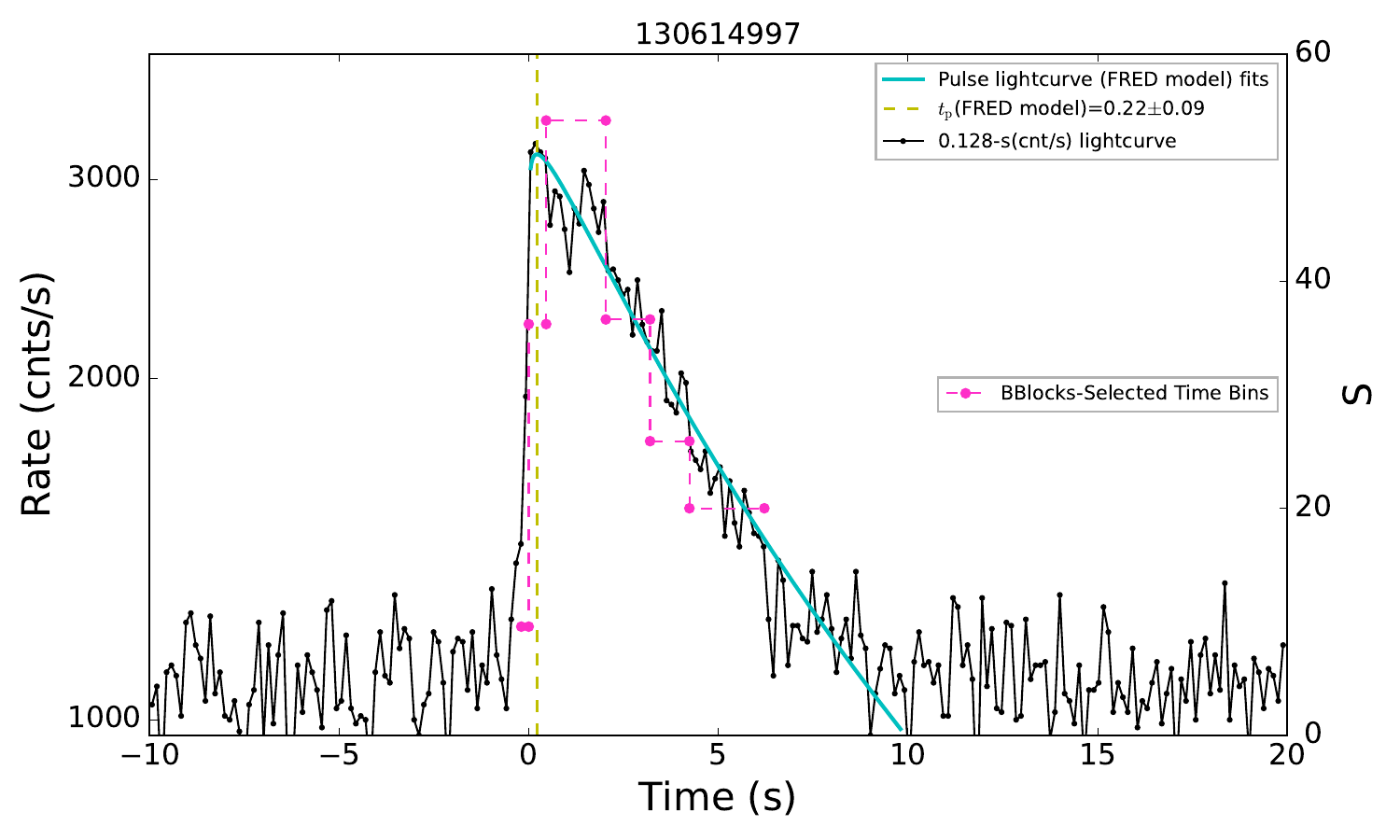}
\includegraphics[width=0.5\hsize,clip]{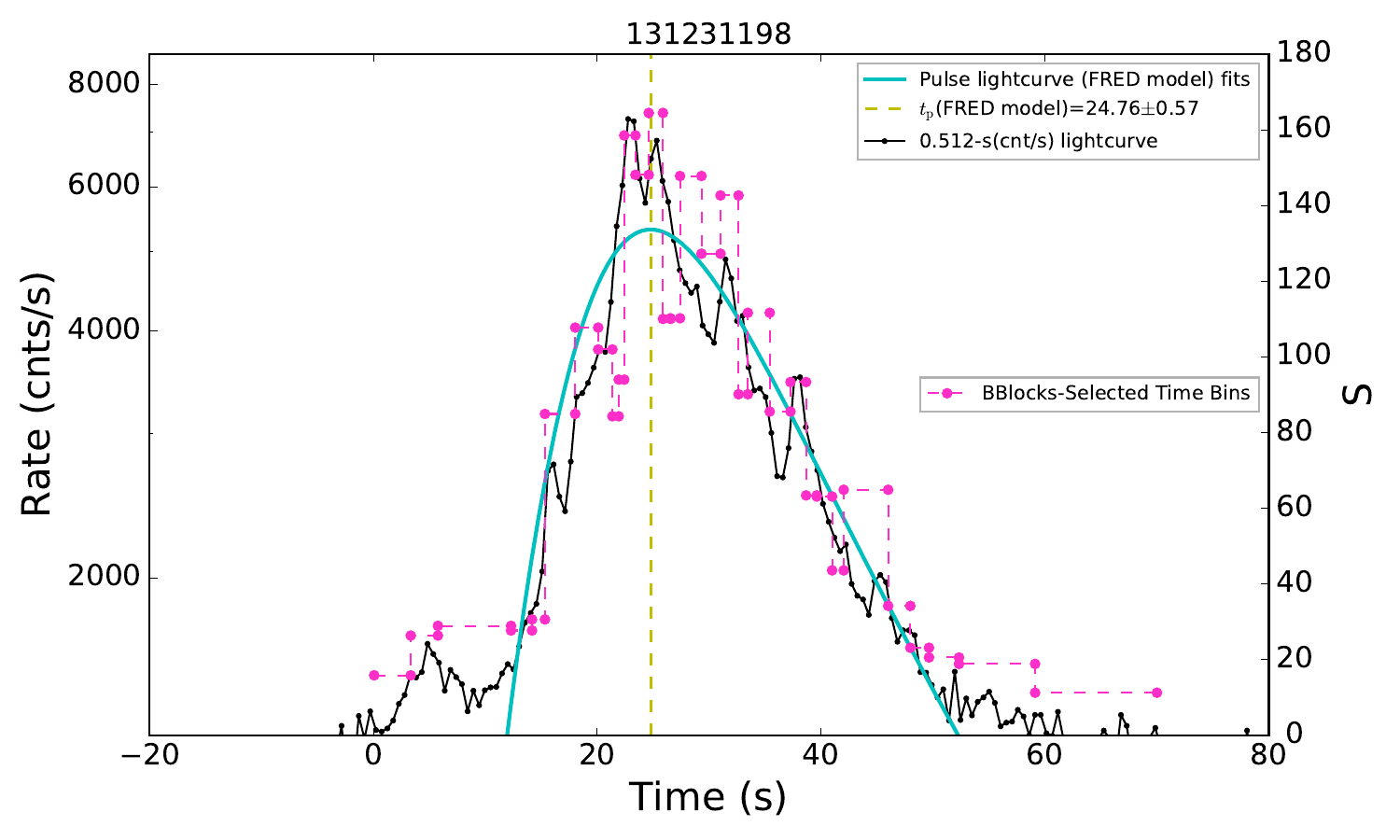}
\includegraphics[width=0.5\hsize,clip]{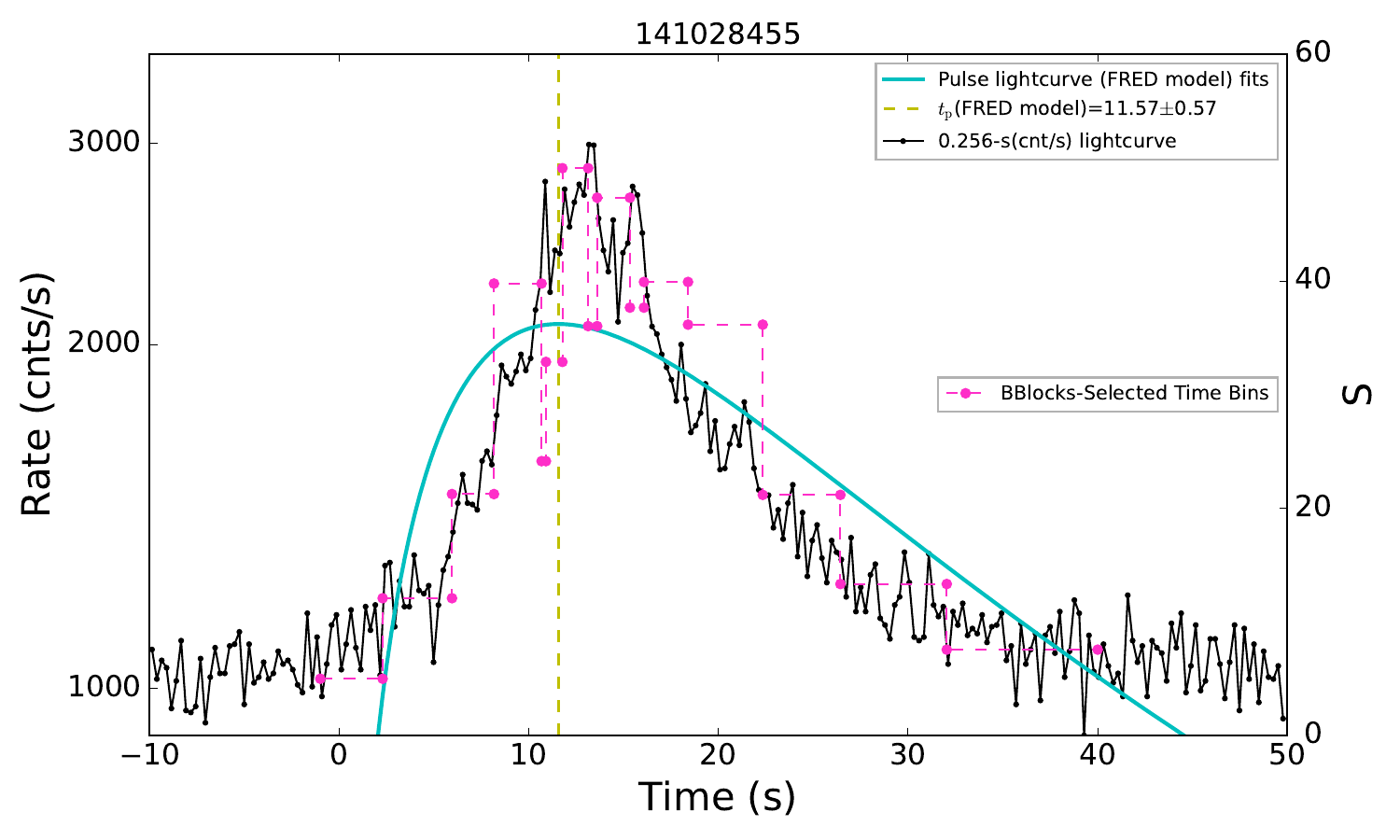}
\includegraphics[width=0.5\hsize,clip]{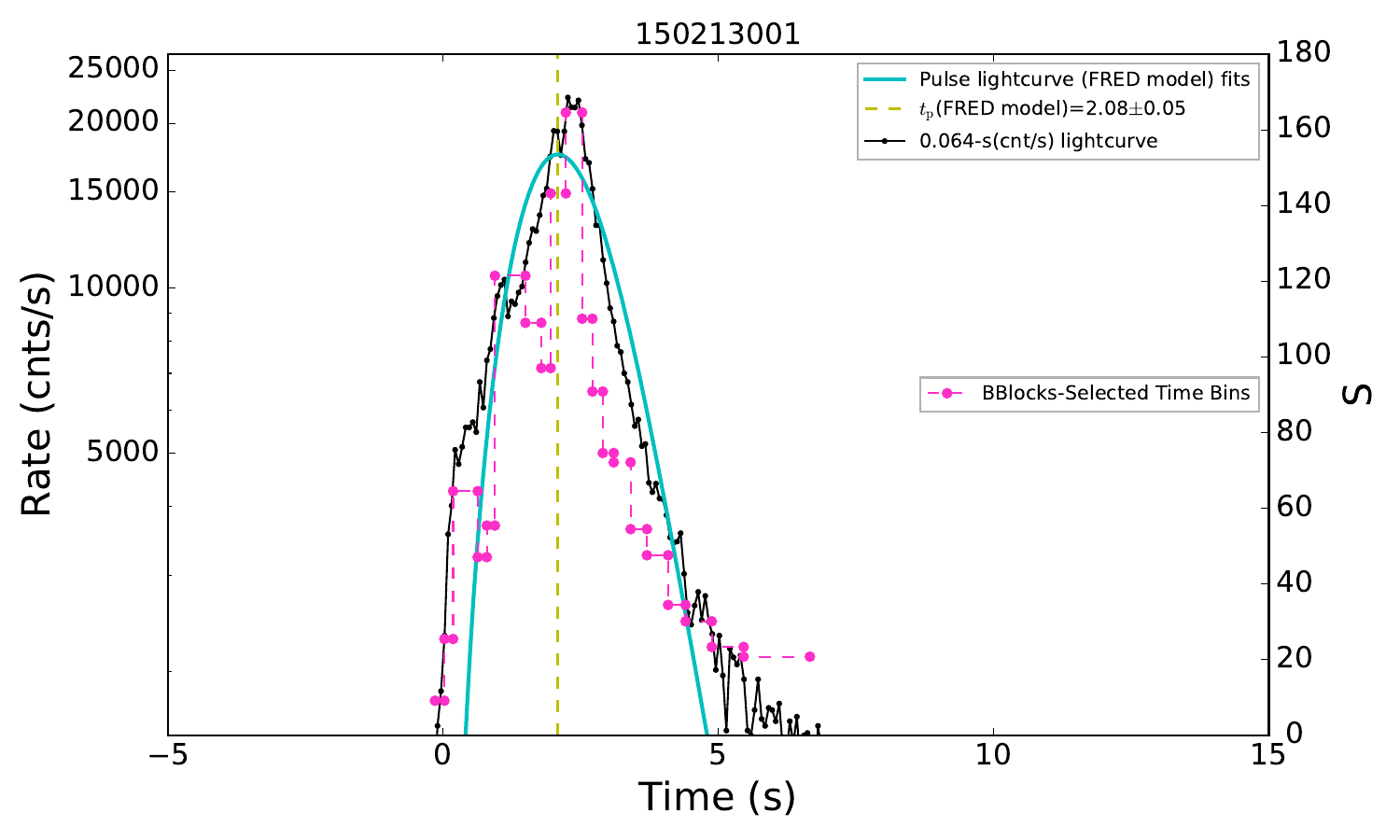}
\includegraphics[width=0.5\hsize,clip]{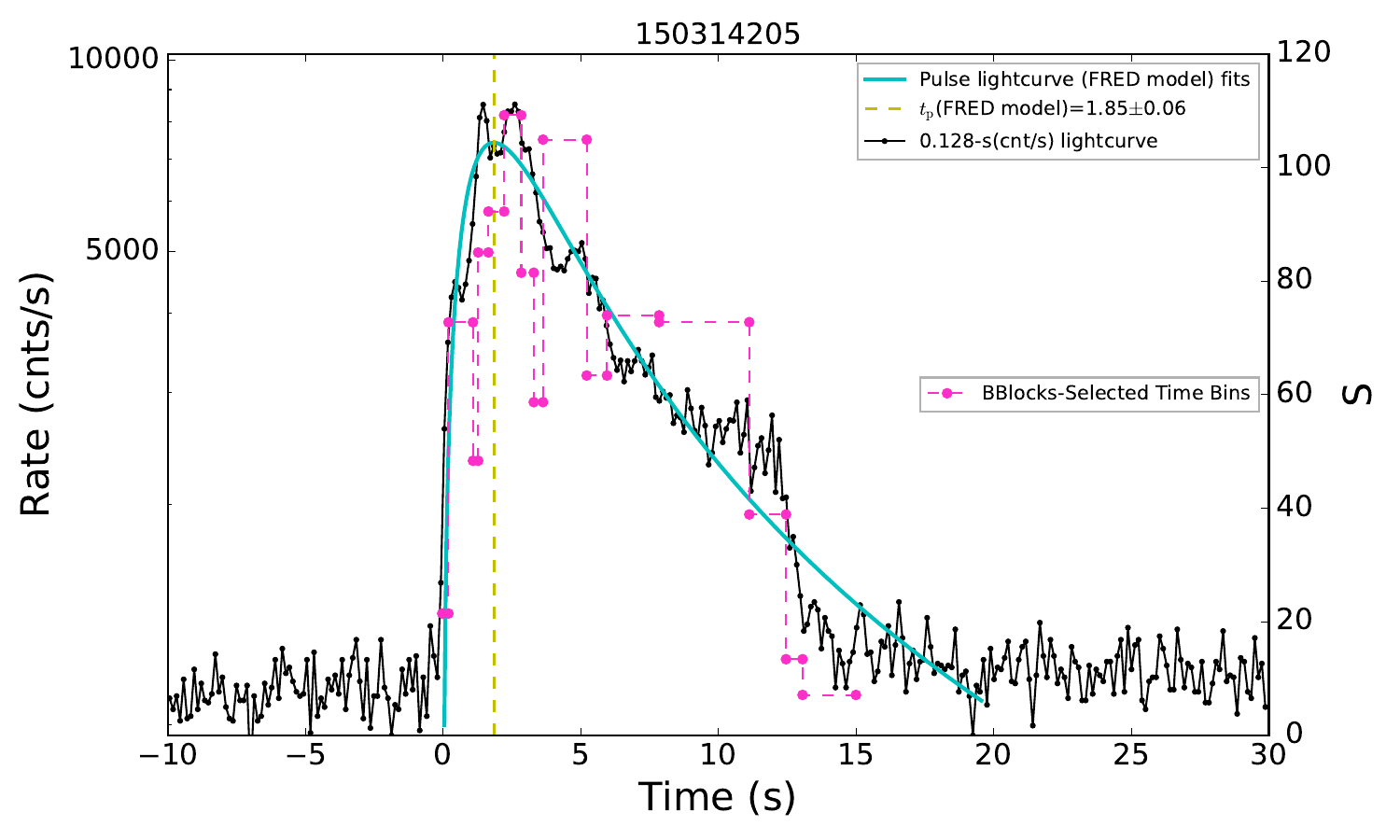}
\includegraphics[width=0.5\hsize,clip]{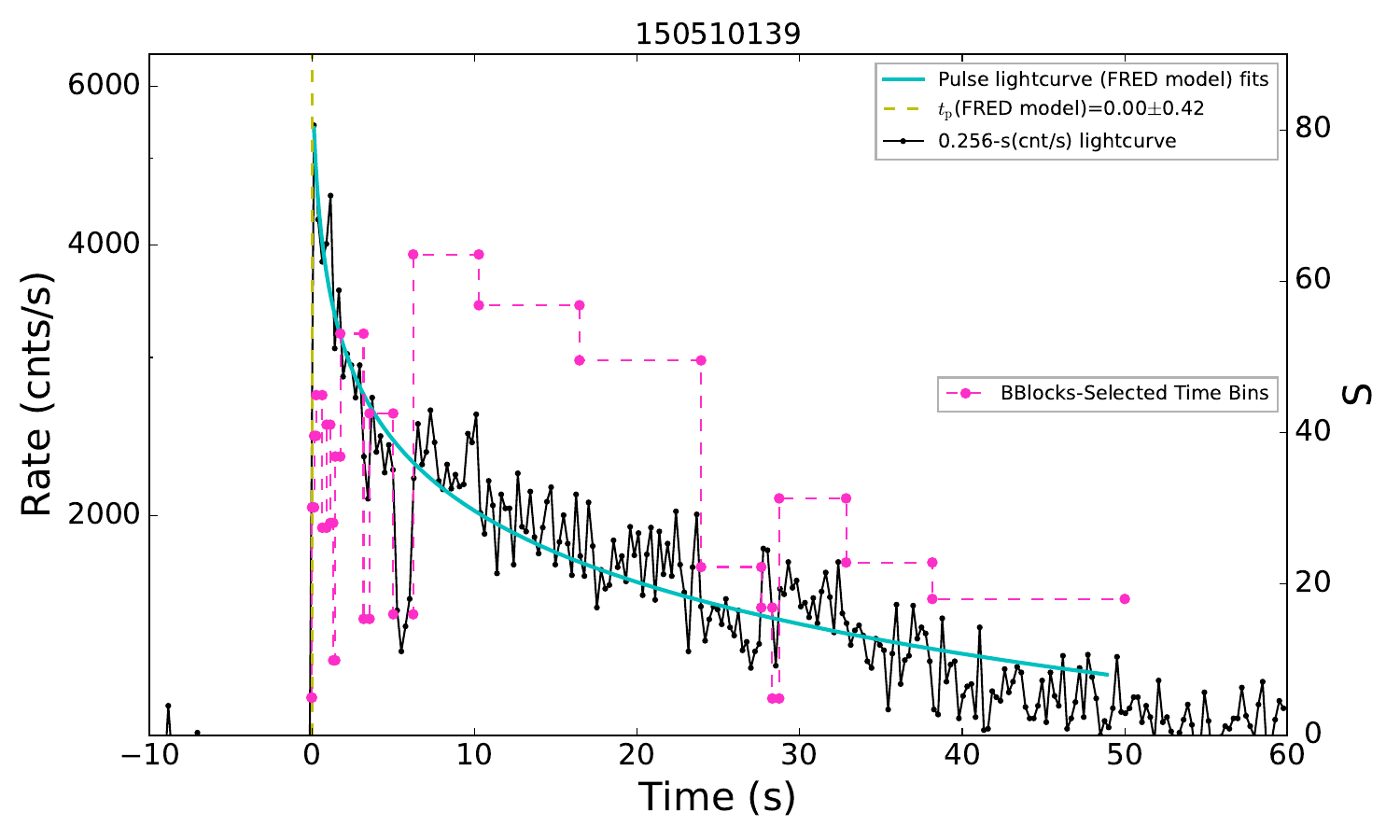}
\center{Fig. \ref{fig:FRED}--- Continued}
\end{figure*}
\begin{figure*}
\includegraphics[width=0.5\hsize,clip]{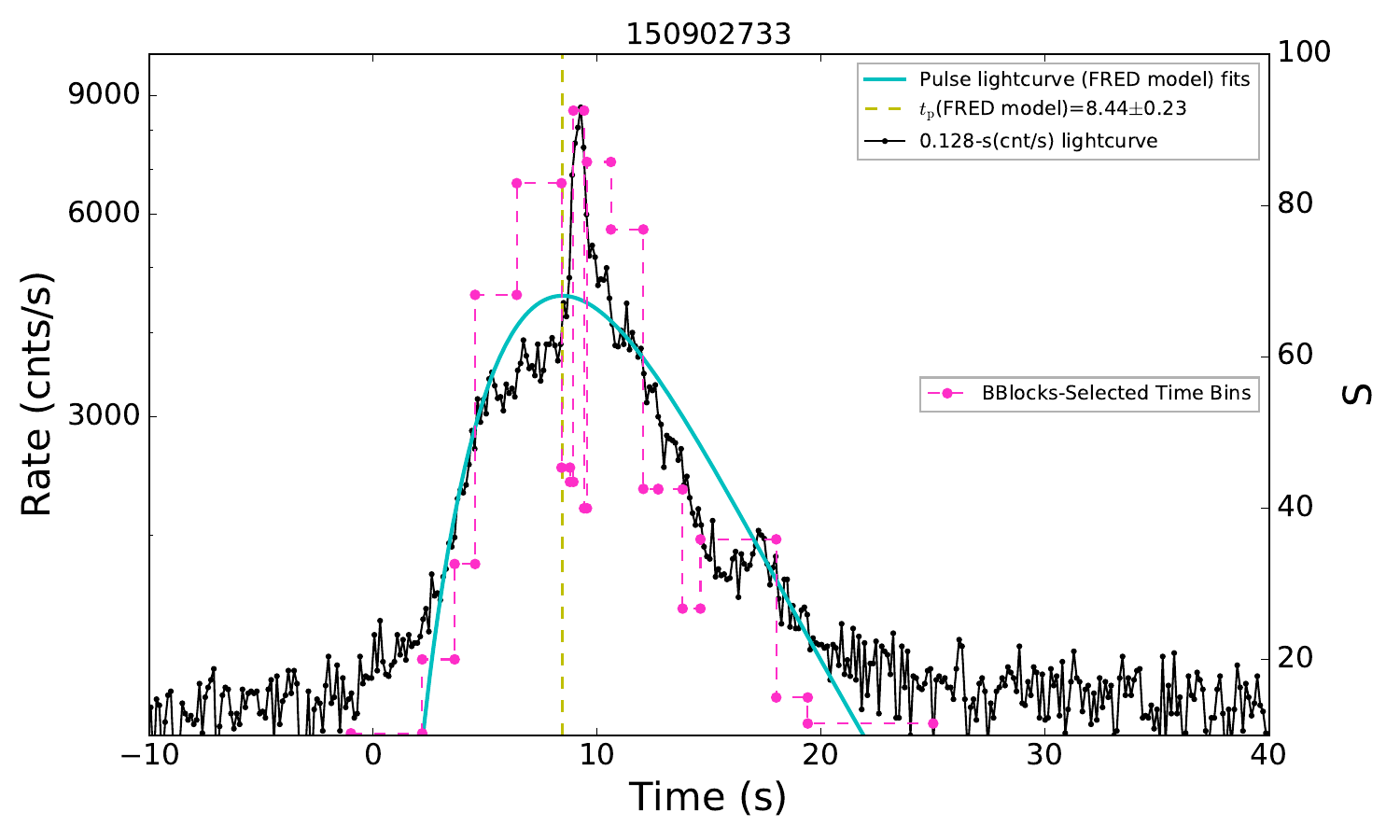}
\includegraphics[width=0.5\hsize,clip]{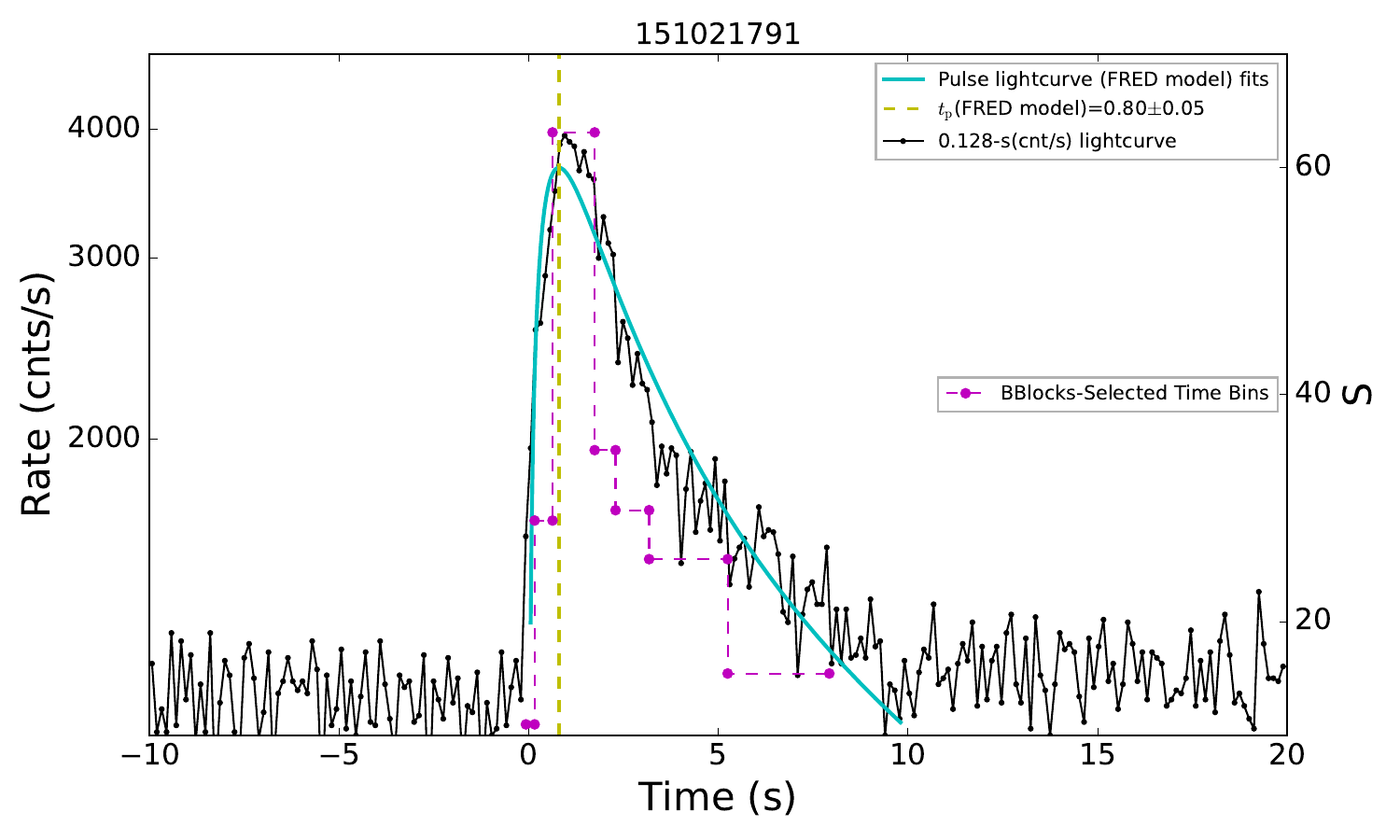}
\includegraphics[width=0.5\hsize,clip]{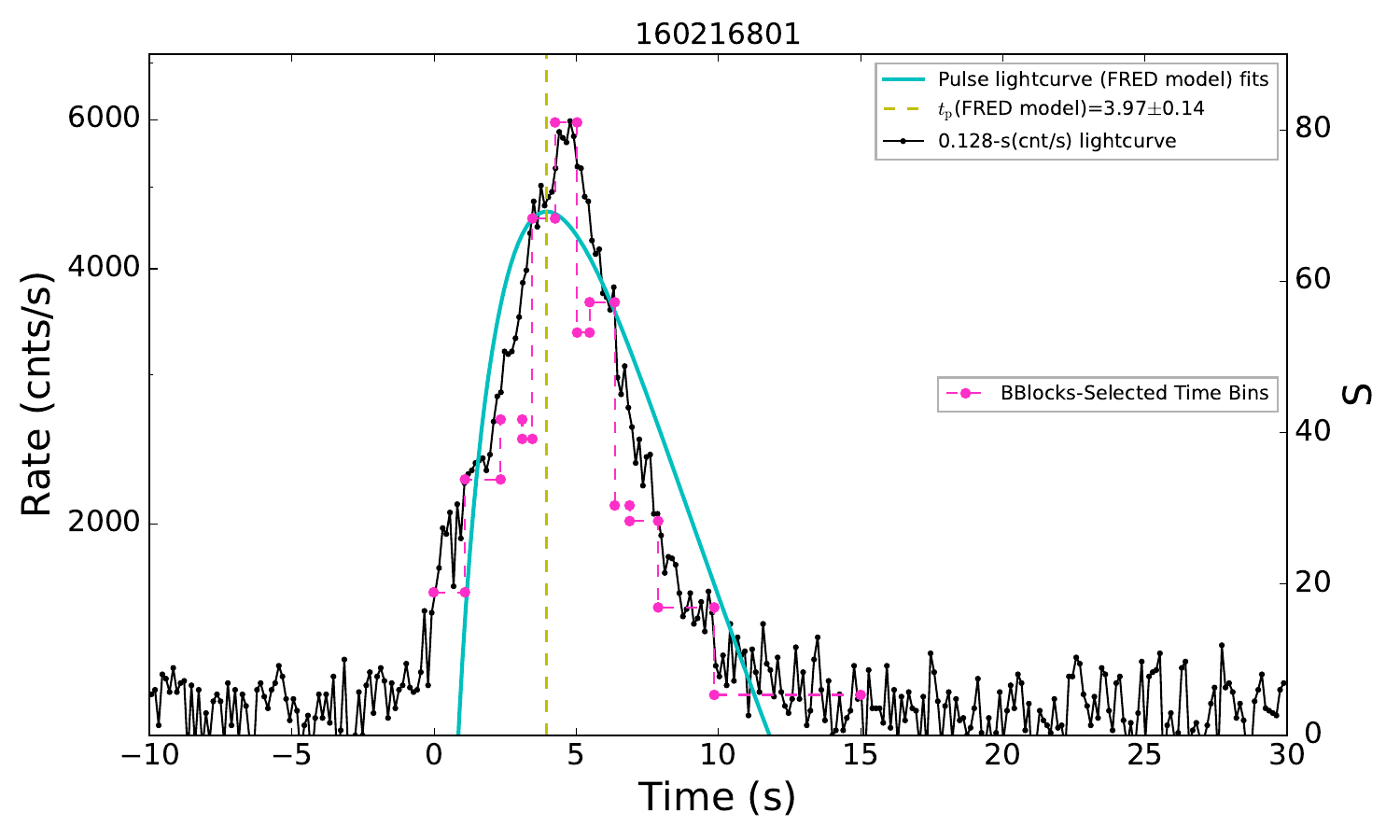}
\includegraphics[width=0.5\hsize,clip]{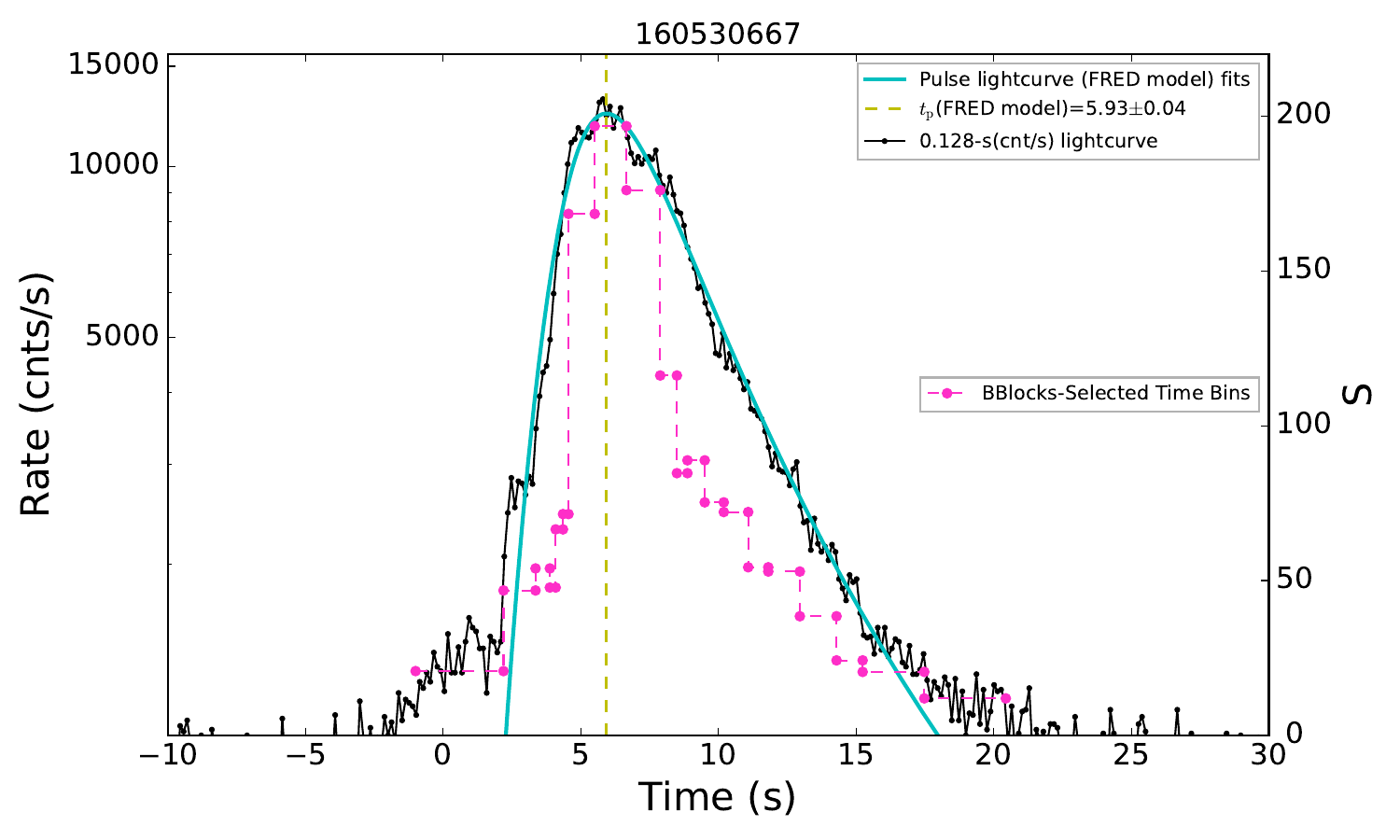}
\includegraphics[width=0.5\hsize,clip]{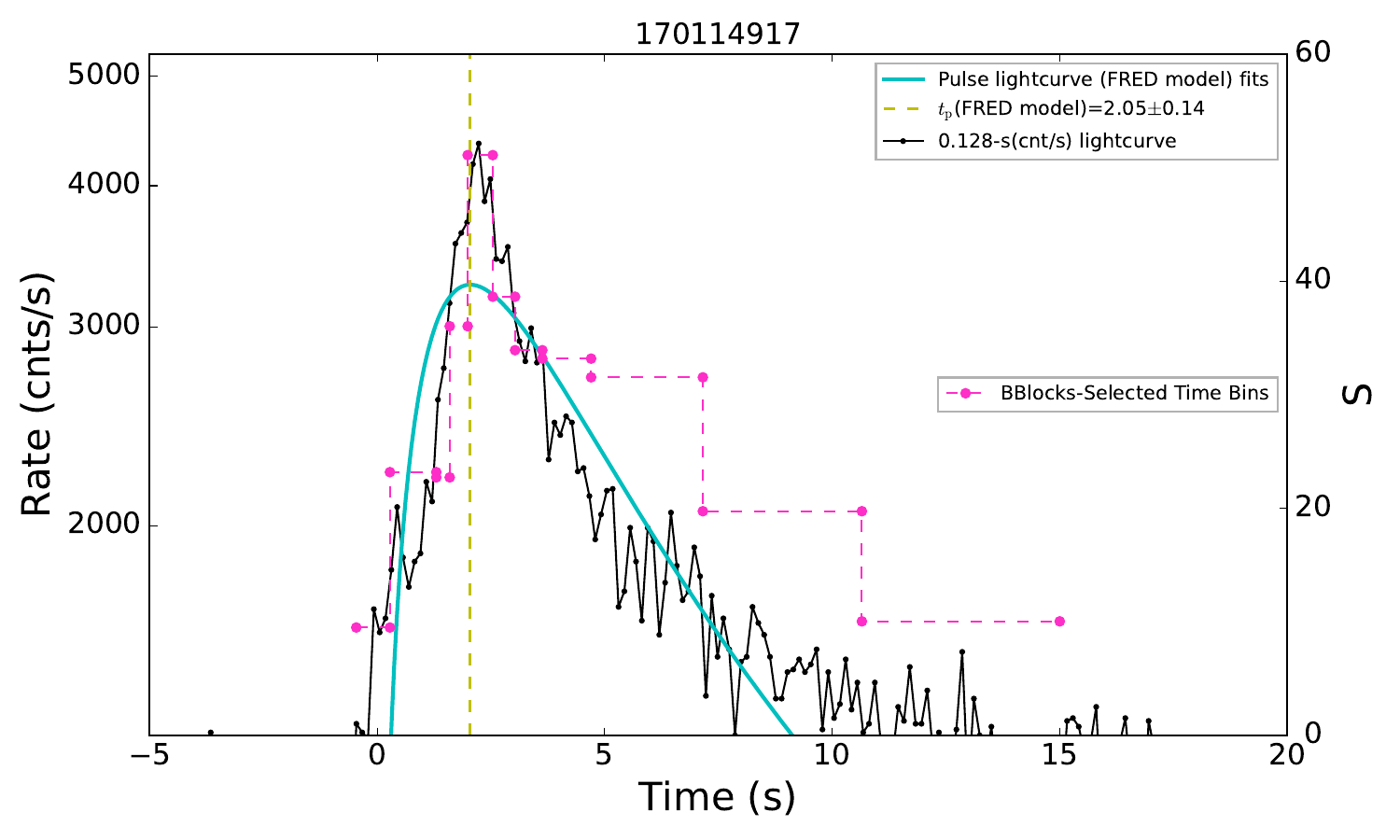}
\includegraphics[width=0.5\hsize,clip]{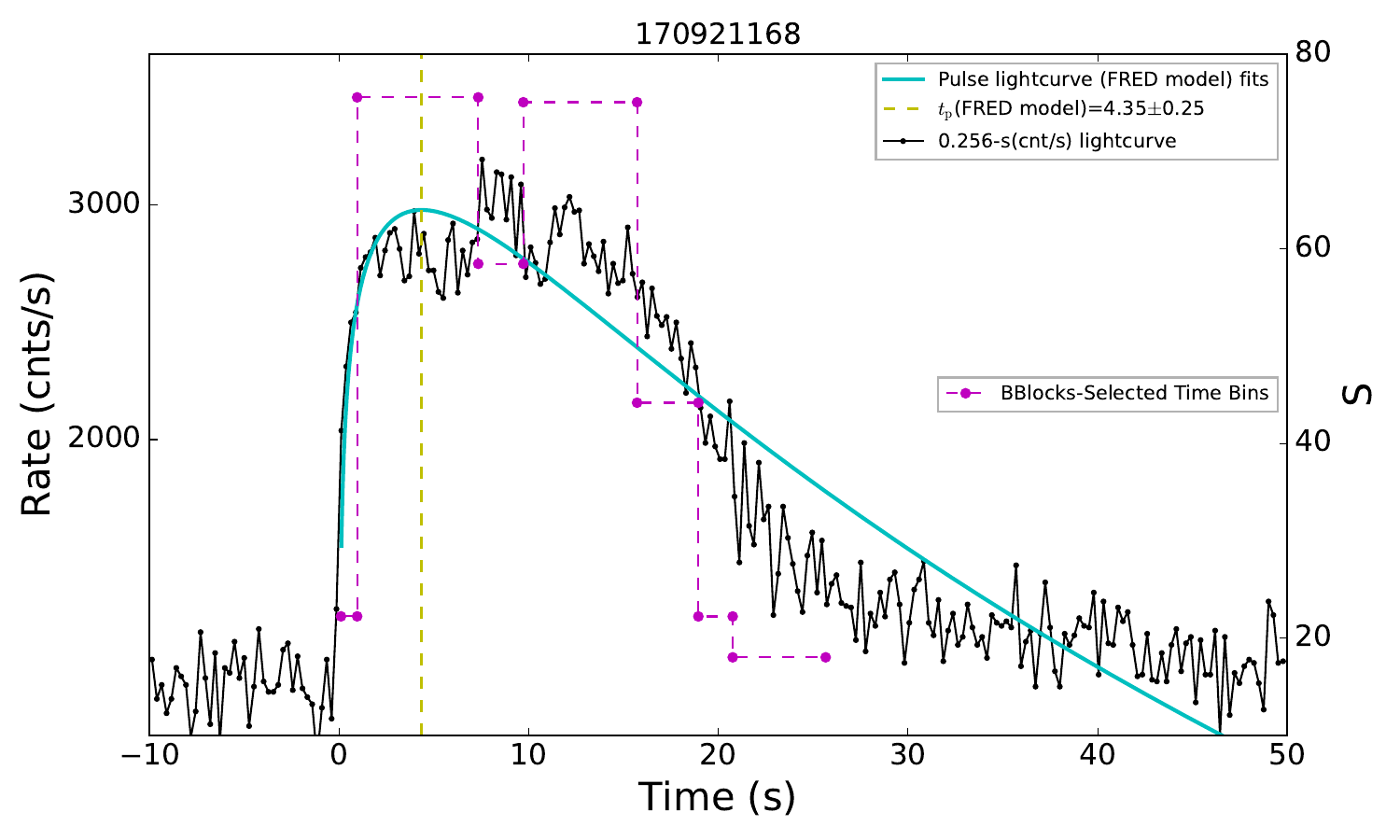}
\includegraphics[width=0.5\hsize,clip]{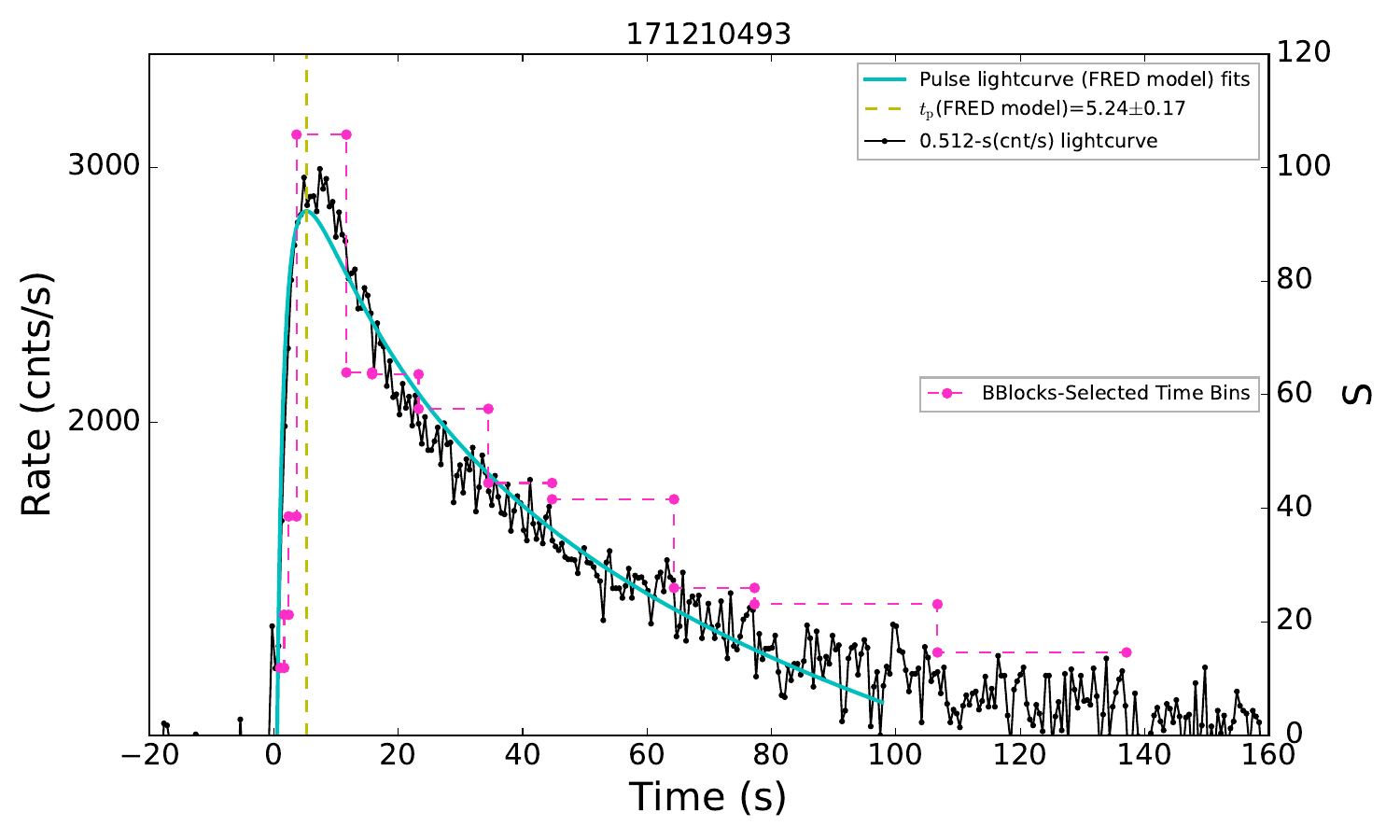}
\includegraphics[width=0.5\hsize,clip]{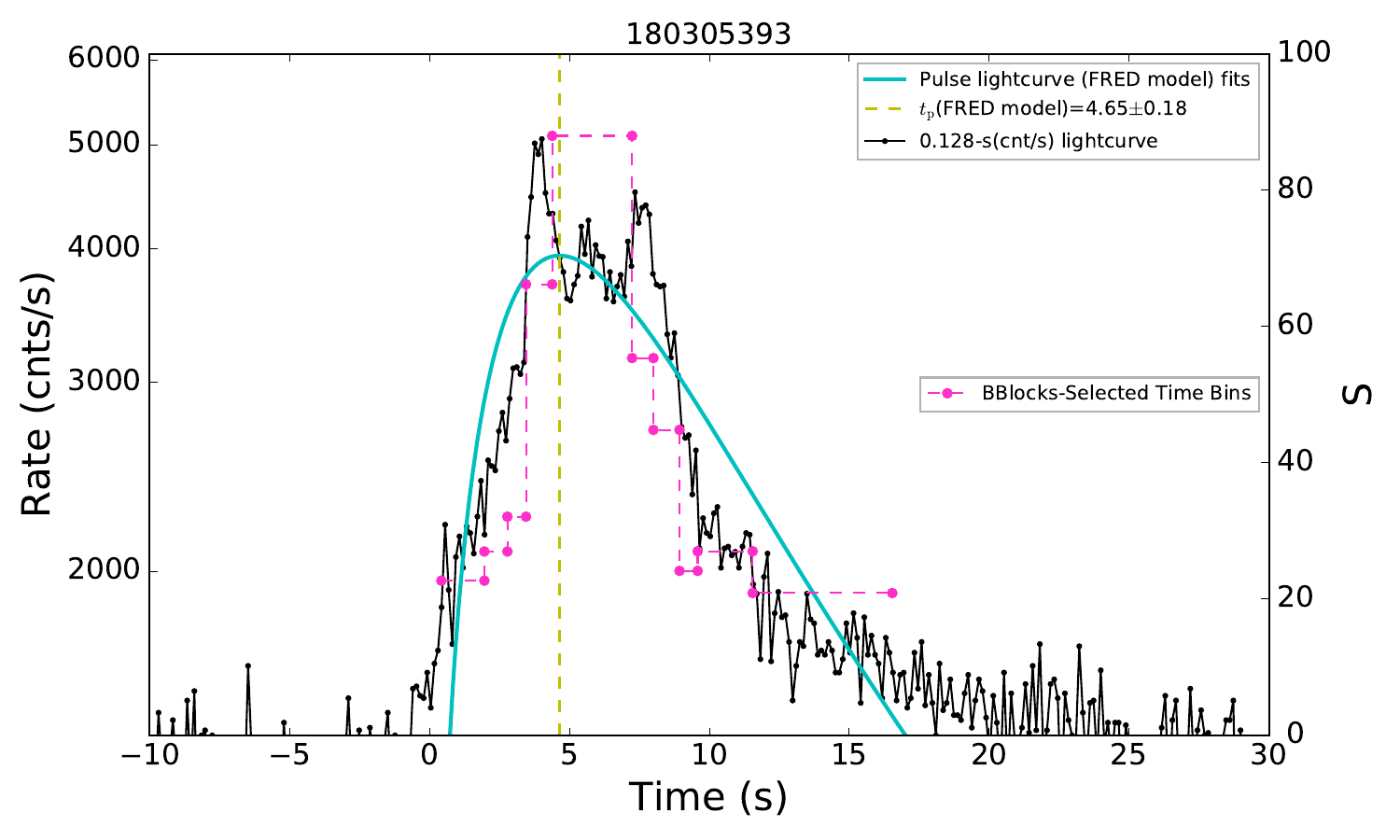}
\center{Fig. \ref{fig:FRED}--- Continued}
\end{figure*}
	
\begin{figure*}
\centering
\includegraphics[width=0.8\hsize,clip]{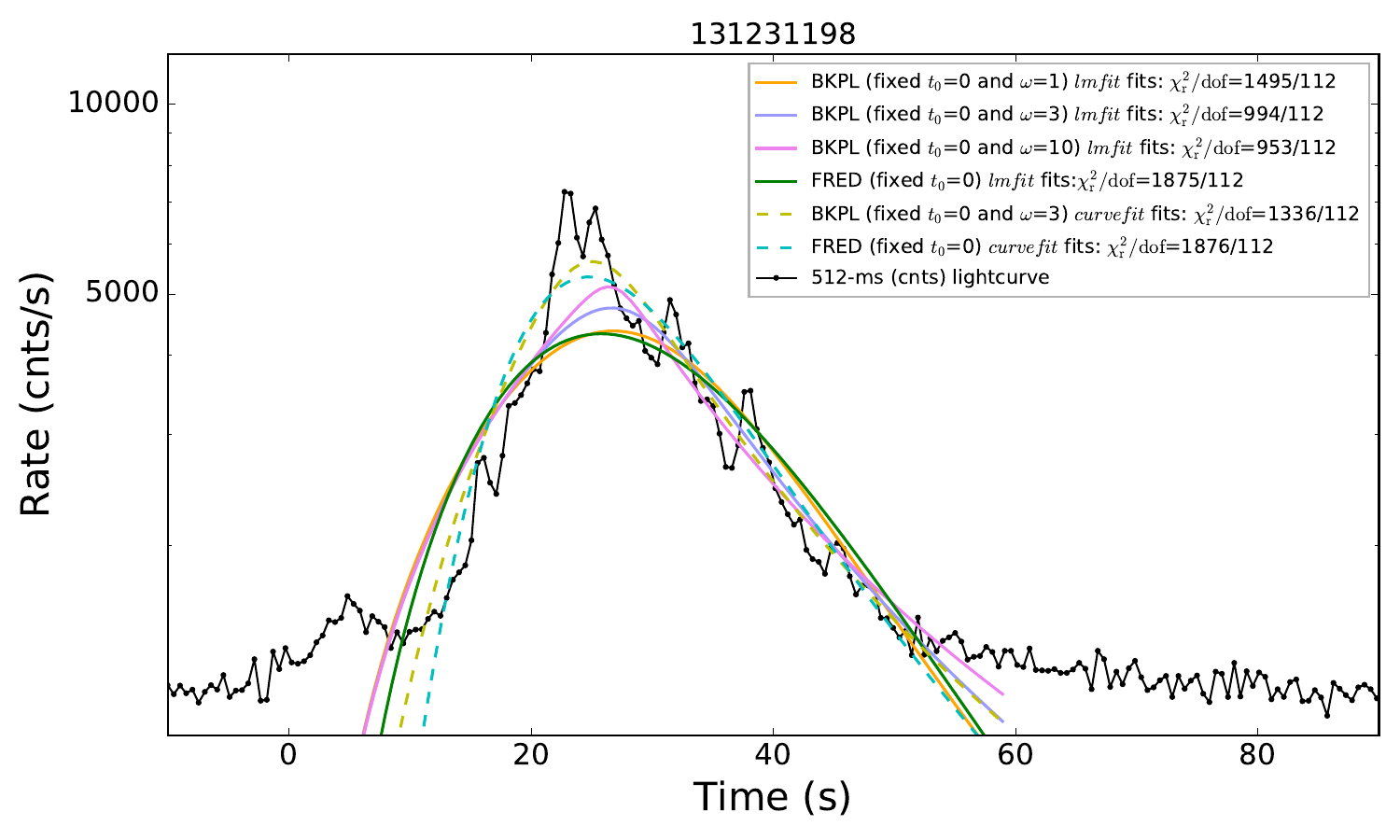}
\caption{Example of the best fits of the count-rate lightcurve for GRB 131231198 with different models (comparing FRED with BKPL) or packages used (comparing $lmfit$ with $scipy.optimize.curve\_fit$) or the same BKPL model with different $\omega$ values (comparing $\omega$=1, $\omega$=3, and $\omega$=10). The points connected by the black solid line represent its 512 ms count-rate lightcurve. Solid curves with different colors indicate the $lmfit$ cases (orange: BKPL model with fixed $\omega$=1; violet: BKPL model with fixed $\omega$=3; pink: BKPL model with fixed $\omega$=10; green: FRED model), while dashed lines indicate the $scipy.optimize.curve\_fit$ cases (yellow: BKPL model with fixed $\omega$=3; and cyan: FRED model). The reduced chi-squared is calculated by assuming its uncertainties with a typical value: 10\% of the values of its data points.}\label{fig:comparison}
\end{figure*}

\clearpage
In addition to the FRED model with a given $t_{0}$, another  five-parameter ($F_{0}$, $t_{\rm b}$, $\alpha_{1}$, $\alpha_{2}$, $\omega$) model, namely, the smoothly broken power law (BKPL), may also be used to characterize the pulse shape:
\begin{equation}
F(t) =  F_0 \, \left[\left(\frac{t+t_{0}}{t_{b}+t_{0}}\right)^{\alpha_1\omega}+\left(\frac{t+t_{0}}{t_{b}+t_{0}}\right)^{\alpha_2\omega}\right]^{-1/\omega},
\label{eq:BKPL}
\end{equation}
where $\alpha_1$, and $\alpha_2$ are the temporal slopes, $t_{\rm b}$ is the break time, $F_{\rm b} = F_0\, 2^{-1/\omega}$ is the flux of the break time; and $\omega$ describes the sharpness of the break. Note that the smaller the $\omega$ parameter, the smoother the break, and it is often fixed as 3. On the other hand, several other similar Python packages (e.g., $scipy.optimize.curve\_fit$) may be also be competent to carry out the current task. Figure \ref{fig:comparison} shows the fit results of the lightcurve of GRB 131231198, compared with the different models (FRED and BKPL) or packages ($lmfit$ and $scipy.optimize.curve\_fit$), or the same model (BKPL) set up with different $\omega$ values ($\omega$=1, $\omega$=3, and $\omega$=10).

\clearpage
\begin{figure*}
\includegraphics[width=0.5\hsize,clip]{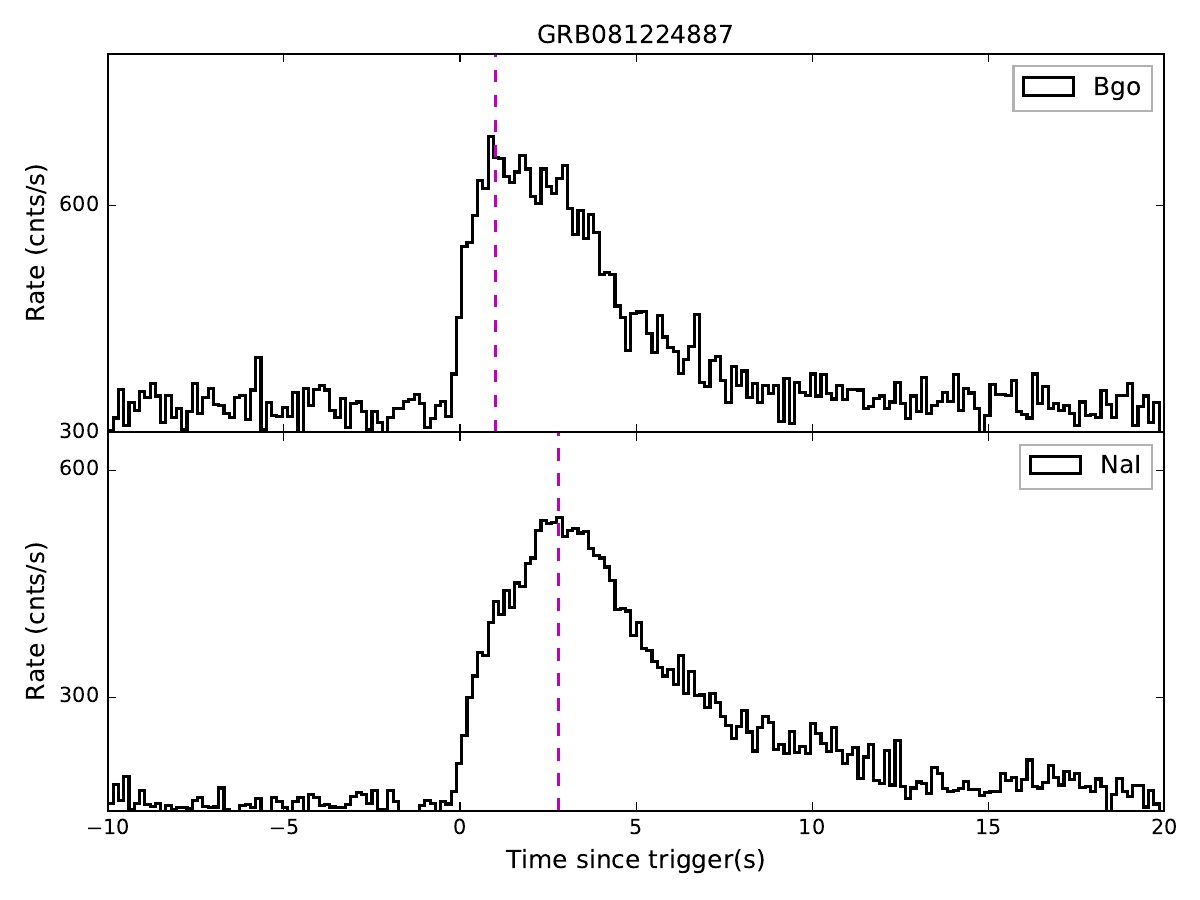}
\includegraphics[width=0.5\hsize,clip]{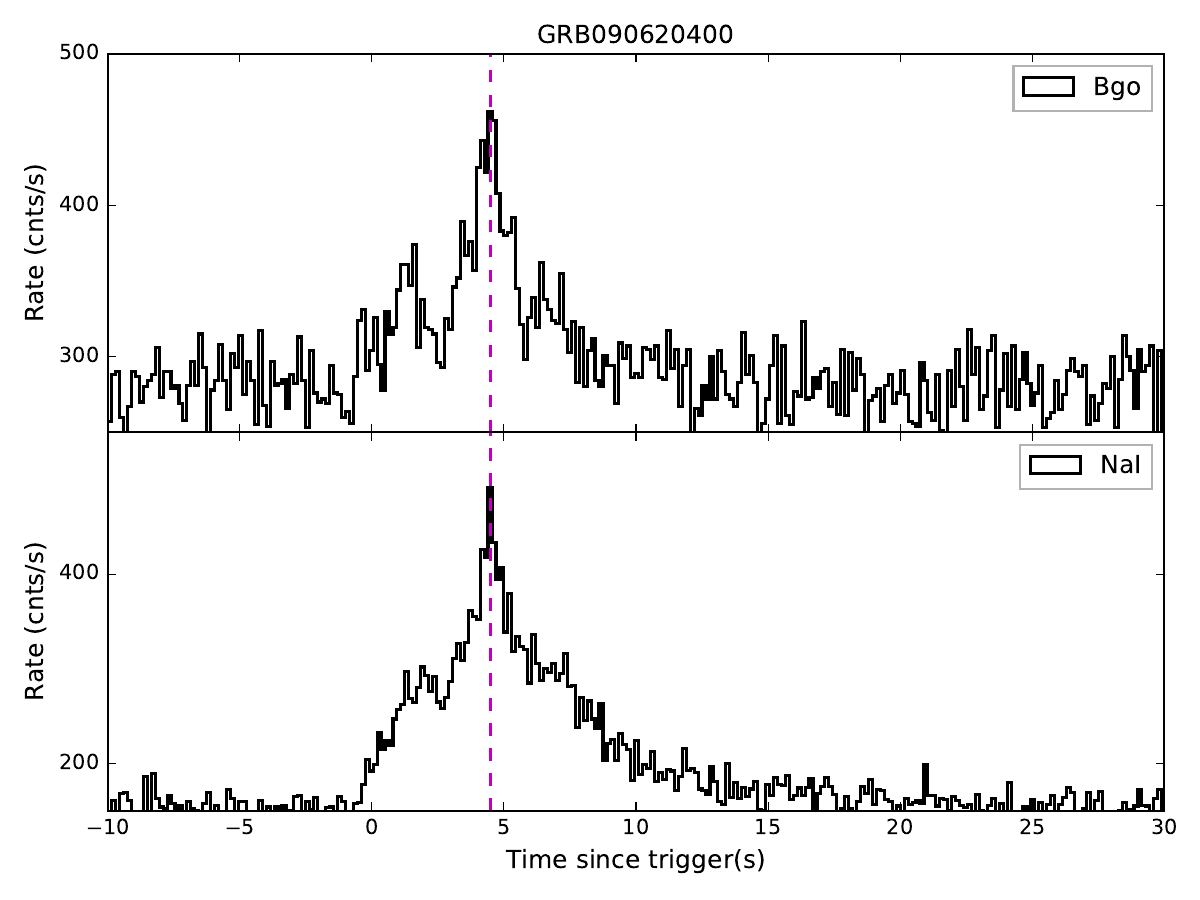}
\includegraphics[width=0.5\hsize,clip]{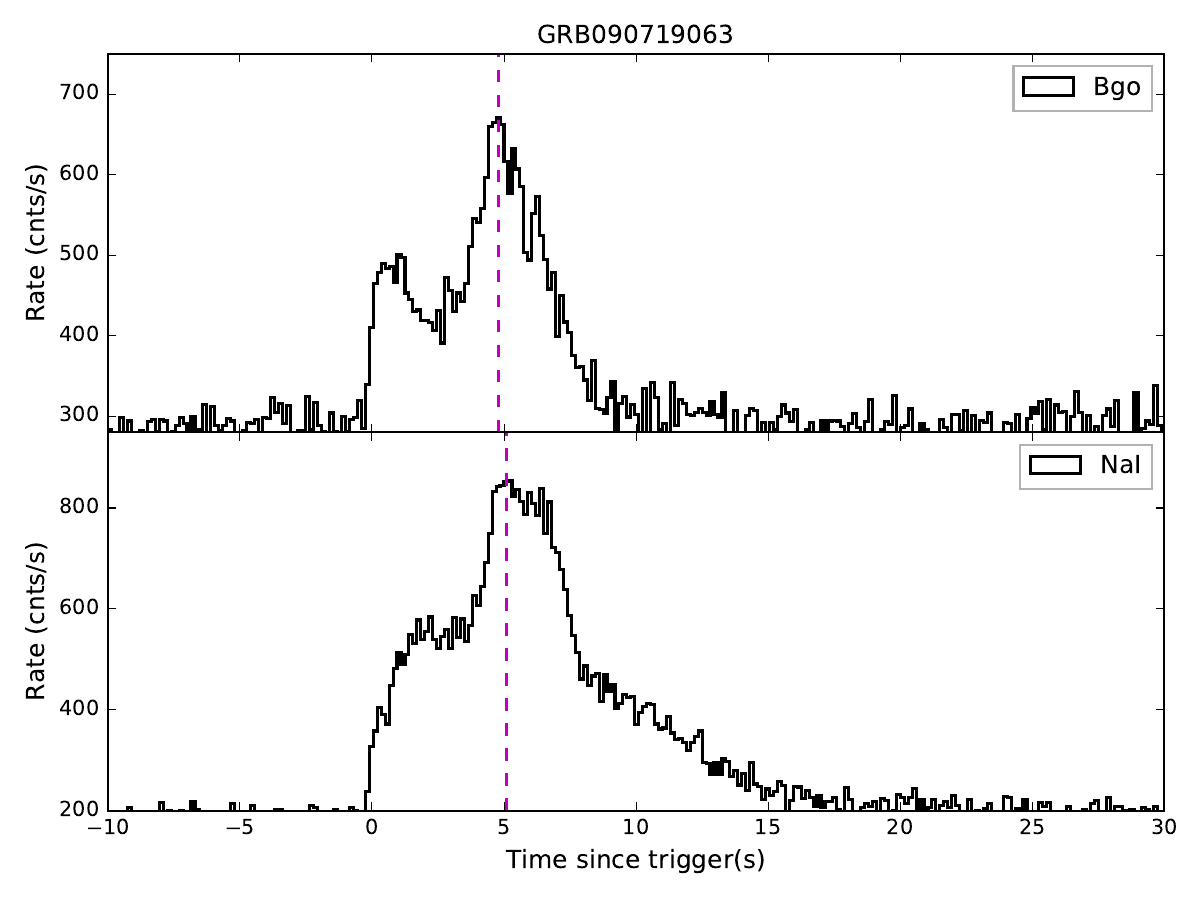}
\includegraphics[width=0.5\hsize,clip]{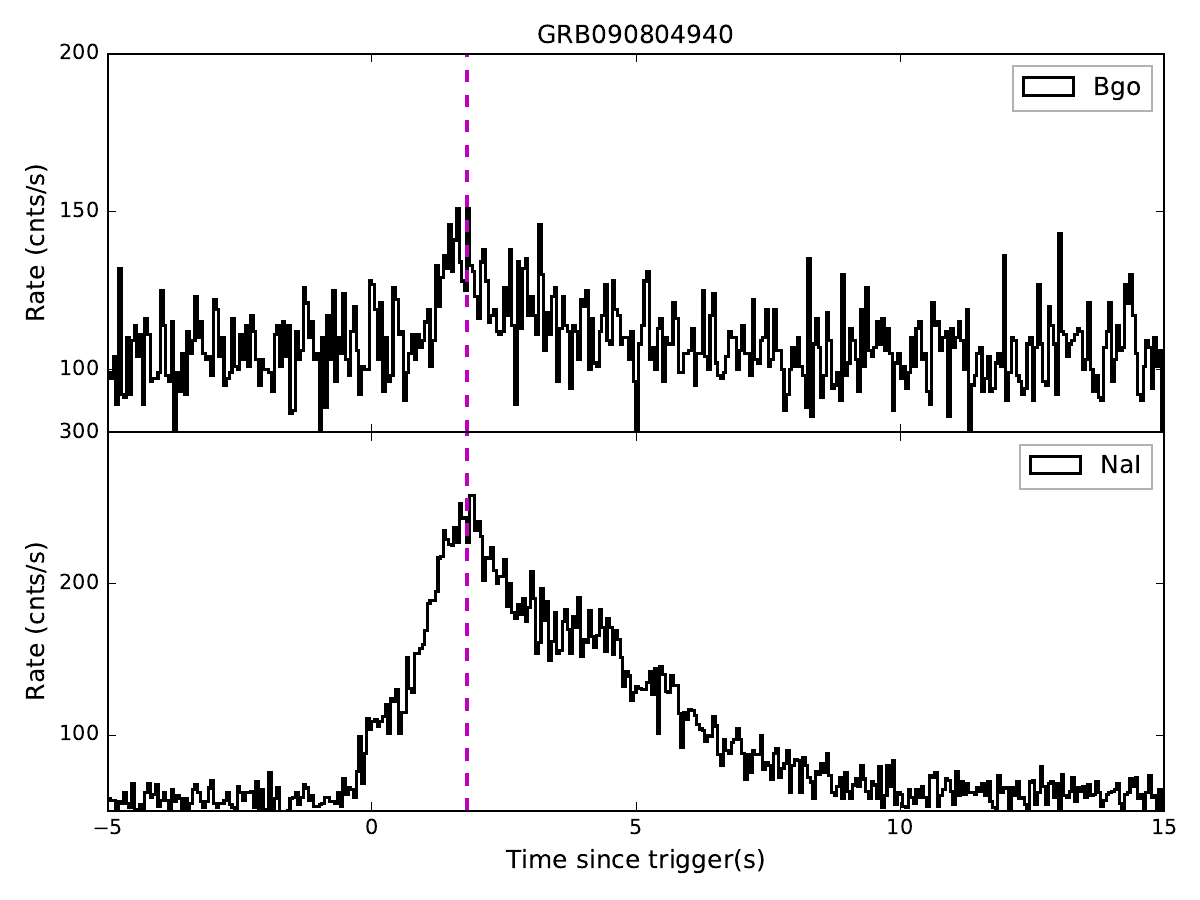}
\includegraphics[width=0.5\hsize,clip]{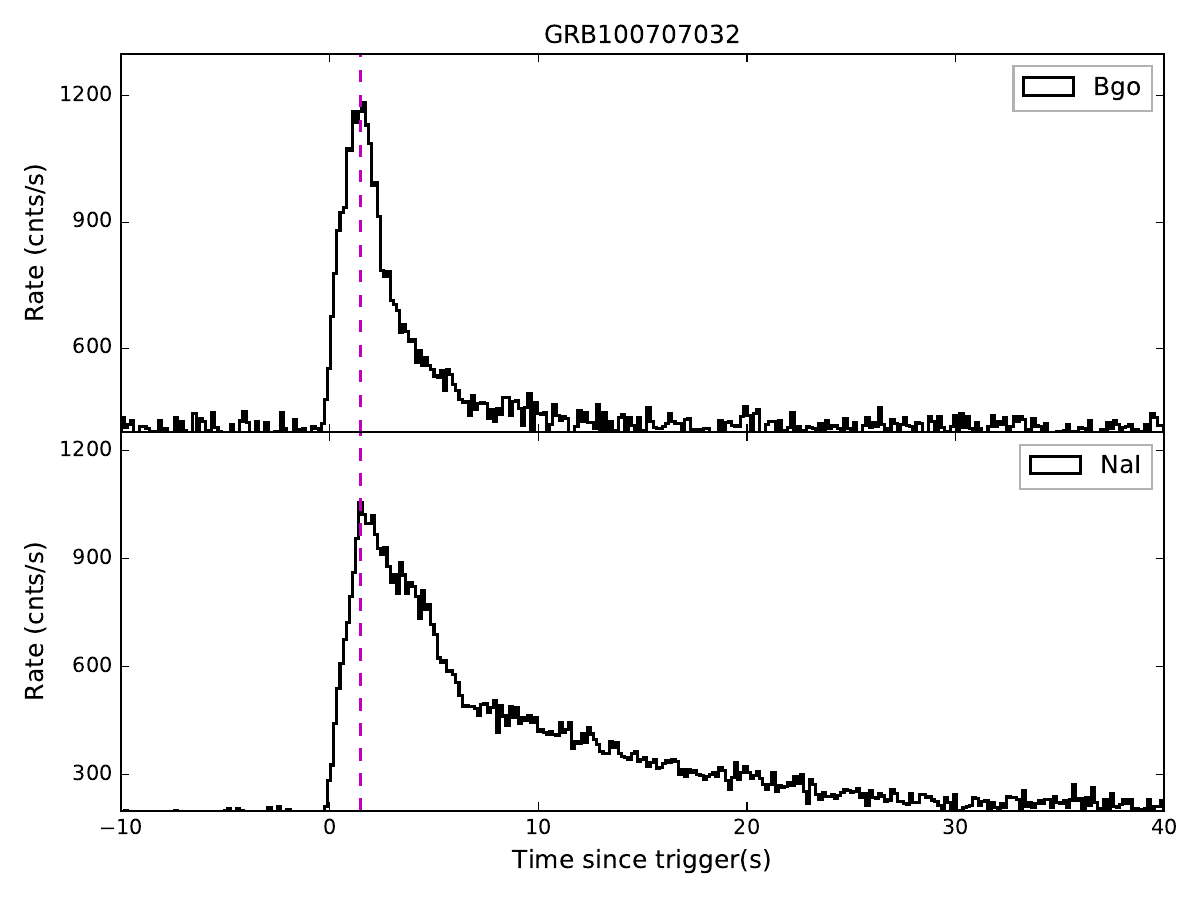}
\includegraphics[width=0.5\hsize,clip]{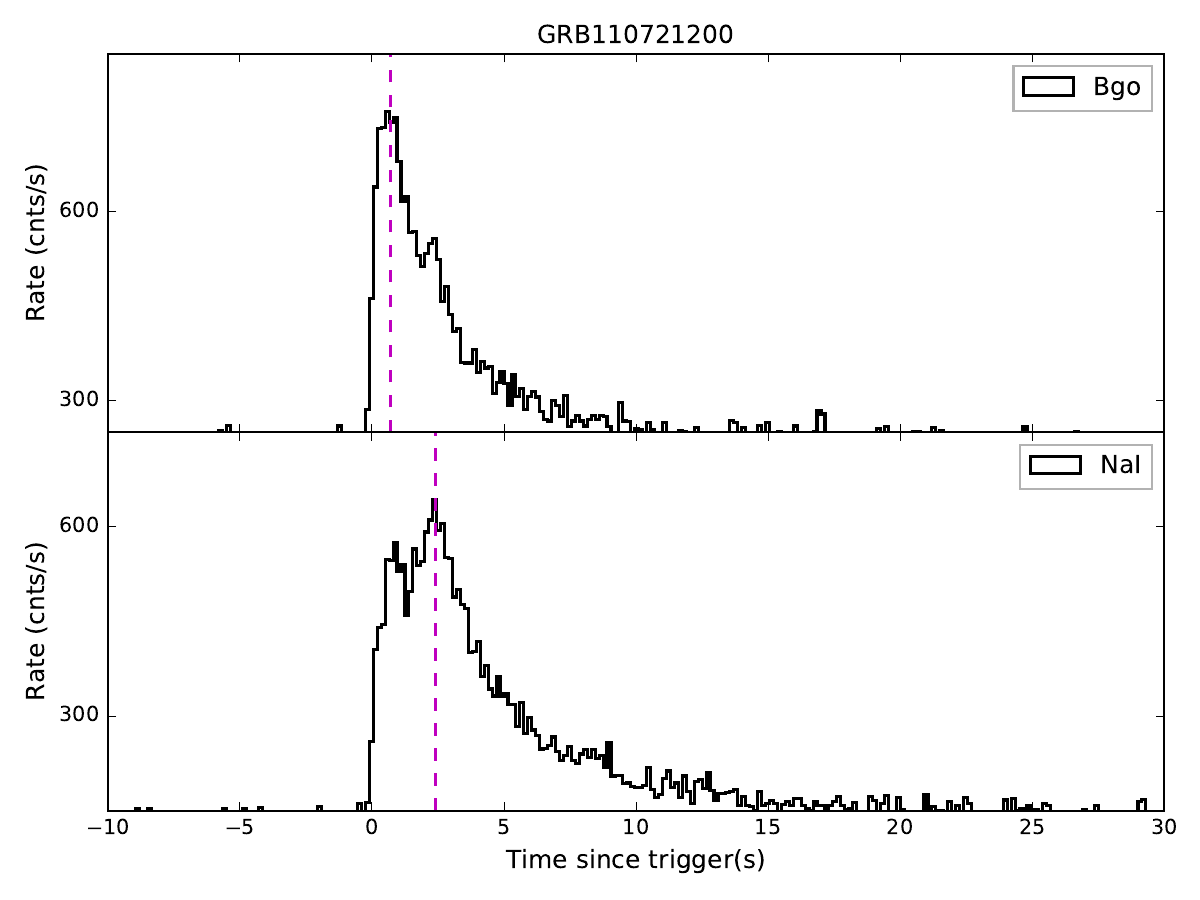}
\caption{Comparison of the count lightcurves for different GBM detectors (NaI and BGO). For each individual burst, the vertical magenta dashed lines are the peak times of two detectors identified by eye by inspecting the flux.}\label{fig:Bgo+NaI_LCs}
\end{figure*}
\begin{figure*}
\includegraphics[width=0.5\hsize,clip]{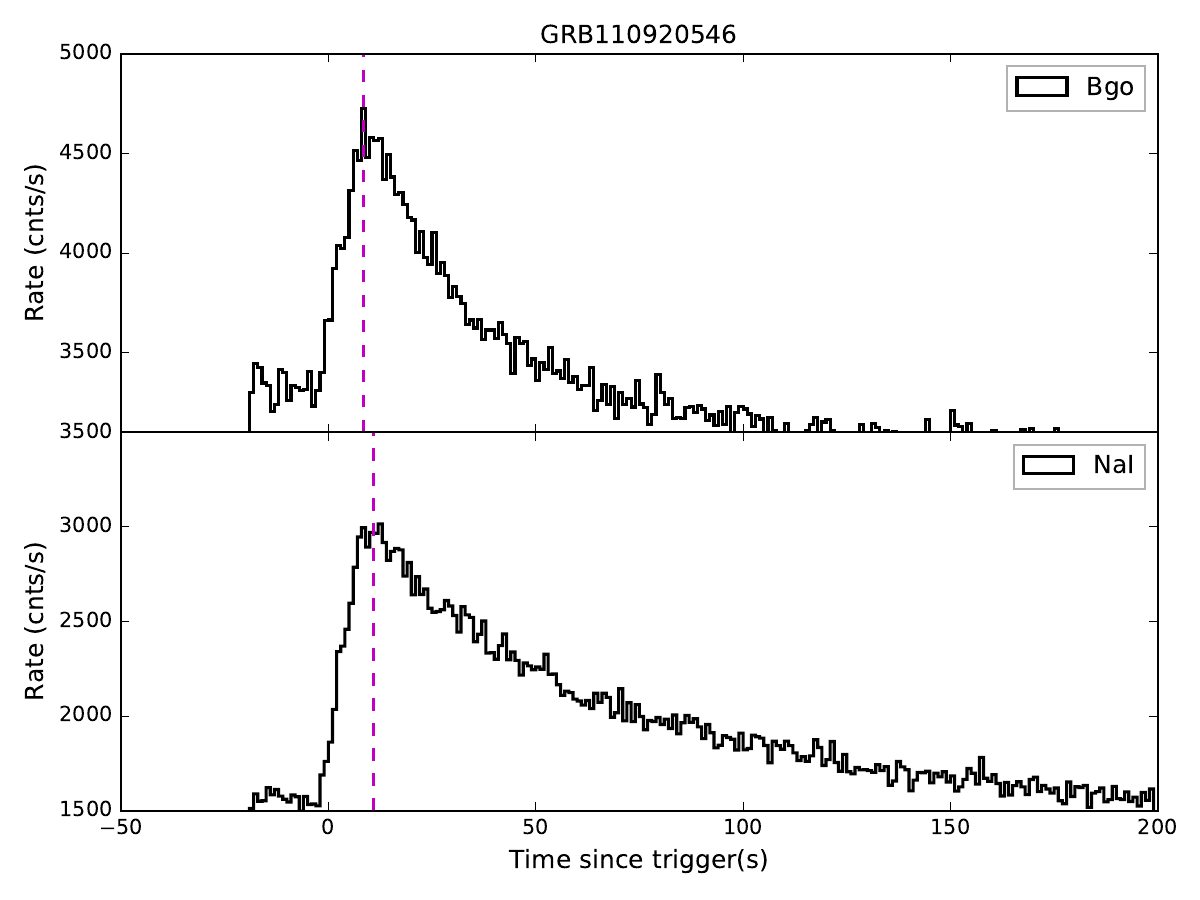}
\includegraphics[width=0.5\hsize,clip]{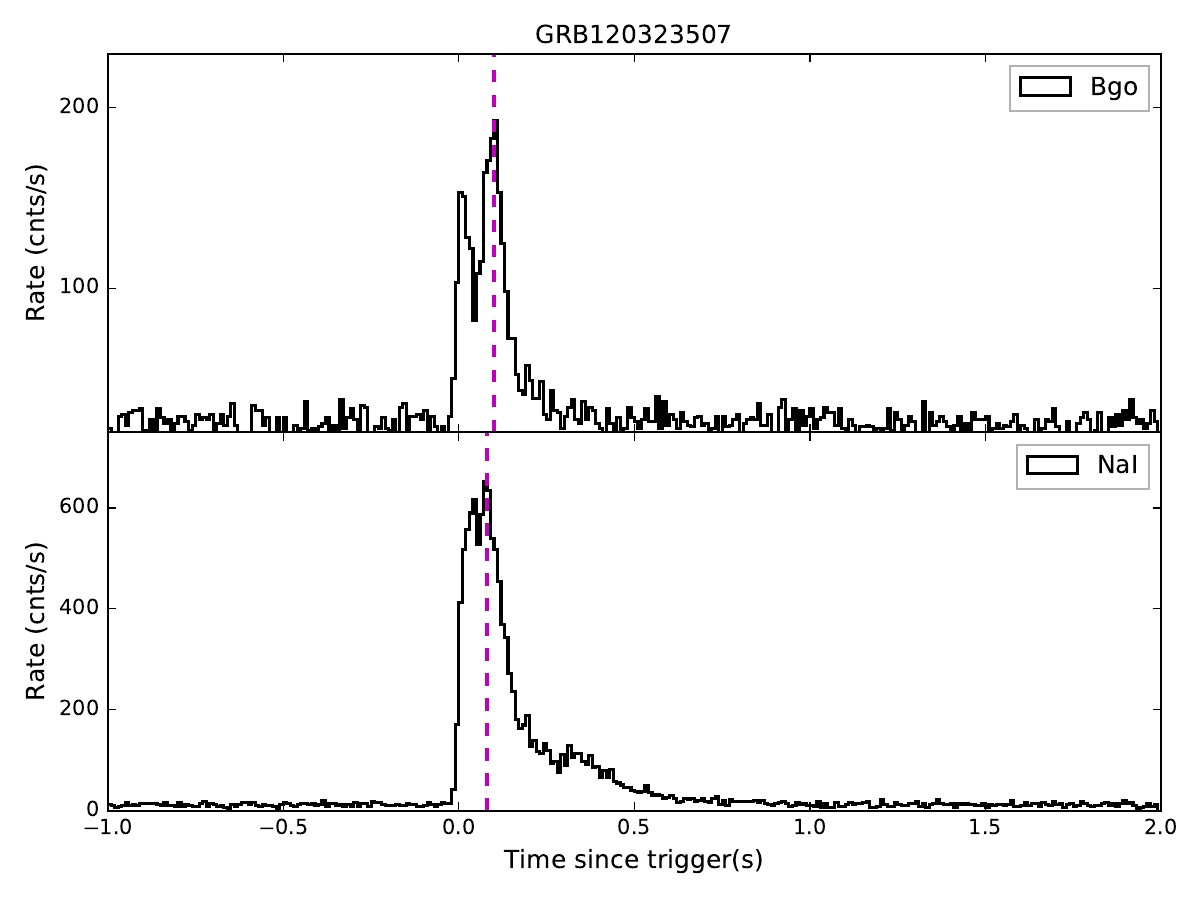}
\includegraphics[width=0.5\hsize,clip]{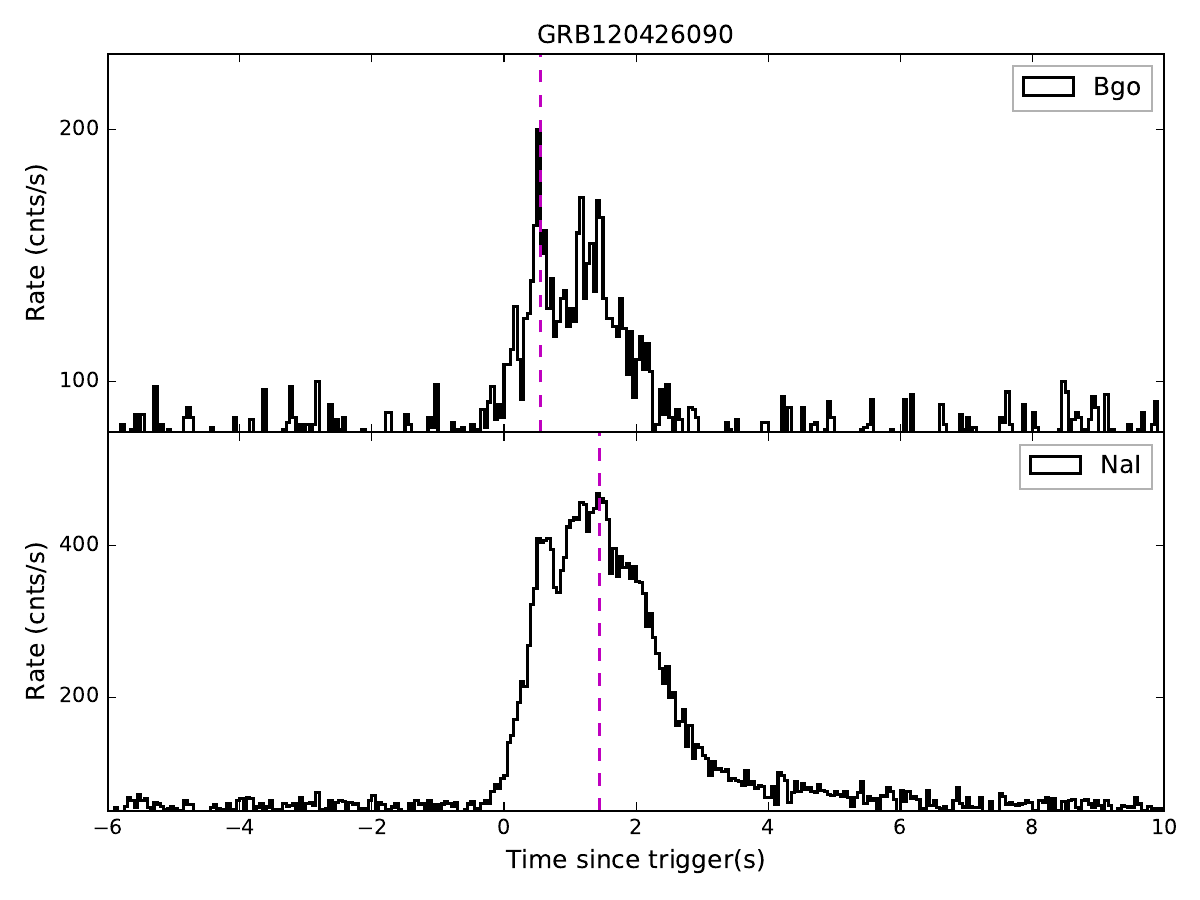}
\includegraphics[width=0.5\hsize,clip]{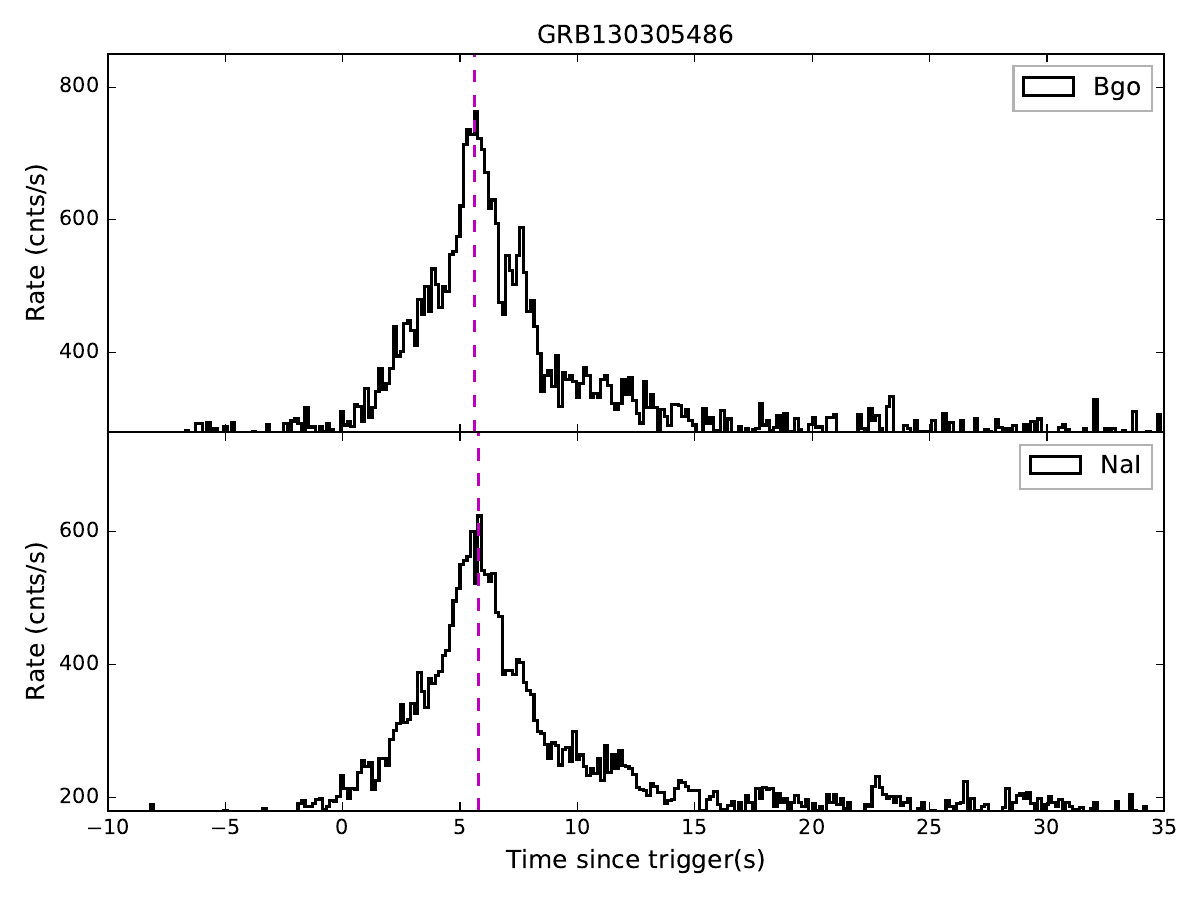}
\includegraphics[width=0.5\hsize,clip]{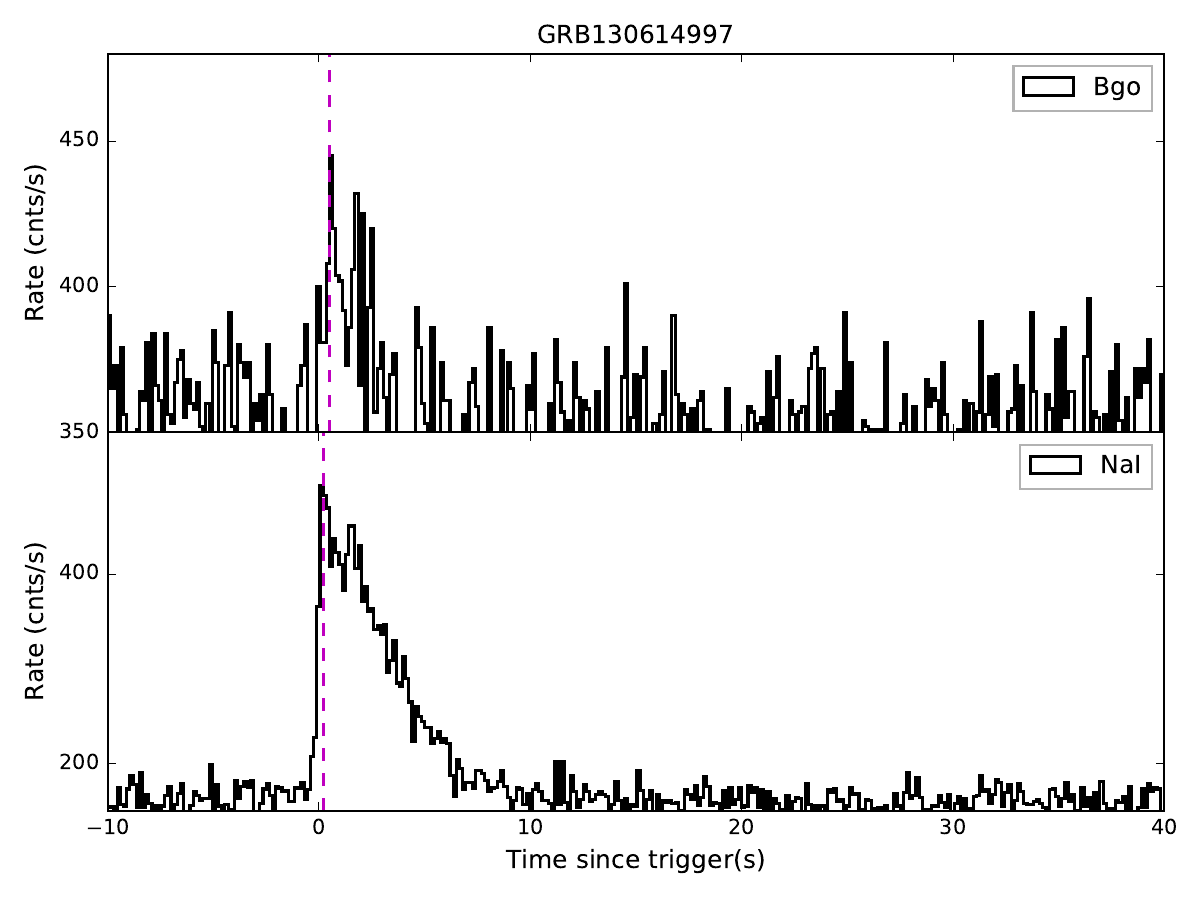}
\includegraphics[width=0.5\hsize,clip]{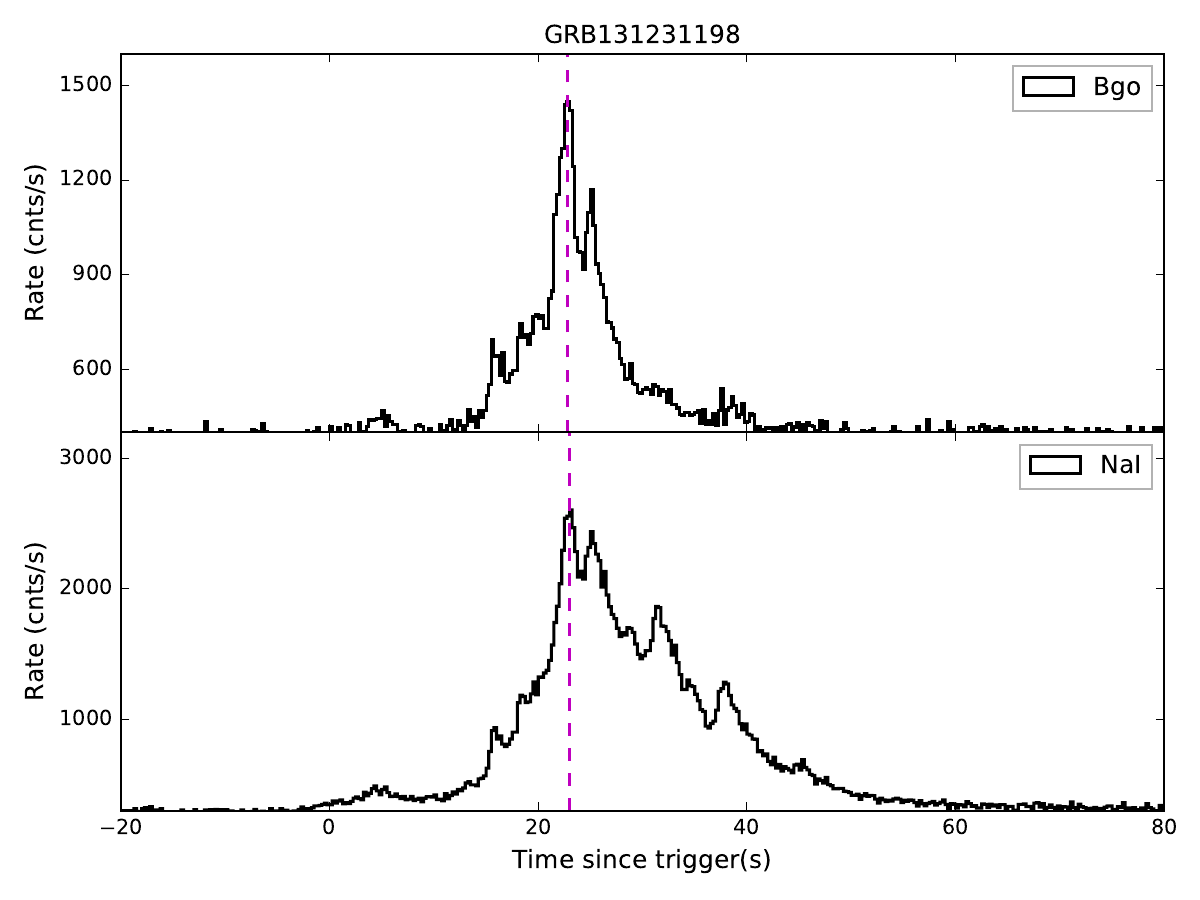}
\center{Fig. \ref{fig:Bgo+NaI_LCs}--- Continued}
\end{figure*}
\begin{figure*}
\includegraphics[width=0.5\hsize,clip]{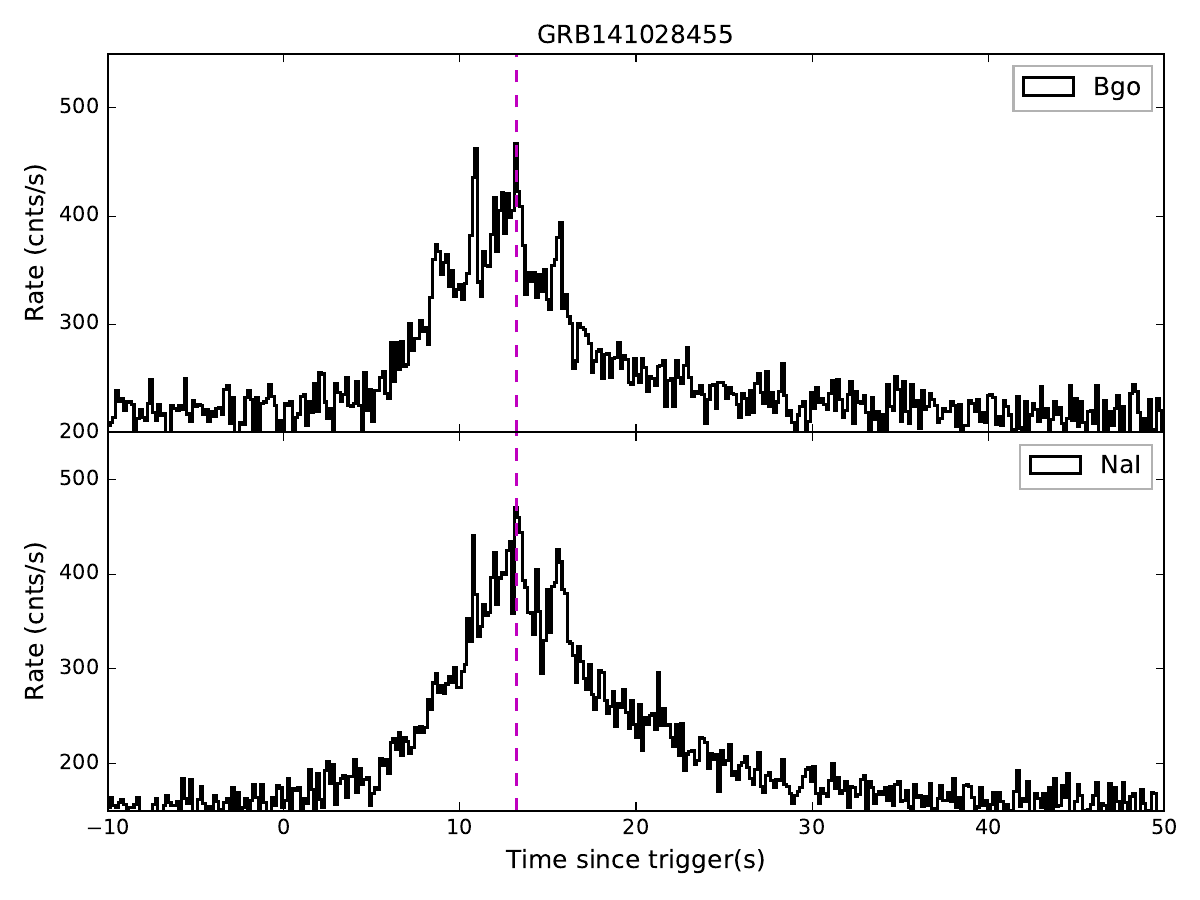}
\includegraphics[width=0.5\hsize,clip]{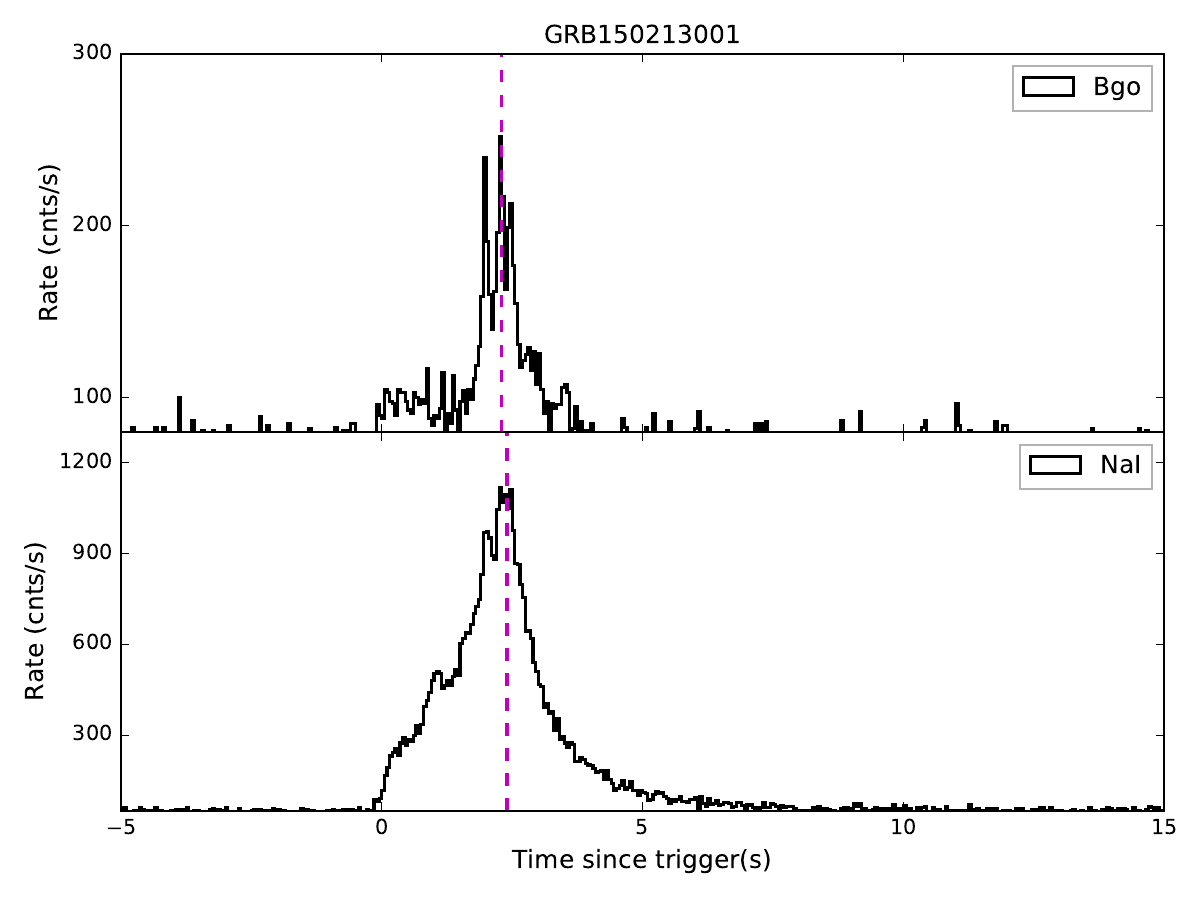}
\includegraphics[width=0.5\hsize,clip]{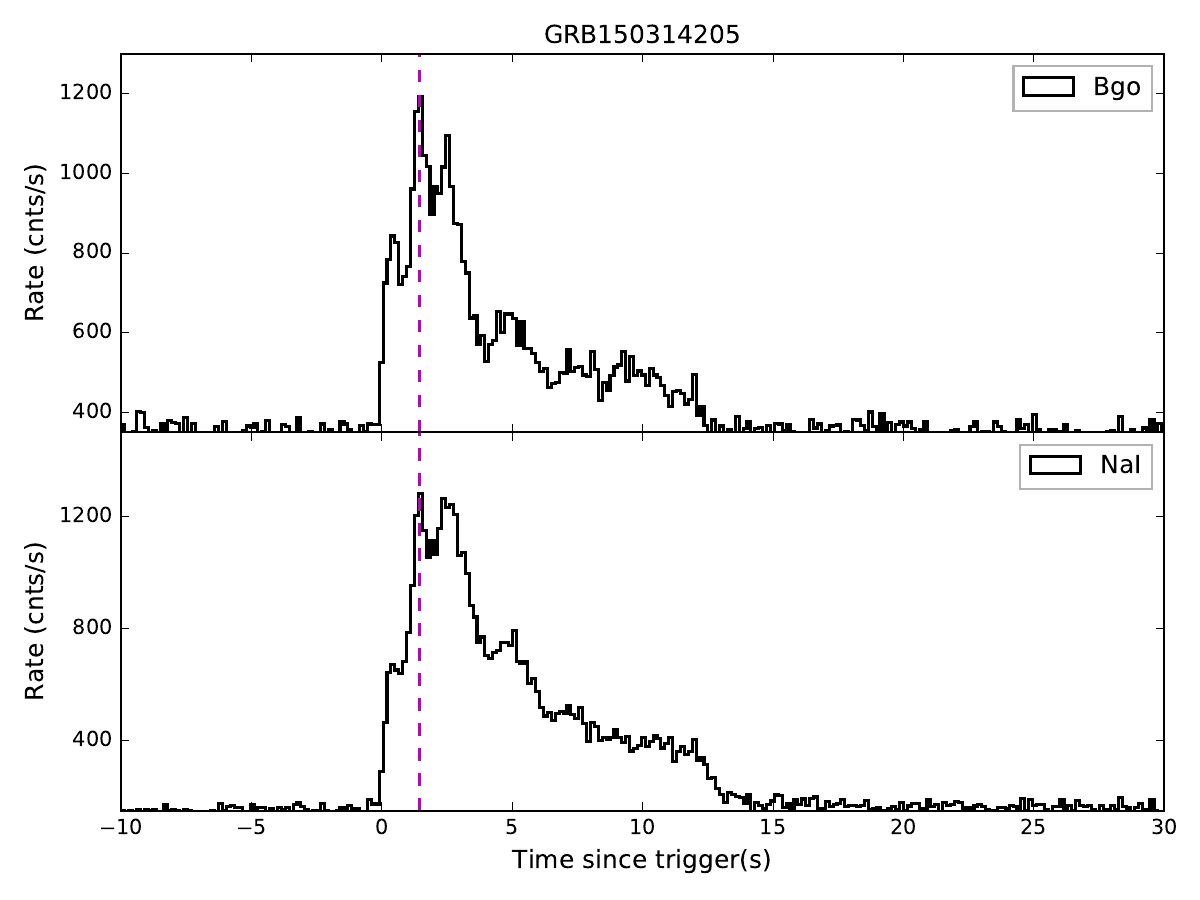}
\includegraphics[width=0.5\hsize,clip]{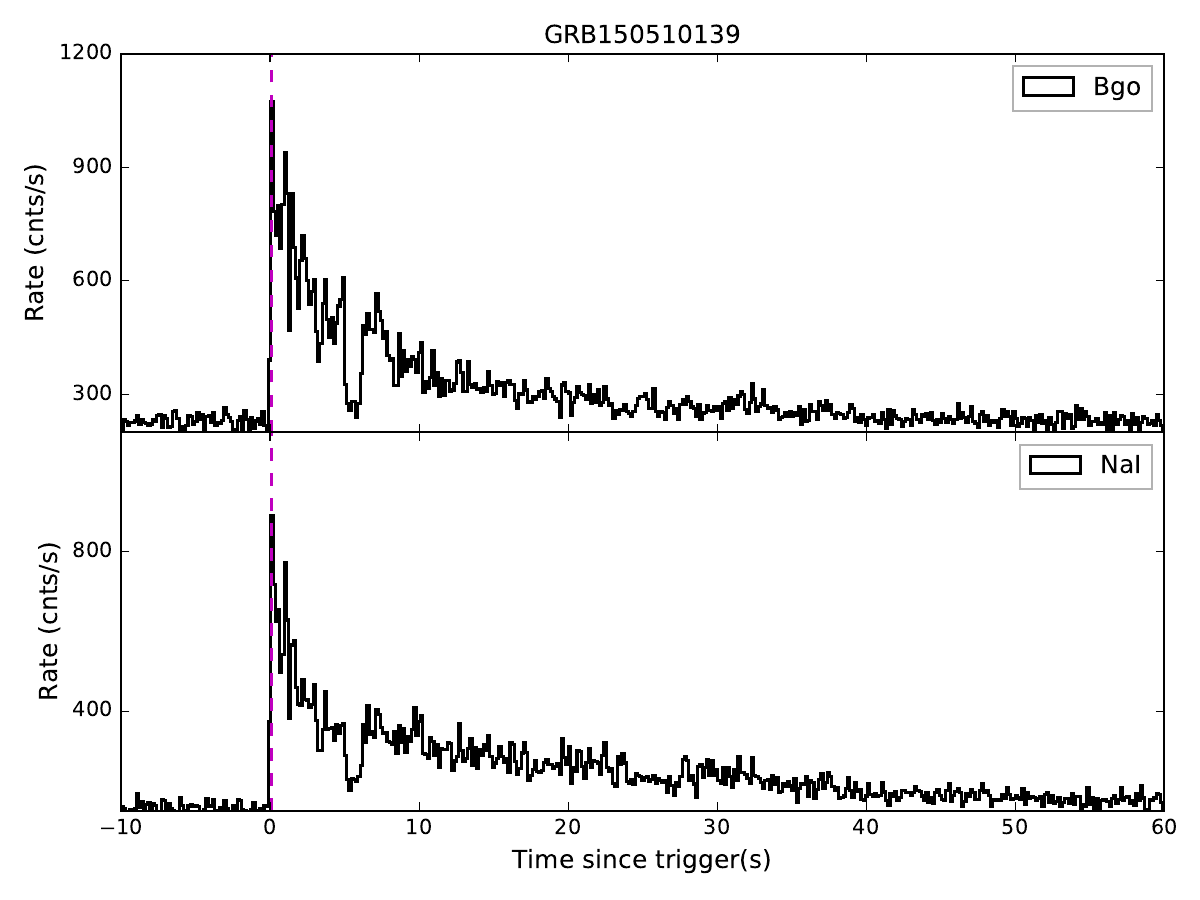}
\includegraphics[width=0.5\hsize,clip]{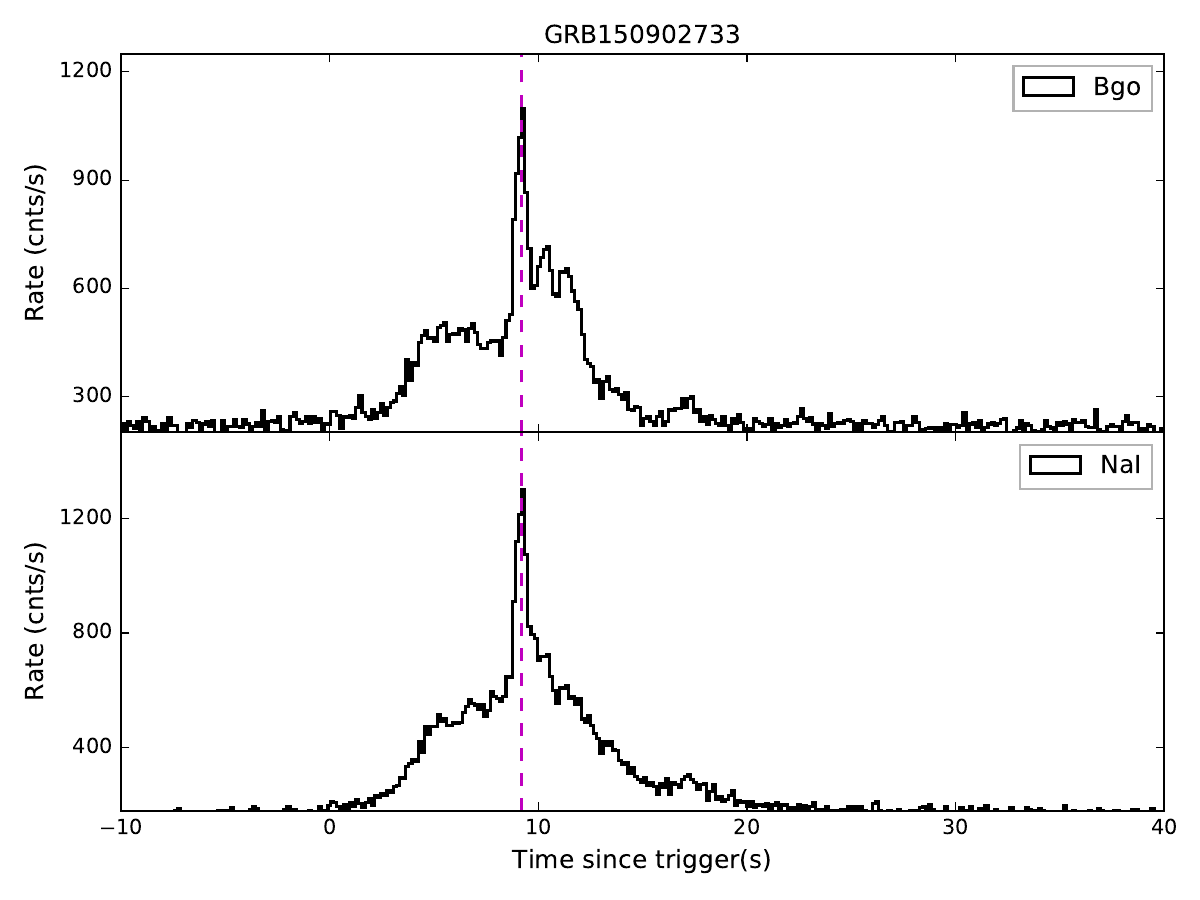}
\includegraphics[width=0.5\hsize,clip]{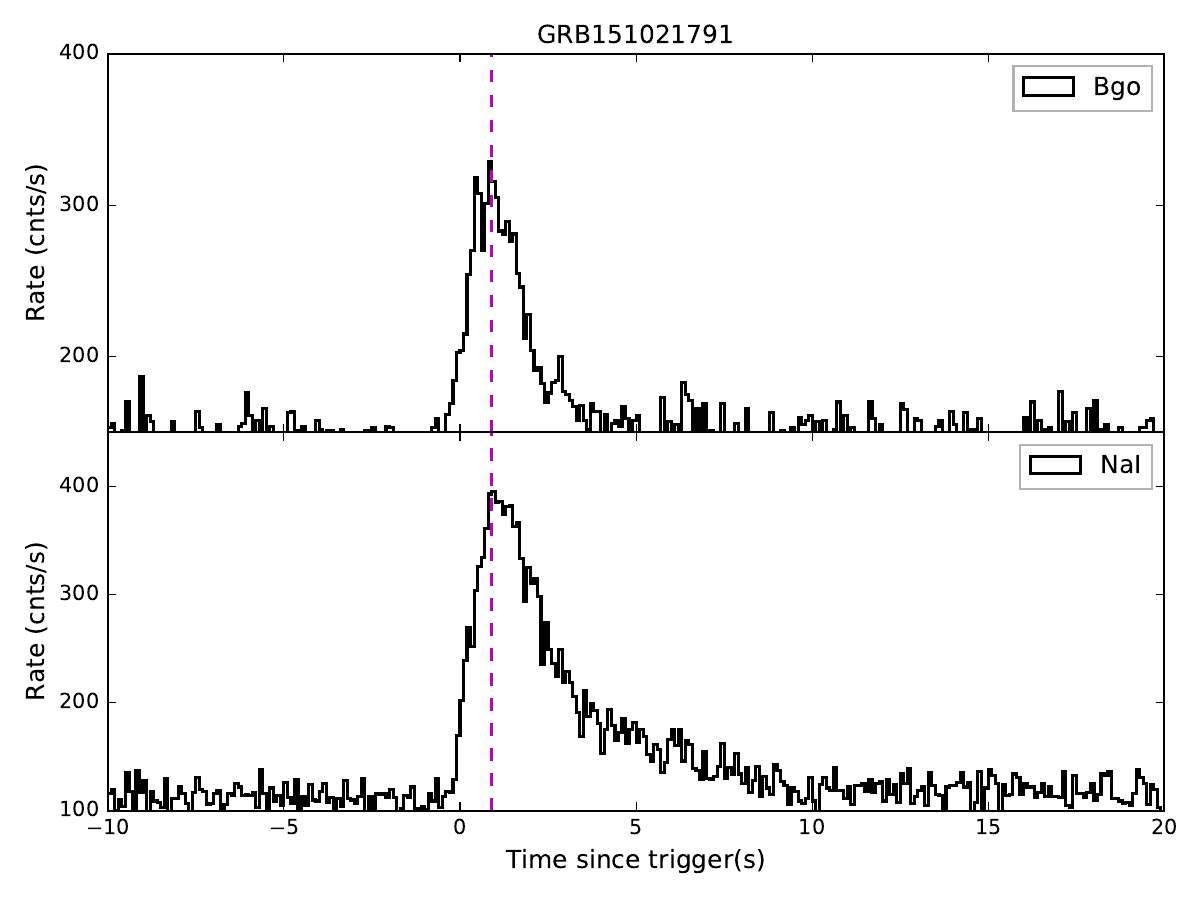}
\center{Fig. \ref{fig:Bgo+NaI_LCs}--- Continued}
\end{figure*}
\begin{figure*}
\includegraphics[width=0.5\hsize,clip]{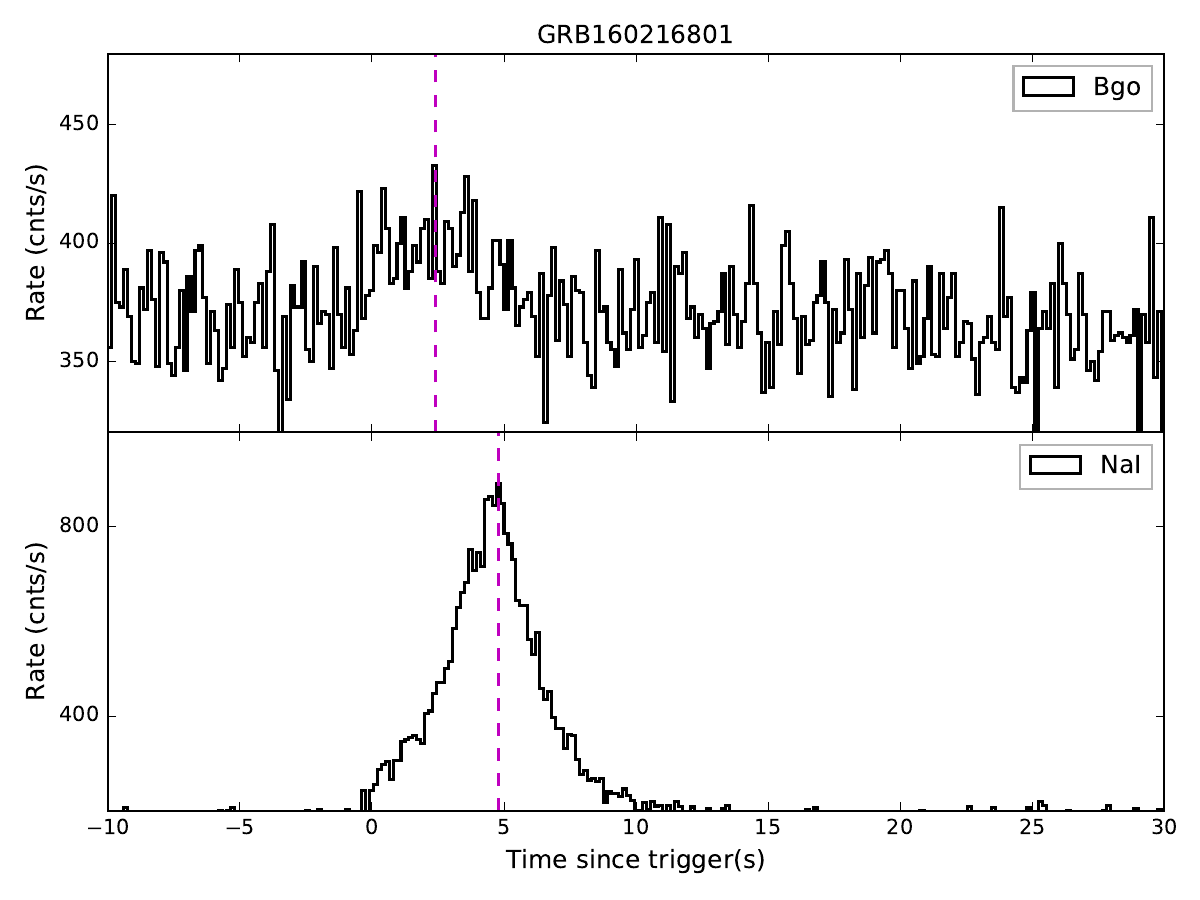}
\includegraphics[width=0.5\hsize,clip]{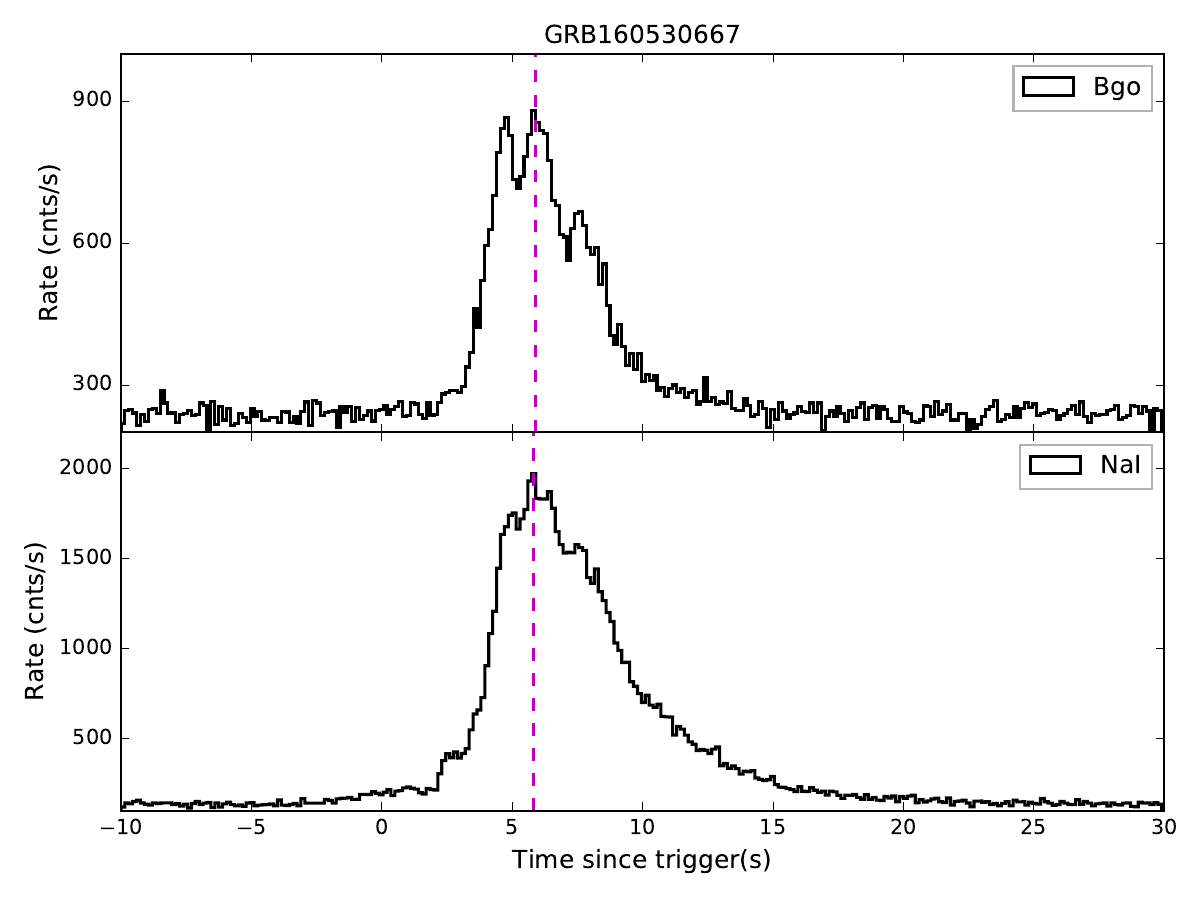}
\includegraphics[width=0.5\hsize,clip]{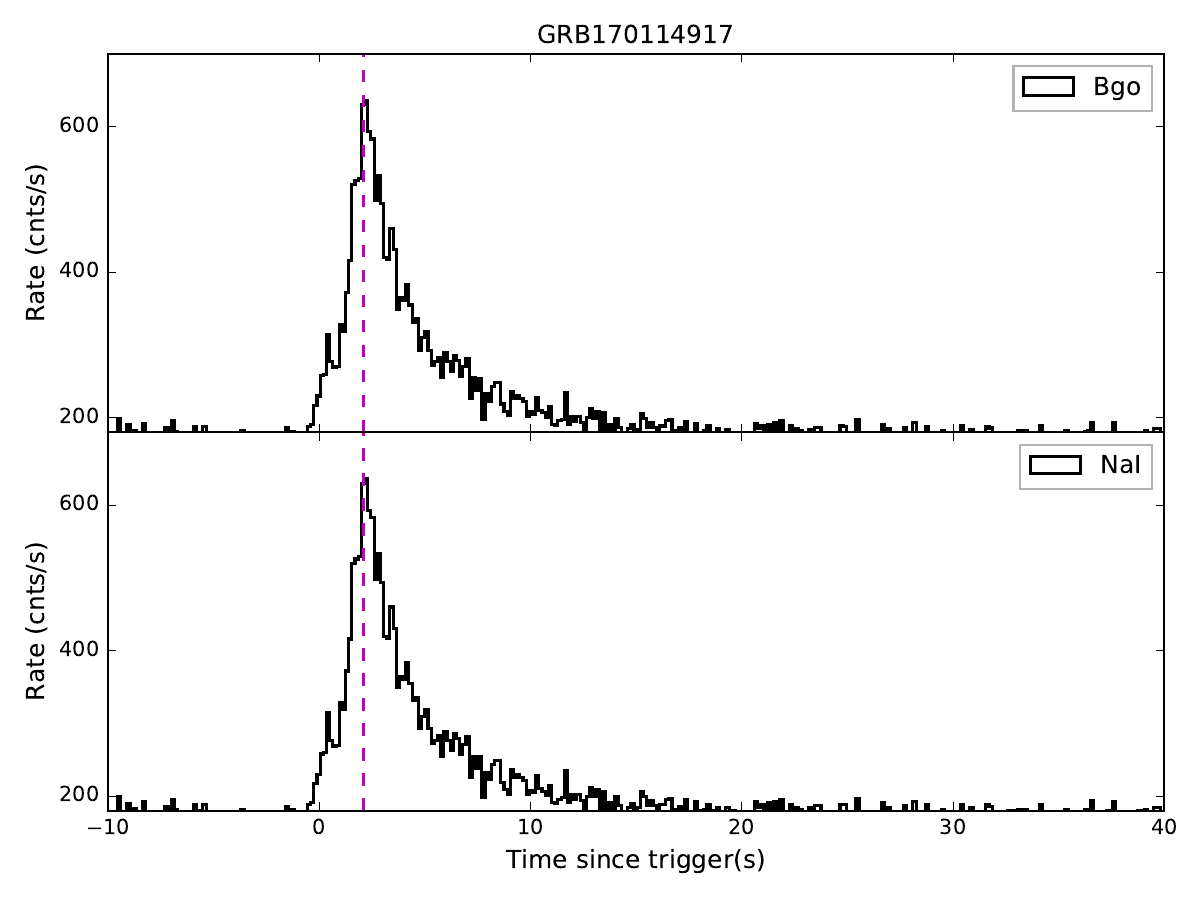}
\includegraphics[width=0.5\hsize,clip]{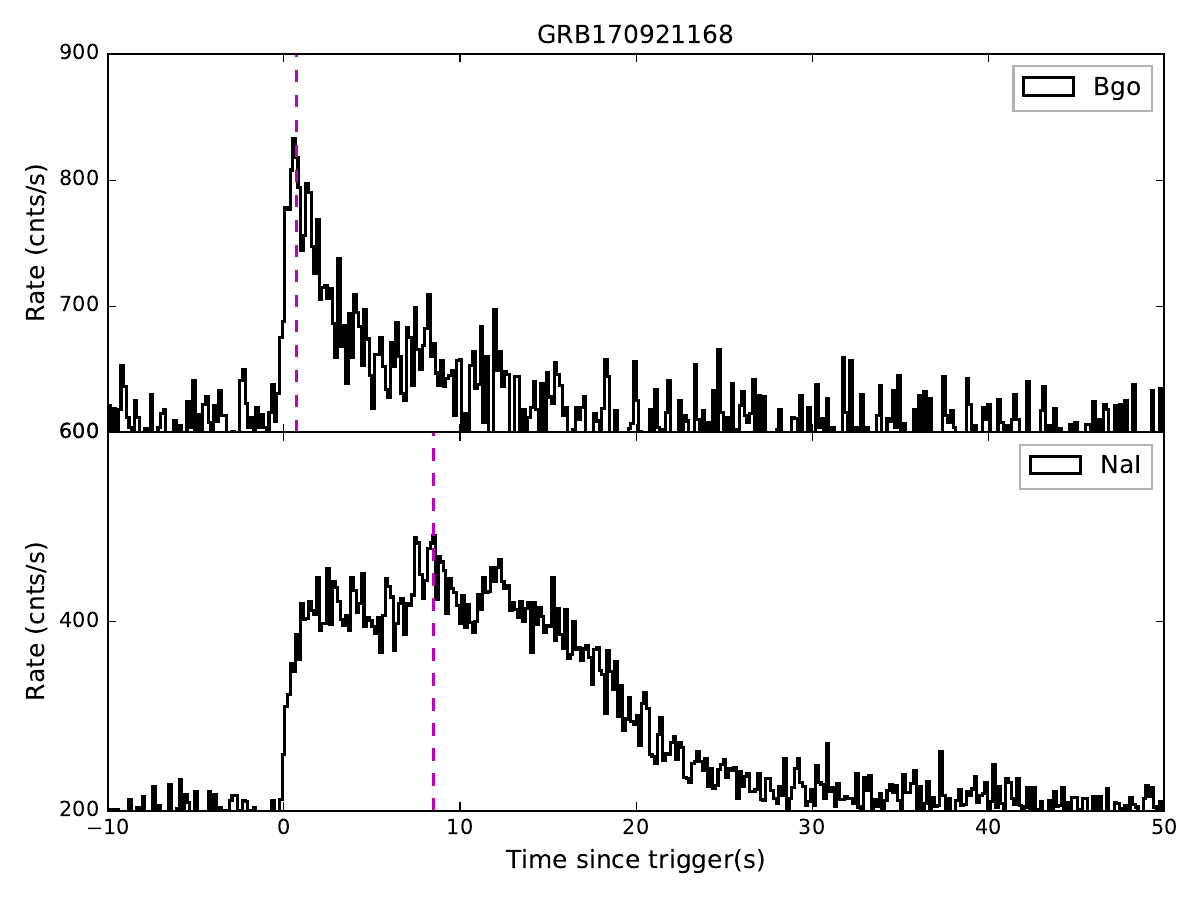}
\includegraphics[width=0.5\hsize,clip]{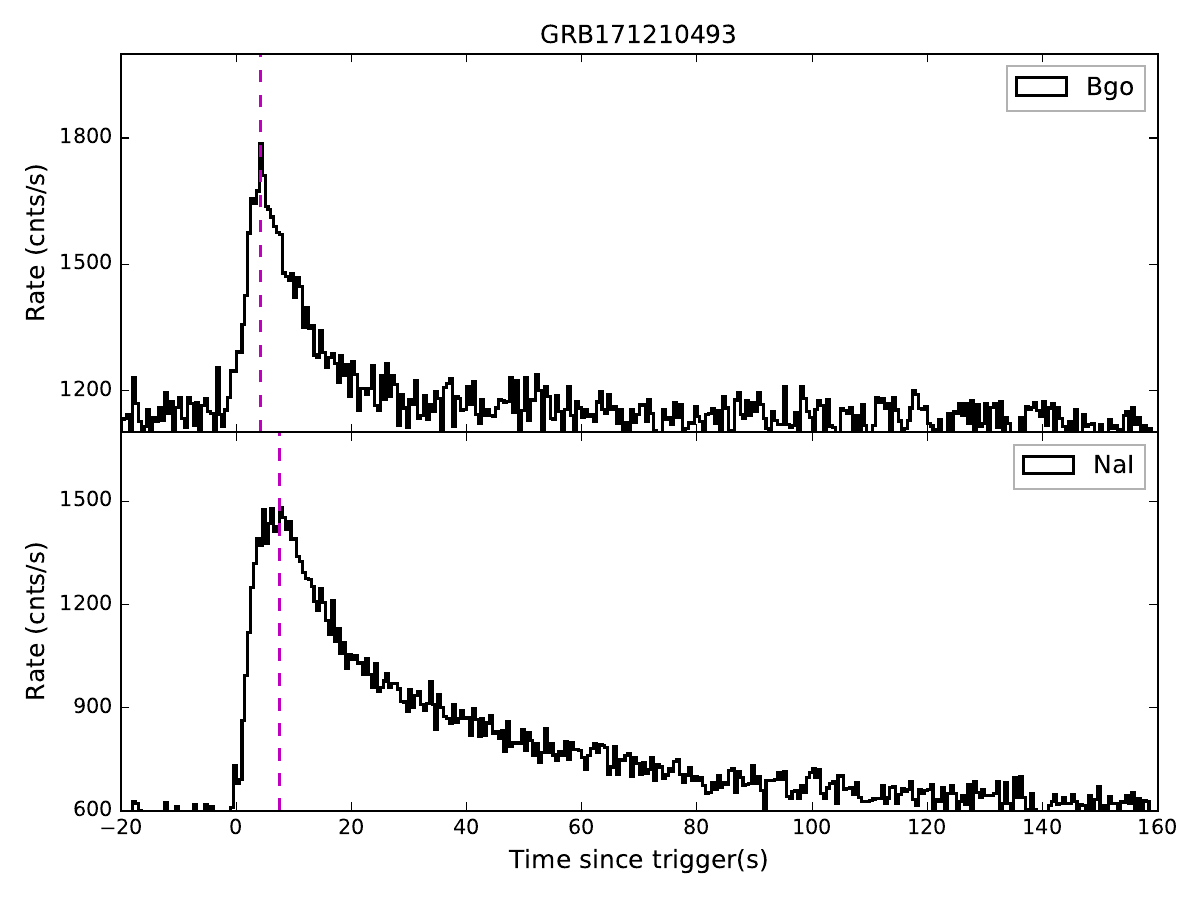}
\includegraphics[width=0.5\hsize,clip]{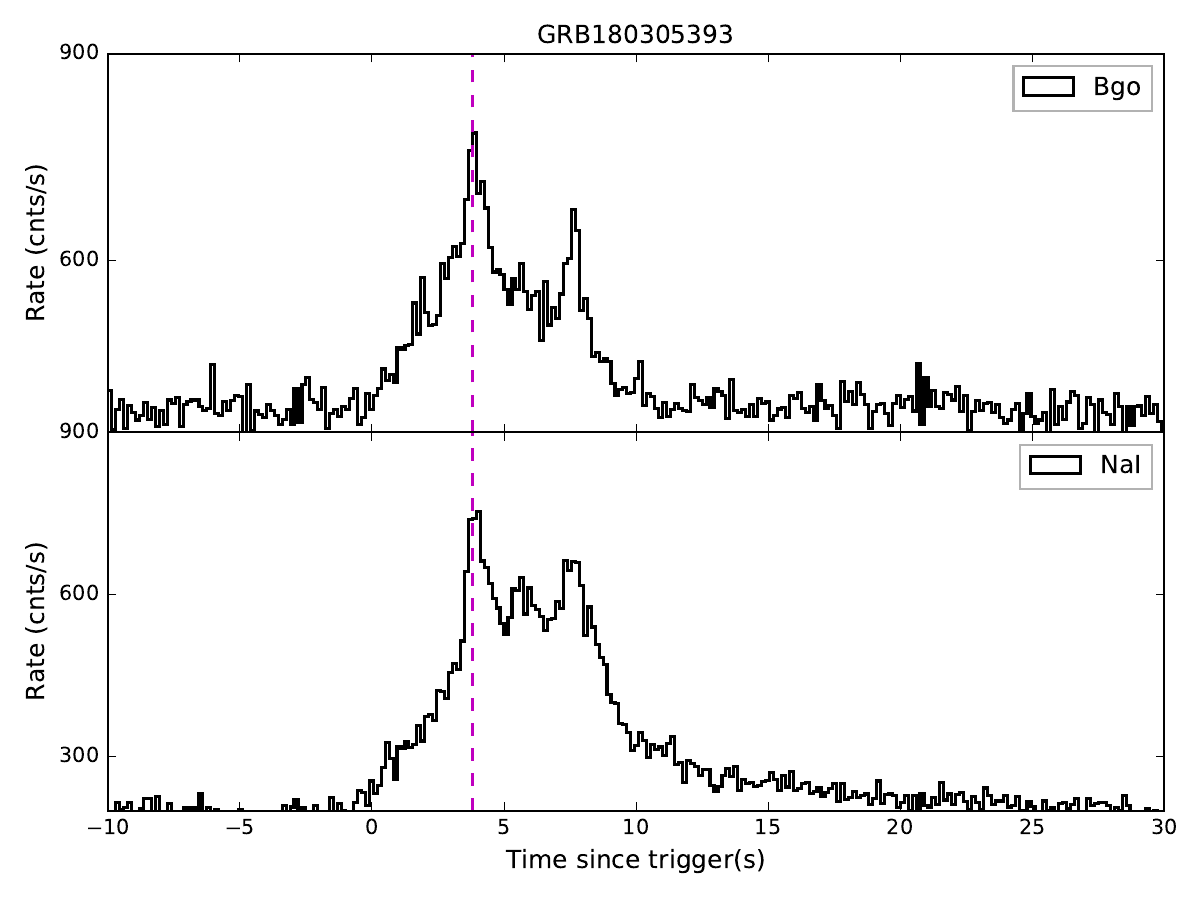}
\center{Fig. \ref{fig:Bgo+NaI_LCs}--- Continued}
\end{figure*}

\clearpage
\begin{figure*}
\begin{center}
\includegraphics[width=0.7\hsize,clip]{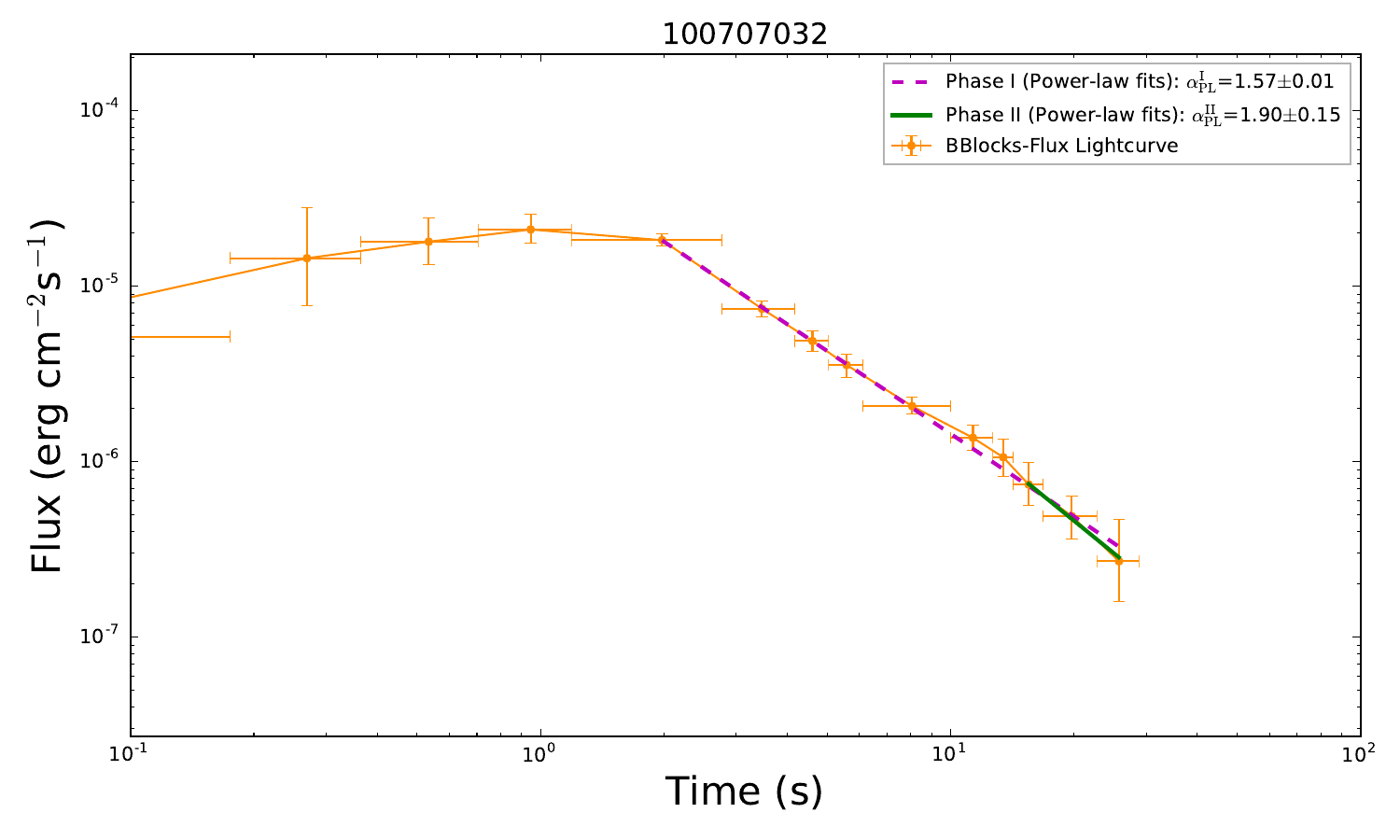}
\end{center}
\caption{Same as Figure \ref{fig:PL} but for the [log(Flux),log(time)] lightcurve. GRB 100707032 is taken as an example.}\label{fig:loglog}
\end{figure*}

\end{document}